\documentclass[twoside,12pt]{article}

\usepackage[toc,titletoc,title]{appendix} 
\usepackage{epsfig}
\usepackage{adjustbox}
\usepackage{graphicx}
\usepackage{amssymb}
\usepackage{amsmath}
\usepackage{tabularx}
\usepackage{multirow}
\usepackage{float}
\usepackage{wasysym,marvosym}
\usepackage{color}
\usepackage{hyperref}
\usepackage{pstricks}
\usepackage{lscape}
\usepackage{rotating}
\usepackage{ulem}

\newcommand{\be}{\begin{equation}}
\newcommand{\ee}{\end{equation}}
\newcommand{\bea}{\begin{eqnarray}}
\newcommand{\eea}{\end{eqnarray}}
\newcommand{\nn}{\nonumber}
\newcommand{\nnn}{\nonumber \\}

\newcommand{\vsigma}{\boldsymbol{\mathbf\sigma}}
\newcommand{\vtau}{\boldsymbol{\mathbf\tau}}
\newcommand{\br}{\boldsymbol{\mathbf r}}

\begin{document}

\title{ \vspace{1cm} Linear Response Theory with finite-range interactions}

\author{D.\ Davesne,$^1$ A.\ Pastore$^2$ J. Navarro$^3$ \\
\\
$^1$Universit\'e de Lyon, F-69003 Lyon, France \\
Institut de Physique des 2 Infinis, CNRS-IN2P3,  \\ 
UMR 5822, Universit\'e Lyon 1, F-69622 Villeurbanne, France \\
$^2$Department of Physics, University of York, Heslington, York, YO10 5DD, UK\\
$^3$IFIC (CSIC University of Valencia), E-46980 Paterna, Spain
}

\maketitle


\begin{abstract}
This review focuses on the calculation of infinite nuclear matter response functions using phenomenological finite-range interactions, equipped or not with tensor terms. These include Gogny and Nakada families, which are commonly used in the litterature. Because of the finite-range, the main technical difficulty stems from the exchange terms of the particle-hole interaction. We first present results based on the so-called Landau and Landau-like approximations of the particle-hole interaction. Then, we review two methods which in principle provide numerically exact response functions. The first one is based on a multipolar expansion of both the particle-hole interaction and the particle-hole propagator and the second one consists in a continued fraction expansion of the response function. The numerical precision can be pushed to any degree of accuracy, but it is actually shown that two or three terms suffice to get converged results. Finally, we apply the formalism to the determination of possible finite-size instabilities induced by a finite-range interaction.
\end{abstract}


\section{Introduction}


The study of quantum systems having a large or even infinite number of fermions is of interest in many physical situations, involving fields as diverse as quantum chemistry, atomic physics, condensed matter, nuclear physics or astrophysics. For instance, liquid $^3$He or conduction electrons in metals are familiar examples of such systems in condensed matter~\cite{fet71,rin80,pin66,lip08,dic08,din18}. In nuclear physics, the concept of infinite nuclear matter as an homogeneous medium made of interacting nucleons is broadly used, because of its relative simplicity and related to the suppression of the boundaries. This idealized system is nevertheless connected to the physics of the inner part of atomic nuclei and also of some regions of compact stars~\cite{cen90,bal99,cen07,cha08}. For these reasons, it is a very useful testing ground for various theories. 

A usual way to get information about a physical system is by means of its response to an external probe~\cite{fet71,rin80}. Some well-known examples among many of physical phenomena that require the knowledge of the response functions are the transition strengths, the inelastic cross-sections, the electron scattering by nuclei~\cite{wes75}, the propagation of neutrinos in nuclear matter~\cite{bur00} and finite nuclei~\cite{vol00} or the study of vibrational modes in finite nuclei~\cite{bortignon2019giant,suh07,har01} and neutron stars~\cite{kha05,bar10,cha13}. When the interaction with the probe is small enough, the response of the system can be calculated in the linear approximation. Its basic ingredients are the particle-hole (ph) interaction and the ph Green's function, whose poles represent the energies of excited states~\cite{dic08}. 

Two main approaches are introduced to get response functions in a non-relativistic frame: either by starting from a bare two-body nucleon-nucleon (NN) interaction, with the possible addition of three-body forces, or by using a phenomenological NN interaction. In the first approach, the many-body problem is treated as exactly as possible, using a variety of methods such as Monte-Carlo simulations~\cite{car15}, Coupled-Cluster~\cite{hag14}, similarity Renormalisation Group~\cite{bog10}, self-consistent Green's function~\cite{dic04} or the Brueckner-Hartree-Fock method~\cite{zuo02,akm98,ter87}. In the second approach, a mean-field approximation~\cite{ben03,gra19} is used  with an interaction whose parameters are adjusted so as to describe selected observables of finite nuclei and some bulk properties of nuclear matter as well. This second approach is the subject of the present review.

Several important informations about nuclear matter can be obtained by analysing the resulting response function. For instance, its singularities at zero energy can be related to a phase transition such as the spinodal one~\cite{duc07} or to instabilities such as the pion condensation \cite{bar73,alb80} or the ferromagnetic transition in neutron stars~\cite{vid84}. Moreover, if these instabilities appear at density values around the saturation density, $\rho_{0}$, they can affect the description of finite nuclei properties. This connection between singularities in the infinite medium and instabilities in atomic nuclei has been investigated in Refs.~\cite{pas12b,hel13,pas15,mar19}. It has been shown that the appearance of such singularities at densities lower than $\simeq 1.3 \rho_0$ (with $\rho_0$ the saturation density) may affect the convergence of the calculations in finite nuclei~\cite{les06,fra12}. To avoid such a problem, the response functions of nuclear matter are currently included into the fitting protocol used to adjust new interactions~\cite{pas12b,pas13,sad12,bec17}.

Historically,  phenomenological nuclear interactions have been limited to central, spin-orbit and density-dependent terms. However, in recent years, the importance of the tensor term has been stressed. Indeed, the bare nucleon-nucleon interaction contains an important tensor part, necessary to reproduce not only  the phase shifts of the nucleon-nucleon scattering, but also the quadrupole moment of the deuteron. Several groups have worked on the introduction of a tensor term in effective interactions and its impact on the ground state properties of finite nuclei has been discussed for example in Refs.~\cite{bro06,les07,ben09B,sag14}. 

Methods to get infinite nuclear matter response functions using the zero-range Skyrme interaction~\cite{sky59} have been reviewed in Ref.~\cite{report}.  The case of finite-range interactions is the subject of the present review, concentrating on the Gogny~\cite{gog75,dec80,rob19} and the Nakada~\cite{nak03,nak20} interactions. Although these are not the only finite-range interactions available in the literature~\cite{rai14,ben17}, they are currently the most commonly used to study finite nuclei at the mean field level~\cite{ben03,rob19,nak20}. 

The most difficult technical aspect in calculating response functions is related to the exchange term of the ph-interaction. This difficulty is overcome if a zero-range interaction is used, and it is thus convenient to briefly summarise some results obtained with the well-known family of zero-range Skyrme interactions. We recall that the standard form of the Skyrme interaction, fixed by Vautherin and Brink~\cite{vau72}, contains central, spin-orbit and density-dependent terms only, with a quadratic momentum dependence which simulates finite-range effects. It has been shown~\cite{gar92} that both its zero-range and its simple momentum dependence allow one to get exact and relatively simple algebraic expressions for the response functions and other related quantities. These results were originally presented for symmetric nuclear matter (SNM), and afterwards extended to asymmetric nuclear matter and pure neutron matter (PNM) including the special case of charge-exchange operators~\cite{bra94,bra95,her96,her97,her99}. They have been applied to a variety of problems, including neutrino transport in neutron stars~\cite{nav99,mar03}. The original form of the Skyrme interaction~\cite{sky59} also contains other terms, namely zero-range tensor terms, which are relevant for the spin and spin-isospin channels.  However, apart from some exploratory studies~\cite{sta77}, these terms have been omitted in most calculations of finite nuclei in recent years, they have have been included either perturbatively to existing central ones~\cite{bri07,col07,bai09,bai09a}, or with a complete refit of the parameters~\cite{bro06,les07,ton83,liu91,she19}.  Again, due to their zero-range and their particular momentum dependence, it is possible to get the exact response functions, although the tensor terms make algebraic expressions rather cumbersome. It has been shown~\cite{report} that the tensor interaction has a very strong impact on the response functions for both spin channels due to spin-orbit coupling. 

Skyrme interactions are reasonably well controlled around the saturation density $\rho_0$ of SNM, for moderate isospin asymmetries and zero temperature. However, it has a certain number of drawbacks. For instance, most Skyrme parametrizations predict that the isospin asymmetry energy becomes negative when the density is increased~\cite{sam04,cen09,dan09,roc11,rep13,blo13}. Consequently, the symmetric system would be unstable at some density beyond the saturation one, preferring a largely asymmetric system made by an excess of either protons or neutrons. Another type of instability refers to the magnetic properties of neutron matter. Most Skyrme interactions predict that even in the absence of a magnetic field a spontaneous magnetisation arises in pure neutron  matter at some critical density~\cite{vid84,kut89,rio05}. Actually, it has been shown~\cite{mar02} that any reasonable Skyrme parametrisation predicts instabilities of nuclear matter beyond some critical density. Another inconvenient of zero-range interactions refers to pairing properties. Indeed, it leads to ultraviolet divergency~\cite{bul02} that needs to be treated with either an explicit regularisation~\cite{gra03} or via a cutoff of the available phase space~\cite{bor06}. Moreover, the UNEDF-SciDAC scientific collaboration~\cite{ber07} has investigated the role of the optimisation procedure on the quality and predictive power of the standard Skyrme functional~\cite{kor10}, concluding  that there is no more room to improve spectroscopic qualities by simply acting on the optimisation procedure~\cite{kor14}. To overcome this inconvenient, higher order momenta terms~\cite{car10} have been considered, namely N2LO and  N3LO Skyrme generalisations, which simulate finite-range effects~\cite{car08,rai11,dav13}. Even if the response function formalism is more complex in this case, it is still possible to obtain analytical results and to use them to avoid finite-size instabilities during the adjustment of the parameters of the interaction~\cite{bec17,bec15}. However, the use of a limited number of momentum powers to simulate finite-range effects may restrict the applicability of Skyrme-like interactions to describe the response functions at high values of the momentum transfer. In principle, all above mentioned difficulties should be removed by using a finite-range interaction.

The simplest way to deal with the exchange term of the ph interaction is to consider the limit of zero transferred momentum by placing the particle and hole momenta over the Fermi sphere. This is named Landau approximation of the ph-interaction. In this case, the form of the ph interaction becomes universal and it is possible to find analytical expressions for the response function~\cite{pas13b}. Since $q=0$ could be a too strong approximation, some authors have tried to keep an explicit momentum dependence only over the direct term and perform a  Landau approximation for the exchange term~\cite{mar05}. Either meson-exchange~\cite{alb80,ose82} or effective~\cite{mar05} interactions have been used in this approach. A general method to obtain response functions consists in using a partial wave expansion of both the ph interaction and the RPA ph propagator~\cite{mar05}. This method leads to a system of coupled  integro-differential equations that need to be solved numerically. Its validity depends on the convergence of the expansion. For all the considered interactions, a few multipoles suffice to get the converged response. As an alternative to the multipolar expansion, some authors have investigated a method based on the continuous fraction (CF) approximation to the RPA response function~\cite{del85,bri87,sch89,pac98,mar08,pac16}. In both cases, the exchange term is treated exactly, and the calculations can be carried out up to any degree of accuracy. Actually, two or three terms suffice to get converged results. The interest of the CF  method is that the formalism is the same for both infinite matter and finite nuclei. The CF convergence can be assessed by comparing with the multipolar expansion for nuclear matter, and the results could be hopefully translated to finite nuclei. 

The plan of this review is the following.  In Sect.~\ref{Sec:formalism} we present the general formalism to calculate the response function in an infinite nuclear system. The relevant quantities are defined, as the ph propagator, the Bethe-Salpeter equation, and the response functions. In Sec.~\ref{Sec:interactions} we present the phenomenological finite-range interactions considered in this review and the connection between finite- and zero-range interactions. The simplest approximation to deal with the exchange term, using either a Landau or a Landau-like ph interactions discussed in Sect.~\ref{Sec:Landau}. Then we review the multipolar expansion method in Sect.~\ref{sec:multipole}, and the continued fraction method in Sect.~\ref{Sec:CF}, to get the response function as two different expansions. In Sect.~\ref{sec:instabilities} we review the use of the response function to detect finite-size instabilities of phenomenological interactions. Finally, we give some concluding remarks in Sect.~\ref{Sec:concluding}. Some useful technical details and formulae are given in appendices. 


\section{Linear Response Formalism} \label{Sec:formalism}


In this section, we present a general description of the formalism used throughout this article to determine nuclear response functions and some related quantities. This formalism 
is based on the Green's function and it has been discussed in great details in several books and articles~\cite{fet71,dic08,mig67}. We summarise it here, mainly to fix the notations and signal some interesting points. Since we shall only consider homogeneous systems, the proper definitions of the strength function and of any related quantities should be understood as divided by a normalisation volume. To get the same quantities per nucleon one has just to divide by the density. Along this paper we shall use units such that $\hbar$ and $c$ are put to 1. 
 
\subsection{\it  Linear response to an external spin-isospin perturbation}

The strength function, also called dynamical structure function or dynamical form factor, is commonly the accessible experimental quantity to describe the response of a system to an external probe. It is defined as
\be
S(\mathbf{q},\omega) = \sum_{n \neq 0} | \langle n | \hat{Q}(\mathbf{q}) | 0 \rangle |^2 \delta(E_{n0} - \hbar \omega) \, ,
\label{strength1}
\ee
where $\omega$ and $\mathbf{q}$ are respectively the energy and momentum transferred by the probe. $|n\rangle$ and $E_n$ are the eigenstates and eigenvalues of the nuclear Hamiltonian ($H |n\rangle = E_n | n \rangle$), and the sum over $n$ includes both discrete and continuum contributions. Finally, $E_{n0}\equiv E_n - E_0$ is the excitation energy and $\hat{Q}$  is a well-chosen operator that couples directly to the desired densities. Typically, experiments determine integrals of $S(\mathbf{q},\omega)$ weighted with a kinematical factor associated with the probe. Therefore, all the physical properties of the system are embodied into the above strength function. 

We shall consider here density fluctuations in the spin-isospin channels  $(\alpha) = (S,I)$ of symmetric nuclear matter (SNM), excited by one-body operators  of the type:
\be
\hat{Q}^{(\alpha)} = \sum_j {\rm e}^{i \mathbf{q} \cdot \mathbf{r}_j} \hat{\Theta}^{(\alpha)}_j \, ,
\label{Q-operator}
\ee
where the index $j$ stands for the $j$th particle, and  
\be
\hat{\Theta}^{(0,0)}_j = \hat{1} \ , \  
\hat{\Theta}^{(1,0)}_j = \hat{\vsigma}_j \ , \ 
\hat{\Theta}^{(0,1)}_j = \hat{\text{\boldmath{$\tau$}}}_j \  \ , \ 
\hat{\Theta}^{(1,1)}_j = \hat{\vsigma}_j \, \hat{\text{\boldmath{$\tau$}}}_j \, ,
\ee
with ${\vsigma}_j$ and {\boldmath{$\hat{\tau}$}}$_j$ being the spin and isospin Pauli matrices. The spin-isospin channel $(\alpha)$ will be explicitly indicated from now on and detailed for each specific system. When necessary, the spin and isospin projections ($M$ and $Q$ respectively) will be also included in $(\alpha)=(S, M; I, Q)$. Most of the formal expressions deduced along this paper are also valid for PNM by simply changing the contents of symbol $(\alpha)=(S, M;n)$. 

If the interaction between the probe and the system is sufficiently weak, the change in the density is proportional to the perturbation induced by the external probe. The factor is the response function, also called dynamical susceptibility, which at first-order perturbation theory can be written as
\be
\chi^{(\alpha)}(\mathbf{q},\omega) = \sum_{n \neq 0}  
\left\{ \frac{|\langle n | \hat{Q}^{(\alpha)} | 0 \rangle|^2 }{\omega - E_{n0} + i \eta} 
- \frac{|\langle n | \hat{Q}^{(\alpha)} | 0 \rangle|^2 }{\omega + E_{n0} + i \eta} \right\} \, ,
\label{response1}
\ee
where $\eta$ is a positive quantity arbitrarily small associated with the adiabatic condition on the external field. Using the relation
\be
\lim_{\eta \to 0^+} \frac{1}{x-a+i \eta} = {\cal P} \frac{1}{x-a} - i \pi \delta(x-a) \, ,
\ee
one can write the following connection between the strength function and the response function
\be
S^{(\alpha)}(\mathbf{q},\omega)  - S^{(\alpha)}(\mathbf{q},- \omega) = - \frac{1}{\pi} {\rm Im} \, \chi^{(\alpha)}(\mathbf{q},\omega) \, .
\ee
At zero temperature, the system is initially in the ground state, and the sole possible effect of the probe is to excite the system, so that $\omega \ge 0$. In that case $S^{(\alpha)}(\mathbf{q},-\omega)$ is identically zero and we have
\be
S^{(\alpha)}(\mathbf{q},\omega)  = - \frac{1}{\pi} {\rm Im} \, \chi^{(\alpha)}(\mathbf{q},\omega) \, .
\label{strength2}
\ee
However, at finite temperature it is possible to transfer energy from the system to the probe, so that negative values of $\omega$ are admissible. We refer the reader to Refs.~\cite{fet71,vau96,report} for details about the formalism to get response functions at finite temperature.
 A more detailed discussion on the properties of the strength function for a system of fermions can be found for example in Refs.\cite{fet71,lip08,dic08}. We recall that the strength function (\ref{strength2}) is per unit volume, and has units MeV$^{-1}$\,fm$^{-3}$. If one is interested in the strength per nucleon, one must divide (\ref{strength2}) by the density and the strength function has units of MeV$^{-1}$.  

For simplicity, we consider that the ground state can be approximated by a Slater determinant, as it happens when we use the mean-field or Hartree-Fock (HF) approximation (the subindex HF will be used to denote functions calculated in this approximation). The ground state is thus a sequence of states filled up to the Fermi energy $\varepsilon_F$ and empty above. Since the action of the external probe on the system (or mathematically of the operator $\hat{Q}$) is to excite the fermions of the system by promoting them in particle states, we shall use the RPA (see~\cite{fet71} for a detailed discussion on this approximation) expressed in terms of ph propagators, whose momentum average will provide the RPA response function. In practice, the steps we will follow are: i) calculate the ph propagator in the mean-field approximation, ii) determine the ph matrix elements of the residual interaction, and iii) write the Bethe-Salpeter (BS) equation for the RPA propagator. The main difficulty to solve exactly that equation stems from the exchange terms of the ph interaction. We shall discuss different approximations employed to deal with such a problem.  

\subsection{\it Particle-hole propagators and linear response} \label{ph-propagators}

We give now a schematic description of the method used to obtain an expression of the strength function defined by Eq.~(\ref{strength2}). An example will be presented in the case of a simple residual interaction at the end of the section. For the sake of simplicity, we first consider a system with only one Fermi surface (as in SNM or in non polarised PNM), so that the isospin index is omitted. In this case, the retarded propagator $G^{HF}$ of a non-interacting ph pair can thus be expressed as
\begin{equation}
G^{HF}(\mathbf{k},\mathbf{q},\omega) = \frac{n(\mathbf{k})-n(\mathbf{q}+\mathbf{k})}{ \omega +\varepsilon(\mathbf{k})- \varepsilon(\mathbf{q}+\mathbf{k})+i \eta} \, ,
 \label{GHF}
 \end{equation}
where $n(\mathbf{k})$ 
is the Fermi-Dirac occupation number, which reduces to the step function $\theta(k_F-k)$ at zero temperature, and 
 $\varepsilon(\mathbf{k})$ is the single-particle energy
\be\label{single-particle}
\varepsilon(\mathbf{k}) = \frac{k^2}{2 m} + U(k) \;,
\ee
$U$ being the mean field. Actually, a parabolic approximation to the mean field is currently employed by most authors, so that the HF propagator has actually the same form as the free propagator when replacing the nucleon mass with the effective mass at the Fermi surface deduced from the mean field, that is 
\be
\varepsilon({\bf k})-\varepsilon({\bf k}+{\bf q}) \to \frac{\hbar^2}{2 m^*}({\bf k}^2-({\bf k}+{\bf q})^2) \, .
\label{sp-diff-m*}
\ee
The validity of this approximation will be discussed in Sect.~\ref{sect:effectivemass}.

The response function is then obtained as
\be
\chi^{HF}(\mathbf{q},\omega) = n_d \int \frac{d^3 \mathbf{k}}{(2 \pi)^3} \, G^{HF}(\mathbf{k},\mathbf{q},\omega) \, ,
\label{chiHF}
\ee
where $n_d$ is the degeneracy factor ($n_d=4$ for SNM and $n_d=2$  for PNM). The function $\chi^{HF}(\mathbf{q},\omega)$ is usually called the Lindhardt function~\cite{lin54}, although strictly speaking the latter refers to the response of a free fermion gas instead of a system of fermions in a mean field. Furthermore, this expression corresponds to the case where no spin or isospin flip occurs and the absence of the $(\alpha)$ index indicates that the result is independent of the spin-isospin channel. 

If we now consider the presence of a residual interaction acting between particle and holes, correlations are taken into account through the correlated ph propagator which is obtained by solving  the Bethe-Salpeter equation~\cite{rin80}
\bea
G^{(\alpha)}_{RPA}(\mathbf{k}_{1},\mathbf{q},\omega) &=& G^{HF}(\mathbf{k}_{1},\mathbf{q},\omega) \nnn
&+& 
G^{HF}(\mathbf{k}_{1},\mathbf{q},\omega) \sum_{(\alpha')} \int \frac{d^{3}\mathbf{k}_{2}}{(2 \pi)^3} \,
V_{ph}^{(\alpha,\alpha')}({\mathbf q}, \mathbf{k}_1, \mathbf{k}_2) G^{(\alpha')}_{RPA}(\mathbf{k}_{2},\mathbf{q},\omega) \, , 
\label{bethe-salpeter}
\eea
where $V_{ph}^{(\alpha,\alpha')}(\mathbf{q}, \mathbf{k}_1, \mathbf{k}_2)$ is the residual interaction matrix element which describes the ph excitations of the system built on a mean-field (Hartree-Fock) ground state. The ph residual interaction links two ph pairs with quantum numbers $(\alpha)$ and $(\alpha')$, and hole momenta $\mathbf{k}_1$ and $\mathbf{k}_2$, respectively. Integrating over $\mathbf{k}_1$ gives the RPA response function
\be
\chi^{(\alpha)}_{RPA}(\mathbf{q},\omega)  = n_d  \int \frac{d^3 \mathbf{k}}{(2 \pi)^3} \, G^{(\alpha)}_{RPA} (\mathbf{k},\mathbf{q},\omega) \, .
\ee

\begin{figure}[tb]
\begin{center}
\includegraphics[width=0.6\textwidth,angle=0]{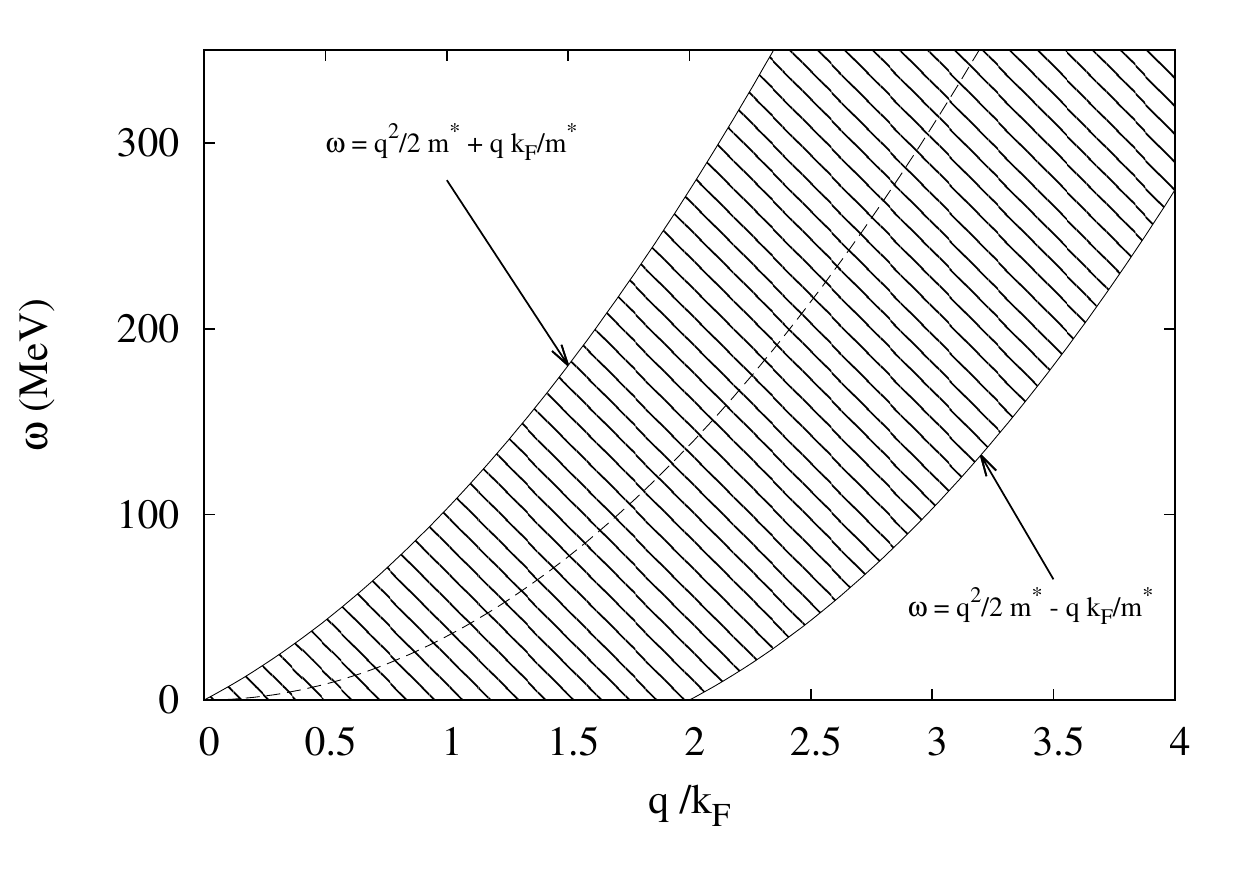}

\caption{Domain of allowed excitation energies  associated with a particle-hole excitation. For illustration, calculations have been done at  $k_F$=1.33~fm$^{-1}$ and $m^*/m$=1.  The dotted line inside this region corresponds to the value $q^2/2m^*$. See text for details. }
\label{excitation}

\end{center}
\end{figure}

At zero temperature, $0\le k \le k_F$ and it is possible to set limits on the available energy space of all possible 1p1h excitations. When a momentum ${\bf q}$ is transferred to a particle with momentum $\mathbf{k}$ in the Fermi sea, the excitation energy can be written as $\omega = (\mathbf{k}+\mathbf{q})^2/2m^* - k^2/2m^*$, assuming the parabolic approximation to the mean field. This energy depends on the relative angle between $\mathbf{k}$ and $\mathbf{q}$, but it lies  inside the shaded region of Fig.~\ref{excitation}. The dotted line corresponds to the value $q^2/2m^*$ (particle initially at rest). This region of 1p1h excitations is the one considered in this review for all RPA calculations. 

\subsection{\it The ph interaction}

\begin{figure}[tb]

\begin{center}
\includegraphics[width=0.5\textwidth,angle=0]{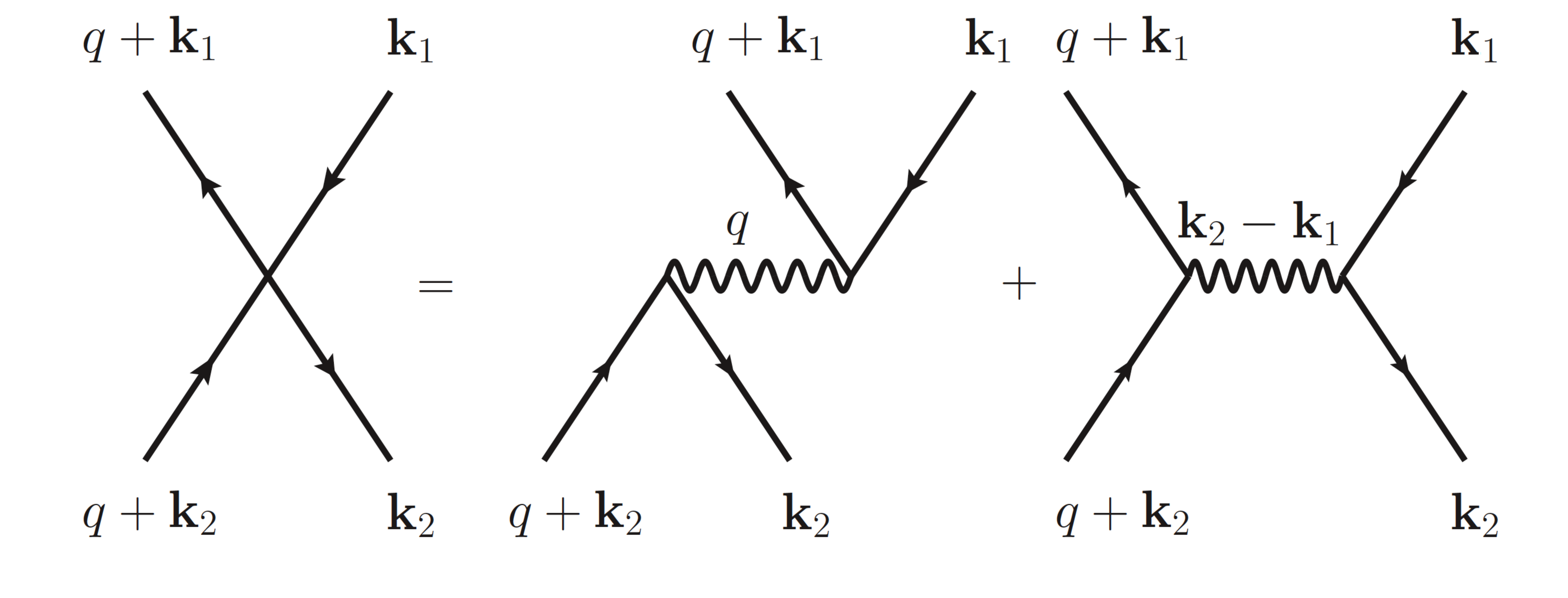}

\caption{Diagrammatic representation of the particle-hole interaction.}
\label{dir:exch}
\end{center}
\end{figure}

Within  the RPA, the excitations of the system result from the residual ph interaction. In the theory of Fermi liquids, the matrix elements of this ph interaction can be easily obtained by the second functional derivative with respect to non-diagonal densities taken at the Hartree-Fock solution~\cite{rin80}
\begin{equation}
\label{secondderivative}
\langle {\bf a'} , {\bf b'} | V_{ph} | {\bf a} , {\bf b} \rangle = \frac{\delta^{2 }E[\rho]}{\delta  \hat{\rho}({\bf a'}, {\bf a})
\delta \hat{\rho}({\bf b'}, {\bf b})} \bigg|_{HF}
\end{equation}
where the symbol $\bf a$ is a shorthand notation for $({\bf x}_{\bf a}, \sigma_{\bf a}, \tau_{\bf a})$, that is the spatial coordinates $\mathbf{x_a}$ and the spin and isospin variables 
$\sigma_{\bf a}$ and $\tau_{\bf a}$. In general, the resulting two-body matrix elements depend on four-momenta at most. However, because of momentum conservation there are actually three independent momenta. Following~\cite{gar92}, we choose them as the initial (final) momentum $\mathbf{k}_1 (\mathbf{k}_2) $ of the holes and the external momentum transfer $\bf{q}$ in the following way: $\bf{k}_{a}'=\bf{k}_1+\bf{q}$, $\bf{k}_a=\bf{k}_1$, $\bf{k}_{b}'=\bf{k}_2$, and $\bf{k}_b=\bf{k}_2+\bf{q}$ as illustrated in Fig.~\ref{dir:exch}. Finally, the matrix elements in the spin-isospin space of the ph interaction may be written as
\be
\langle {\bf q} + {\bf k}_1, {\bf k}_2; \alpha | V_{ph} | {\bf k}_1, {\bf q} + {\bf k}_2; \alpha' \rangle
= V^{(\alpha, \alpha')}_{ph}({\bf q}, {\bf k}_1, {\bf k}_2)\;,
\label{ME-Vph}
\ee 
as anticipated in Eq.~(\ref{bethe-salpeter}). From Fig.~\ref{dir:exch}, we see that the ph interaction is the sum of a direct contribution, which only depends on the transferred momentum $\mathbf{q}$, and an exchange one, which depends on the relative momentum $\mathbf{k}_{12} \equiv \mathbf{k}_1-\mathbf{k}_2$. 

\begin{figure}[tb]
\begin{center}
\begin{center}
\includegraphics[width=0.43\textwidth,angle=0]{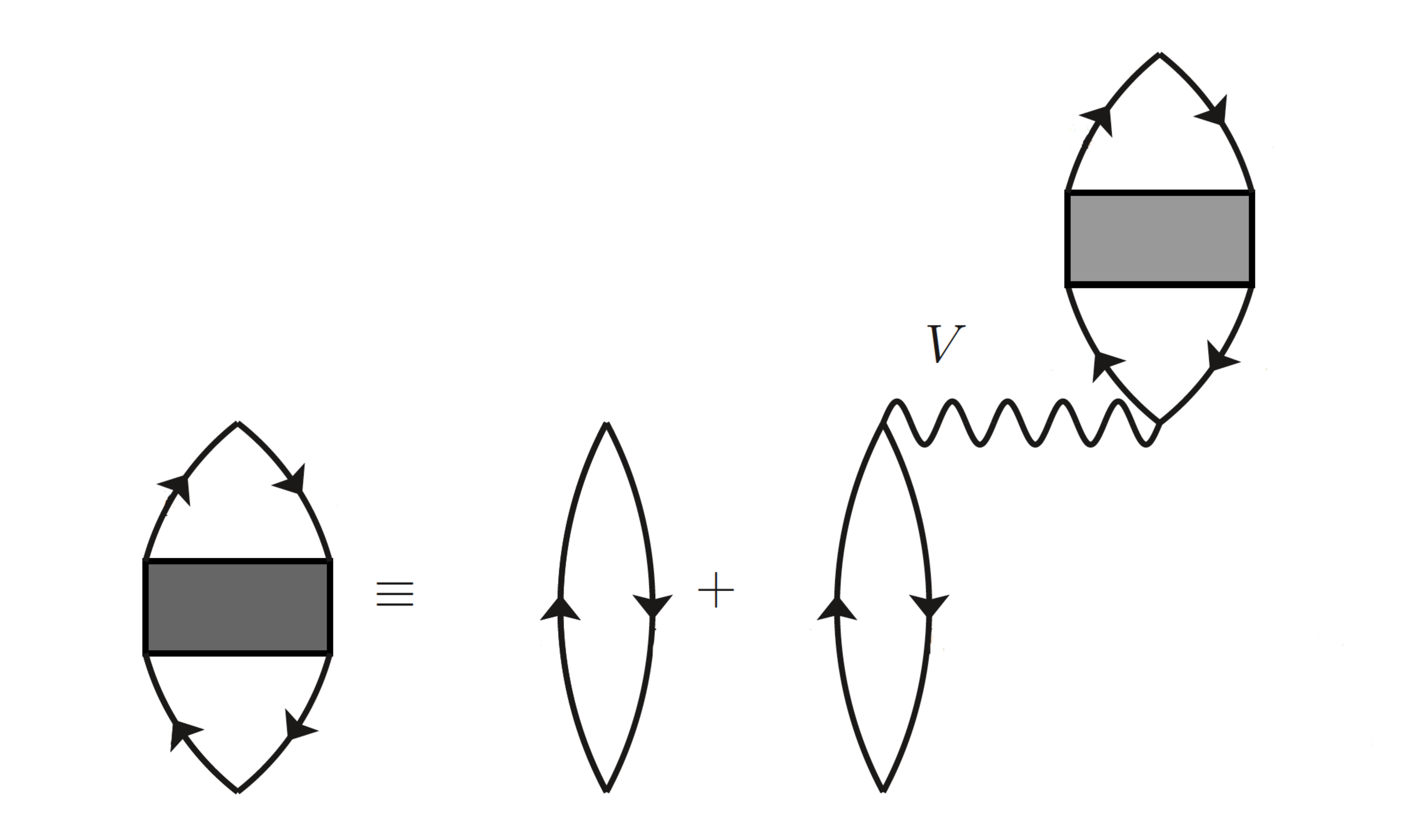}
\end{center}
\caption{Diagrammatic representation of the Bethe-Salpeter equation for the response function restricted to the direct term.}
\label{ring}
\end{center}
\end{figure}

One can thus recast the previous expression as
\be
V^{(\alpha, \alpha')}_{ph}({\bf q}, {\bf k}_1, {\bf k}_2) = V^{(\alpha, \alpha')}_D(\mathbf{q})  + V^{(\alpha, \alpha')}_E(\mathbf{k}_{12}) \;.
\label{res}
\ee
\noindent The phenomenological interactions considered here contain an explicit density-dependent part, but it is worth noticing that this part only contributes to the direct term.
Actually, the exchange term represents the main difficulty in solving the  Bethe-Salpeter equation. When treated in some approximation so that its ${\bf k}_{12}$ dependence simplifies, the resolution of Eq.~(\ref{bethe-salpeter}) becomes feasible. For instance, assuming no ${\bf k}_{12}$ dependence at all, and the direct term diagonal in the spin-isospin space, one can immediately solve the BS equation so that response function takes the simple form 
\begin{equation}
\chi^{(\alpha)}_{RPA}(q,\omega)=\frac{\chi^{HF}(q,\omega)}{1-V_D^{(\alpha)}(q) \chi^{HF}(q,\omega)}
\, .
\label{ring2}
\end{equation}
This is the so-called ring approximation, diagrammatically represented in Fig.~\ref{ring}.  As a matter of terminology, it is worth keeping in mind that in condensed matter textbooks, such an approximation is indicated as ``RPA'', whereas ``extended RPA'' refers to what it is called here  RPA, that is, including the full residual interaction. 


Some general properties of the RPA response function can be deduced from Eq.~(\ref{ring2}). The residual interaction not only modifies the distribution strength, but it can also give rise to excitations outside the energy domain illustrated in Fig.~\ref{excitation} where  ${\rm Im}\, \chi^{(\alpha)}_{RPA}(q,\omega) = 0$. Indeed, the response has a pole when the denominator of Eq.~(\ref{ring2}) vanishes:  $1-V_{ph}^{(\alpha)} \chi^{HF} = 0$. This defines a particular collective excitation~\cite{fet71}, also called zero sound, for historical reasons~\cite{alb82b}. Increasing the value of $q$, the excitation energy eventually crosses the upper ph edge and the excitation is absorbed into the continuum.  

In the case of a standard Skyrme interaction the strategy for obtaining the response function is very simple~\cite{gar92,mar06,dav09,pas12,report}. In momentum space, the ph interaction can be written as a polynomial form in the relative hole momenta, containing only terms of the form $1 \, , \, k_i^2 \, , k_i Y_{1,\mu}(\hat{k}_i)$, and $k_1Y_{1,\mu}(\hat{k}_1) k_2 Y_{1,\mu'}(\hat{k}_2)$. The idea is thus to multiply the Bethe-Salpeter equation Eq.~(\ref{bethe-salpeter}) with appropriate monomials in ${\bf{k}}_i$ and integrate over momentum, thus obtaining a closed set of algebraic equations for the momentum averages of the ph propagator $\langle G^{(\alpha)}_{RPA}\rangle$, 
$\langle k^2 G^{(\alpha)}_{RPA}\rangle$, $\langle k Y_1 ^0 G^{(\alpha)}_{RPA}\rangle$,
$\langle k^2 |Y_1 ^1| ^2 G^{(\alpha)}_{RPA}\rangle$ and $\langle k^2 |Y_1 ^0| ^2 G^{(\alpha)}_{RPA}\rangle$. 
Depending on the ph interaction, one can therefore obtain either a single equation or a set of equations coupling the different spin-isospin channels. In principle the system can be analytically solved, although for some systems and/or excitation operators, the  expressions are cumbersome and the numerical approach is preferable.

It is worth mentioning that an essential assumption used to solve algebraically the Bethe-Salpeter equation is that the momentum dependence of the single particle states is given by Eq.~(\ref{single-particle}), {\it i.e.} there is no explicit momentum dependence in the effective mass $m^*$. This is the case for a standard Skyrme interaction, but it is no longer true when dealing with Skyrme N2LO or N3LO interactions or in general finite-range interactions. Taking explicitly into account the full momentum dependence of the effective mass would require solving the problem numerically from the start.  The problem has been discussed in  Ref~\cite{bec15} for the Skyrme N2LO interaction:  by evaluating $m^*$ at the Fermi momentum and thus removing the explicit momentum dependence of the effective mass it is then possible to find an analytical expression for the response function. By doing that, a violation has been observed for the energy weighted sum rule of at most 10\% for large values of the transferred momentum ($\simeq 2k_F)$. We will come back to this point in Sect.~\ref{sect:effectivemass}.

\section{Phenomenological finite-range interactions} \label{Sec:interactions}

There are two popular types of finite-range interactions being currently used to describe nuclear structure, namely Gogny~\cite{gog75} and Nakada~\cite{nak03} interactions. Both types can be cast in the following general form
\be\label{eq.int}
V(\mathbf{r_1},\mathbf{r}_2) = V_{C}(\mathbf{r_1},\mathbf{r}_2) + V_{DD}(\mathbf{r_1},\mathbf{r}_2) + V_{SO}(\mathbf{r_1},\mathbf{r}_2) + V_{T}(\mathbf{r_1},\mathbf{r}_2) \;,
\ee
which is a sum of central, density-dependent, spin-orbit and tensor terms. A formal difference between these interactions  lies in the finite-range form factor, consisting of a superposition of either Gaussians or Yukawians, respectively. In both cases, the density-dependent term is a zero-range interaction of the same form as the Skyrme interaction.

Gogny's interaction was originally conceived to describe pairing correlations using the same central interaction for both particle-particle (pp) and ph channels, thus avoiding the introduction of two separated potentials as required when dealing with the Skyrme interaction. It is a purely phenomenological interaction containing a central part, made as a superposition of two Gaussians, plus a density-dependent  and a spin-orbit term both similar to those present in the Skyrme interaction. The parameters are fitted so as to reproduce selected (pseudo)-observables in both finite nuclei and nuclear matter as well. Nowadays, it is currently used to describe many aspects of nuclear structure at the Hartee-Fock-Bogoliubov (HFB) level, or even beyond mean field~\cite{rod07}. The merits and drawbacks of this interaction have been reviewed in Refs.~\cite{per14, egi16,rob19}.

Nakada interaction is a semi-realistic one, actually based on the so-called Michigan three range Yukawa (M3Y) interaction~\cite{ber77}. Originally, the M3Y interaction was conceived for inelastic nucleon-nucleus scattering studies, and its parameters were fitted to G-matrix results derived from realistic interactions. However, it provides unsatisfactory results in finite nuclei, particularly for saturation and spin-orbit splittings. It was then generalized by Nakada~\cite{nak03}, starting from a M3Y parametrisation which fits the G-matrix obtained from the Paris-potential~\cite{ana83}. It includes finite-range central, spin-orbit and tensor terms plus a Skyrme-like density-dependent term. Some of its parameters were fitted to nuclear structure data, while keeping others unchanged. In particular, the long range part of the interaction has been kept, so that the main features of the one pion exchange potential (OPEP) are asymptotically fulfilled.  Afterwards, improving the fit of nuclear observables has resulted in the construction of several parameterisations.  A recent review of this interaction can be found in Ref.~\cite{nak20}. In the following we discuss in more detail the structure of each of these terms for both interactions. 

For both families, the general residual ph interaction matrix elements is written as a sum
\be
V_{ph}^{(\alpha,\alpha')}(k_1,k_2,q) = V_{C \, ph}^{(\alpha,\alpha')} + V_{DD \, ph}^{(\alpha,\alpha')} +
V_{SO \, ph}^{(\alpha,\alpha')} + V_{T \, ph}^{(\alpha,\alpha')} 
\label{Vph-general}
\ee
of central, density dependent, spin-orbit and tensor parts respectively. 

\subsection{\it Gogny interaction}

The various contributions to Eq.~(\ref{eq.int}) for the Gogny interaction read
\bea
V_{C}(\mathbf{r_1},\mathbf{r}_2) &=& \sum_{i=1,2} \left( W_i + B_i P_{\sigma} - H_i P_{\tau} - M_i P_{\sigma} P_{\tau} \right) 
{\rm e}^{-r_{12}^2 / \mu_i^2} \;, \\
V_{DD}(\mathbf{r_1},\mathbf{r}_2) &=&  t_3 \left( 1 + x_3 P_{\sigma} \right) \rho^{\gamma} \delta(\mathbf{r_1}-\mathbf{r}_2)\,,\\
V_{SO}(\mathbf{r_1},\mathbf{r}_2) &=&  i W_0 (\bf{k'} \wedge \bf{k}) \cdot ({\vsigma}_1+{\vsigma}_2 )\delta(\mathbf{r_1}-\mathbf{r}_2) \,.
\label{V-gogny-1}
\eea
\noindent $P_{\sigma}, P_{\tau}$ are the standard spin and isospin exchange operators. The values of $\mu_i$ are fixed beforehand to simulate short- and long-range terms, typically between 0.5 and 0.8 fm for the shorter and 1-1.2 fm for the longer.  The sum over range indices  will be omitted in the following to alleviate the notations. The density power $\gamma$ is also fixed, usually to the value $1/3$. The remaining parameters are adjusted on some selected properties of finite nuclei and SNM according to the adopted fitting protocol~\cite{dec80,cha08b,gor09a}.

Originally, no tensor terms were considered. However, some authors have recently investigated the possibility of equipping the Gogny interaction with a finite-range tensor. In Ref.~\cite{ots06} a finite-range Gaussian tensor isospin term was added to the D1S interaction, with a full refit of the parameters, leading to the so-called GT2 interaction. However, the resulting parametrisation lead to strong instabilities in the infinite medium~\cite{pac16}, thus making the interaction not suitable for calculations in atomic nuclei. In Refs.~\cite{co11,ang11}, an isospin tensor term was perturbatively added to D1S and D1M interactions. It was taken from the Argonne AV18 interaction~\cite{wir95}, with a regularisation factor. The resulting interactions were labelled D1ST and D1MT. Later on, the same authors~\cite{ang12,ang16} used a Gaussian tensor form as suggested in Ref.~\cite{oni78}
\be
V_{T}(\mathbf{r_1},\mathbf{r}_2) = \left( V_{T1} + V_{T2} P_{\tau} \right) {\rm e}^{-\frac{1}{4} r_{12}^2 \mu^2} S_T({\hat{\bf r}}_{12}) \;,
\label{V-gogny-2}
\ee
where $S_T$ is the usual tensor operator 
\be\label{eq:tensor}
S_T({\hat{\bf r}}_{12})  =  3 ({\vsigma}_1 \cdot {\hat{\bf r}}_{12})  ({\vsigma}_2 \cdot {\hat{\bf r}}_{12}) -  ({\vsigma}_1 \cdot {\vsigma}_2) \;.
\ee
Again, it was introduced perturbatively by simultaneously adjusting the spin-orbit and the tensor terms on the shell structure of selected nuclei, without touching the central part terms. Some stable parameterisations, labelled D1ST2a and D1ST2b, suitable for systematic calculations in atomic nuclei have been obtained. 

Finally, it is worth mentioning that in Ref.~\cite{cha15a} a finite-range version of the density-dependent term has been suggested. This new density dependent term has the form
\begin{eqnarray}\label{DD:fr}
V_{DD,fr}(\mathbf{r_1},\mathbf{r}_2) &=&(W_3+B_3P_\sigma-H_3P_\tau-M_3P_\sigma P_\tau) \,
\frac{e^{- r_{12}^2/\mu_3^2}}{(\mu_3 \sqrt{\pi})^3} \, 
\frac{\rho^\alpha(r_1)+\rho^\alpha(r_2)}{2}\;.
\end{eqnarray}
The main reason was to improve the reproduction of the four spin/isospin channels of the equation of state (see next section). The resulting parametrisation is called D2, but, to our knowledge, no systematic calculations have been implemented with this parametrization.

\subsubsection{\it Mean field}

\begin{figure}[!h]
\begin{center}
\includegraphics[width=0.43\textwidth,angle=0]{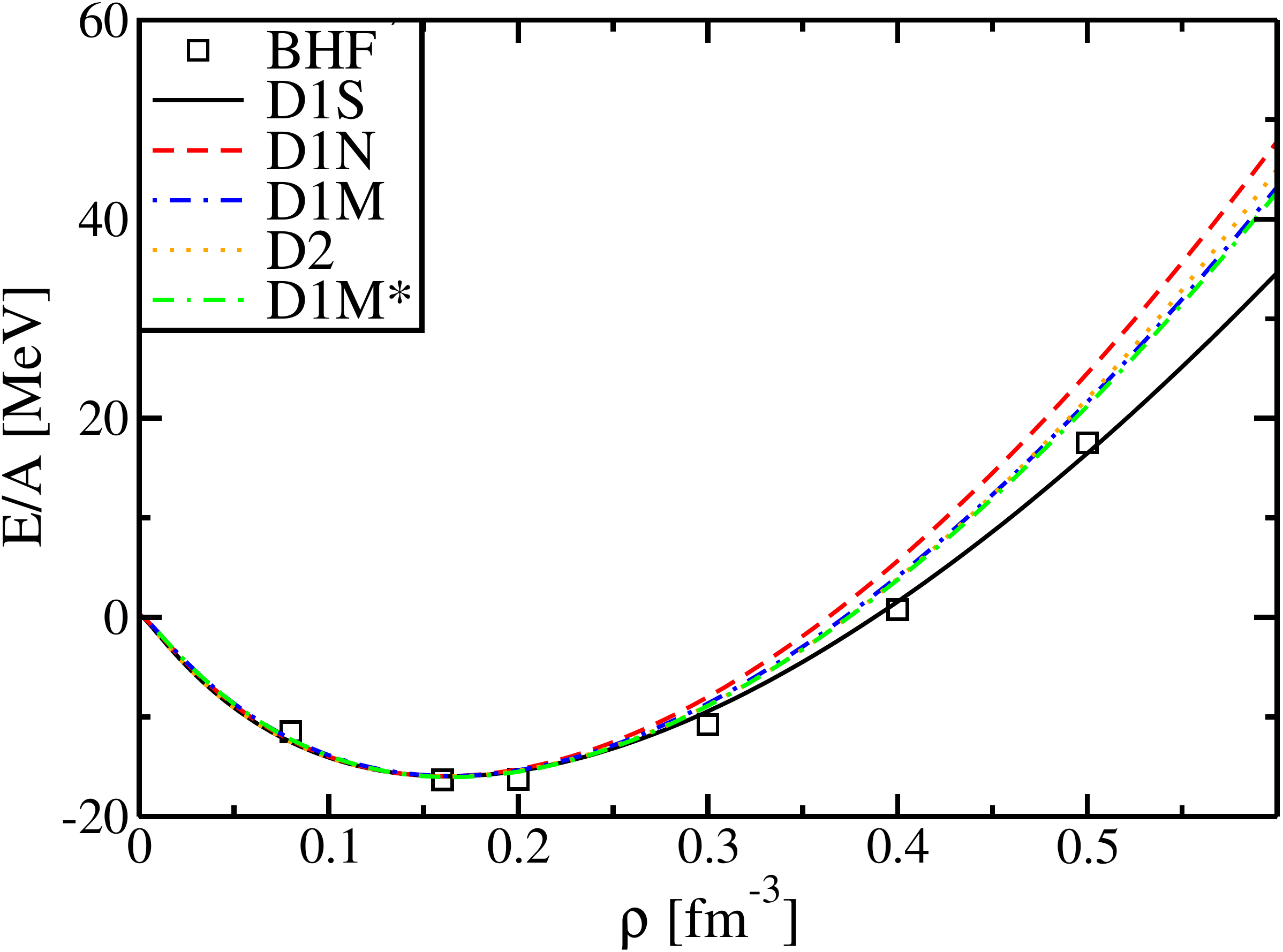}
\end{center}
\caption{EoS in SNM for the various Gogny interactions. The squares correspond to BHF calculations. Taken from Ref.~\cite{dav16b}.}
\label{fig1-Gogny-EA}
\end{figure}

In the Hartree-Fock approach, it is possible to find an analytical solution for ground state properties of infinite matter. For instance, in Fig.~\ref{fig1-Gogny-EA}, the equation of state (EoS) is plotted as a function of the density for SNM with D1S~\cite{dec80}, D1N~\cite{cha08b}, D1M~\cite{gor09a}, D1M*~\cite{gon18} and D2~\cite{cha15a} interactions. As a comparison, on the same figure are also displayed the Brueckner-Hartree-Fock (BHF) results~\cite{bal97}, based on the Argonne AV14 nuclear interaction plus the Urbana model for the three-body interaction. 

We see that the various interactions have a quite similar behaviour up to twice the saturation density, then the results tend to differ. For completeness, we  report in Tab.~\ref{tab:gogny-SNM} other relevant properties of selected Gogny parametrizations around saturation density $\rho_0$ as the energy per particle at saturation  $\left.E/A\right|_0$, the compressibility $K_0$ and the symmetry energy $J$. We refer the reader to Ref.~\cite{sel14} for a more detailed discussion on the SNM properties of Gogny interactions.
\begin{table}[h]
\begin{center}
\begin{tabular}{c|ccccc}
\hline 
 & $m^*/m$ & $\rho_0$ [fm$^{-3}$]& $\left.E/A\right|_0 $ [MeV] & $K_0$ [MeV]& $J$ [MeV]\\
\hline
D1M* & 0.746 &0.165 & -16.06 & 225.4 & 30.25 \\
 D1M & 0.747 & 0.165 & -16.02 & 225.0 & 28.55\\
 D1N & 0.747 & 0.161 & -15.96 & 225.6 & 29.60 \\
 D1S & 0.697 & 0.163 & -16.01 & 202.9 & 31.13 \\
 D2 & 0.738 & 0.163 & -16.00 & 209.3 & 31.13\\
\hline
\end{tabular}
\caption{SNM properties as predicted by selected Gogny parametrizations. Adapted from Ref.~\cite{gon18}.}
\label{tab:gogny-SNM}
\end{center}
\end{table}

\begin{figure}[!h]
\begin{center}
\includegraphics[width=0.43\textwidth,angle=0]{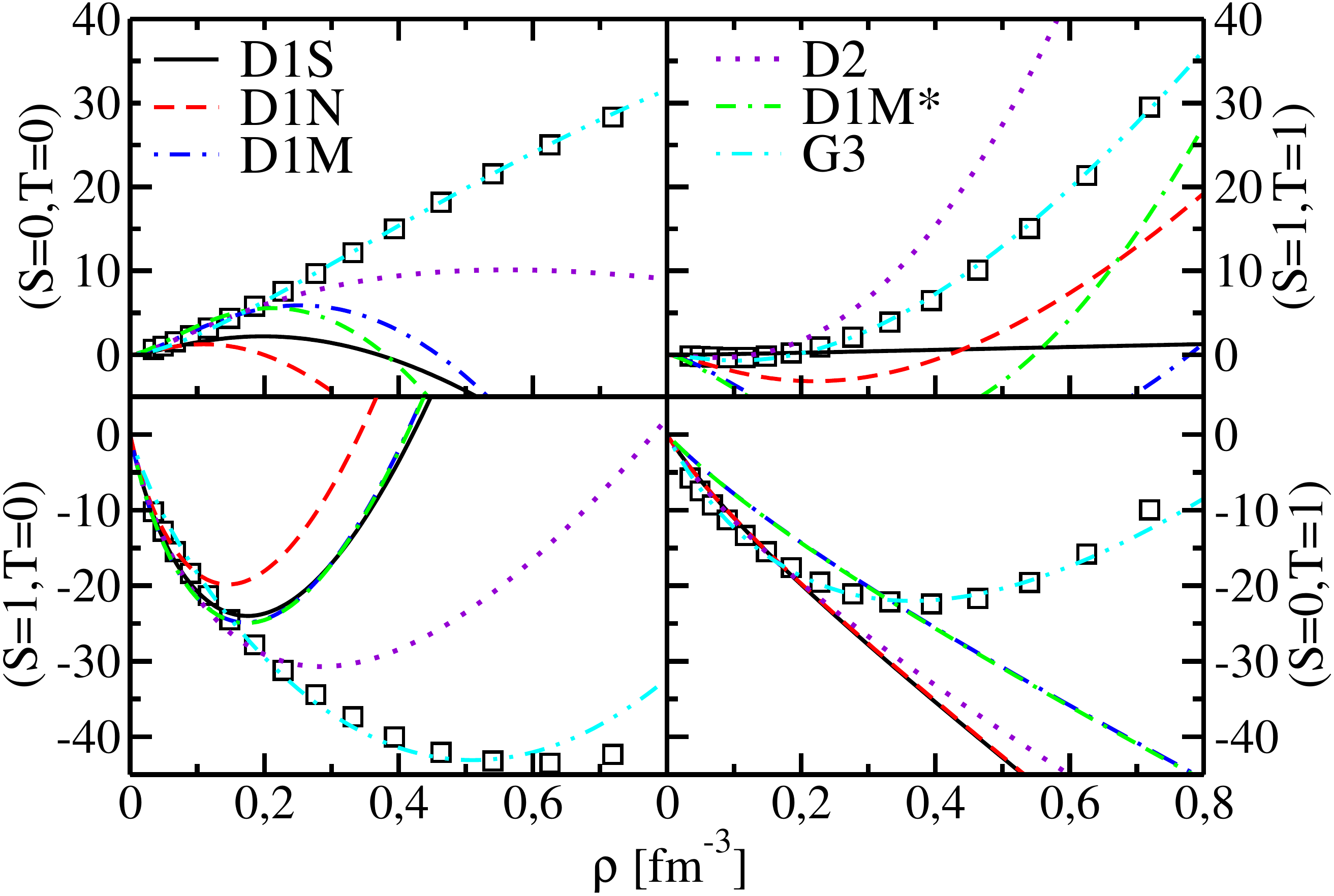}
\end{center}
\caption{Spin/Isospin decomposition of the EoS for various Gogny interactions. The dots correspond to BHF calculations. Adapted from Ref.~\cite{dav16b}.}
\label{fig1-Gogny-ST}
\end{figure}

In Fig.~\ref{fig1-Gogny-ST}, we  illustrate the spin-isospin decomposition of the SNM EoS~\cite{les06} for the Gogny parametrizations of Tab.~\ref{tab:gogny-SNM} together with the BHF results~\cite{bal97}. 
As already mentioned, the D2 interaction has been equipped with the finite-range density-dependent term given in Eq.~(\ref{DD:fr}) to improve on the reproduction of the (S,T) channels as compared to BHF results. We can see that the agreement between mean-field and BHF results is thus better than the other parametrizations, but restricted to densities below $\simeq 1.5$ the saturation density.

To improve the agreement at higher values of the density, the inclusion of a third Gaussian to the original Gogny interaction has been suggested in Ref.~\cite{dav17b}. Up to now, no parametrization with three gaussians which takes into account finite nuclei constraints has been obtained yet: the interaction G3, based on D1S, has been adjusted only on infinite matter. However, the inclusion of a third gaussian seems to contain the extra required flexibility and one can observe in Fig.~\ref{fig1-Gogny-ST} a much better agreement with realistic BHF results. See discussion in Ref.~\cite{dav16b}.

\begin{figure}[!h]
\begin{center}
\includegraphics[width=0.43\textwidth,angle=0]{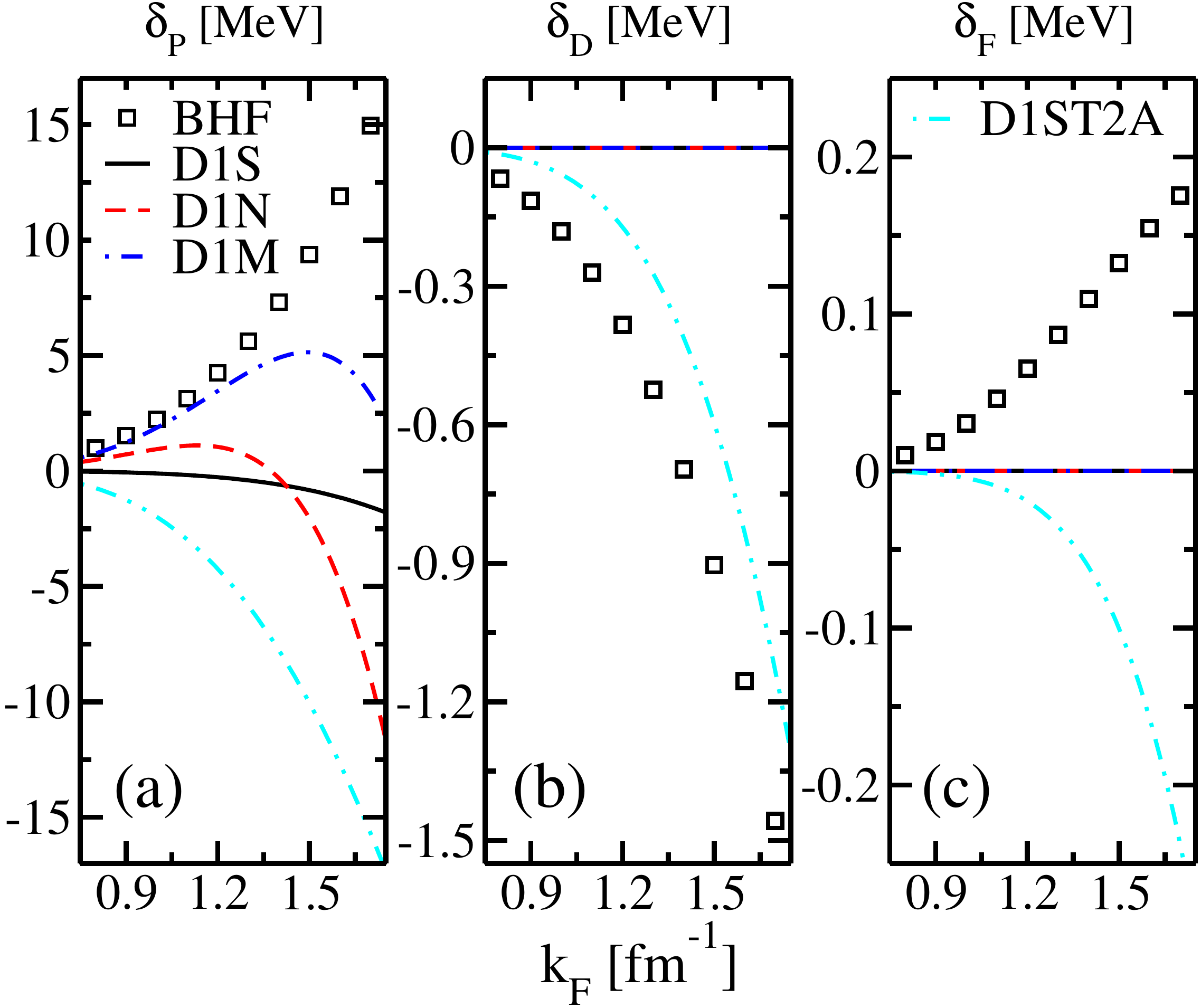}
\end{center}
\caption{Difference of partial waves for BHF results (dots) and some Gogny parametrizations (lines). Taken from Ref.~\cite{dav16b}.}
\label{fig-gogny-deltas}
\end{figure}

Only the central and density-dependent terms are tested in the $(S,T)$ decomposition of the EoS since at the mean field level, neither the tensor nor the spin-orbit contribute~\cite{vid11}. However, as discussed in Refs.~\cite{dav15a,dav16b}, it is possible to test these non-central terms as well, by inspecting the following partial wave decomposition of the total EoS
\begin{eqnarray}\label{partial}
\frac{E}{A}=\frac{3}{5} \frac{\hbar^2}{2m}k_F^2 +\sum_{JLS}\mathcal{V}\left(^{2S+1}L_J \right) \;,
\end{eqnarray}
\noindent where $\vec{J}=\vec{L}+\vec{S}$ is the total angular momentum, $\vec{L}$ is the orbital angular momentum and $\vec{S}$ is the total spin of a pair of particles. To better disentangle the effects related to spin-orbit and tensor, in Ref.~\cite{dav16b} the energy difference of some specific partial waves was considered 
\begin{eqnarray}
\delta_P =\frac{1}{3}\mathcal{V}\left(^3P_1 \right)-\mathcal{V}\left(^3P_0 \right)\label{partial:p} \;, \\
\delta_D =\frac{1}{5}\mathcal{V}\left(^3D_2 \right)-\frac{1}{7}\mathcal{V}\left(^3D_3 \right)\label{partial:d} \;, \\
\delta_F =\frac{1}{7}\mathcal{V}\left(^3F_3 \right)-\frac{1}{9}\mathcal{V}\left(^3F_4 \right) \label{partial:f} \;.
\end{eqnarray}
In principle, some other combinations are possible, but they are contaminated by extra parity mixing which is absent in the HF formalism. The behaviour of these energy differences is displayed in Fig.~\ref{fig-gogny-deltas} for BHF calculations and some of the Gogny parametrizations considered in the previous section. Since there is no tensor in these parametrizations, the contribution is identically zero for $\delta_D$ and $\delta_F$, the spin-orbit term giving a non zero contribution to $\delta_P$ only. Therefore, the parametrization D1ST2a~\cite{ang11} which contains a tensor has also been included. However, we see in Fig.~\ref{fig-gogny-deltas} that this tensor term, introduced perturbatively via an adjustment on some selected properties of spherical nuclei gives the wrong sign in infinite matter when compared to BHF calculations.

As discussed in Ref.~\cite{dav16c}, it is thus not clear whether the constraints on the tensor coming from finite nuclei are compatible with the ones coming from {\it ab initio} calculations. However it is neither clear if such a difference is relevant and has any impact on calculations of properties of finite nuclei. More investigation in this direction is clearly required before drawing strong conclusions.

\subsubsection{\it ph interaction}

The different terms entering Eq.~(\ref{Vph-general}) read 
\bea
V_{C \, ph}^{(\alpha,\alpha')} &=& \delta_{\alpha,\alpha'} \left\{ D^{(\alpha)} F_{C}(q) + E^{(\alpha)} F_{C}(\bf{k}_{12}) \right\} \;, \label{Appb:central}\\
V_{DD \, ph}^{(\alpha,\alpha')} &=& \delta_{\alpha,\alpha'} R^{(\alpha)} \;, \label{Appb:density}\\
V_{SO \, ph}^{(\alpha,\alpha')} &=& q X^{(I)} \left\{ M'(k_{12})^1_{M'}\delta_{S',1}\delta_{S,0}+M(k_{12})^1_{-M} \delta_{S,1}\delta_{S',0}  \right\} \delta_{I,I'}\delta_{Q,Q'} \;, \label{Appb:so} \\
  V_{T\,ph}^{(\alpha,\alpha')} & = & \delta_{S,1}\delta_{S',1}\delta_{I,I'}\delta_{Q,Q'} \left\{ 
  D_T^{(I)}   F_T (q) q^2 \left[3\delta_{M,0}\delta_{M',0} -\delta_{M,M'} \right]  \right. \nonumber \\ 
& & \hskip 2.2 true cm \left. 
+  E_T^{(I)}  F_{T}(k_{12}) 
\left[3(-1)^{M}(k_{12})^{1}_{-M}(k_{12})^{1}_{M'} -k_{12}^2\delta_{M,M'}\right] \right\} \;, \label{Appb:ten}
\end{eqnarray}
where $\left(k_{12}\right)^{1}_{M}$ is a rank-1 tensor defined as $\left( k_{12}\right)^{1}_{M} = \sqrt{\frac{4 \pi}{3}} \left( Y_{1M}(\hat{k}_1) - Y_{1M}(\hat{k}_2) \right)$, $Y_{1M}$ being the usual spherical harmonic. The coefficients $D^{(\alpha)}, E^{(\alpha)}, R^{(\alpha)}$ are combinations of the central and density-dependent parameters, given in Tab.~\ref{gogny-central} for both SNM and PNM. The coefficients $X^{(I)}, D_T^{(I)}, E_T^{(I)}$ are combinations of the non-central parameters. They only depend on the isospin index, and their expressions are given in Tab.~\ref{gogny-noncentral}. The notation $D$ and $E$ refers to the direct and exchange contributions in each case. As the Gogny spin-orbit interaction is zero-range, both contributions merge in a single combination $X^{(I)}$. The functions $ F_{C,T}$ corresponds to the Fourier transform of the radial functions, whose expressions are 
\bea
F_{C}(k) &=& {\rm e}^{-k^2 \mu^2/4} \label{FC:gogny} \;, \\
F_{T}(k) & = & \frac{1}{4 k^5} \left(  6\pi \hbox{Erf}\left(\frac{k\mu}{2}\right)-\sqrt{\pi}k\mu (6+k^2\mu^2)e^{-k^2\mu^2/4} \right)\;.  
\eea

\begin{table}[h]
\begin{center}
\begin{adjustbox}{max width=\textwidth}
\begin{tabular}{ccccc}
\hline 
&&&& \\[-3mm]
SNM & $(S,I)$ & $D^{(S,I)}$ & $E^{(S,I)}$  & $R^{(S,I)}$  \\
\hline
& (0,0) & $\pi^{3/2}\mu^3 (4W+2B-2H-M)$ & $\pi^{3/2}\mu^3 (-W-2B+2H+4M)$  & $\frac{3}{2}(\gamma +1)(\gamma+2)t_3 \rho^\gamma$ \\
& (0,1) & $ -\pi^{3/2}\mu^3 (2H+M)$ & $-\pi^{3/2}\mu^3 (W+2B) $ & $ -(1+2x_3)t_3 \rho^\gamma$ \\
& (1,0) &  $\pi^{3/2}\mu^3 (2B-M)$  & $\pi^{3/2}\mu^3 (-W+2H)$  & $ -(1-2x_3)t_3 \rho^\gamma$  \\
& (1,1) &   $-\pi^{3/2}\mu^3 M$ &   $-\pi^{3/2}\mu^3 W$ & $-t_3 \rho^\gamma$ \\
\hline
PNM & $(S)$ & $D^{(S;n)}$ & $E^{(S;n)}$ & $ R^{(S;n)}$ \\
\hline
 & (0) & $\pi^{3/2}\mu^3 (2W+B-2H-M)$ & $\pi^{3/2}\mu^3 (-W-2B+H+2M)$ & $\frac{1}{2}(\gamma +1)(\gamma+2) (1-x_3) t_3  \rho^\gamma$ \\
 & (1) &   $\pi^{3/2}\mu^3 (B-M)$   & $\pi^{3/2}\mu^3 (-W+H)$ & $ -(1-x_3) t_3 \rho^\gamma$ \\
 \hline
\end{tabular}
\end{adjustbox}
\end{center}
\caption{Contributions from the direct, exchange and density dependent terms (\ref{Appb:central}, \ref{Appb:density}) to the Gogny ph matrix elements.}
\label{gogny-central}
\end{table}

\begin{table}[h]
\begin{center}
\begin{tabular}{ccccc}
\hline
&&&& \\[-3mm]
SNM & $(I)$ & $X^{(I)}$  & $D_T^{(I)}$ & $E_T^{(I)}$  \\
\hline
& $(0)$ &  $- 3 W_0$ & $- 8 \pi (2 V_{T1}+V_{T2})$ & $8 \pi (V_{T1}+2V_{T2})$ \\
& $(1)$ &  $ - W_0$ & $-8 \pi V_{T2} $ & $8 \pi V_{T1}$  \\
\hline
PNM & & $X^{(n)}$  & $D_T^{(n)}$ & $E_T^{(n)}$  \\
\hline
& & $-2 W_0$ & $- 8 \pi (V_{T1}+V_{T2})$ &  $8 \pi (V_{T1}+V_{T2})$ \\
\hline
\end{tabular}
\end{center}
\caption{Contributions from the non-central terms (\ref{Appb:so}, \ref{Appb:ten}) to the Gogny ph matrix elements.}
\label{gogny-noncentral}
\end{table}

For completeness, we also give the ph interaction for the extra D2 finite-range density-dependent term
\bea
V_{DD,fr \, ph}^{(\alpha,\alpha')} &=& \rho^\gamma D_3^{(\alpha)}\int d\mathbf{r} e^{i\mathbf{q} \mathbf{r}}G(r)-\rho^\gamma E_3^{(\alpha)}\int d\mathbf{r} e^{i\mathbf{k}_{12} \mathbf{r}}G(r)\nonumber\\
&+&\gamma \rho^\gamma D_3^{(0,0)}\int d\mathbf{r} (1+e^{i\mathbf{q} \mathbf{r}})G(r) \nonumber \\
&-&\gamma \rho^\gamma D_3^{(0,0)}\int d\mathbf{r}\frac{3j_1(k_Fr)}{k_Fr}G(r) \frac{1}{2}\left(e^{-i\mathbf{k}_1\mathbf{r}}+e^{-i\mathbf{k}_2\mathbf{r}} \right) (1+e^{i\mathbf{q} \mathbf{r}})\nonumber\\
&+&\frac{1}{2}\gamma(\gamma-1) D_3^{(0,0)}\int d\mathbf{r} G(r)
-\frac{1}{2}\gamma(\gamma-1) E_3^{(0,0)}\int d\mathbf{r} G(r)\left( \frac{3j_1(k_Fr)}{k_Fr}\right)^2\;,
\eea
where we used $G(r)= \frac{e^{-{r^{2}}/{\mu_3^2}}}{\pi^{3/2}\mu_3^3}$ to simplify the notation and  $j_1$ is the standard spherical Bessel function. The coefficients $D_3^{(\alpha)}$, $E_3^{(\alpha)}$ are given by the same combinations of Tab.~\ref{gogny-central} for the central terms, but using the parameters of the density-dependent interaction.  

One can see in Eq.~(\ref{Appb:so}) that the spin-orbit interaction mixes both $S=0$ and 1 channels. The tensor interaction instead  acts on the $S=1$ channel, as shown in Eq.~(\ref{Appb:ten}), However, due the coupling induced by the spin-orbit interaction, it has also effects in the $S=0$ channels. Finally, notice also that the matrix elements of the ph interaction are diagonal in the isospin, which justifies the isospin convention used here to simultaneously deal with both SNM and PNM. These comments applies of course to any ph interaction with the general structure given in Eq.~(\ref{Vph-general}).

\subsection{\it Nakada interaction}

The various contributions to the Nakada interaction read
\bea
V_C(\mathbf{r}_1,\mathbf{r}_2) &=& \sum_{i=1,3} \left( t_i^{(SE)} P_{SE} + t_i^{(TE)} P_{TE} + t_i^{(SO)} P_{SO} + t_i^{(TO)} P_{TO}  \right) \frac{{\rm e}^{-\mu_i r_{12} }}{\mu_ir_{12}}  \label{m3y:interactionC} \;, \\
V_{SO}(\mathbf{r}_1,\mathbf{r}_2) &=& \sum_{i=1,2} \left(  t_i^{(LSE)} P_{TE} + t_i^{(LSO)} P_{TO} \right) \frac{{\rm e}^{-\mu_i r_{12}}}{\mu_i r_{12}} \left( {\bf r}_{12} \wedge {\bf p}_{12} \right) \cdot ({\bf s}_1+ {\bf s}_2) \label{m3y:interactionSO} \;, \\
V_T(\mathbf{r}_1,\mathbf{r}_2) &=& \sum_{i=1,2} \left(  t_i^{(TNE)} P_{TE} + t_i^{(TNO)} P_{TO} \right) \frac{{\rm e}^{-\mu_i r_{12}}}{\mu_i r_{12}} r_{12}^2 S_T({\hat{\bf r}}_{12}) 
\label{m3y:interactionT} \;,
\eea \label{m3y:interaction}
where the projectors on the different singlet-triplet spin and odd-even space pair states are given by
\bea
&& P_{SE} = \frac{1}{4} (1-P^{\sigma}) (1+P^{\tau}) \, , \, P_{TE} = \frac{1}{4} (1+P^{\sigma}) (1-P^{\tau}) \label{proj1} \;, \\
&& P_{SO} = \frac{1}{4} (1-P^{\sigma}) (1-P^{\tau})  \, , \,  P_{TO} = \frac{1}{4} (1+P^{\sigma}) (1+P^{\tau}) \label{proj2} \;.
\eea
$S_{T}(\mathbf{r}_{12})$ is the tensor operator defined in Eq.~(\ref{eq:tensor}) and ${\bf s}_i = {\vsigma}_i/2$. The Nakada interaction contains three different ranges in the central part, each of them corresponding to a specific meson mass (790, 490 and 140 MeV respectively). The third Yukawian, corresponding to the longest range, has been adjusted on matrix elements of the one pion-exchange potential (OPEP) and has been kept unchanged in all existing parametrizations. Also notice that the Nakada interaction contains finite-range form factors in all terms apart from the density-dependent one. In the first M3Y-P1 and M3Y-P2 parameterisations~\cite{nak03}, this term was identical to the standard Skyrme one, but later on it  was rewritten~\cite{nak10} as the sum of two zero-range terms, one for each even channel, as
\begin{eqnarray}\label{eq:dd:nak2}
V_{DD}(\mathbf{r}_1,\mathbf{r}_2) &=& \left[ t^{(SE)}_\rho P_{SE} \rho^{\gamma^{(SE)}}+ t^{(TE)}_\rho P_{TE} \rho^{\gamma^{(TE)}}\right] \delta({\bf r}_{12})\;. \label{m3y:interactionDD} 
\end{eqnarray}
Following discussion done in Ref.~\cite{nak08}, the presence of a density dependent term is important to obtain reasonable properties of infinite matter at saturation and a reasonable value for the effective mass~\cite{dav18}. The parameters in the triplet-even (TE) channels of the density dependent term have been adjusted to obtain a value of the nuclear incompressibility close to the accepted values~\cite{dut12}, the singlet-even (SE) channel gives little contribution to $K_0$ and it is then adjusted to reproduce pairing properties of PNM.

  Finally, we notice that in Ref.~\cite{nak15} a density dependent term is also added to enrich the spin-orbit one. We will not consider such a case here. Again, to simplify the notation, the sum over range indices, as well as the indices themselves, will be omitted.

\subsubsection{\it Mean field}\label{mf:nakada}

\begin{figure}[!h]
\begin{center}
\includegraphics[width=0.43\textwidth,angle=0]{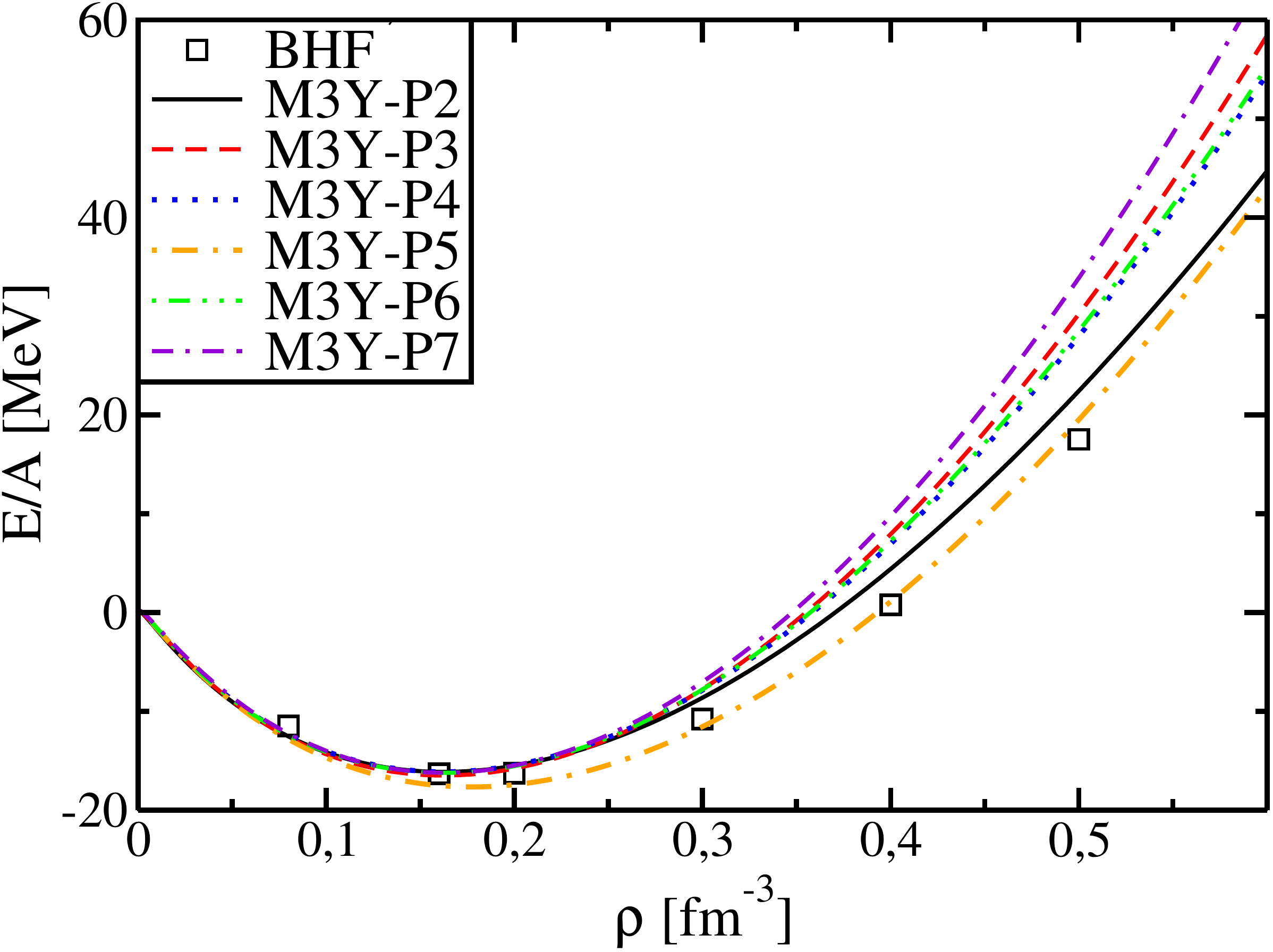}
\end{center}
\caption{EoS in SNM for the various Nakada interactions. The squares correspond to BHF calculations. Taken from Ref.~\cite{dav16b}.}
\label{fig-M3Y-EA}
\end{figure}

We start by briefly summarising infinite matter properties obtained using some selected Nakada parameterisations. As an example, the main SNM quantities obtained with them are given in Tab.~\ref{tab:M3Y-SNM}  and in Fig.~\ref{fig-M3Y-EA}, we plot the corresponding equations of state in SNM as well as the BHF results~\cite{bal97}. As for the Gogny case, the different parametrizations provide roughly the same features of SNM. This is due to a fine tuning of the bulk part (the only part that contributes to SNM) except for the long-range part which is kept unchanged.

\begin{table}[h]
\begin{center}
\begin{tabular}{c|ccccc}
\hline 
 & $m^*/m$ & $\rho_0$ [fm$^{-3}$]& $\left.E/A\right|_{\rho=\rho_0}$ [MeV] & $K_0$ [MeV]& $J$ [MeV]\\
\hline
M3Y-P2~\cite{nak03} &0.652&0.162& -16.14 & 220.4 & 30.61\\
M3Y-P3~\cite{nak08} &0.658&0.162& -16.51 & 245.8 & 29.75\\
M3Y-P4~\cite{nak08} &0.665&0.162& -16.13 & 235.3 & 28.71\\
M3Y-P5~\cite{nak08} &0.629&0.162& -16.12 & 235.6 & 29.59\\
M3Y-P6~\cite{nak13} &0.596&0.162& -16.24 & 239.7 & 32.14\\
M3Y-P7~\cite{nak13} &0.589&0.162& -16.22 & 254.7 & 31.74\\
\hline
\end{tabular}
\end{center}
\caption{SNM properties as predicted by selected Nakada interaction}
\label{tab:M3Y-SNM}
\end{table}

\begin{figure}[!h]
\begin{center}
\includegraphics[width=0.43\textwidth,angle=0]{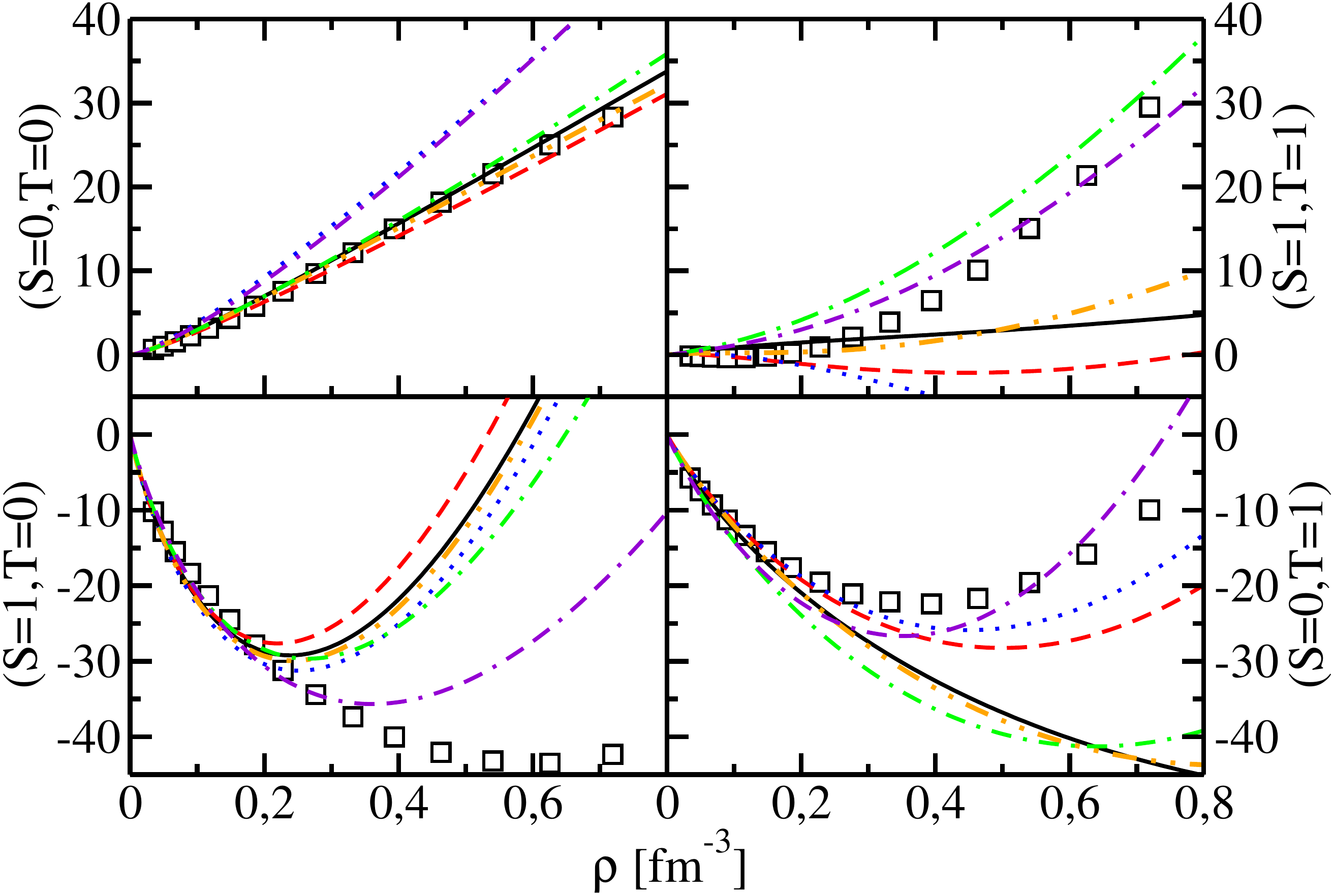}
\end{center}
\caption{Spin/isospin decomposition of the EoS for various Nakada interactions. Same symbols as in Fig.~\ref{fig-M3Y-EA}. Adapted from Ref.~\cite{dav16b}.}
\label{fig-M3Y-ST}
\end{figure}

The spin/isospin decomposition of the EoS is also displayed in Fig.~\ref{fig-M3Y-ST}  for the various Nakada interactions. As compared to Gogny results displayed in Fig.~\ref{fig1-Gogny-ST}, one can notice that on average the reproduction of the BHF results is more satisfactory. Most likely the presence of three ranges provides more flexibility to better reproduce the $(S,T)$ channels.

\begin{figure}[!h]
\begin{center}
\includegraphics[width=0.43\textwidth,angle=0]{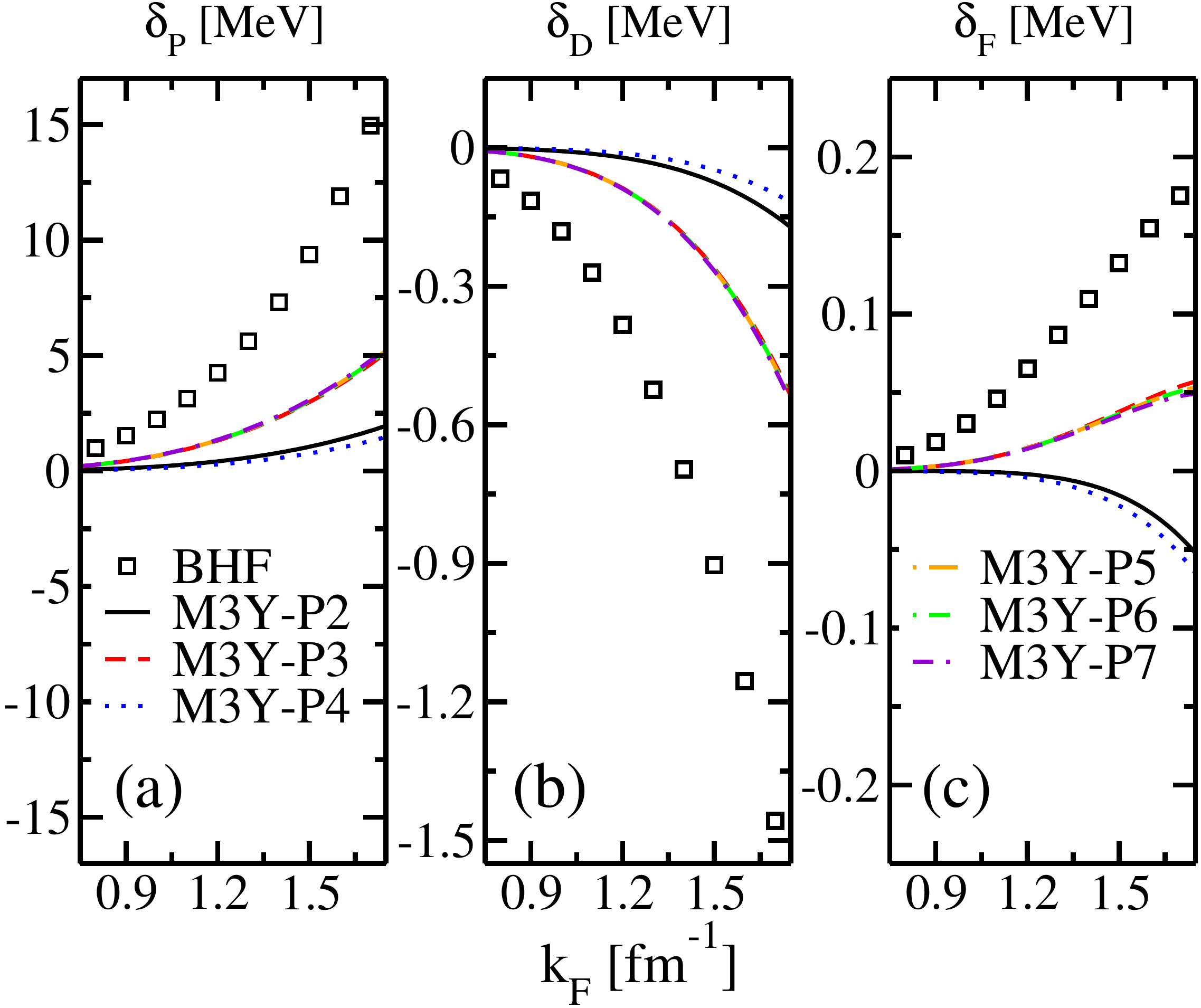}
\end{center}
\caption{Difference of partial waves for BHF results (dots) and Nakada interactions (lines). Taken from Ref.~\cite{dav16b}.}
\label{fig-M3Y-deltas}
\end{figure}

The differences $\delta_P,\delta_D$ and $\delta_F$, introduced in Eqs.~(\ref{partial:p}-\ref{partial:f}), are displayed in Fig.~\ref{fig-M3Y-deltas}. Contrary to the Gogny case, the spin-orbit and the tensor both contribute to the $P, D$ and $F$ waves since the spin-orbit has now a finite-range. We first notice that the tensor contribution of the Nakada interaction is weaker than a realistic tensor, but has the correct sign, contrary to the Gogny case. The tensor terms of M3Y-P3, P5, P6 and P7 are identical, while they have been switched off in M3Y-P4 and have a considerable reduced strength in M3Y-P2. Indeed, we can see on Fig.~\ref{fig-M3Y-deltas} that M3Y-P3, P5, P6 and P7 give the same results and that M3Y-P2 and M3Y-P4 roughly coincide (and correspond to the contribution of the spin-orbit alone).

\subsubsection{\it ph interaction}

The ph interaction as defined in Eq.~(\ref{ME-Vph}) read 
\bea
V_{C \, ph}^{(\alpha,\alpha')} &=& \delta_{\alpha,\alpha'} \left\{ D^{(\alpha)} F_C(q) +E^{(\alpha)} F_C({\bf k}_{12}) 
\right\} \;, \label{central-Nakada} \\
V_{DD \, ph}^{(\alpha,\alpha')} &=& \delta_{\alpha,\alpha'} R^{(\alpha)}  \;,
\label{density-Nakada} \\
V_{SO \, ph}^{(\alpha,\alpha')} & = & q \delta_{I,I'}\delta_{Q,Q'}\left\{ D_{SO}^{(I)} F_{SO}(q) 
+  E_{SO}^{(I)} F_{SO}(k_{12})  \right\} \nonumber \\
& & \hskip 2 true cm 
\left\{ M'(k_{12})^1_{M'}\delta_{S',1}\delta_{S,0}+M(k_{12})^1_{-M} \delta_{S,1}\delta_{S',0}  \right\} \;, \label{so-Nakada}\\
V_{T ph}^{(\alpha,\alpha')} & = & \delta_{S,1}\delta_{S',1}\delta_{Q,Q'} \left\{  D_T^{(I)} F_{T}(q) q^2 \left[3\delta_{M,0}\delta_{M',0} -\delta_{M,M'}\right] \right. \nnn
& & \left. 
+ E_T^{(I)}  F_T(k_{12}) \left[3(-1)^{M}(k_{12})^{1}_{-M}(k_{12})^{1}_{M'} -k_{12}^2\delta_{M,M'}\right]  \right\}
\;. \label{ten-Nakada}
\end{eqnarray}

The notation is similar to that given for Gogny interaction, except that the Nakada spin-orbit term is finite-range and consequently, there are now distinct direct and exchange contributions. 
The coefficients $D^{(\alpha)}$, $E^{(\alpha)}$, and $R^{(\alpha)}$ are combinations of the central parameters, and are given in Tab.~\ref{nakada-central} for both SNM and PNM. 
Notice that in Tab.~\ref{nakada-central}, we provide the expressions of the ph interaction for the density dependent term given in Eq.~(\ref{eq:dd:nak2}). If the Gogny-like density dependent is used, the expressions can be taken from Tab.~\ref{gogny-central}.  The coefficients $D_{SO}^{(I)}$,  $E_{SO}^{(I)}$, $D_T^{(I)}$ and $E_T^{(I)}$ are combinations of the non-central parameters and are given in Tab.~\ref{nakada-noncentral}. As for Gogny, a sum over  range indices is to be understood.

\begin{table}[h]
\begin{center}
\begin{adjustbox}{max width=\textwidth}
\begin{tabular}{ccccc}
\hline
& &&& \\[-3mm]
SNM & $(S,I)$ & $D^{(S,I)}$ & $E^{(S,I)}$ & $R^{(S,I)}$ \\
\hline
& (0,0) & $\frac{\pi}{\mu} (3 t^{(SE)}+t^{(SO)}+3t^{(TE)}+9t^{(TO)})$ & $\frac{\pi}{\mu} (3 t^{(SE)}-t^{(SO)}+3t^{(TE)}-9t^{(TO)})$ & $+\frac{3}{4}(\gamma^{(SE)}+1)(\gamma^{(SE)}+2)t_\rho^{(SE)}\rho^{\gamma^{(SE)}}$ \\
         &                                                              &                                    &                                      & $+\frac{3}{4}(\gamma^{(TE)}+1)(\gamma^{(TE)}+2)t_\rho^{(TE)}\rho^{\gamma^{(TE)}}$  \\
& (0,1) & $\frac{\pi}{\mu} (t^{(SE)}-t^{(SO)}-3t^{(TE)}+3t^{(TO)})$ &  $\frac{\pi}{\mu} (t^{(SE)}+t^{(SO)}-3t^{(TE)}-3t^{(TO)})$ & $\frac{1}{2}t_\rho^{(SE)}\rho^{\gamma^{(SE)}}-\frac{3}{2}t_\rho^{(TE)}\rho^{\gamma^{(TE)}}$\\
& (1,0) & $\frac{\pi}{\mu} (-3 t^{(SE)}-t^{(SO)}+t^{(TE)}+3t^{(TO)})$ & $\frac{\pi}{\mu} (-3 t^{(SE)}+t^{(SO)}+t^{(TE)}-3t^{(TO)})$ & $-\frac{3}{2}t_\rho^{(SE)}\rho^{\gamma^{(SE)}}+\frac{1}{2}t_\rho^{(TE)}\rho^{\gamma^{(TE)}}$ \\  
& (1,1) &   $\frac{\pi}{\mu} (-t^{(SE)}+t^{(SO)}-t^{(TE)}+t^{(TO)})$  &   $\frac{\pi}{\mu} (-t^{(SE)}-t^{(SO)}-t^{(TE)}-t^{(TO)})$  &  $-\frac{1}{2}t_\rho^{(SE)}\rho^{\gamma^{(SE)}}-\frac{1}{2}t_\rho^{(TE)}\rho^{\gamma^{(TE)}}$\\
\hline
PNM & $(S)$ & $D^{(S;n)}$ & $E^{(S;n)}$ & $R^{(S;n)}$ \\
\hline
& $(0)$ & $\frac{2 \pi}{\mu} (t^{(SE)}+3t^{(TO)})$ & $\frac{2\pi}{\mu} ( t^{(SE)}-3t^{(TO)})$ & $\frac{1}{2}(\gamma^{(SE)}+1)(\gamma^{(SE)}+2)t_\rho^{(SE)}\rho^{\gamma^{(SE)}}$\\
& $(1)$ & $\frac{2\pi}{\mu} (-  t^{(SE)}+t^{(TO)})$ & $-\frac{2\pi}{\mu} ( t^{(SE)}+t^{(TO)})$ & 
$- t_\rho^{(SE)}\rho^{\gamma^{(SE)}}$ \\
\hline
\end{tabular}
\end{adjustbox}
\end{center}
\caption{Contributions from the direct, exchange and density dependent terms to the ph matrix elements for the central terms of the Nakada interaction.}
\label{nakada-central}
\end{table}

\begin{table}[h]
\begin{center}
\begin{tabular}{cccccc}
\hline
&&&&& \\[-3mm]
SNM & $(I)$ & $D_{SO}^{(I)}$ & $E_{SO}^{(I)}$ & $D_T^{(I)}$ & $E_T^{(I)}$  \\
\hline
& $(0)$ & $\frac{\pi}{2} (t^{(LSE)}+3t^{(LSO)})$ & $ \frac{\pi}{2} (- t^{(LSE)}+3t^{(LSO)})$ & $-4 \pi (t^{(TNE)}+3 t^{(TNO)})$ & $4 \pi (- t^{(TNE)}+3t^{(TNO)})$ \\
& $(1)$ & $ \frac{\pi}{2} (- t^{(LSE)}+t^{(LSO)})$ & $\frac{\pi}{2} (t^{(LSE)}+t^{(LSO)})$ & $-4 \pi ( - t^{(TNE)}+t^{(TNO)})$ & $ 4 \pi (t^{(TNE)}+t^{(TNO)})$ \\
\hline
PNM & & $D_{SO}^{(n)}$ & $E_{SO}^{(n)}$ & $D_T^{(n)}$ & $E_T^{(n)}$ \\
\hline
          & & $ \pi t^{(LSO)}$ & $ \pi t^{(LSO)}$ & $- 8 \pi t^{(TNO)}$ & $8 \pi t^{(TNO)}$ \\
\hline
\end{tabular}
\end{center}
\caption{Coefficients entering the Nakada non-central parts of the ph interaction}
\label{nakada-noncentral}
\end{table}

The functions $F_C, F_{SO}, F_{TN}$ are obtained via a Fourier transform and they read
\bea
F_C(q) &=& \frac{1}{q^2+\mu^2} \label{FC:nakada} \;,\\
F_{SO}(q) &=& \frac{2}{\mu ( q^2+\mu^2)^2}\;,\\
F_{T}(q) &=& \frac{8}{\mu (q^2+\mu^2)^3}\;.
\eea

\subsection{\it Comparing the tensor interactions}\label{sec:comp:tens}
\begin{figure}[!h]
\begin{center}
\includegraphics[width=0.4\textwidth,angle=0]{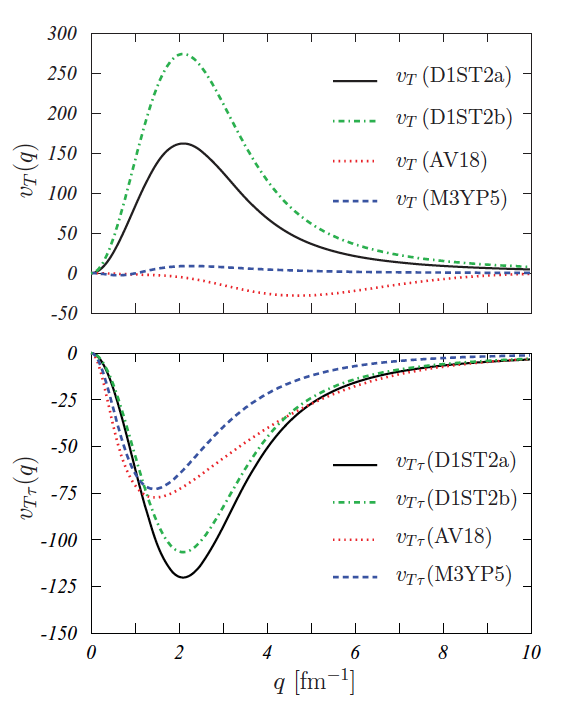}
\end{center}
\caption{Tensor isoscalar (upper panel) and isovector (lower panel) of Gogny interactions D1ST2a and D1ST2b, and Nakada M3Y-P5, compared with the analogous terms of the realistic AV18 interaction. Taken from Ref.~\cite{ang12}.}
\label{anguiano-fig3}
\end{figure}

The difference between the Gogny and Nakada tensor has previously been illustrated in Figs. \ref{fig-gogny-deltas}-\ref{fig-M3Y-deltas}, by performing a partial wave decomposition of the EoS. It is now interesting to compare the radial parts of the tensor interactions. In Fig.~\ref{anguiano-fig3} the Fourier transform of the radial form factors are plotted for three typical effective interactions, namely  Gogny D1ST2a and D1ST2b~\cite{ang12} and Nakada M3Y-P5~\cite{nak08}, and compared with the realistic AV18 interaction~\cite{wir95}. One can immediately see that both isoscalar and isovector  AV18 tensor terms are attractive. Instead, the effective interactions are repulsive in the isoscalar tensor channel, except for a small attraction presented by M3Y-P5 for $q< 1$~fm$^{-1}$. 

The intensity of the radial part is also very different for the different interactions, in particular we observe that the Nakada interaction is reasonable close to the AV18 result in the isovector channel, while the two Gogny interactions lead to $\approx50$\% stronger intensity.

In view of the results presented in the following, it is very instructive to analyse in more detail the isoscalar channel: we observe that the Nakada tensor is very weak in this channel, even weaker than the one given by the AV18 interaction. On the contrary the Gogny tensor leads to a very strong intensity, even larger (in absolute scale) than the one observed in the isovector channel. The apparent discrepancy between the AV18 tensor and the Gogny one was also discussed in Ref.~\cite{dav16c} by comparing the partial wave decomposition of the EoS with the one obtained using {\it ab-initio} methods.
For a more detail comparison between the two types of tensor, one can also inspect the corresponding Landau parameters as illustrated in Tab.~\ref{land-param-SNM}.

The Gogny-like tensor contains two parameters, which have been fitted  to the energy difference between the $1f_{5/2}$ and $1f_{7/2}$ single particle neutron states in $^{48}$Ca, and to the energy of the first $0^-$ state in the $^{16}$O nucleus. The parameters have been adjusted together with the spin-orbit term~\cite{gra13}, but leaving unchanged the parameters of the central term. Nakada tensor parameters are first fitted to the microscopic interaction, and then multiplied by a reduction factor fixed so as to reproduce the single-particle level ordering for $^{208}$Pb. 

 At present a systematic analysis of the impact of finite-range tensor on nuclear observables as done for the case of  Skyrme interactions~\cite{les07,bai09a,bai10,cao10,min13} is still missing. It is thus not possible to conclude which is the most adapted form of tensor to be used in this case of finite-range interactions and more importantly its strength.

\subsection{\it Connection with zero-range interactions}

\begin{figure}[!h]
\begin{center}
\includegraphics[width=0.4\textwidth,angle=0]{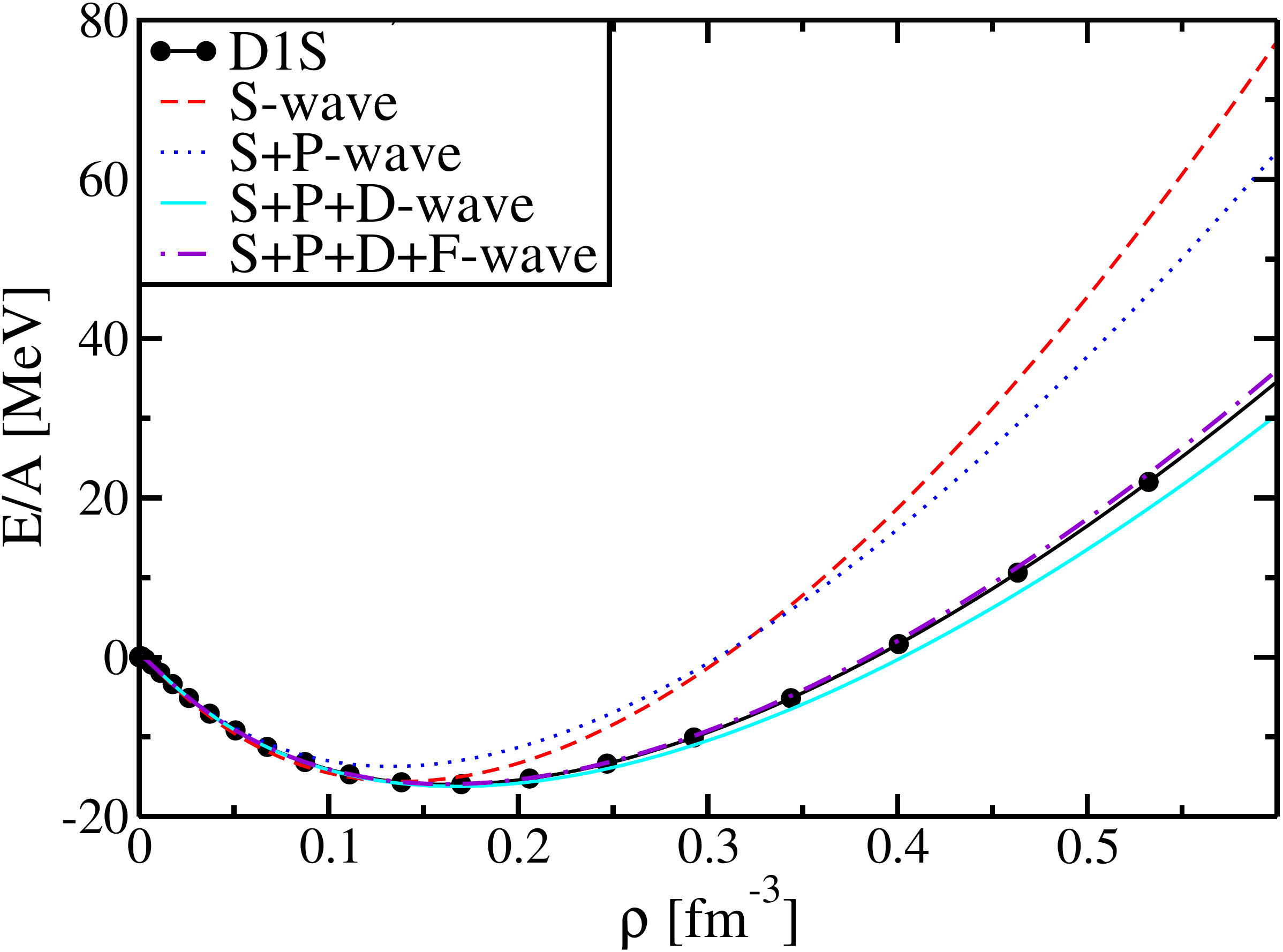}
\includegraphics[width=0.4\textwidth,angle=0]{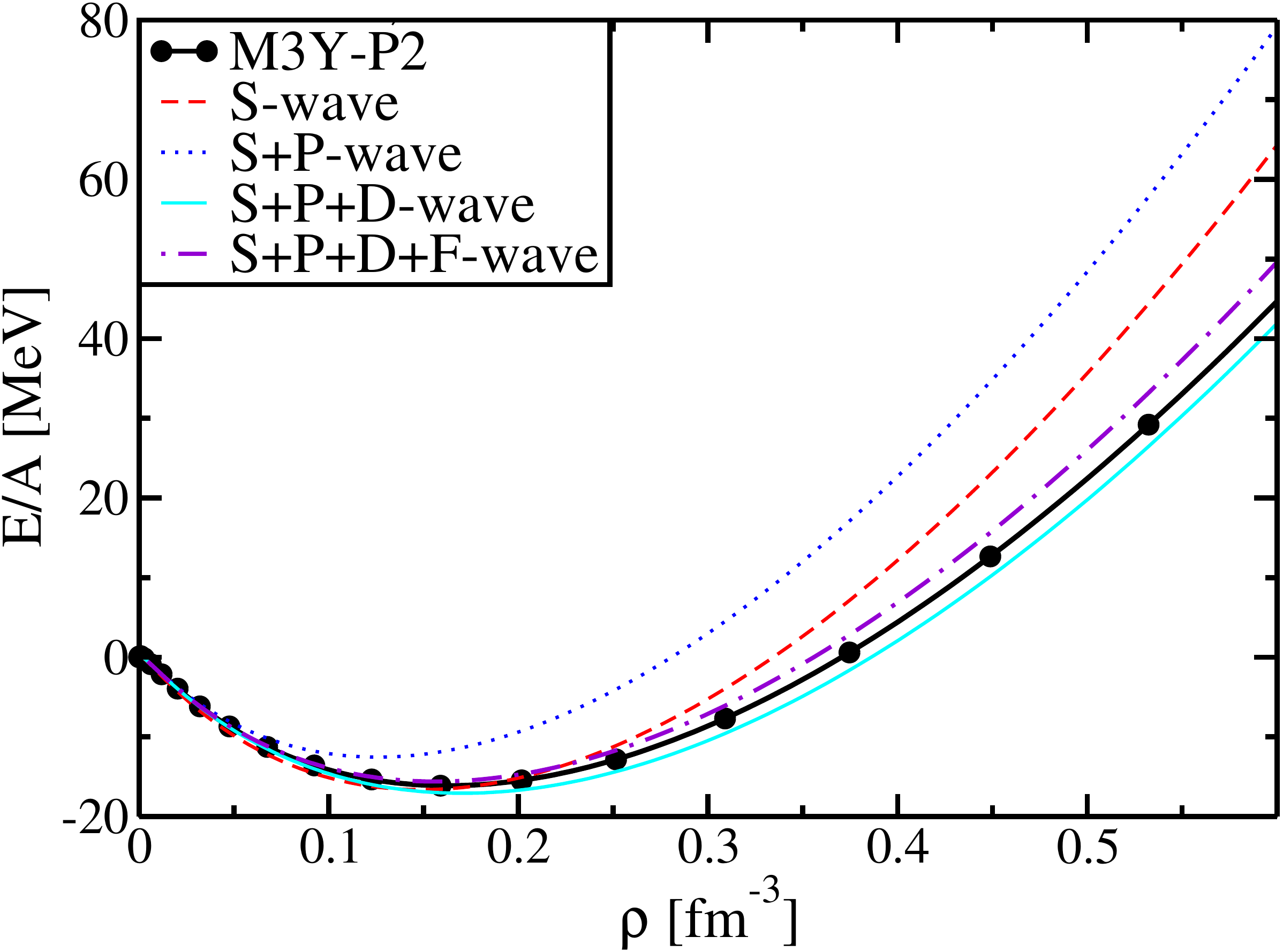}
\end{center}
\caption{EoS in SNM for the complete interaction (solid line with dots) and sum of different partial waves (lines). Taken from Ref.~\cite{dav16b}.}
\label{fig1-D1S_JLS}
\end{figure}

It is interesting to analyse the zero-range limit of the finite-range interactions considered in this review. Following the original Skyrme idea~\cite{sky59}, it has been shown in Ref.~\cite{dav16b} how it is possible to expand in momentum space {\it any} finite-range interaction and truncate such an expansion at a given order. By considering only second order, one recovers the form of the  standard Skyrme interaction~\cite{per04} and by considering the higher order momenta one obtains new zero-range interactions named N$\ell$LO, with $\ell=2,3,\dots$ ~\cite{rai11}. Furthermore, by means of a partial-wave decomposition of the EoS of SNM~\cite{bec15} of a finite-range interaction, one observes that the main contributions to the ground state arise from S- and P- waves, thus proving that the Skyrme interaction is able with its simplicity to grasp the main physical features of a more complex finite-range NN interaction. To get an insight on the convergence properties of the partial-wave decomposition in Eq.~(\ref{partial}), in Fig.~\ref{fig1-D1S_JLS} are compared the exact EoS of SNM and the one obtained by truncating it at various partial waves for interactions D1S and M3Y-P2. 

It can be seen that up to densities $\simeq 2 \rho_0$, the S, P and D partial waves give a satisfactory convergence. When the F wave is included, two different scenarios emerge. In the case of Gogny interaction, a full convergence is reached even at higher densities, but the convergence is not yet totally reached in the Nakada case. This result is related to the relatively short range character of Gogny ranges as compared to the Nakada ones. However, we can safely claim that the contributions up to $D$ wave provide a very good approximation up to 2 times the saturation density. Following this idea, D waves have been added to the original Skyrme interaction~\cite{bec15,car08,dav13,rai14,dav14,dav15a,dav16} and a parametrisation at order N2LO, suitable for calculations in finite nuclei~\cite{bec17}, has been obtained. Generally speaking, this D wave (and even F wave) terms can be generated by a momentum expansion of  Gogny, M3Y or {\it any} finite-range interactions up to the sixth order. Such an expansion definitely shows that the resulting series matches exactly with the extended N3LO Skyrme interaction derived in Refs.~\cite{rai11,kor13}. The {\it{standard}} Skyrme interaction reads
\begin{eqnarray}
V_{C}(\mathbf{r}_1,\mathbf{r}_2)&=&t_0(1+x_0P_{\sigma})\delta(\mathbf{r})+\frac{1}{2}t_1(1+x_1P_{\sigma})\left[\mathbf{k}^{'2}\delta(\mathbf{r})+)\delta(\mathbf{r})\mathbf{k}^2 \right]+t_2(1+x_2P_{\sigma})\left[\mathbf{k}'\cdot \delta(\mathbf{r})\mathbf{k} \right] \;, \label{Skyrme:interactionC} \\
V_{DD}(\mathbf{r_1},\mathbf{r}_2) &=&  \frac{1}{6}t_3 \left( 1 + x_3 P_{\sigma} \right) \rho^{\gamma} \delta(\mathbf{r_1}-\mathbf{r}_2)\,,\\
V_{SO}(\mathbf{r}_1,\mathbf{r}_2)&=&iW_0({\vsigma}_1+{\vsigma}_2)\cdot\left[ \mathbf{k}' \wedge \delta(\mathbf{r})\mathbf{k}\right] \;. \label{Skyrme:interactionSO} 
\end{eqnarray}
By investigating the various terms, we  clearly recognise the original S and P wave contributions of Ref.~\cite{sky59}. Some parameterisations include also a zero-range tensor term of the form
\begin{eqnarray}
V_T(\mathbf{r}_1,\mathbf{r}_2) &=& \frac{1}{2}t_eT_e(\mathbf{k}',\mathbf{k})+\frac{1}{2}t_oT_o(\mathbf{k}',\mathbf{k}) \;. \label{Skyrme:interactionT}
\end{eqnarray}
The tensor operators  $T_e$ and $T_o$, are respectively even and odd under parity transformations and are defined in the appendix~\ref{app:tens}.
In Tab.~\ref{skyrme-from-finite}, we give as an illustration the Skyrme parameters in terms of Gogny and Nakada ones when these finite-range interactions are expanded up to second order in momenta.  No density-dependent term appear at this stage as it is originated from a three-body interaction term~\cite{vau72}. In practice it is added by hand phenomenologically. 
It has been shown in Ref.~\cite{dav13} that the terms of finite-range spin-orbit expansion are not gauge-invariant, apart from the first one, which represents the standard Skyrme spin-orbit term. Therefore, the finite-range spin-orbit interaction seems to be in conflict with the continuity equation. 

\begin{table}[h]
\begin{center}

\begin{tabular}{ccc}
\hline
Skyrme & Gogny & Nakada \\
\hline
$t_0$       & $\pi^{3/2} \mu^3 (W+M)$                      &  $\frac{2 \pi}{\mu^3} (t^{(SE)}+t^{(TE)})$   \\
$t_0 x_0$ & $\pi^{3/2} \mu^3 (B+H)$                      &  $\frac{2 \pi}{\mu^3} (- t^{(SE)}+t^{(TE)})$ \\
$t_1$        & $- \frac{1}{2} \pi^{3/2} \mu^5 (W+M)$  & $-  \frac{4 \pi}{\mu^5} (t^{(SE)}+t^{(TE)})$ \\
$t_1 x_1$ & $- \frac{1}{2} \pi^{3/2} \mu^5 (B+H)$    & $\frac{4 \pi}{\mu^5} (t^{(SE)}-t^{(TE)})$ \\
$t_2$        &  $\frac{1}{2} \pi^{3/2} \mu^5 (W-M)$    & $\frac{4 \pi}{\mu^5} (t^{(SO)}+t^{(TO)})$ \\
$t_2 x_2$ & $ \frac{1}{2} \pi^{3/2} \mu^5 (B-H)$      & $\frac{4 \pi}{\mu^5} (- t^{(SO)}+t^{(TO)})$ \\
$W_o$     & $W_o$                                                  & $- \frac{\pi}{\mu^5} t^{(LSO)}$ \\
$t_e$        & $\frac{\pi \sqrt{\pi}}{5} \mu^5 (V_1+V_2)$   & $- \frac{64 \pi}{\mu^7} t^{(TNE)}$ \\
$t_o$         & $ \frac{\pi \sqrt{\pi}}{5} \mu^5 (- V_1+V_2)$ & $ \frac{64 \pi}{\mu^7} t^{(TNO)}$ \\
\hline
\end{tabular}
\end{center}
\caption{Skyrme parameters deduced from the momentum expansion of Gogny and Nakada interactions. A sum over the ranges is to be understood.}
\label{skyrme-from-finite}
\end{table}
It should be stressed that the Skyrme parameters deduced in this way from a finite-range interaction only contain a part of these interactions. As a consequence, the resulting Skyrme interaction cannot produce reliable results. Only for the $t_0$ parameter one obtains values similar to those currently found in a genuine Skyrme interaction, whose parameters are directly adjusted to experimental data.

Since the focus of this review is on finite-range interactions, we refer the reader to Ref~\cite{dut12} for a detailed discussion on SNM properties of zero-range (Skyrme) interactions, and to Ref.~\cite{pas15} for the response functions in infinite nuclear matter. We provide here only the expression of the matrix elements of the Skyrme ph interaction since, as stated above, it can be formally seen as a special limit case. It will be used explicitly to test the methodology developed in the next sections to determine finite-range response functions. One obtains
\begin{eqnarray}
V_{C;DD \, ph}^{(\alpha,\alpha')} &=& \delta_{\alpha,\alpha'}\left( W_1^{(\alpha)}(q)+W_2^{(\alpha)}(\mathbf{k}_1-\mathbf{k}_2)^2\right)\;,\\
V_{SO \, ph}^{(\alpha,\alpha')} &=& q X^{(I)} \left\{ M'(k_{12})^1_{M'}\delta_{S',1}\delta_{S,0}+M(k_{12})^1_{-M} \delta_{S,1}\delta_{S',0}  \right\} \delta_{I,I'}\delta_{Q,Q'} \;,\\
 V_{T \, ph}^{(\alpha,\alpha')} &=& \delta_{S,1}\delta_{S',1}\delta_{I,I'}\delta_{Q,Q'}
 \left\{ Z_1^{(I)} q^2 \left( 3 \delta_{M,0} \delta_{M',0} - \delta_{M,M'} \right) \right. \nonumber \\
 && \left. \hspace{3cm} + Z_2^{(I)} 
\left[ 3(-)^M (\mathbf{k}_{12})_{-M}^1 (\mathbf{k}_{12})_{M'}^1- \mathbf{k}_{12}^2\delta_{M,M'}\right] \right\}\;.
\end{eqnarray}
The various coefficients entering the above equations are given in Tabs.~\ref{skyrme-central} and \ref{skyrme-noncentral}, for both SNM and PNM.

\begin{table}[h]
\begin{center}
\begin{adjustbox}{max width=\textwidth}
\begin{tabular}{cccc}
\hline
&&& \\[-3mm]
SNM & $(S,I)$ & $W_1^{(S,I)}$ & $W_2^{(S,I)}$ \\
\hline
&(0,0) & $3t_0+\frac{1}{4}(\gamma+1)(\gamma+2)t_3\rho^\gamma+\frac{3}{4}t_1q^2 -\frac{1}{4} (5+4x_2)t_2q^2$ & $\frac{3}{4}t_1+\frac{1}{4}(5+4x_2)t_2$\\
&(0,1) & $-(1+2x_0)t_0-\frac{1}{6}(1+2x_3)t_3\rho^\gamma -\frac{1}{4}(1+2x_1)t_1q^2-\frac{1}{4}(1+2x_2)t_2q^2$ & $-\frac{1}{4}(1+2x_1)t_1+\frac{1}{4}(1+2x_2)t_2$\\
&(1,0) & $-(1-2x_0)t_0-\frac{1}{6}(1-2x_3)t_3\rho^\gamma -\frac{1}{4}(1-2x_1)t_1q^2-\frac{1}{4}(1+2x_2)t_2q^2$& $-\frac{1}{4}(1-2x_1)t_1+\frac{1}{4}(1+2x_2)t_2$\\
&(1,1) & $-t_0-\frac{1}{6}t_3\rho^\gamma-\frac{1}{4}t_1q^2-\frac{1}{4}t_2q^2$ & $-\frac{1}{4}t_1+\frac{1}{4}t_2$\\
\hline 
PNM & $(S)$ & $W_1^{(S;n)}$ & $W_2^{(S;n)}$ \\
\hline
 & (0) & $(1-x_0) t_0+\frac{1}{12}(\gamma+1)(\gamma+2)(1-x_3) t_3\rho^\gamma+\frac{1}{4}(1-x_1) t_1q^2 -\frac{1}{4} (1+x_2)t_2q^2$ & $\frac{1}{4}(1-x_1)t_1+\frac{3}{4}(1+x_2)t_2$ \\
 & (1) & $-(1-x_0)t_0-\frac{1}{6}(1-x_3)t_3\rho^\gamma -\frac{1}{4}(1-x_1)t_1q^2-\frac{1}{4}(1+x_2)t_2q^2$ & $-\frac{1}{4}(1-x_1)t_1+\frac{1}{4}(1+x_2)t_2$\\
 \hline
\end{tabular}
\end{adjustbox}
\end{center}
\caption{Coefficients of the Skyrme ph interaction in the various spin/isospin channels}
\label{skyrme-central}
\end{table}

\begin{table}[h]
\begin{center}
\begin{tabular}{ccccc}
\hline
&&&& \\[-3mm]
SNM & $(I)$ & $X^{(I)}$  & $Z_1^{(I)}$ & $Z_2^{(I)}$  \\
\hline
& $(0)$ &  $- 3 W_0$ & $\frac{1}{2} (t_e-3t_o)$ & $\frac{1}{2} ( t_e+3t_o)$ \\
& $(1)$ &  $ - W_0$ & $-\frac{1}{2} (t_e+t_o)$ & $-\frac{1}{2} ( t_e-t_o)$  \\
\hline
PNM & & $X^{(n)}$  & $Z_1^{(n)}$ & $Z_2^{(n)}$  \\
\hline
&& $-2 W_0$ & $-t_o$ & $t_o$ \\
\hline
\end{tabular}
\end{center}
\caption{Coefficients entering the non-central parts of the Skyrme ph interaction}
\label{skyrme-noncentral}
\end{table}

\subsection{\it Approximations to the effective mass and the response function} \label{sect:effectivemass}

\begin{figure}[!h]
\begin{center}
\includegraphics[width=0.4\textwidth,angle=0]{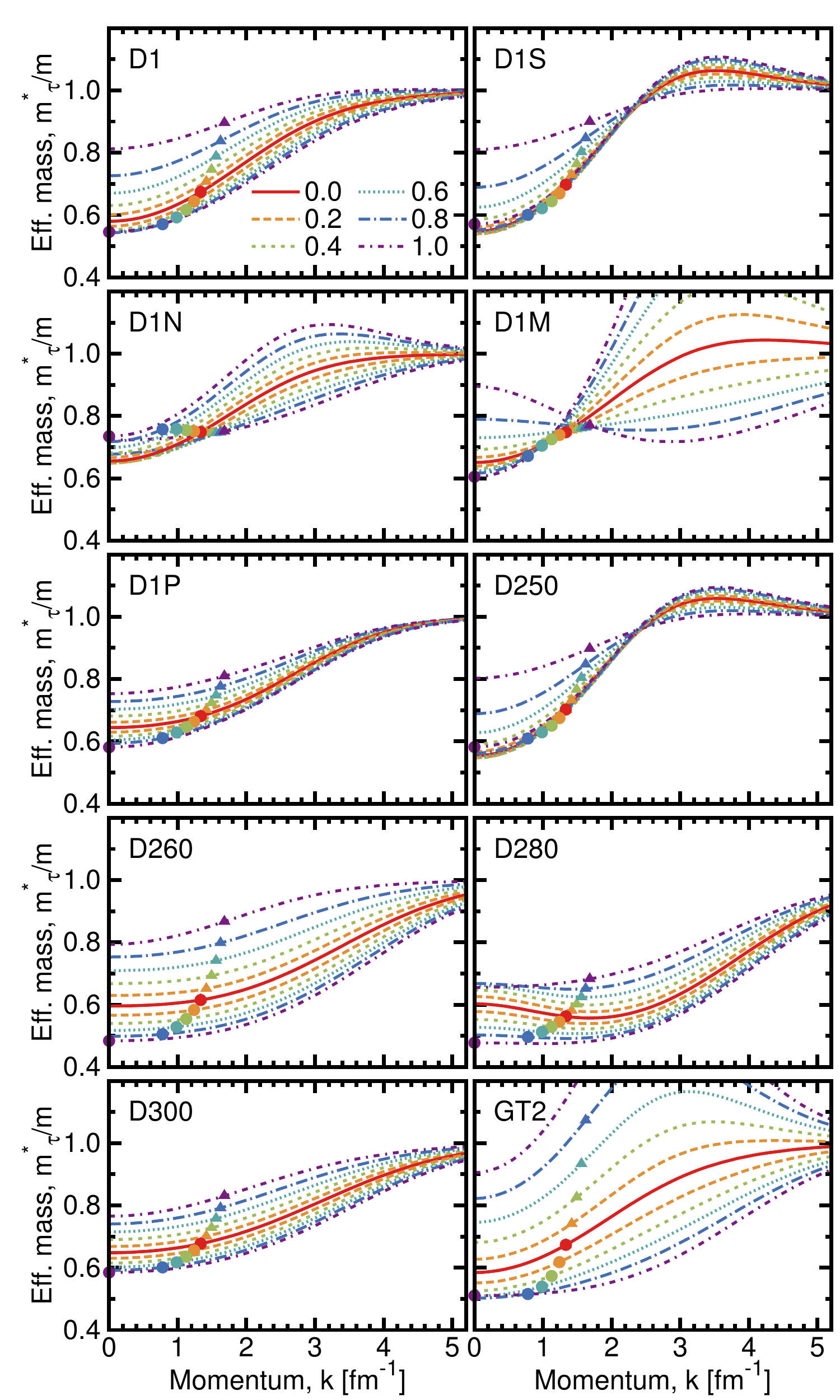}
\end{center}
\vspace{-.15in}
\caption{Effective masses of neutrons (triangles) and protons (cercles) as a function of momentum for
six different isospin asymmetries, calculated for several Gogny-like interactions. Symmetric nuclear matter corresponds to the solid lines. The symbols denote the values of the effective mass at the respective Fermi momentum, assuming a density value $\rho=0.16$\,fm$^{-3}$. Taken from Ref.~\cite{sel14}.}
\label{arnau-fig4}
\end{figure}

We now turn to the approximation to the mean field mentioned in Sect.~\ref{ph-propagators}. The denominator of the HF propagator (\ref{GHF}) contains the difference $\varepsilon({\bf k})-\varepsilon({\bf q}+{\bf k})$,
where $\varepsilon({\bf k})= \frac{\hbar^2}{2m} k^2 + U({\bf k})$ is the single-particle energy, $U$ being the mean-field. The effective mass is defined as
\be
\frac{m}{m^*} = 1 + \frac{m}{\hbar^2} \frac{1}{k} \frac{\partial U(k)}{\partial k} \,.
\label{eff-mass}
\ee
As the momentum dependence of the Skyrme mean field is quadratic, it can be absorbed into the kinetic energy through a constant effective mass. However, in the case of a finite-range interaction one has to deal with an effective mass $m^*(k)$ depending on the momentum $k$. 
Such a momentum dependence has been analysed in Ref.~\cite{sel14} for a set of Gogny-like interactions and several isospin asymmetries. The results are shown in Fig.~\ref{arnau-fig4}, and include interactions D1~\cite{dec80} and D250, D260, D280, and D300~\cite{bla95}, besides those previously encountered. 

One can see that the behaviour of $m^*(k)$ as a function of $k$ is rather disparate. In some cases, the effective mass steadily increases, saturating at $m^* \simeq m$ at large momenta. In other cases, $m^*(k)$ has a maximum (a minimum in one case), and then goes to the limit value of the bare mass. In all cases, defining an approximate constant effective mass is restricted to small values of $k$ only. 

\begin{figure}[!h]
\begin{center}
\includegraphics[width=0.4\textwidth,angle=0]{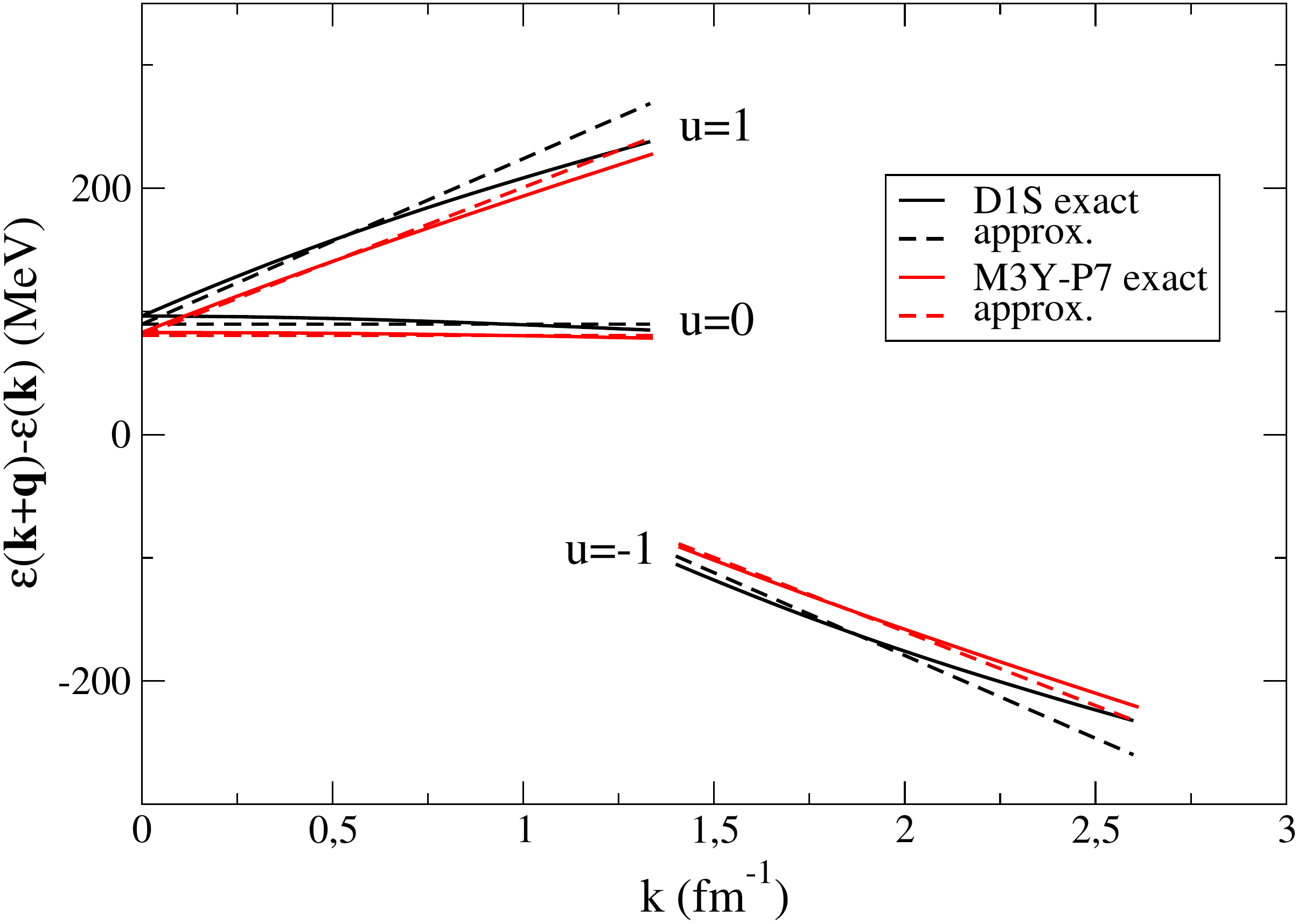}
\end{center}
\vspace{-.15in}
\caption{$\varepsilon({\bf q}+{\bf k})- \varepsilon({\bf k})$ calculated exactly (solid lines) and with the parabolic approximation (dashed lines) for interactions D1S and M3Y-P7 for three values of $u={\hat{\bf k}} \cdot {\hat {\bf q}}$ at the transferred momentum $q=k_F$. }
\label{fig-Diff-energies}
\end{figure}

The momentum dependence of the effective mass implies some numerical complications in calculating the HF response function, in particular in the determination of the poles. This is the reason why a parabolic approximation of the mean field is largely used. However, instead of using the simple Taylor expansion 
$U(k) \simeq U(0) + \left( \partial U(k)/\partial k^2 \right)_{k=0} \, k^2$, which is only valid for small values of $k$, it was shown in Ref.~\cite{jerome} that the approximation 
\be
U(k) \simeq U(0) + \left( \frac{\partial U(k)}{\partial k^2} \right)_{k=k_F}  {\hskip -7 mm} k^2 \quad \;,
\label{Uk-parabole}
\ee
or using the value of the effective mass at the Fermi surface, reasonably reproduces the mean field for values of $k$ up to $\simeq 2 k_F$. The analysis of Ref.~\cite{jerome} was based on the specific D1 interaction, but the conclusion seems to hold in general. Since the kinetic energy dominates at high values of the momenta, we must consider directly the single-particle energy difference $\varepsilon({\bf k})-\varepsilon({\bf q}+{\bf k})$. It is plotted in Fig.~\ref{fig-Diff-energies} for three values of $u={\hat{\bf k}} \cdot {\hat {\bf q}}$ at the transferred momentum $q=k_F$. The plot is limited to the integration domain $k$ fixed by the numerator of Eq.~(\ref{GHF}). One can therefore see that this approximation works also well for D1S interaction and even better for M3Y-P7. 

\begin{figure}[!h]
\begin{center}
\includegraphics[width=0.5\textwidth,angle=0]{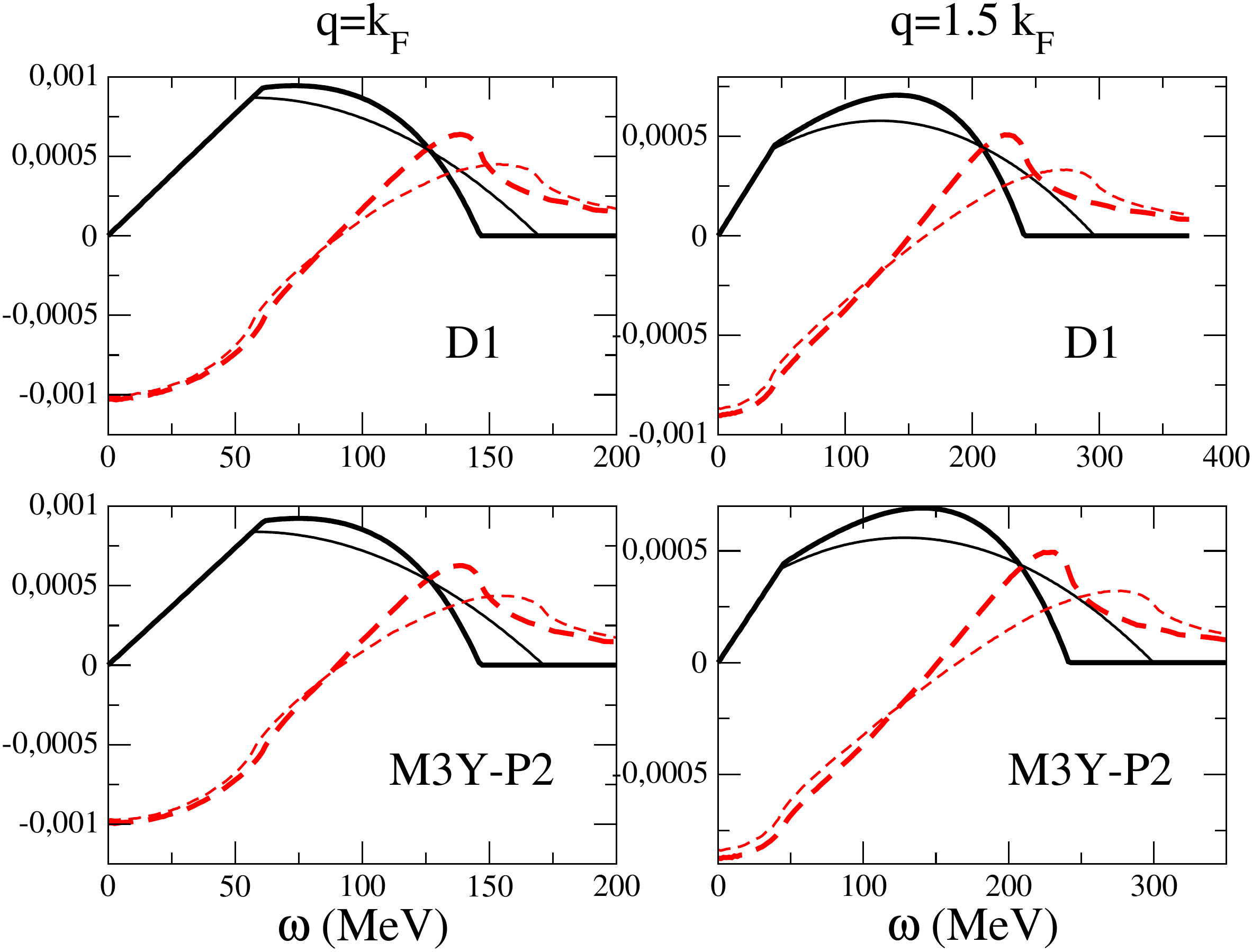}
\end{center}
\vspace{-.15in}
\caption{Hartree Fock response functions calculated exactly (thick lines) and with the approximation $m^*(k) \simeq m^*(k_F)$ (light lines). Solid and dashed lines refer to imaginary and real parts respectively. }
\label{fig-Pi-effmass}
\end{figure}

\begin{figure}[!h]
\begin{center}
\includegraphics[width=0.45\textwidth,angle=0]{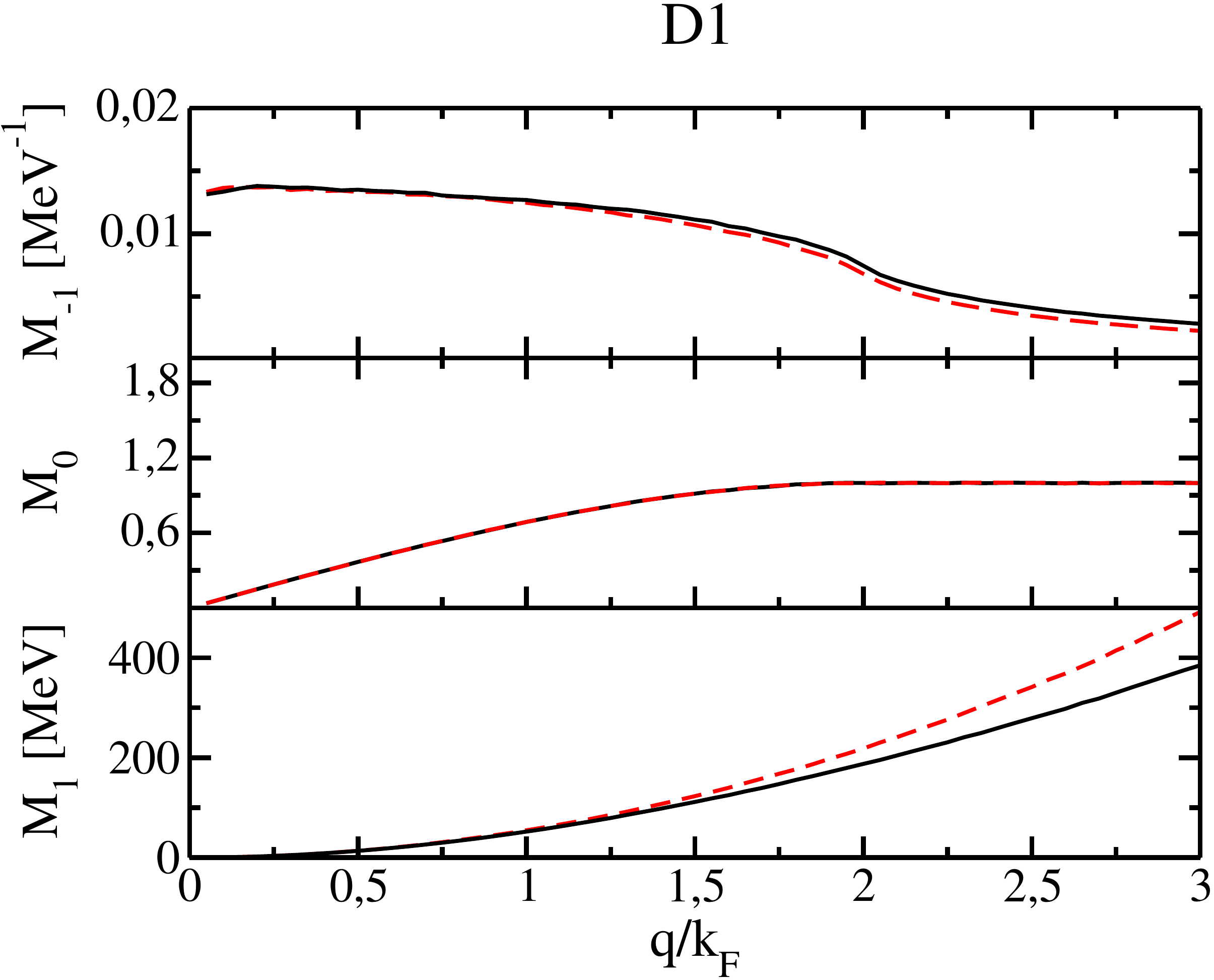}
\includegraphics[width=0.45\textwidth,angle=0]{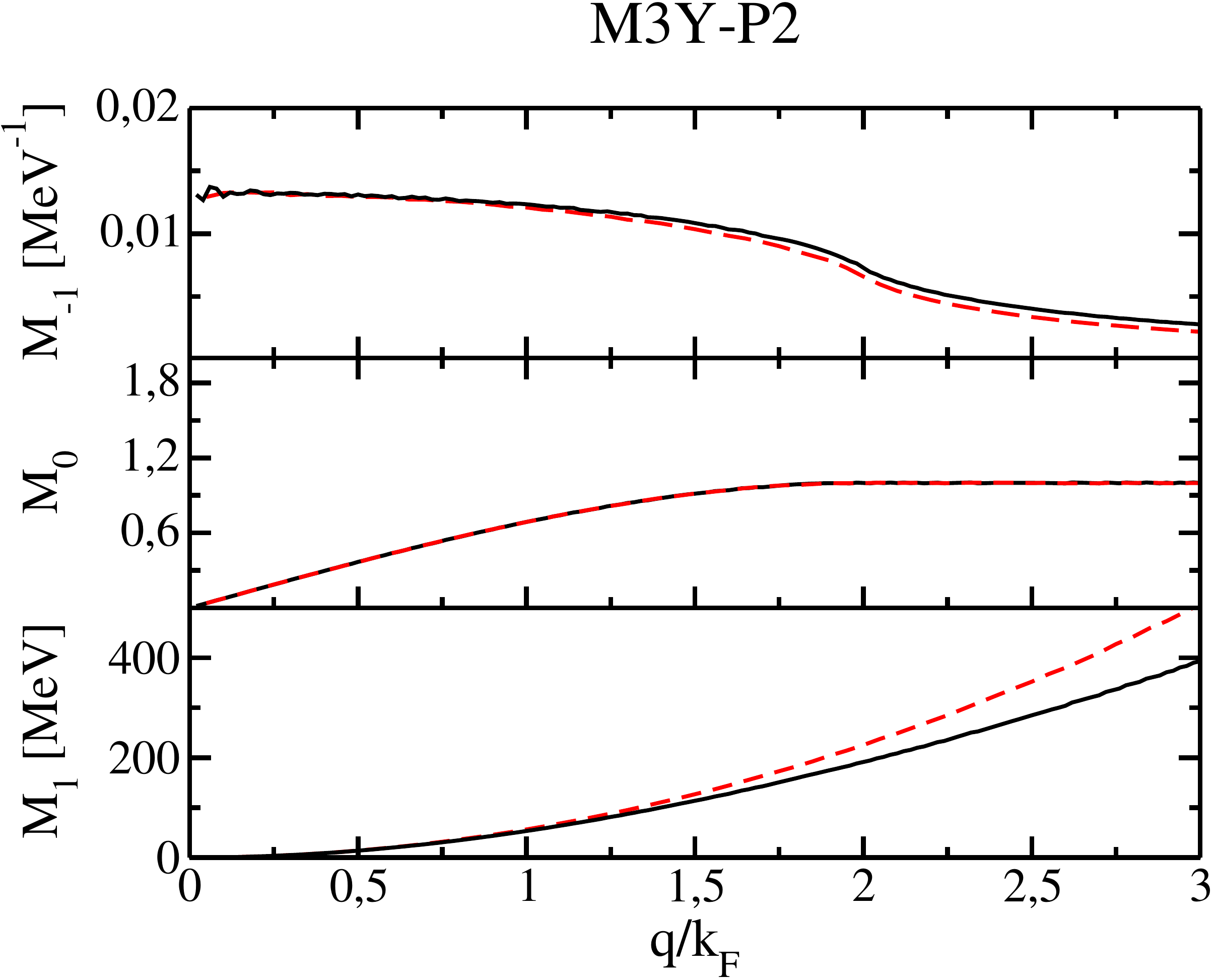}
\end{center}
\caption{Hartree Fock energy weighted sum rules at saturation density calculated exactly (solid lines) and with the approximation $m^*(k) \simeq m^*(k_F)$ (dashed lines). See text for details.}
\label{fig-SR-effmass}
\end{figure}

It should be stressed that the interest of such an approximation is neither for the mean field nor for the effective mass, but merely concerns the resulting HF response itself. We thus plot the real and imaginary parts of the HF response in Fig.~\ref{fig-Pi-effmass} at two values of the transferred momentum ($q=k_F$ and $1.5 k_F$) for interactions D1 and M3Y-P2. One can see that using the effective mass $m^*(k_F)$ produces some differences as compared with the exact response, resulting in a redistribution to the high energy region. 
To better quantify it, we consider the sum rules
\begin{eqnarray}
M_n(q,\rho)=\frac{1}{\pi \rho }\int_{0}^{\infty}  \omega^n Im \chi_{HF}(q,\omega) d\omega \;.
\end{eqnarray}
In Fig.~\ref{fig-SR-effmass} are compared the exact and approximate HF sum rules with $n=-1,0,1$ for interactions D1 and M3Y-P2. We see that the parabolic approximation (dashed lines) reproduces very nicely the exact $M_{-1}(q,\rho)$ and $M_0(q,\rho)$ sum rules (solid lines), while a sizeable deviation for $M_1$ beyond $q\approx 2 k_F$ is observed.  Concerning the $M_1$ sum rule, this kind of observation has already appeared in an extension of the Skyrme potential called N2LO~\cite{bec15}. It was concluded that the applicability of the model is questionable when the discrepancy is too important ($\simeq 10\%$). From Fig.~\ref{fig-SR-effmass} we can draw an important conclusion which will be used in Sect.~\ref{sec:instabilities}, where unphysical instabilities of finite-range interactions are discussed in terms of the strengths functions at zero energy. Since such strengths are proportional to the RPA $M_{-1}(q,\rho)$ sum rule, we can expect the parabolic approximation provide reasonably results in that respect.

All in all, the parabolic approximation is reasonably accurate for values of the transferred momentum $q$ smaller than $\simeq 2 k_F$. In Ref.~\cite{pac98} a  ``bi-parabolic" approximation was suggested for higher values of $q$. 
The idea is to fit separately the particle and the hole parts of the mean field involved in the HF propagator (\ref{GHF}), restricting the fit to the range of momenta actually involved in the integration to get the HF response. Such an approach was used for calculations of the quasi-elastic nuclear response. We are not going to enter into more details, because this specific approximation has been applied  to situations where the values of the transferred momentum are significantly higher than those currently explored by the works reviewed here, which are all based on the single parabolic approximation.


\section{Landau approximations to the ph interaction} \label{Sec:Landau}


The Landau's theory~\cite{lan57a,lan57b,lan59,pin66,mig67,abr59,bay91} encompasses the basic properties of Fermi liquids. It is a simple method largely used to calculate response functions of nuclear matter and in some cases, finite nuclei as well~\cite{kha02}. In this approach, the excitations of a strongly interacting normal Fermi system are described in terms of weakly interacting quasiparticles -or ph excitations- which are long-lived only near the Fermi surface. This quasiparticle interaction is solely characterised by a set of Landau parameters, which can be obtained from phenomenological or realistic interactions. The so-called Landau limit consists in taking $q \to 0$, but keeping the quotient $\omega/q$ fixed so that the response functions depend only on the dimensionless variable $\omega/(v_F q)$, where $v_F$ is the Fermi velocity. As an illustrative example, one can quote Gogny and Padjen~\cite{gog77} who calculated the SNM response function in the Landau limit with parameters derived from the Gogny parameterisation D1~\cite{gog75}.

\subsection{\it Landau limit}

In this limit ($q \to 0$), the particles are restricted to be at the surface of their respective Fermi sphere ($|\mathbf{k_1}| = |\mathbf{k_2}| = k_F$) so that the interaction only depends on the relative angle $\theta$ between vectors ${\bf k}_1$ and ${\bf k}_2$. The standard form of the SNM nuclear ph interaction in the Landau limit is then given as
\bea
\label{landau-ph}
V_{Landau}&=& f(\theta) + f'(\theta) ({\vtau}_1 \cdot {\vtau}_2) 
+ g(\theta) ({\vsigma}_1 \cdot {\vsigma}_2)
+ g'(\theta) ({\vtau}_1 \cdot {\vtau}_2) \, ({\vsigma}_1 \cdot {\vsigma}_2) \nnn 
& & \quad\quad\quad + \left[ h(\theta) + h'(\theta)  \, ({\vtau}_1 \cdot {\vtau}_2 ) \right]  \frac{{\mathbf k}_{12}^{2}}{k_{F}^{2}} \, S_T(\hat{\mathbf{k}}_{12}).
\eea
Although the factor ${\mathbf k}_{12}^{2}/k_F^2$ in the tensor term must be properly written as $2 (1 - \cos \theta )$, we keep this form for the time being and postpone the discussion of this term.  
Since a spin-orbit ph interaction, either zero or finite-range, is proportional to the transferred momentum~\cite{dav16b}, it gives no contribution in the Landau limit. The great advantage of a description \`a la Landau is that all the expressions obtained from this interaction are completely general. The set of usual Landau functions  \cite{bac79,bac80,fri81,hae82,fuj87} $f,f',g, g'$ then represent the central ph interaction in the spin-isospin ph spaces $(0,0), (0,1), (1,0)$ and $(1,1)$ respectively, while $h$ and $h'$ correspond to the tensor part in the spin $S=1$ and isospin spaces $I=0,1$, respectively. As a short notation for these functions, we shall write them as $f^{(\alpha)}(\theta)$ for the central components and $h^{(\alpha)}(\theta)$ for the tensor ones. Since the isospin matrix elements of Eq.~(\ref{landau-ph}) do not depend on the isospin projection $Q$, the spin-isospin label $\alpha$ will actually stand for $=(S,M,I)$. We note that the same notation can be also used for PNM, in which case the symbol $\alpha$ refers only to the spin indices $(S,M;n)$. 

Such functions are expanded in terms of Legendre polynomials 
\bea
f^{(\alpha)}(\theta) = \sum_{\ell}  f^{(\alpha)}_{\ell} \, P_{\ell}(\cos \theta)\; ,\\
h^{(\alpha)}(\theta) = \sum_{\ell}  h^{(\alpha)}_{\ell} \, P_{\ell}(\cos \theta) \;, 
\eea
and the coefficients are called Landau parameters. The matrix elements of the ph interaction to be plugged into the Bethe-Salpeter equation Eq.~(\ref{bethe-salpeter}) can therefore be written as
\bea
\label{LAN-Vph-ME}
 V_{ph}^{(\alpha,\alpha')}/n_d &=& \delta_{\alpha,\alpha'} \sum_{\ell}   f^{(\alpha)}_{\ell} 
P_{\ell}( \cos \theta ) 
 + \delta_{S,1} \delta_{S,S'} \delta_{I,I'} \sum_{\ell} h^{(\alpha)}_{\ell}  
P_{\ell}( \cos \theta ) 
S_T^{(M,M')}( {\mathbf {\hat k}_1} , {\mathbf {\hat k}_2}  ) \, , \quad 
\eea
where $n_d$ is the degeneracy number, coming from the calculation of the matrix elements in the spin-isospin spaces, and
\bea\label{eq:tensore}
S_T^{(M,M')}( {\mathbf {\hat k}_1} , {\mathbf {\hat k}_2}  ) =  
3  (-)^M \left( \hat{k}_{12}\right)^{(1)}_{-M} \left( \hat{k}_{12}\right)^{(1)}_{M'}
 - 2 \, \left( 1- \cos \theta \right) \, \delta_{M,M'}\,.
\eea
Introducing the density of states at the Fermi surface $N(0) = n_d \, m^* k_F / (2 \pi^2)$, one usually defines dimensionless Landau parameters as $F^{(\alpha)}_{\ell} \equiv N(0) f^{\alpha)}_{\ell}$ which provide dimensionless measures of the strength of the ph interaction on the Fermi surface. The stability  of the spherical Fermi surface against small deformations can then be expressed in terms of some inequalities for the dimensionless Landau parameters~\cite{bac79}, as we shall discuss later on.

Let us now come back to the tensor term. It has been written in Eq.~(\ref{landau-ph}) following the conventional definition~\cite{bac79,dab76}. However, some authors~\cite{ols04,sch04,ben13,hol13} have defined it without the factor ${\bf k}^2_{12}/k_F^2$ because it leads to a faster convergence~\cite{ols04}, in the sense that the absolute value of parameters $h_{\ell}, h'_{\ell}$ decreases as $\ell$ increases. Although the physical information contained in the ph interaction is the same in both cases, the Landau parameters are different, because of the extra $2 (1 - \cos \theta)$ factor entering the conventional definition. Both sets of parameters are actually connected through a recurrence relation~\cite{ols04,dav15}. The form (\ref{landau-ph}) is the most usually considered, and it is well adapted to the method we will discuss later on to calculate the response function. We thus consider only this form in the following.

However, Eq.~(\ref{landau-ph}) does not include the most general ph interaction. Indeed, as Schwenk and Friman~\cite{sch04} have pointed out, in the many-body medium the presence of the Fermi sea defines a preferred frame. Thus, two more non-central components must be included in the ph interaction that explicitly depend on the center-of-mass momentum $\hat{\bf P}_{12}$. These have the form
\bea
K_{12}(\hat{\bf P}_{12}) &=& 3 (\hat{\bf P}_{12} \cdot {\vsigma}_1) (\hat{\bf P}_{12} \cdot {\vsigma}_2) - ({\vsigma}_1 \cdot {\vsigma}_2) \;,\\
A_{12}(\hat{\bf k}_{12}, \hat{\bf P}_{12}) &=& ({\vsigma}_1 \cdot \hat{\bf P}_{12}) ({\vsigma}_2 \cdot \hat{\bf k}_{12}) - ({\sigma}_1 \cdot \hat{\bf k}_{12}) ({\vsigma}_2 \cdot \hat{\bf P}_{12}) \;,
\eea
which are designed as center-of-mass tensor and cross-vector interactions, respectively. The latter arises at second order in perturbation theory from the coupling of spin-orbit terms in the free-space interaction with any other non-spin-orbit term~\cite{hol13}.
Consequently, and with the same provisos previously mentioned concerning the ${\bf k}^2_{12}/k_F^2$ factor, one should add to Eq.~(\ref{LAN-Vph-ME}) the terms
\be
 \delta_{S,1} \delta_{S,S'} \delta_{I,I'} \sum_{\ell} \left[ k^{(\alpha)}_{\ell} K_T^{M.M'}(\hat{\bf{k}}_1, \hat{\bf{k}}_2)  
+ l^{(\alpha)}_{\ell} A^{M,M'}_T(\hat{\bf{k}}_1, \hat{\bf{k}}_2)  \right] P_{\ell}(\cos \theta) \;,
\label{non-central}
\ee
where $k_{\ell}, l_{\ell}$ are new Landau parameters related to these interactions, and where we have defined
\bea
K_T^{M.M'}(\hat{\bf{k}}_1, \hat{\bf{k}}_2) &=& 3 (-)^M (k_{12})^{(1)}_{-M} (k_{12})^{(1)}_{M'} - 2 (1+ \cos \theta) \delta_{M,M'}, \\
A^{M,M'}_T(\hat{\bf{k}}_1, \hat{\bf{k}}_2) &=& \frac{8 \pi}{3} \left[ Y^*_{1,M}(\hat{\bf{k}}_1) Y_{1,M'}(\hat{\bf{k}}_2) -
Y_{1,M'}(\hat{\bf{k}}_1) Y^*_{1,M}(\hat{\bf{k}}_2) \right].
\eea

For completeness, we mention that other non-central terms have been considered by Fujita and Quader~\cite{fuj87} in a study of electron systems and heavy fermions. These new terms were deduced from a general ph interaction invariant under a combined rotation in spin and orbital spaces. But as far as we know, no attempt to study such terms in nuclear systems has been made. 

\begin{figure}[!h]
\begin{center}
\includegraphics[width=0.363\textwidth,angle=0]{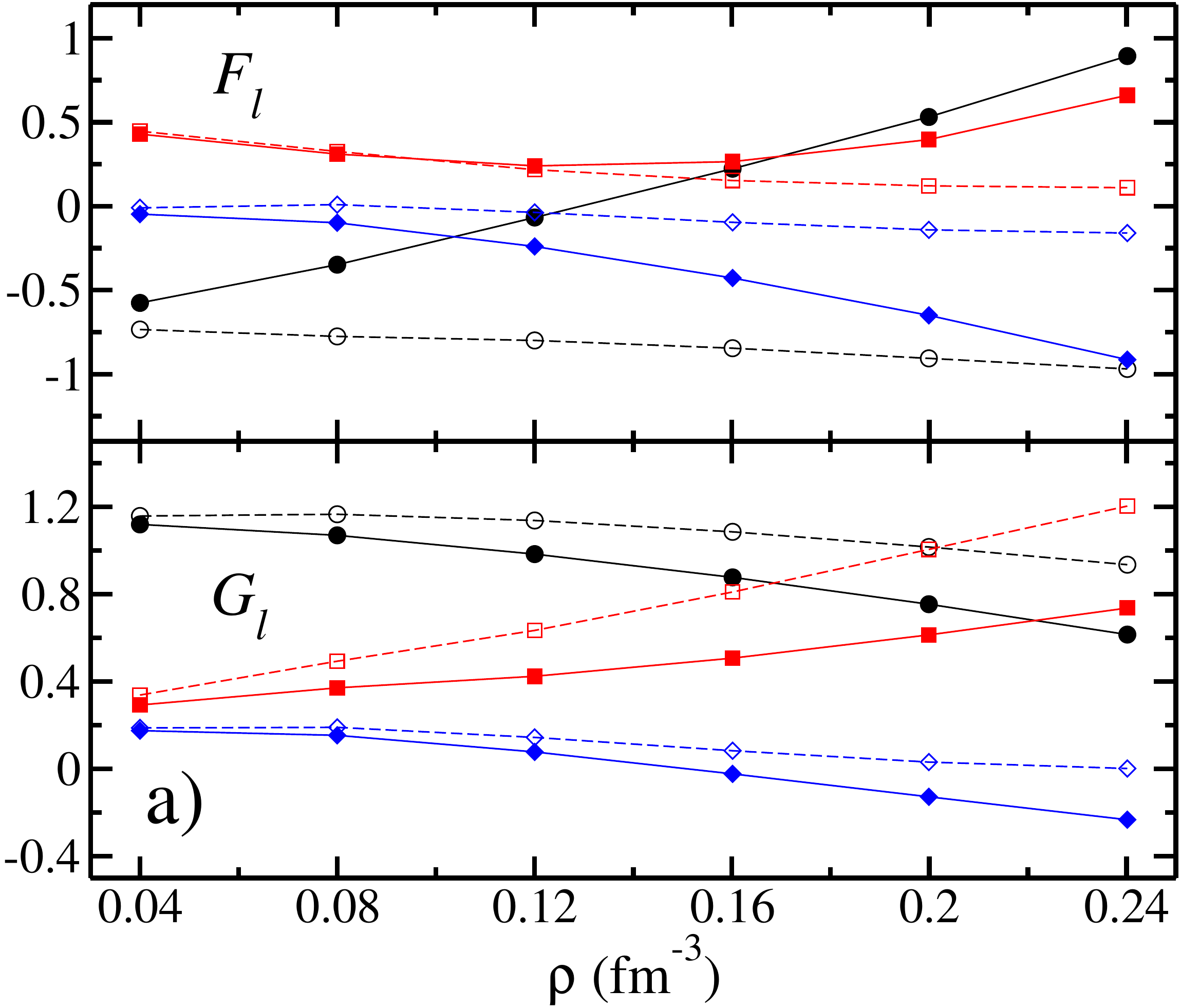}
\includegraphics[width=0.35\textwidth,angle=0]{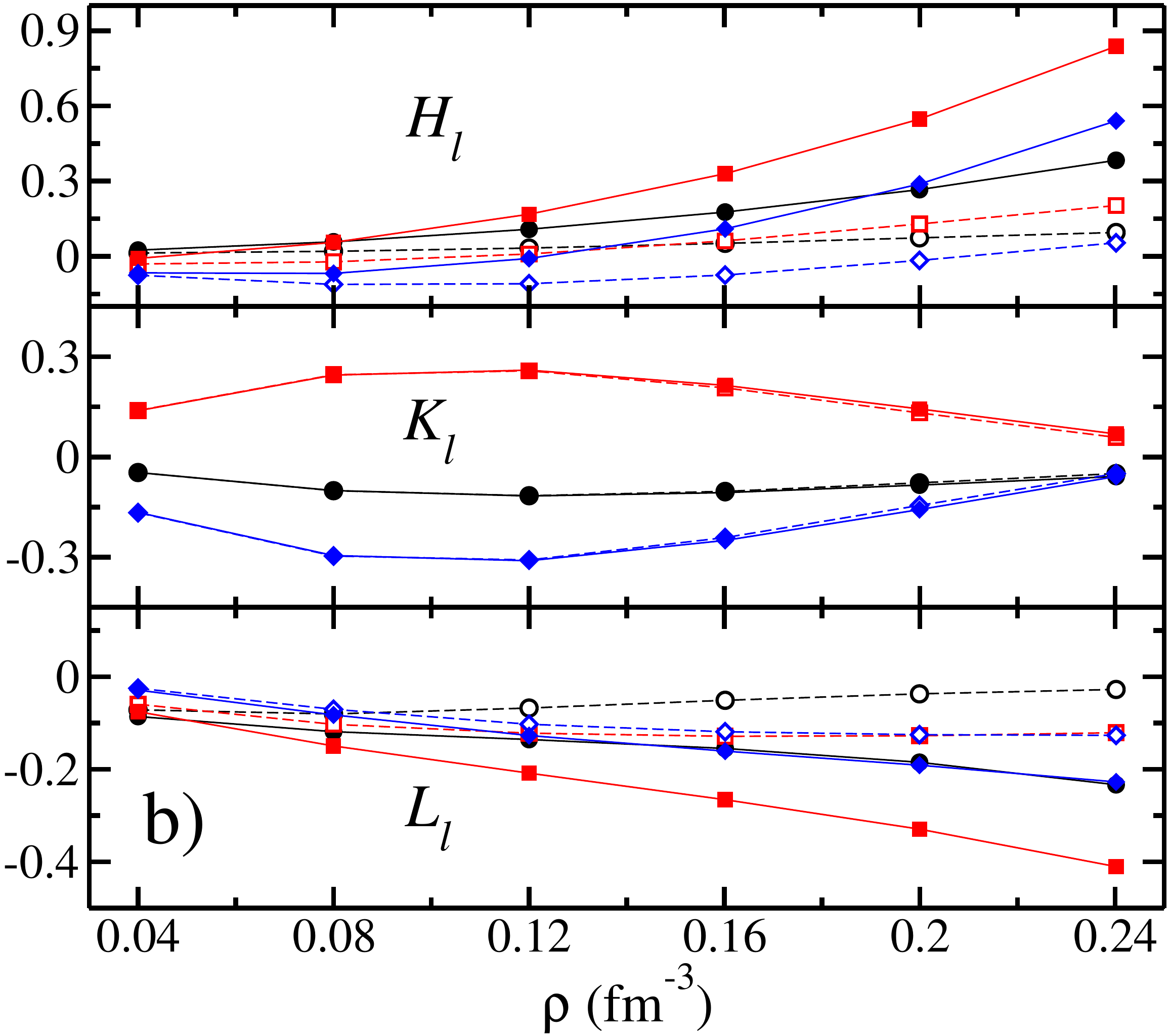}
\end{center}
\vspace{-.15in}
\caption{Left panel (a): central PNM Landau parameters $F_{\ell}$ and
$G_{\ell}$ associated with the chiral nuclear force  at the cutoff scale $\Lambda=450$
MeV as a function of the density of the system. Right panel (b): same but for the non-central
Landau parameters $ H_{\ell}$, $ K_{\ell}$ and $ L_{\ell}$.
Black circles, red squares and blue diamonds correspond to multipoles $\ell=0, 1, 2$, respectively.
Dashed lines and open symbols refer to results obtained when only two-body interactions
are included, while solid lines and symbols refer to the full two- and three-body interactions. Taken from Ref.~\cite{dav15}.}
\label{fig1-FGcut450}
\end{figure}

Some microscopic calculations of Landau parameters have been done~\cite{hol13,hol11,hol12} in the framework of chiral effective-field theory. They include the first and second-order perturbative contributions from two-body forces as well as the leading term from the chiral three-nucleon force. It was found that such a scheme lead to a good description of the bulk equilibrium properties of SNM, including the compressibility modulus and the symmetry energy. The same scheme was also employed to obtain the PNM Landau parameters in a large range of density values in Ref.~\cite{dav15}. The results depend on the value of the momentum-space cutoff in the chiral nuclear interaction, but it was checked that the qualitative features of the response functions are insensitive to the cutoff value. In Fig.~\ref{fig1-FGcut450} are displayed the dimensionless Landau parameters as a function of the density for a cutoff  value $\Lambda=450$ MeV. One can see that non-central parameters are of the same order of magnitude as the central ones. Therefore, based on their absolute magnitude alone, it is not possible to discard them when calculating the response functions. 

\subsubsection{\it Response functions}
\label{resp-landau-analytical}

The method for obtaining the RPA response functions with the ph interaction given by Eqs. (\ref{LAN-Vph-ME}) and (\ref{non-central}) has been discussed in~\cite{pas14a,report} and we recall it here briefly. The first step is to understand how the ph interaction depends on $\hat{\bf k}_1$ and $\hat{\bf k}_2$. Let us, for instance, consider the central components. They depend on Legendre polynomials $P_\ell(\cos \theta)$, that is, some linear combinations of $(\hat{\bf k}_1 \cdot \hat{\bf k}_2)^n$. In turn, these powers are represented by combinations of the products 
\bea
Y^*_{1, \mu_1}(\hat{\bf k}_1) \dots Y^*_{1, \mu_n}(\hat{\bf k}_1) Y_{1, \mu_1}(\hat{\bf k}_2) \dots Y_{1, \mu_n}(\hat{\bf k}_2)\, . \nonumber
\eea
 Similarly, the tensor components additionally contain terms like 
 \bea
 Y^*_{1, M}(\hat{\bf k}_1)  Y_{1, M'}(\hat{\bf k}_1), \, Y^*_{1, M}(\hat{\bf k}_1)  Y_{1, M'}(\hat{\bf k}_2), \, Y^*_{1, M}(\hat{\bf k}_2)  Y_{1, M'}(\hat{\bf k}_1), \, Y^*_{1, M}(\hat{\bf k}_2)  Y_{1, M'}(\hat{\bf k}_2), \dots \nonumber
 \eea 
 Taking the momentum average of the Bethe-Salpeter equation in Eq.~(\ref{bethe-salpeter}), the response function appears, by definition, in the left hand side. One then realises immediately that this momentum average $\langle G^{(\alpha)}_{RPA}\rangle$ is coupled to other averages of the ph propagator containing a number of spherical harmonics  $\langle Y_{1, \mu_1}(\hat{\bf k}_2) \dots Y_{1, \mu_n}(\hat{\bf k}_2)G^{(\alpha)}_{RPA} \rangle$. Therefore, one multiplies the BS equations by these products of spherical harmonics and integrates over the momenta thus getting a new equation for each of these averages, until one ends up with a closed system of coupled linear equations for these unknown functions. The coefficients depend only on the Landau parameters and some momentum averages of the HF propagator, which can be determined separately.

As an example, let us restrict the ph interaction to its central components with $\ell=0, 1,2$. In this case the BS equations can be solved in a simple way and  the response function can be written as 
\be\label{chirpa:landau}
\chi_{RPA}^{(\alpha)}(q,\omega) = \frac{\chi^{HF}(q,\omega)}{1 - W^{(\alpha)} \,  \chi^{HF}(q,\omega)} \;,
\ee
\noindent where
\be
\label{W-landau}
W^{(\alpha)} = f^{(\alpha)}_0 -\frac{1}{2} f^{(\alpha)}_2 + \nu^2 \, 
\frac{f^{(\alpha)}_1 + \left(\frac{27}{8} \nu^2 + \frac{3}{10} F^{(\alpha)}_1 \right) f^{(\alpha)}_2}{\left(1+ \frac{1}{3} F^{(\alpha)}_1 \right) \left( 1 + \frac{1}{8} \left[ - 9 \nu^2 + \frac{12}{5} \right] F^{(\alpha)}_ 2 \right)} \;,
\ee
where $\nu = \omega m^*/(q k_F)$ is a dimensionless quantity. The inclusion of tensor components lead to a more complex algebraic system, for which one gets too cumbersome expressions in the general case. However, it is relatively easy to obtain an analytical expression in the limit $q=0$ and $\omega=0$, that is, to the static susceptibility. This susceptibility is important in the sense that it is a thermodynamical quantity and can be derived from numerical methods. It was given in Ref.~\cite{nav13} for the ph interaction in Eq.~(\ref{LAN-Vph-ME}), and generalized in Ref.~\cite{dav15} to include the non-central terms of Eq.~(\ref{non-central}). For SNM and the $S=1$ channel, it has been written as 
\be
\frac{\chi_{HF}(0,0)}{\chi_{RPA}^{(M,I)}(0,0)} = 1 + G^{(I)}_0 - \frac{T_1}{T_2} \;,
\ee
with
\bea
T_1 &=& 2 \, \left(H^{(I)}_0 - \frac{2}{3} H^{(I)}_1 + \frac{1}{5} H^{(I)}_2  +K^{(I)}_0 + \frac{2}{3} K^{(I)}_1 + \frac{1}{5} K^{(I)}_2 \right)^2 \\
T_2 &=& 1+\frac{1}{5} G^{(I)}_2-\frac{7}{15} H^{(I)}_1 + \frac{2}{5} H^{(I)}_2 - \frac{3}{35} H^{(I)}_3 \nnn
&& \quad + \frac{7}{35} K^{(I)}_1 + \frac{2}{5} K^{(I)}_2 + \frac{3}{35} K^{(I)}_3 + \frac{2}{5} L^{(I)}_1 - \frac{6}{35} L^{(I)}_3 \, .
\label{suscept}
\eea

Notice that this expression is independent of the spin projection $M$, {\it i.e.} the spin susceptibility is identical for the longitudinal ($M=0$) and the transverse ($M=1$) channels. Obviously, it reduces to the familiar spin susceptibility of a Fermi liquid when no tensor terms are considered. These formulae are also valid for PNM, by replacing $G^{(I)}_{\ell}, H^{(I)}_{\ell}...$ with the corresponding PNM parameters and in the $S=0$ channel, by replacing $G^{(I)}_{\ell}$ with $F^{(I)}_{\ell}$ and dropping the tensor parameters. It is worth noticing that Olsson et al.~\cite{ols04} have also deduced the static susceptibility by solving the usual Landau equation, instead of the method based on the BS equation for the propagator. These authors included the non-central terms of Eq.(\ref{non-central}), but without the prefactor ${\bf k}_{12}^2/k_F^2$ in the definition of the tensor operators. Both results are in agreement, but the parameters $L_{\ell}$ entered as an infinite sum. If the prefactor was included instead, the contribution of Landau parameters with $\ell>3$ would have canceled out exactly, as pointed out in Ref.~\cite{nav13}.

\begin{figure}[!h]
\begin{center}
\includegraphics[angle=0,width=0.4\textwidth]{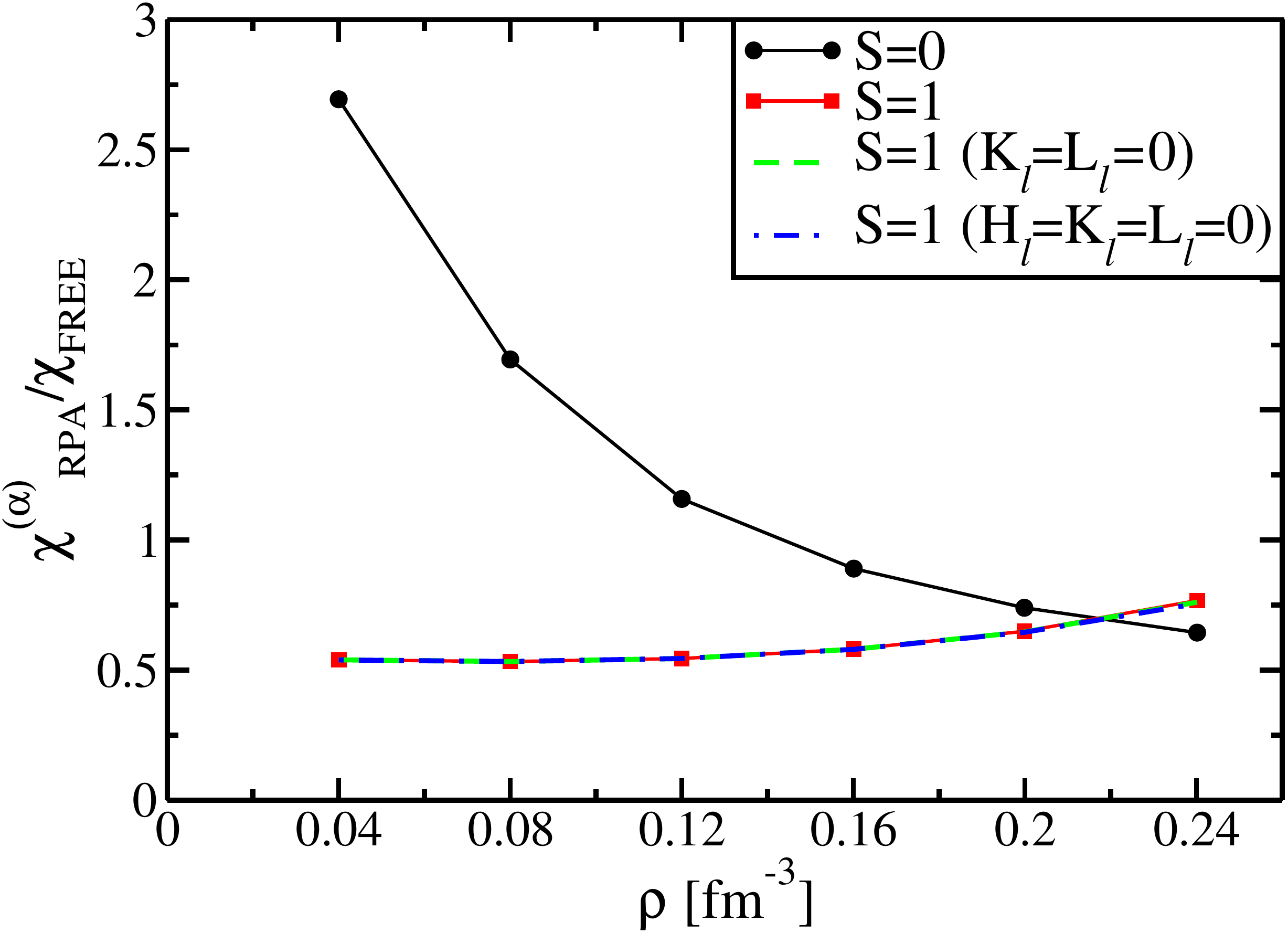}
\end{center}
\caption{Static spin susceptibility of PNM from the chiral nuclear potential for $\Lambda=450$
MeV as a function of the density. Taken from Ref.~\cite{dav15}.}
\label{fig:static}
\end{figure}

In Fig.~\ref{fig:static}, is plotted the PNM static spin susceptibility as a function of density, as obtained with the Landau parameters of Fig.\ref{fig1-FGcut450}. To discern the importance of the tensor parameters, the spin susceptibility has also been calculated by dropping them or keeping only $H_\ell$. One can  deduce that the tensor contribution is negligible as compared to the contribution of the $G_0$ term. Actually, as signalled in Ref.~\cite{dav15}, there is an approximate cancellation between the combination of parameters $H_\ell$ and $K_\ell$ which results in $T_1 \simeq 0$. It follows that the non-central terms play almost no role in the static spin susceptibility.

\begin{figure}[H]
\begin{center}
\includegraphics[angle=0,width=0.4\textwidth]{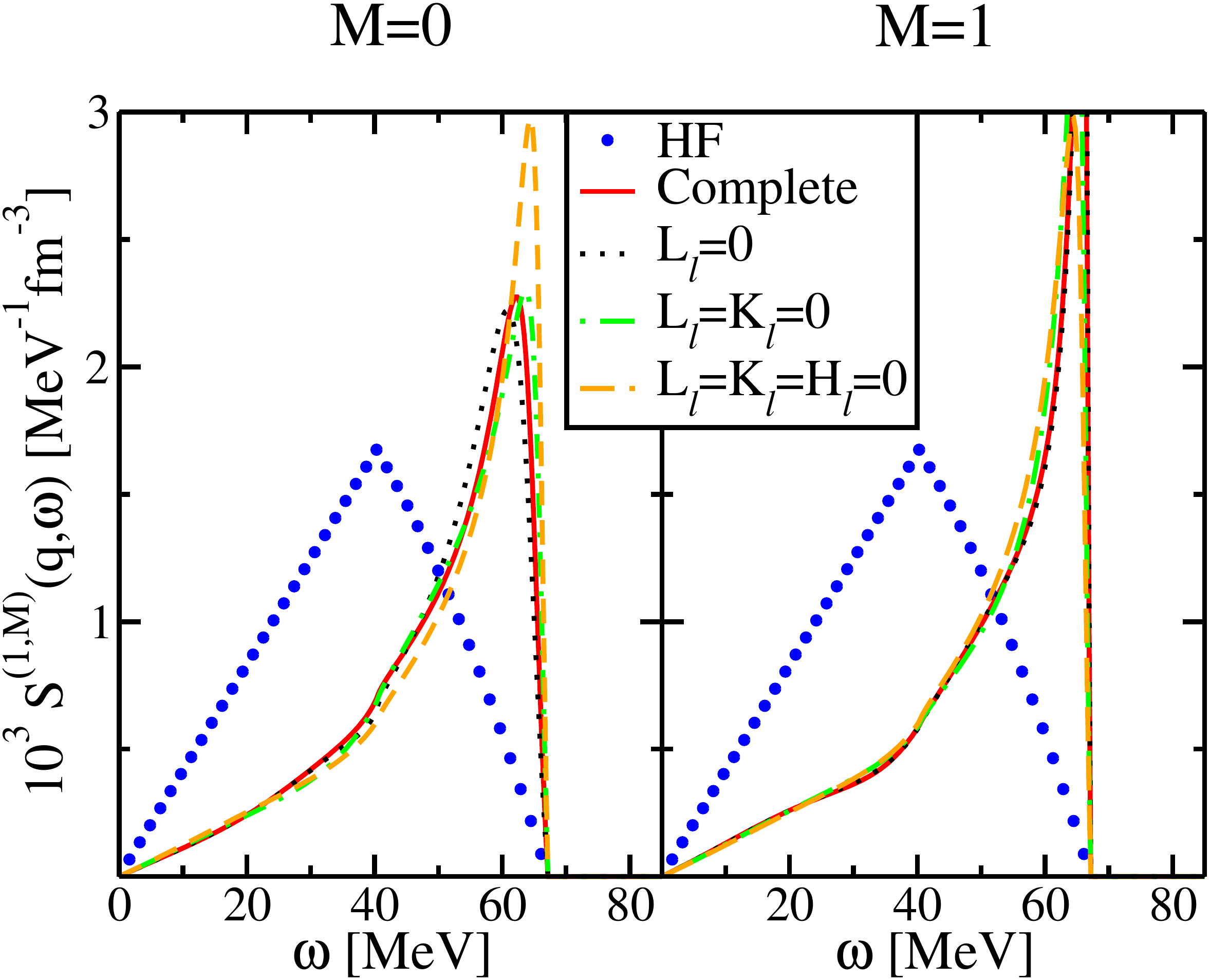}
\end{center}
\caption{PNM strength function at $k^{}_{F}=1.68$~fm$^{-3}$ and $q/k_{F}=0.5$
in the $S=1$ channel calculated at $l_{max}=3$ with and without the different contributions
of the tensor terms. Adapted from Ref.~\cite{dav15}}
\label{fig:hkl}
\end{figure}

\begin{figure}[H]
\begin{center}
\includegraphics[angle=0,width=0.4\textwidth]{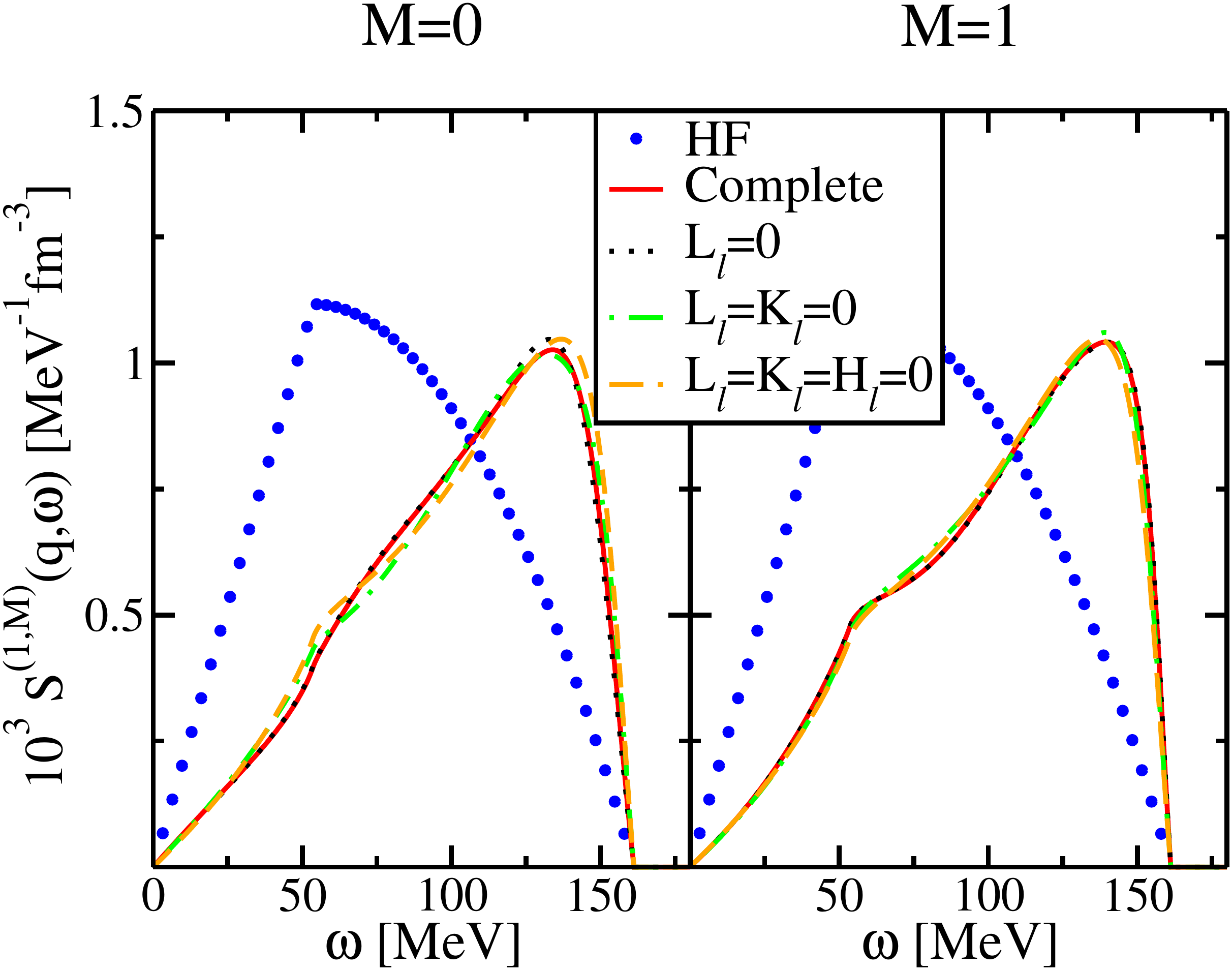}
\includegraphics[angle=0,width=0.4\textwidth]{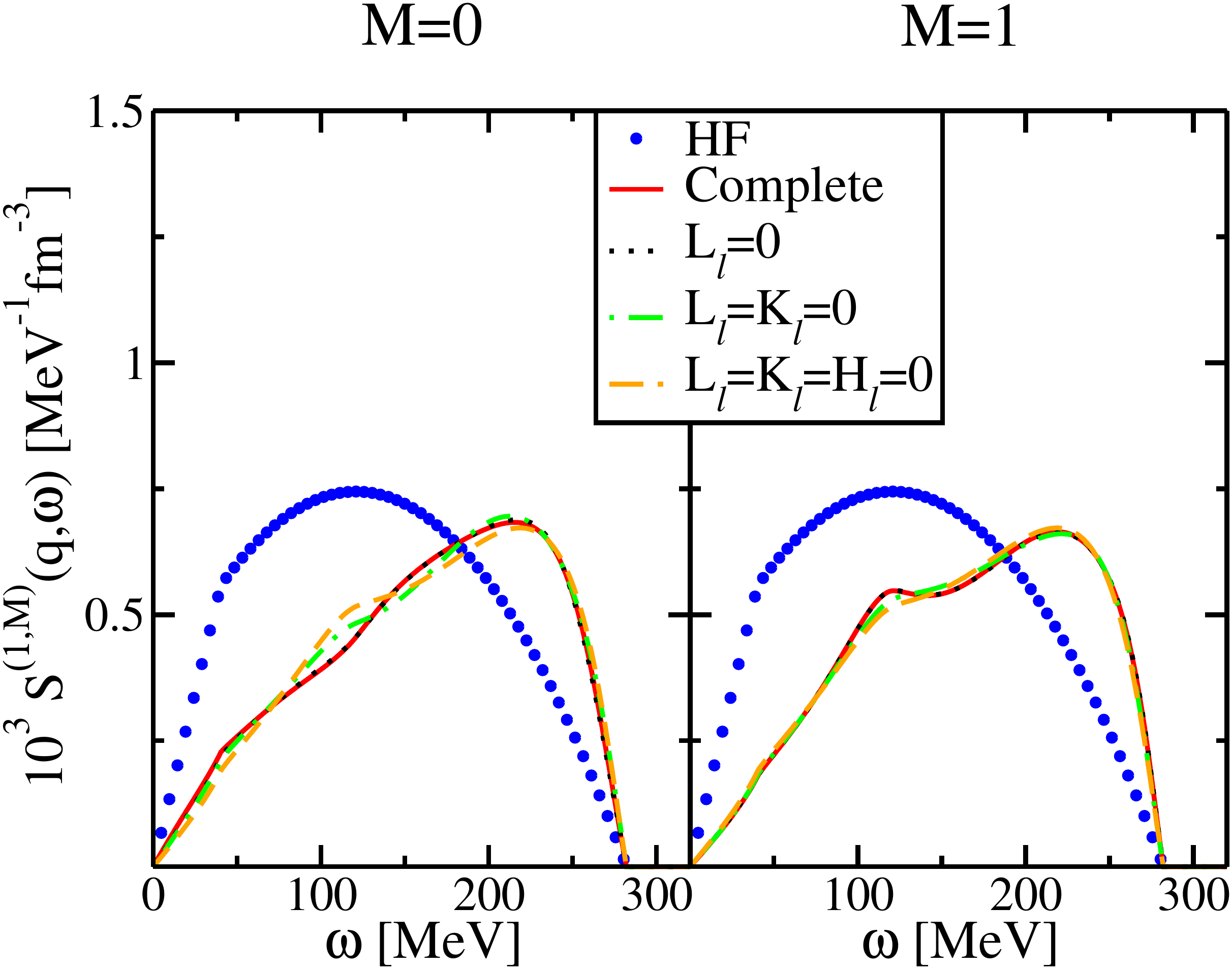}
\end{center}
\caption{Same as Fig.\ref{fig:hkl} for $q/k_F=1$ (left panel) and $q/k_F=1.5$ (right panel) }
\label{fig:hkl2}
\end{figure}

The PNM strength functions based on the Landau parameters given in Fig.~\ref{fig1-FGcut450} are displayed in Fig.~\ref{fig:hkl} as a function of energy for $k_{F}=1.68$~fm$^{-3}$ and $q/k_{F}=0.5$. The ph interaction includes up to $\ell=3$ parameters. The HF response is also plotted as a reference. To see the effect of non-central terms, calculations have been performed in three additional cases, by ignoring $L_{\ell}$, both $L_{\ell}$ and $K_{\ell}$ and all of them. One can see that the contribution of tensor terms $L_\ell$ and $K_\ell$ is negligible in both $M=0$ and 1 channels, while the term $H_\ell$ reduces the peak of the response and slightly shifts its position towards lower energies. In conclusion, the new non-central terms can be safely ignored. This is confirmed by Fig.~\ref{fig:hkl2} where the same curves are depicted but with a higher transferred momentum.

\subsubsection{\it Landau parameters from Gogny and Nakada interactions}

For a given effective two-body interaction taken in the Landau limit, it is always possible to cast the ph interaction in the form of Eq.~(\ref{landau-ph}). The interaction has thus a {\it universal} form and it is completely characterised by the Landau parameters $F_\ell,G_\ell,H_\ell$ (the parameters $L_\ell$ and $K_\ell$ will be neglected in the following). Explicit expressions of Landau parameters for Gogny and Nakada interactions have been done in Ref.~\cite{report}. In Tab.~\ref{land-param-SNM}, we thus report their values at saturation density for M3Y-P2 and  D1MT interactions. Notice that the latter is a modified version of D1M~\cite{gor09a} with a tensor term added perturbatively.

 \begin{table}[h!]
 \caption{Landau parameters in SNM and PNM at density $\rho=0.16$ fm$^{-3}$ for D1MT and M3Y-P2 interactions.}
\begin{center}
\begin{tabular}{c|cccccc|ccc}
\hline
\multicolumn{10}{c}{Gogny D1MT \cite{ang11}}\\
\hline
&\multicolumn{6}{c|}{SNM} & \multicolumn{3}{c}{PNM}\\
\hline
$\ell$ & $F_{\ell}$ & $F'_{\ell}$ & $G_{\ell}$ & $G_{\ell}'$ & $H_{\ell}$ & $H_{\ell}'$&  $F^{(n)}_{\ell}$ & $G^{(n)}_{\ell}$ & $H^{(n)}_{\ell}$\\
\hline
0 & -0.310 & 0.724 & -0.037 & 0.731 & 0.306 & -0.102&  -0.560 &0.465  & 0.136 \\
1& -0.756 & 0.397 & -0.348 & 0.624& 0.587 & -0.196 &  -0.690 &  0.029&0.300\\
2& -0.293 & 0.615 & 0.471 & -0.237 & 0.557 & -0.185 &   0.283& 0.226 &0.328\\
3 &-0.055 & 0.130 & 0.108 & -0.060 &   -      &  -        &    0.120  & 0.077 & -     \\
\hline
\multicolumn{10}{c}{M3Y-P2 \cite{nak03}}\\
\hline
&\multicolumn{6}{c|}{SNM} & \multicolumn{3}{c}{PNM}\\
\hline
$\ell$ & $F_{\ell}$ & $F'_{\ell}$ & $G_{\ell}$ & $G_{\ell}'$ & $H_{\ell}$ & $H_{\ell}'$&  $F^{(n)}_{\ell}$ & $G^{(n)}_{\ell}$ & $H^{(n)}_{\ell}$ \\
\hline
0 &  -0.383 & 0.620 &  0.112 &  1.010  & 0.043& -0.015 & -0.564& 0.830&0.026\\
1& -1.039 & 0.632 & 0.272 & 0.200 &0.063 & -0.019      &    -0.362 &0.391&0.047\\
2& -0.433 & 0.243 & 0.161 & 0.040 &0.047 & -0.013       &   -0.174&0.223&0.044\\
3& -0.208 & 0.094 & 0.077 & -0.002&     -      &    -        & -0.106 &0.0972& - \\
\hline
\end{tabular}
\label{land-param-SNM}
\end{center}
\end{table}

Landau parameters have a physical meaning and can therefore be related to some observables at saturation, such as the compressibility or the density of states, via very simple linear relations. It is thus possible to compare the parameters directly. First, we observe that the $F_\ell$ parameters have roughly the same order of magnitude and sign. This is due to the fact that both interactions have been constrained using properties of non-polarised matter. A different conclusion stands for the $G_\ell$ parameters that are linked to spin properties which have not been constrained. One observes much stronger differences in that case. A possible way to better constrain these terms would be to use spin-isospin excitations as discussed in Refs.~\cite{gia81,ben02,roc12}. This is important since these terms may play a minor role in determining the ground state properties of atomic nuclei, but they may play a major role in determining the structure of the excited spectrum. Similar conclusions hold for the $H_\ell$ terms, where one observes significant differences.

As already mentioned, another important information we can extract by observing Landau parameters, concerns the stability of the Fermi surface against small deformations. This stability can be expressed in terms of general inequalities satisfied by Landau parameters. These inequalities impose a stringent test for the bare or phenomenological interactions used to calculate the Landau parameters. For instance, the central parameters must satisfy~\cite{mig67} 
\be
F_{\ell}^{(\alpha)} > - (2 \ell +1)\;.
\ee
As shown in Ref.~\cite{bac79}, the inclusion of tensor components in the ph interaction produces a coupling between the spin-dependent parts, that is, $G^{(I)}$ terms become coupled to $H^{(I)}$ ones. Compact expressions can be obtained considering states with good ph angular momentum $J$, so that the potential part of the free energy for a given $J$ is a $2 \times 2$ matrix with $\ell, \ell'$ values $J \pm 1$. In that case, the stability criterion is given by the condition that these matrices have positive eigenvalues~\cite{cao10}. When $\ell = \ell'$, the matrices are diagonal, and one gets a single stability criterion for each possible value of $J$. We collect here the resulting stability criteria for the lower possible values of $\ell$ and $J$. The first diagonal matrices correspond to the values $\ell=1$, $\ell'=1$, and one gets
\bea
&& 1 + \frac{1}{3} G^{(I)}_1 - \frac{10}{3} H^{(I)}_0 + \frac{4}{3} H^{(I)}_1 - \frac{2}{15} H^{(I)}_2 > 0 \;, \label{Landau:insta1}\\
&& 1 + \frac{1}{3} G^{(I)}_1 + \frac{5}{3} H^{(I)}_0 - \frac{2}{3} H^{(I)}_1 +  \frac{1}{15} H^{(I)}_2 > 0 \;, \label{Landau:insta2} \\
&& 1 + \frac{1}{3} G^{(I)}_1 - \frac{1}{3} H^{(I)}_0 + \frac{2}{15} H^{(I)}_1 - \frac{1}{75} H^{(I)}_2 > 0 \;,  \label{Landau:insta3}
\eea
for $J= 0^-, 1^-$ and $2^-$, respectively. The diagonal $\ell=2, \ell'=2$ case gives
\bea
&& 1 + \frac{1}{5} G^{(I)}_2 - \frac{7}{15} H^{(I)}_1 + \frac{2}{15} H^{(I)}_2 - \frac{3}{35} H^{(I)}_3 > 0 \;,  \label{Landau:insta4}\\
&& 1 + \frac{1}{5} G^{(I)}_2 + \frac{7}{15} H^{(I)}_1 - \frac{2}{5} H^{(I)}_2 + \frac{3}{35} H^{(I)}_3 > 0 \;, \label{Landau:insta5} \\
&& 1 + \frac{1}{5} G^{(I)}_2 - \frac{2}{15} H^{(I)}_1 + \frac{4}{35} H^{(I)}_2 - \frac{6}{245} H^{(I)}_3 > 0  \label{Landau:insta6}\;, 
\eea
for $J=1^+, 2^+$ and $3^+$, respectively. Obviously, when all tensor parameters are put to zero, one recovers the familiar stability conditions $G^{(I)}_{\ell} > - (2 \ell+1)$.

\begin{figure}
\begin{center}
\includegraphics[width=0.4 \textwidth,angle=0]{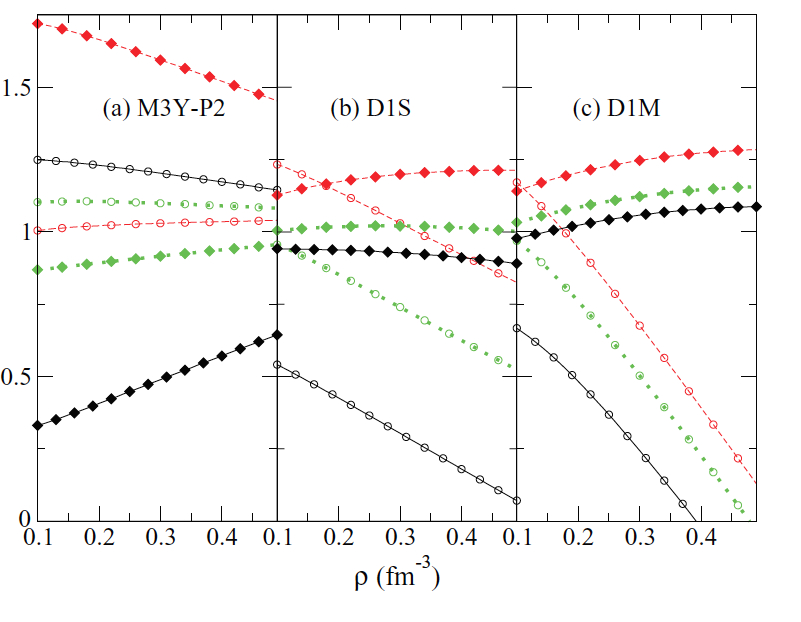}
\end{center}
\caption{(Color online). Instability inequalities of Eqs. (\ref{Landau:insta1}-\ref{Landau:insta6}) for the singlet isospin channel as a function of density for symmetric nuclear matter obtained for the finite-range interactions: (a) M3Y-P2, (b) Gogny D1S and (c) Gogny D1M.  Solid, dashed and dotted lines all with empty circles correspond to $\ell =1,J=0^-,1^-,2^-$ respectively. Solid, dashed and dotted lines all with full diamonds correspond to $\ell =2,J=1^+,2^+,3^+$. The effects of the tensor force are included in all cases. Taken from Ref.~\cite{nav13}.}
\label{fig:Landau:skyrme:instabilities}
\end{figure}

In Fig.~\ref{fig:Landau:skyrme:instabilities}, we show the density dependence of the diagonal stability conditions in the isoscalar channel, corresponding to $\ell=1$, for  
$J=0^-, 1^-, 2^-$, Eqs.~(\ref{Landau:insta1}-\ref{Landau:insta3}) and to $\ell=2$ for $J=1^+, 2^+, 3^+$, Eqs.~(\ref{Landau:insta4}-\ref{Landau:insta6}) for SNM. Figure~\ref{fig:Landau:skyrme:instabilities}(a) shows the results for M3Y-P2 parametrization and Figs.~\ref{fig:Landau:skyrme:instabilities}(b) and \ref{fig:Landau:skyrme:instabilities}(c) correspond to D1S and D1M parametrizations, respectively, supplemented with a tensor force as discussed in Ref.~\cite{ang12}.
 The stability conditions $\ell = 1$ are well respected for all interactions in the density range considered in the figure. However, the $\ell = 2$ stability condition has a negative slope as a function of density. In particular for D1M and for $\ell = 2, J=0^-$ (empty circles joined by solid line), there is a signal of an instability with a critical density close to $\rho=0.4$ fm$^{-3}$. It is worth mentioning that in the case of the isovector channel the stability conditions are all well respected in the range of densities explored. See Ref.~\cite{nav13} for a more detailed discussion.

\subsubsection{\it RPA response function}
\label{subsec:landau}
\begin{figure}[!h]
\begin{center}
\includegraphics[width=0.5\textwidth,angle=0]{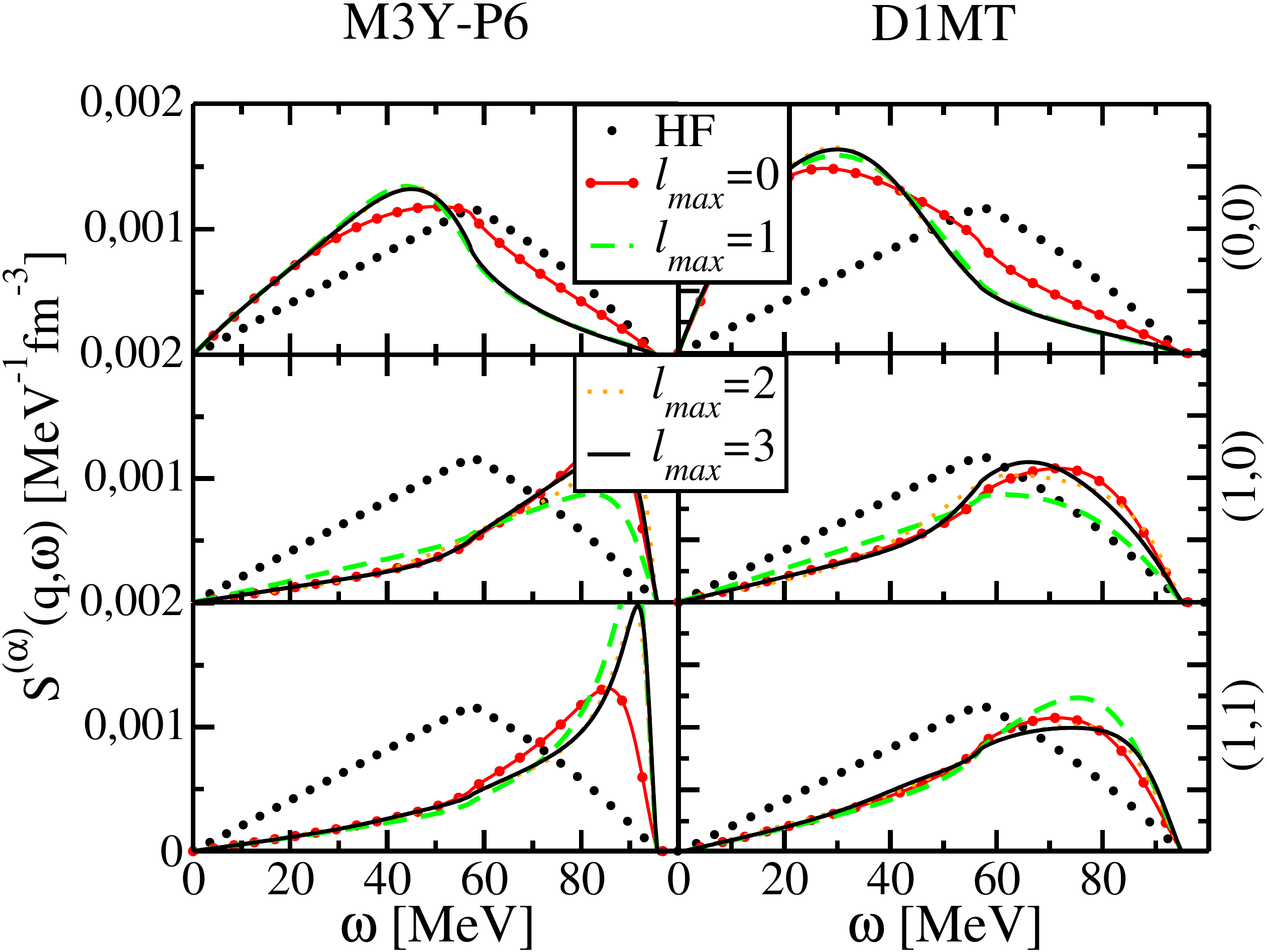}
\end{center}
\caption{Strength functions in pure neutron matter for interactions M3Y-P6 and D1MT at density $\rho=0.16$ fm$^{-3}$ and $q=0.5 k_F$. The HF strengths are also plotted (dots) as a reference. Adapted  from Ref.~\cite{pas14a}. }
\label{fig:cov:gognyD1MT:pnm}
\end{figure}

The RPA response function using a Landau ph interaction as given by Eq.~(\ref{LAN-Vph-ME}) has been calculated in  Ref.~\cite{pas13b}. The resulting BS equation is solved using the same symbolic procedure described in the previous section. The main difference is that one requires intermediate integrals of the form $\alpha_i(q,\omega)=\langle \cos^i \theta G_{HF}\rangle$, whose expressions are given in Ref.~\cite{pas13b}.  In Fig.~\ref{fig:cov:gognyD1MT:pnm},  we present strength functions for PNM at density $\rho=0.16$ fm$^{-3}$ and $q=0.5k_F$ using D1MT and M3Y-P6. Although in the Landau limit the $S=0$ and $S=1$ are not coupled together, we still presents the results in this channel for completeness. Since we deal with finite-range interactions, the number of partial waves and thus the number of Landau parameter is infinite. Hopefully not all of them contribute equally to the response functions. In Fig.~\ref{fig:cov:gognyD1MT:pnm}, the calculations are thus performed as a function of the maximum number of partial waves included in the calculation: we clearly see that in all channels, the convergence is achieved for $\ell_{max}= 2$.
 It is interesting to observe that the major discrepancy  between the two families of curves is obtained for the $S=1$ channel. This is not surprising since it is usually the least constrained during the optimisation procedure.

\begin{figure}[!h]
\begin{center}
\includegraphics[width=0.5\textwidth,angle=0]{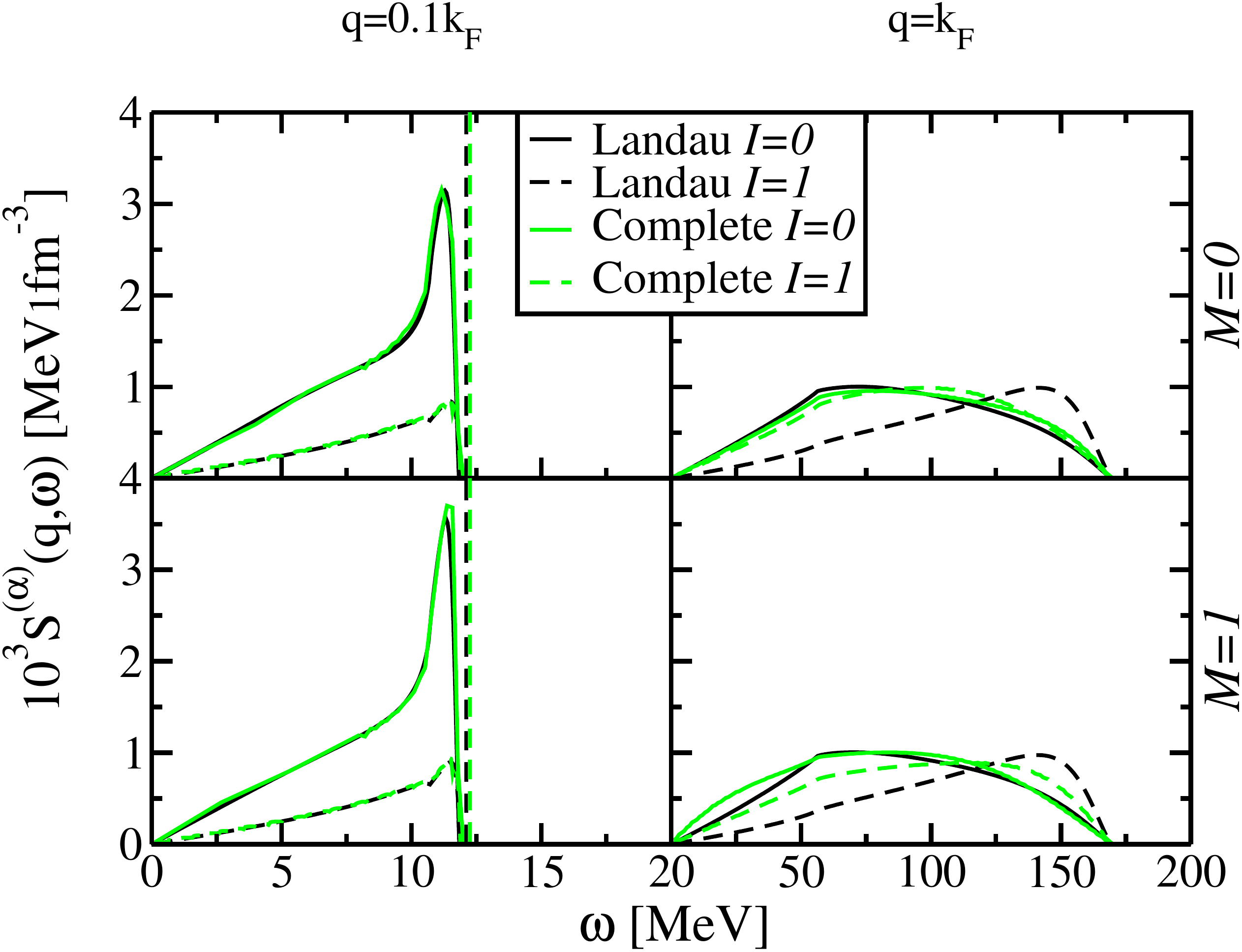}
\end{center}
\caption{Strength function in the $S=1$ channel using M3Y-P2 calculated using both the Landau approximation and full multipolar expansion at two values of transferred momentum $q/k_F=0.1,1$ at $\rho=\rho_0$.}
\label{fig:cov:M3YP2:copmpare:Landau}
\end{figure}

In order to have an insight of the quality of the Landau approach, it is interesting to compare these results with the numerically exact strengths which are presented in Sec.~\ref{sec:multipole}. For instance, we show in Fig.~\ref{fig:cov:M3YP2:copmpare:Landau} the strength function obtained with Landau parameters and $\ell_{max}=2$ with the complete response function for two values of the transferred momentum $q$. Since in the Landau limit, the spin-orbit term is exactly zero, there is no coupling between the S=0 and S=1 channel and as such we just illustrate the differences in the S=1 channel where the tensor acts.
We observe that for low values of transferred momentum, the response functions obtained used the full response or the Landau approximation are essentially on top of each other. The agreement breaks down  by increasing the transferred momentum  and for $q=k_F$ we observe some  differences especially in the isovector channel.

\subsection{\it Beyond Landau approximation}

The strict Landau limit means $q=0$, and consequently the response functions  deduced in the previous subsection should be applied only to small transferred momenta. A possible way to overcome this issue, but keeping the simplicity of the calculation, has been suggested in Ref.~\cite{mar05}.  The idea is to keep the full $q$-dependence of the direct term $V_D^{(\alpha,\alpha')}(q)$ of the ph interaction Eq.~(\ref{res}) and make the Landau approximation on the exchange term $V_E^{(\alpha,\alpha')}({\bf k}_1-{\bf k}_2)$ only. In other words, the hole momenta are restricted to the Fermi surface so that the ph interaction only depends on $q$ and $\cos \theta$, where $\theta$ is the angle between vectors $\hat{k}_1$ and $ \hat{k}_2$. In Ref.~\cite{mar05} this approach was dubbed LAFET, for Landau Approximation For Exchange Term. It is worth noticing that, from a somehow different perspective, a similar ph interaction has also been employed long ago in studies about pion propagation in nuclear matter~\cite{mey81}. We shall briefly describe first this approach concerning pions in nuclear matter and then show that LAFET can be somehow considered as a generalisation of it.

\subsubsection{\it Pionic modes and spin-isospin responses}

Spin-isospin excitations are the ideal arena to observe pionic effects in nuclei~\cite{ose82,ericson-weise}. Migdal~\cite{mig78} suggested the possibility of a pion condensation in nuclear matter. The effect was predicted for densities typical of neutron stars, that is much higher that the nuclear saturation density, so that it is not expected to occur in ordinary nuclei. However, some authors~\cite{gyu77,eri78,tok79} suggested that precursor phenomena could be seen in nuclei as a pole in the pion propagator at energies close to zero. In particular, the authors of Ref.~\cite{alb82} proposed investigating to the nuclear response functions in the different spin-isospin channels as a possible source of evidence of the precursor phenomena. 

\begin{figure}[!h]
\begin{center}
\includegraphics[width=0.4\textwidth,angle=0]{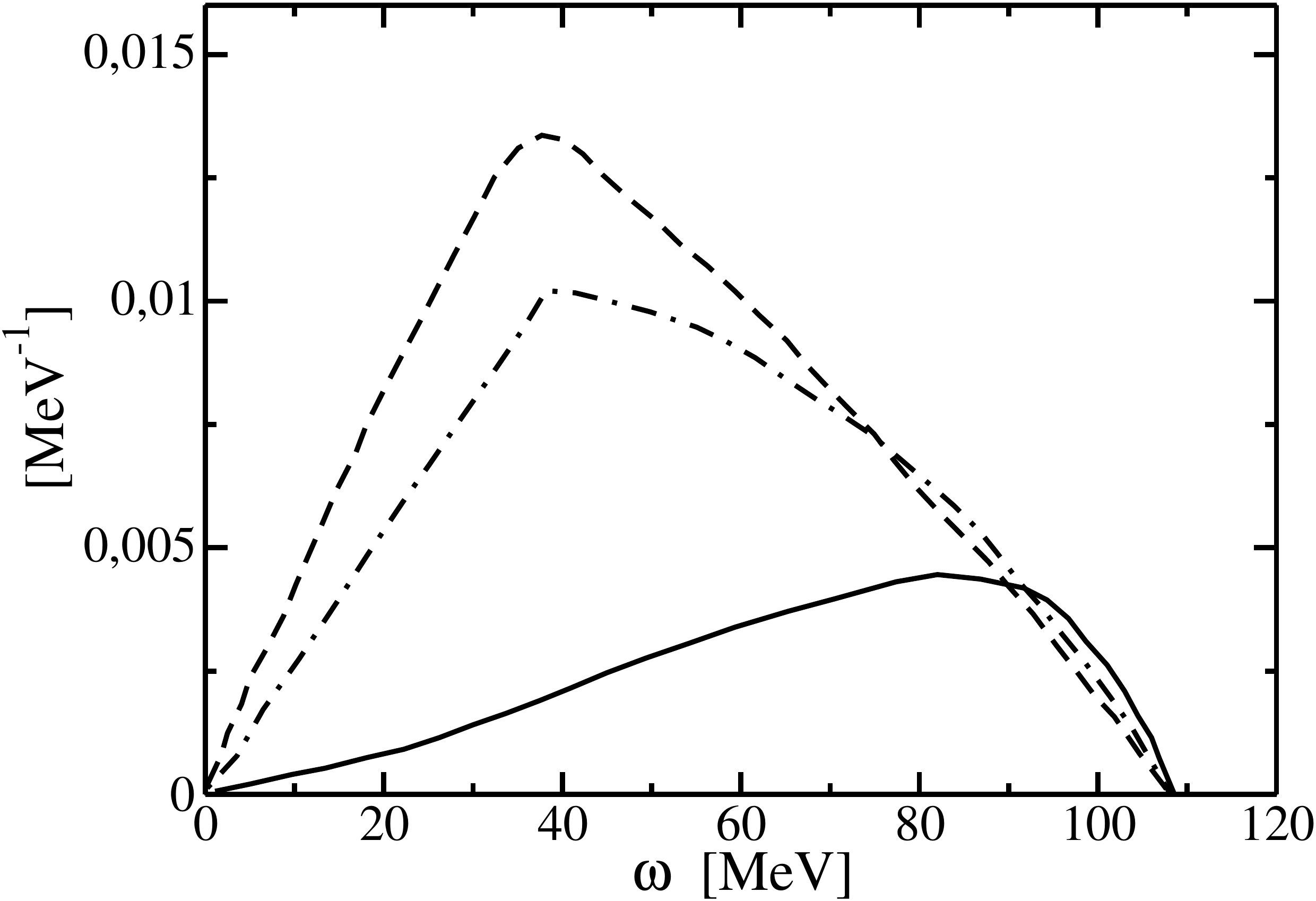}
\end{center}
\caption{The longitudinal (dashed line) and transverse (continuous line) strengths at $q=1.3$\,fm$^{-1}$. The HF strength function with $m^*=m$ is also shown for comparison (dot-dashed line). Adapted from Ref.~\cite{alb82}. }
\label{alberico-fig3}
\end{figure}

The ph interaction usually used to obtain these responses is given by the exchange of pion and rho mesons, which acts on the longitudinal and transverse channels, respectively. Shorter range effects are simulated by adding a phenomenological contact interaction of intensity $g'$:
\be
V_{ph} =  \frac{f_{\pi}^2}{m_{\pi}^2} \bigg\{ \frac{({\vsigma}_1 \cdot {\bf q}) ({\vsigma}_2 \cdot {\bf q})}{\omega^2-q^2-m_{\pi}^2} \, \Gamma_{\pi}^2+ C_{\rho} \frac{({\vsigma}_1 \wedge {\bf q}) ({\vsigma}_2 \wedge {\bf q})}{\omega^2-q^2-m_{\rho}^2} \, \Gamma_{\rho}^2 + g' ({\vsigma}_1 \cdot {\vsigma}_2) \bigg\} \, ({\vtau}_1 \cdot {\vtau}_2) \, ,
\label{pi-rho}
\ee
where $C_{\rho}= (f_{\rho}^2/ m_{\rho}^2) \, (m_{\pi}^2/f_{\pi}^2)$ and 
\be
\Gamma_{\pi,\rho}= \frac{\Lambda_{\pi,\rho}^2-m_{\pi,\rho}^2}{\Lambda_{\pi,\rho}^2-\omega^2+q^2}
\ee
 are the monopole factors of the $\pi N$,  $\rho N$ vertex. The zero-range term corresponds formally to the Landau parameter $g'_0$,  whose value was adjusted to $G'_0 \simeq 0.6-0.7$. A monopole form factor is usually also included in the first two terms. 

The precursor phenomena of pion condensation could appear around the so-called quasi-elastic peak. The authors of Ref.~\cite{alb82} observed that the longitudinal response is softened and enhanced with respect to the free Fermi gas, while the transverse response in quenched and hardened. In Fig.~\ref{alberico-fig3} are displayed their results for the strength function per nucleon at $q=1.3$\,fm$^{-1}$,  using the pion-exchange interaction of Eq.~(\ref{pi-rho}) with $g'=0.7$.

Some authors~\cite{spe80} have also considered an empirical $q$-dependence on parameter $g'$. Tensor effects were also discussed by adding an empirical zero-range interaction $h' S_T(\hat{\bf k}_{12}) ({\vtau}_1 \cdot {\vtau}_2)$, whose intensity is related to the Landau parameter $h'_0$. Some debates emerged on the physical origin of these parameters. For instance, Dickhoff \cite{dic83} argued against a short-range origin of them, stating instead that they are simply related to the exchange terms. The question has also been raised whether these parameters are universal or if one should consider different parameters corresponding to the $NN$, $N\pi$,... interactions. These topics are far from the scope of the present work, and we refer the interested reader to Ref.~\cite{ich06} for a general discussion about them. Our aim is simply to stress that the finite-range ph interaction employed in these works contains the direct term plus a monopole ($\ell=0$) Landau parameter pertinent to the specific spin-isospin mode and allows to obtain the response functions in the ring approximation. In fact, Landau parameters with $\ell>0$ can also be easily included in this method, thus relating $g'$ to the exchange term as suggested by Dickhoff. This is precisely the viewpoint we are going to discuss now.

\subsubsection{\it Landau approximation for exchange term}

Let us first consider a ph interaction restricted to its central and density-dependent terms, that is Eqs.~(\ref{Appb:central}-\ref{Appb:density}) or (\ref{central-Nakada}-\ref{density-Nakada}). As already done in the Landau approximation, it can be expanded on a series of Legendre polynomials as
\be
V^{(\alpha,\alpha')}_{LAFET} = \delta_{\alpha,\alpha'} n_d \sum_{\ell} f_{\ell}^{(\alpha)} P_{\ell}(\cos \theta) 
\, ,
\label{lafet-central}
\ee
where
\be
n_d f_\ell^{(\alpha)}=\delta_{\ell,0} \left\{ D^{(\alpha)} F_C(q)+R^{(\alpha)} \right\} + E^{(\alpha)}J_\ell^C\;,
\ee
with
\be
J_\ell^C = \frac{2 \ell+1}{2} \int_{-1 }^1 {\rm d}x F_C\left(\sqrt{2 k_F^2(1-x)}\right) \, P_{\ell}(x)\;.
\ee
As the direct term only contributes to the $\ell=0$ term, this means that we are dealing with a generalized Landau parameter $f^{(\alpha)}_0(q)$ depending on $q$. Actually, such a generalisation has been employed in the calculation of the response function of liquid $^3$He and related issues~\cite{pin87,wei91,gar92,bar93,barr96}. 

For the Gogny interaction, with $F_C$ given in Eq.~(\ref{FC:gogny}), the functions $J_{\ell}^C$ are obtained from the recurrence relation
\be
J_{\ell+1}^C = \frac{2 \ell+3}{z^2} \left\{ \frac{z^2}{2 \ell-1} J_{\ell-1}^C - 2 J_{\ell}^C \right\}\;,
\ee
with $z=k_F\mu$, and
\begin{eqnarray}
J_0^C&=&\frac{1}{z^2}\left[ 1-e^{-z^2}\right]\;,\\
J_1^C&=&\frac{3}{z^2}\left[ \left(1-\frac{2}{z^2} \right)+\left( 1+\frac{2}{z^2}\right)e^{-z^2}\right]\;.
\end{eqnarray}

Similarly, for the Nakada interaction, the function $F_C$ is given in Eq.~(\ref{FC:nakada}) and the recurrence relation for the functions $J_{\ell}^C$ is
\be
J_{\ell+1}^C = \frac{2 \ell+3}{\ell+1} \frac{1}{2 z^2} 
\left\{ (1+2z^2) J_{\ell}^C - \frac{\ell}{2 \ell-1} \, 2 z^2 J_{\ell-1}^C \right\} \;,
\ee
with $z=k_F/\mu$, and 
\begin{eqnarray}
J_0^C &=& \frac{1}{4 \mu^2 z^2} \ln(1+4 z^2) \\
J_1^C &=& \frac{3}{8 \mu^2 z^2} \left\{ -4 z^2 + (1+2z^2) \ln(1+4 z^2) \right\} \;.
\end{eqnarray}
Keep in mind that an implicit sum over the ranges should be understood in these formulae for both types of interaction.

\begin{figure}[!h]
\begin{center}
\includegraphics[width=0.45\textwidth,angle=0]{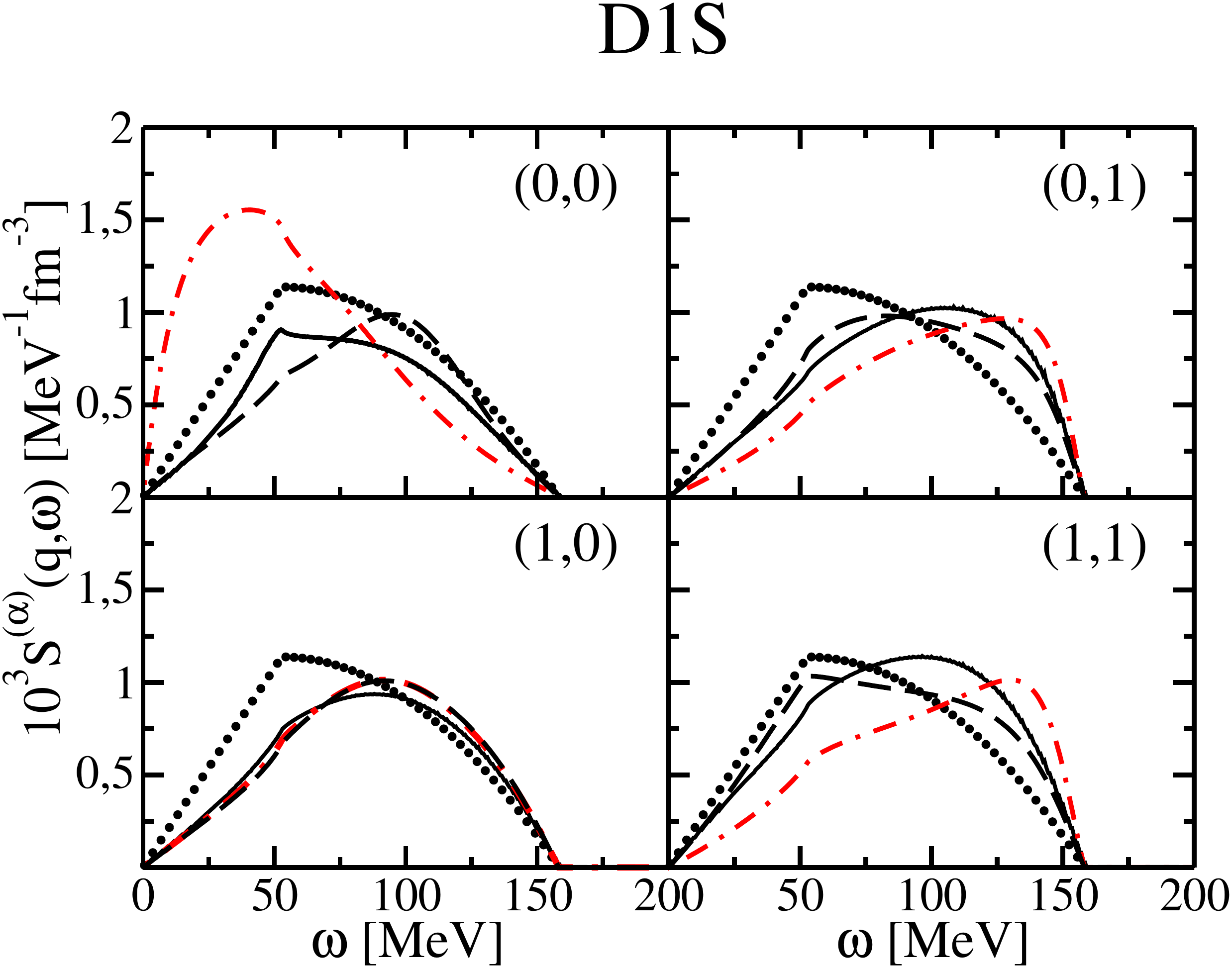}
\includegraphics[width=0.45\textwidth,angle=0]{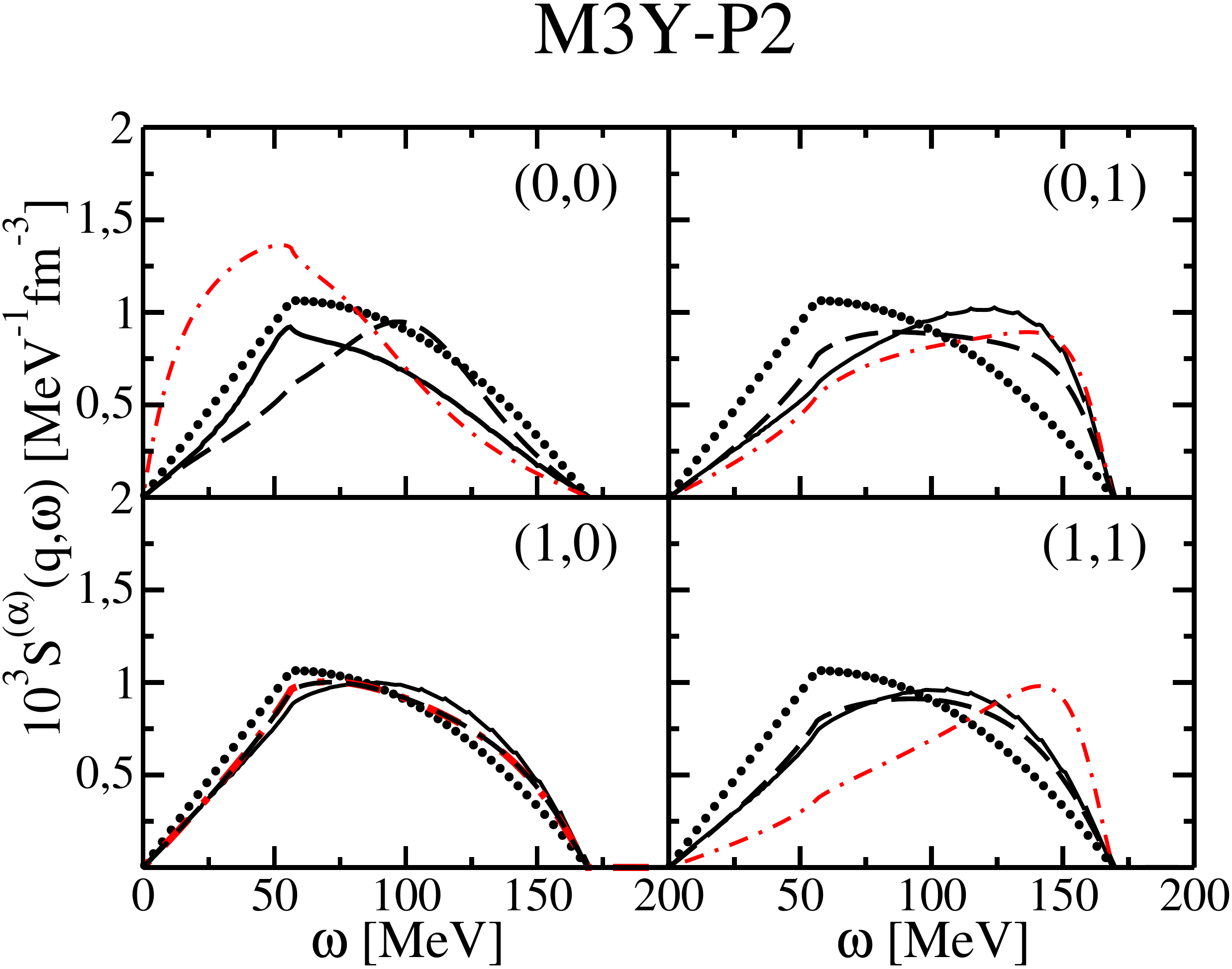}
\end{center}
\caption{Strength functions $S^{S,T}(q,\omega)$ (dashed lines) of D1S interaction and M3Y-P2 interaction (central term only) for $q/k_F=1$ at saturation density. For comparison are also plotted the Landau (dot-dashed lines, the HF (dotted lines) and the numerically converged (solid lines) strengths. Adapted from Ref.~\cite{mar05}.}
\label{Fig:LAFET:originale}
\end{figure}

In Fig.~\ref{Fig:LAFET:originale}, are displayed the LAFET strength functions at saturation density, based on interactions D1S and M3Y-P2 (only central part). For comparison, the (converged) Landau and HF strengths are also displayed. The solid curves in all $(S,T)$ panels correspond to the numerically exact results which are discussed in Sect.~\ref{sec:multipole}. 
LAFET and Landau results include up to $\ell=3$. Since at small transferred momentum, the differences between LAFET and Landau approximation are small, the value $q=k_F$ has been chosen in these figures.  One can see that, with the exception of channel (1,0), LAFET strengths differ significatively from the Landau ones. Actually, LAFET results are closer to the exact ones, in particular for the $(0,0)$ channel. Hence, the LAFET is indeed a very simple extension of the Landau approximation at momentum transfers where the latter does not give reliable results. In principle, one does not expect reliable LAFET results for higher values of $q$. However, LAFET can be useful for extensive calculations when the numerical effort required by exact calculations becomes heavy.

\section{A multipolar expansion}\label{sec:multipole}

We present now a general method to solve directly the Bethe-Salpeter equation in the three-dimensional momentum space, proposed in Ref.~\cite{mar05}. It consists in expanding the Green's functions and the ph interaction on a complete basis of spherical harmonics, thus transforming Eq.~(\ref{bethe-salpeter}) into a set of coupled integral equations on the momentum modulus variable. The interest is that this is an exact method, with no approximations, provided the expansion is pushed up to the required degree of precision/convergence. 

\subsection{\it The method}

The multipolar expansion of the HF Green's function is simple because it has no dependence on the azimuthal angle $\varphi_1$ of the vector ${\bf k}_1$, so that only the $M=0$ component appears in the expansion:
\be
G^{HF}(q, \omega, {\bf k}_1) = \sum_L G^{HF}_{L}(q, \omega, k_1) Y_{L0}(\hat{k}_1) \;.
\label{mult-GHF}
\ee
In Ref.~\cite{mar05} a ph-interaction containing only central and density-dependent components was considered to illustrate the method. In that case, the exchange term depends on the modulus $| {\bf k}_1-{\bf k}_2|$ and therefore its expansion is as follows
\be
V^{(\alpha,\alpha')}_{C;DD \; ph}(q, {\bf k}_1, {\bf k}_2) =  \delta(\alpha, \alpha') \sum_{LM} V^{(\alpha)}_L(q, k_1, k_2) Y^*_{LM}(\hat{k}_1) Y_{LM}(\hat{k}_2) \;.
\label{mult-VphC}
\ee
Inserting these multipolar expansions onto the Bethe-Salpeter equation~(\ref{bethe-salpeter}) one immediately verifies that the expansion of the RPA Green's function only contains $M=0$ components
\be
G^{(\alpha)}(q, \omega, {\bf k}_1) = \sum_L G^{(\alpha)}_{L}(q, \omega, k_1) Y_{L0}(\hat{k}_1) \;.
\ee

When the ph-interaction includes non-central terms, $M \neq 0$ components appear in the matrix elements of both the ph interaction and the RPA Green's function. Their multipolar expansions are thus generalized as
\be\label{multi:res:int}
V^{(\alpha,\alpha')}_{ph}(\mathbf{k_1},\mathbf{k_2})=\sum_{L,M,L',M'}F^{(\alpha,\alpha')}_{LM;L'M'}(k_1,k_2)Y_{LM}(\hat{k}_1)Y^*_{L'M'}(\hat{k}_2)\;,
\ee
and 
\be
G^{(\alpha)}(q,\omega,\mathbf{k}_1)=\sum_L G^{(\alpha)}_{LM}(q,\omega,{k}_1)Y_{LM}(\hat{k}_1) \;.
\ee
Replacing the different pieces of the initial Bethe-Salpeter equation with their multipolar expansion and integrating over the angles $(\theta_1, \phi_1)$ one gets the following system of coupled integral equations for the RPA multipoles
\bea
G^{(\alpha)}_{\ell m}(q,\omega,k_1) & = & G^{HF}_{\ell}(q,\omega,k_1)\delta_{m,0}+\sum_{L,L',\ell',m'} G^{HF}_L(q,\omega,k_1)\frac{\hat{\ell}\hat{L}\hat{L}'}{\sqrt{4\pi}} \left(\begin{array}{ccc} 
\ell&L&L'\\
0&0&0
\end{array} \right) \left(\begin{array}{ccc} 
\ell&L&L'\\
-m&0&m
\end{array} \right) \nonumber\\
& & \quad \int \sum_{\alpha'}\frac{d k_2 k^2_2}{(2\pi)^3}F^{(\alpha,\alpha')}_{L'm;\ell' m'}(k_1,k_2)(-1)^{m}G_{\ell' m'}^{(\alpha')}(q,\omega,k_2) \, ,
\label{full-set}
\eea
where 
\be
F^{(\alpha,\alpha')}_{LM;L'M'}(k_1,k_2,q)=\int d\hat{k}_1d\hat{k}_2 Y^*_{LM}(\hat{k}_1)Y_{L'M'}(\hat{k}_2)V^{(\alpha,\alpha')}_{ph}(\mathbf{k_1},\mathbf{k_2},q) \;,
\label{eq:dvptV}
\ee
and we    used $3j$ coefficients and the standard notation $\hat{J}=\sqrt{2J+1}$. 
The complete expressions of these matrix elements are given in Appendix~\ref{app:B} for Skyrme, Gogny and Nakada interactions. The central and density-dependent terms of the interaction led to diagonal matrix elements in the indices $\alpha, L, M$.  Including spin-orbit and tensor terms lead to non-diagonal components, in particular there are the couplings $L=L'\pm1$ and $L=L'\pm2$

Finally, the response function is given by the momentum integral
\be
\chi^{(\alpha)}(q,\omega) = \frac{\sqrt{4 \pi}}{(2 \pi)^3} n_d \int G^{(\alpha)}_{00}(q, \omega, k) \, k^2 {\rm d}k \;, 
\ee
so that only the $\ell=0$ multipole of the RPA Green's function is required. However, one has to solve the full system of coupled equations, because the interaction couples different multipoles. In practice, a complete calculation implies the choice of a cutoff value $\ell_{max}$ for the summations on angular momenta and a grid of points in momentum space to transform the integrals into discrete sums. The system (\ref{full-set}) can be finally solved by using a matrix inversion whose size is typically 1800 x 1800.

We mention here two useful tests for the numerical matrix inversion, both based on a zero-range ph interaction. The first consists in using a Skyrme interaction, which has only three possible values for $\ell = 0, 1$ and 2, the latter entering only if a tensor interaction is included. The response function can be analytically obtained if non-central interaction terms are excluded, while a relatively simple numerical procedure has to be applied otherwise. The second test is based on the Landau approximation of a finite-range interaction, since we can limit the number of multipoles up to a precise order. The response function can be obtained as explained in Sect.~\ref{subsec:landau}. In both cases, one has an independent way to obtain the response function. See Ref.~\cite{report} for the details.

\subsection{\it Convergence}\label{sec:convergence}

\begin{figure}[!h]
\begin{center}
\includegraphics[width=0.8\textwidth,angle=0]{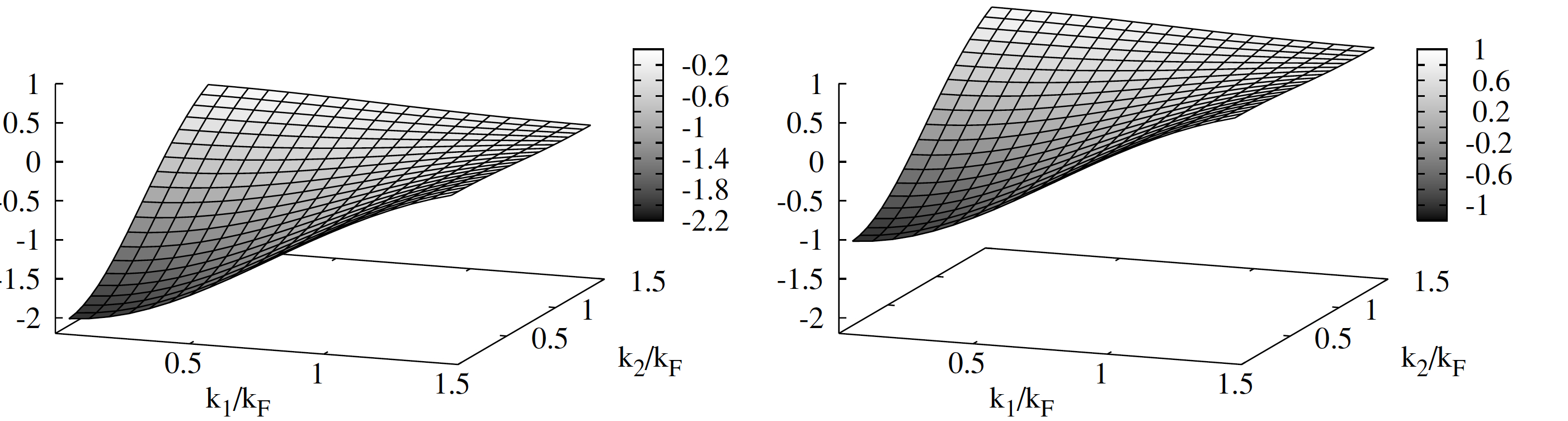}
\end{center}
\caption{The dimensionless function $\tilde{V}^{0,0}_{0,C}(q,k_1,k_2)$ of the Gogny D1S interaction as defined in Eq.~(\ref{eq:mono:dimensionless}) as a function of the momenta $k_1/k_F$ and $k_2/k_F$ at $\rho=\rho_0$. Left  and right panels are for $q=0.1 k_F$ and $q=k_F$, respectively. Taken from Ref.~\cite{mar06}.}
\label{fig1-Gogny-monopole}
\end{figure}

\begin{figure}[!h]
\begin{center}
\includegraphics[width=0.8\textwidth,angle=0]{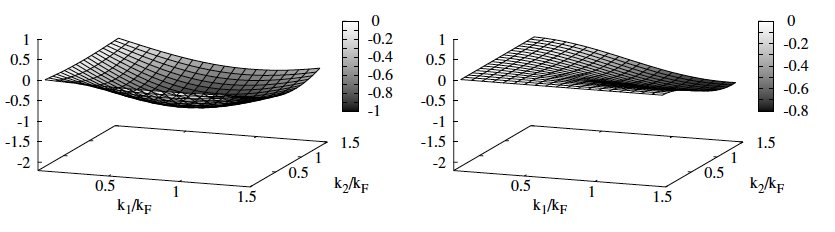}
\end{center}
\caption{The dimensionless functions $\tilde{V}^{0,0}_{L,C}(q,k_1,k_2)$ of the Gogny D1S interaction as defined in Eq.~(\ref{eq:mono:dimensionless}) as a function of the momenta $k_1/k_F$ and $k_2/k_F$ at $\rho=\rho_0$. The left panel represents the case $L=1$ and the right one $L=2$. Taken from Ref.~\cite{mar06}.}
\label{fig2-Gogny-L12}
\end{figure}

\begin{figure}[!h]
\begin{center}
\includegraphics[width=0.32\textwidth,angle=0]{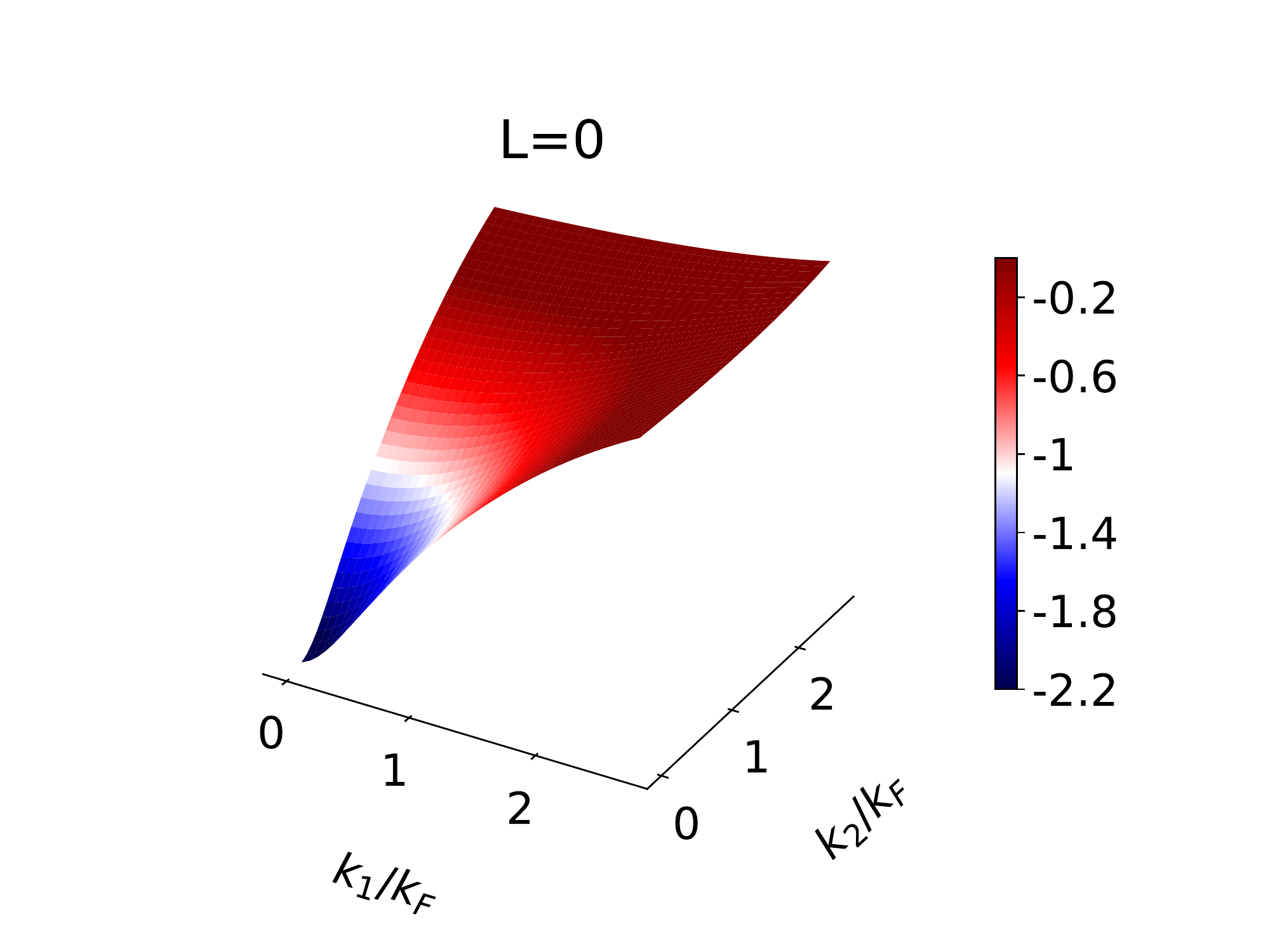}
\includegraphics[width=0.32\textwidth,angle=0]{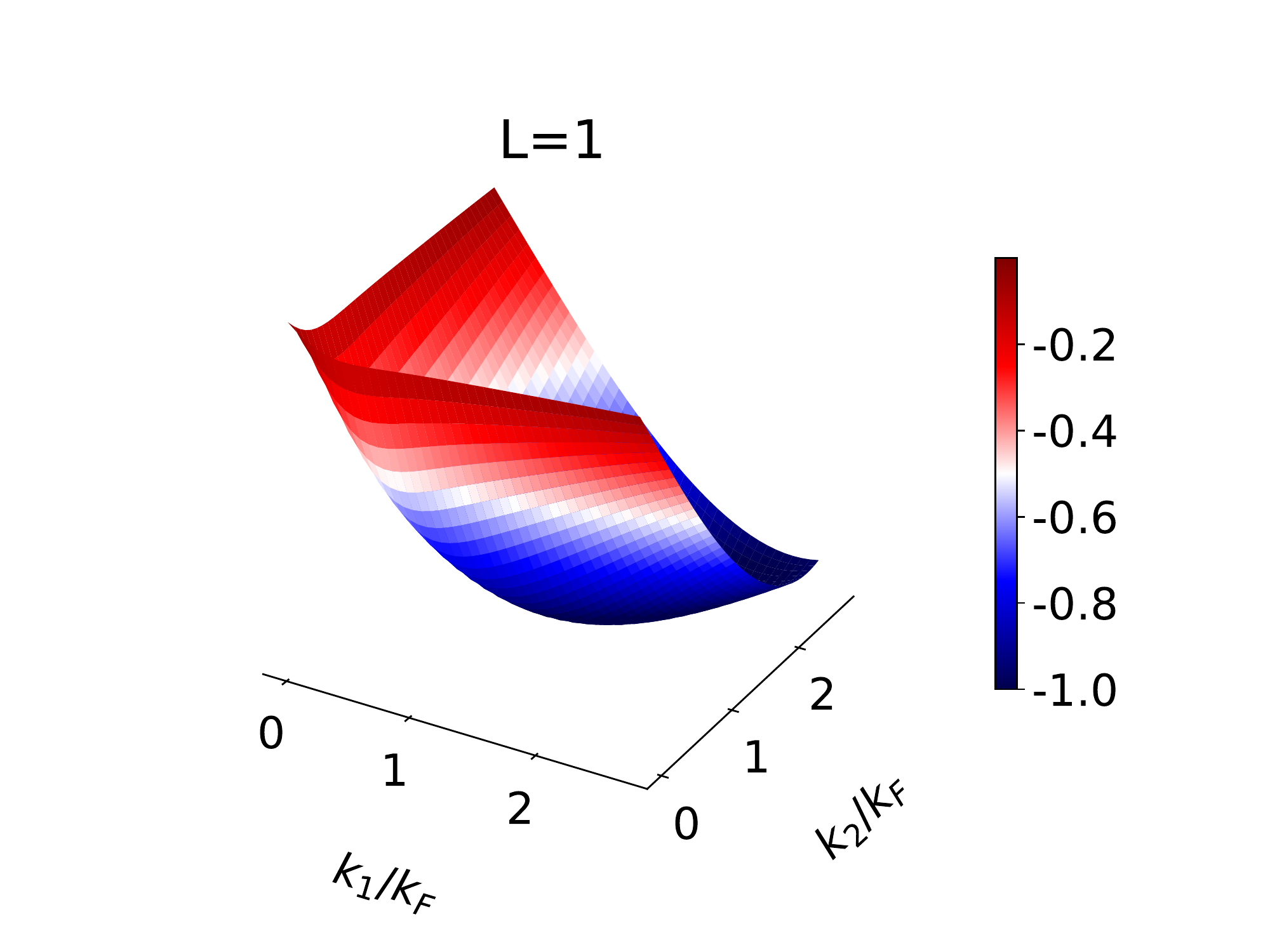}
\includegraphics[width=0.32\textwidth,angle=0]{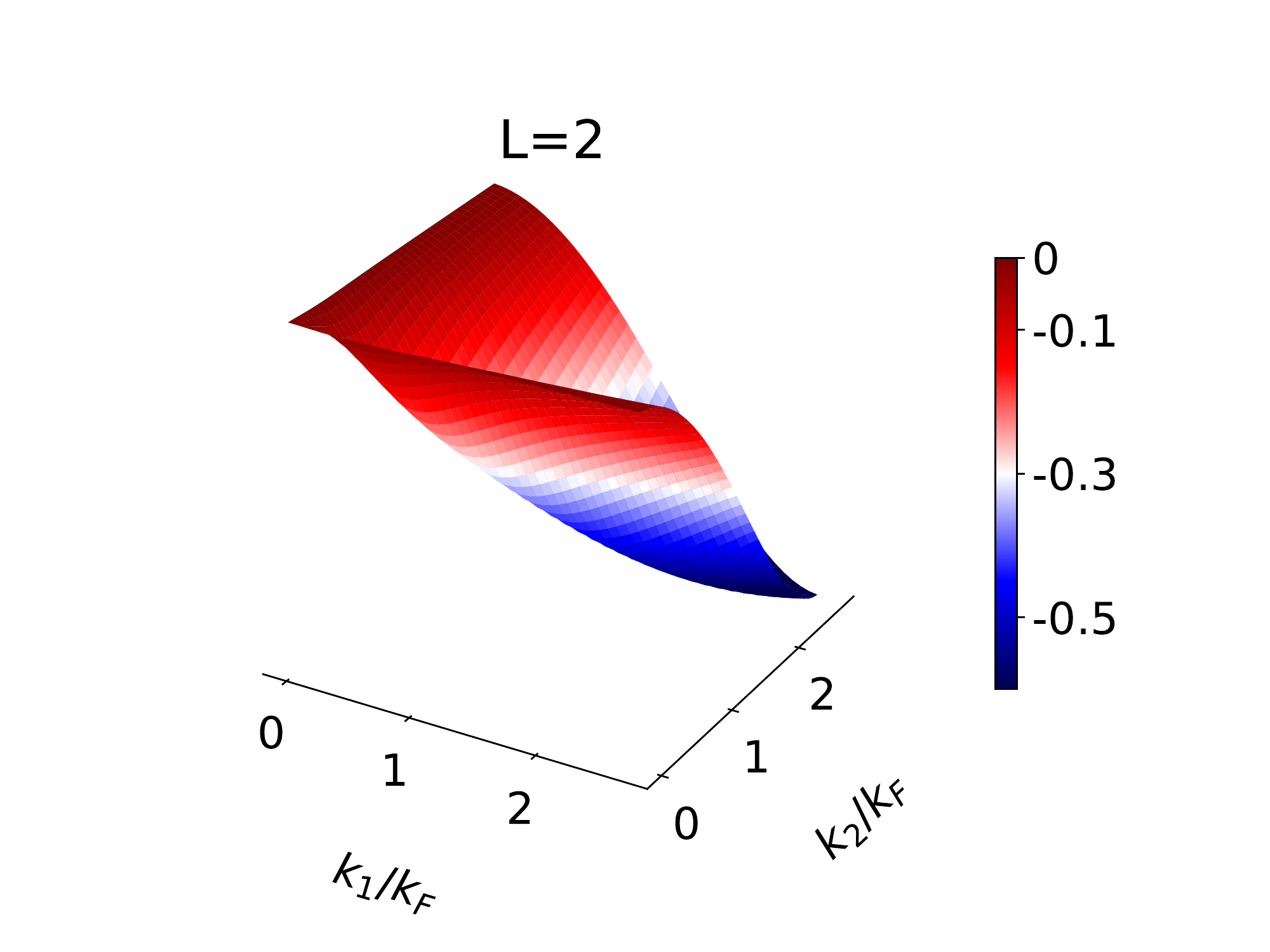}
\end{center}
\caption{The dimensionless functions $\tilde{V}^{0,0}_{L,C}(q,k_1,k_2)$ of the Nakada M3Y-P2 interaction as defined in Eq.~(\ref{eq:mono:dimensionless}) as a function of the momenta $k_1/k_F$ and $k_2/k_F$ at $\rho=\rho_0$. The left panel represents the case $L=0$, the central panel $L=1$ and the right one $L=2$. The transferred momentum is $q=0$ in all cases.}
\label{fig1-M3Y-monopole}
\end{figure}

The convergence of the method depends on the values of the interaction matrix elements (\ref{eq:dvptV}) in the momentum space $(q, k_1, k_2)$. This issue was analysed in Ref.~\cite{mar06} for the case of an interaction with only central and density-dependent terms, by defining the dimensionless functions
\begin{equation}
\tilde{V}^{\alpha,\alpha}_{\ell,C}(q,k_1,k_2)=\frac{2\ell+1}{4\pi }\frac{m^* k_F}{2\pi^2}
F^{(\alpha,\alpha)}_{\ell 0;\ell 0}(k_1,k_2,q) \, ,
\label{eq:mono:dimensionless}
\end{equation}
which, for the values $q=0, k_1=k_F, k_2=k_F$ gives the familiar dimensionless Landau parameters.
In Fig.~\ref{fig1-Gogny-monopole} are plotted the $\ell=0$ functions at saturation density in the plane $(k_1/k_F, k_2/k_F)$ of hole momenta, for two values of the transferred momentum $q$, namely $0.1 k_F$ and $k_F$.  The spin-isospin channel is $(0,0)$ and the interaction is D1S. One can appreciate that both surfaces simply differ by an overall translation. This can be understood from a glance at the explicit expressions of the multipoles given in Appendix \ref{app:B}. Indeed, the $q$-dependence enters as an additive term, independent of $k_1$ and $k_2$, which for the interaction considered produces an overall repulsion. The $\ell \neq 0$ functions are instead independent of $q$, and they become smaller as the value of $\ell$ is increased, as is displayed in Fig.~\ref{fig2-Gogny-L12} for the same D1S interaction at saturation density. Therefore, the convergence of this expansion does not depend on the value of $q$. The M3Y-P2 interaction shows a similar global behaviour, as displayed in Fig.~\ref{fig1-M3Y-monopole} for the values $\ell=0, 1, 2$ and zero transferred momentum. 

Although the figures refer to the $(0,0)$ channel, one gets similar results for the other $(S,I)$ channels.  
In conclusion, one has to introduce a cutoff $L_{max}$ to solve Eq.~(\ref{full-set}) and the previous figures suggest that a relatively small value of $L_{max}$ will suffice to obtain converged results, as far as non-central terms are ignored in the ph-interaction. We will show now that including such terms does not dramatically modify the convergence rate.

\subsection{\it Results}

\begin{figure}[!h]
\begin{center}
\includegraphics[width=0.3\textwidth,angle=0]{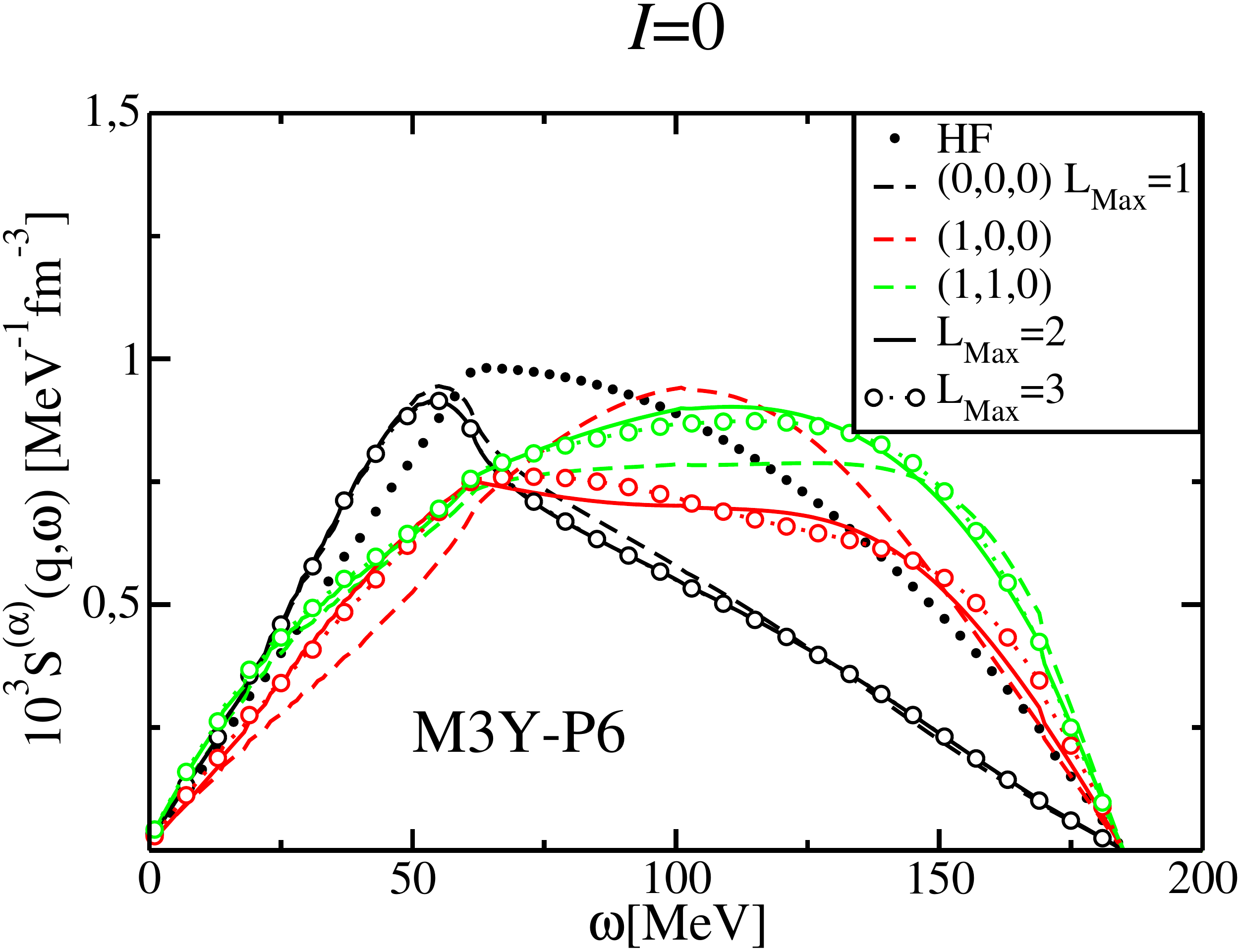}
\includegraphics[width=0.3\textwidth,angle=0]{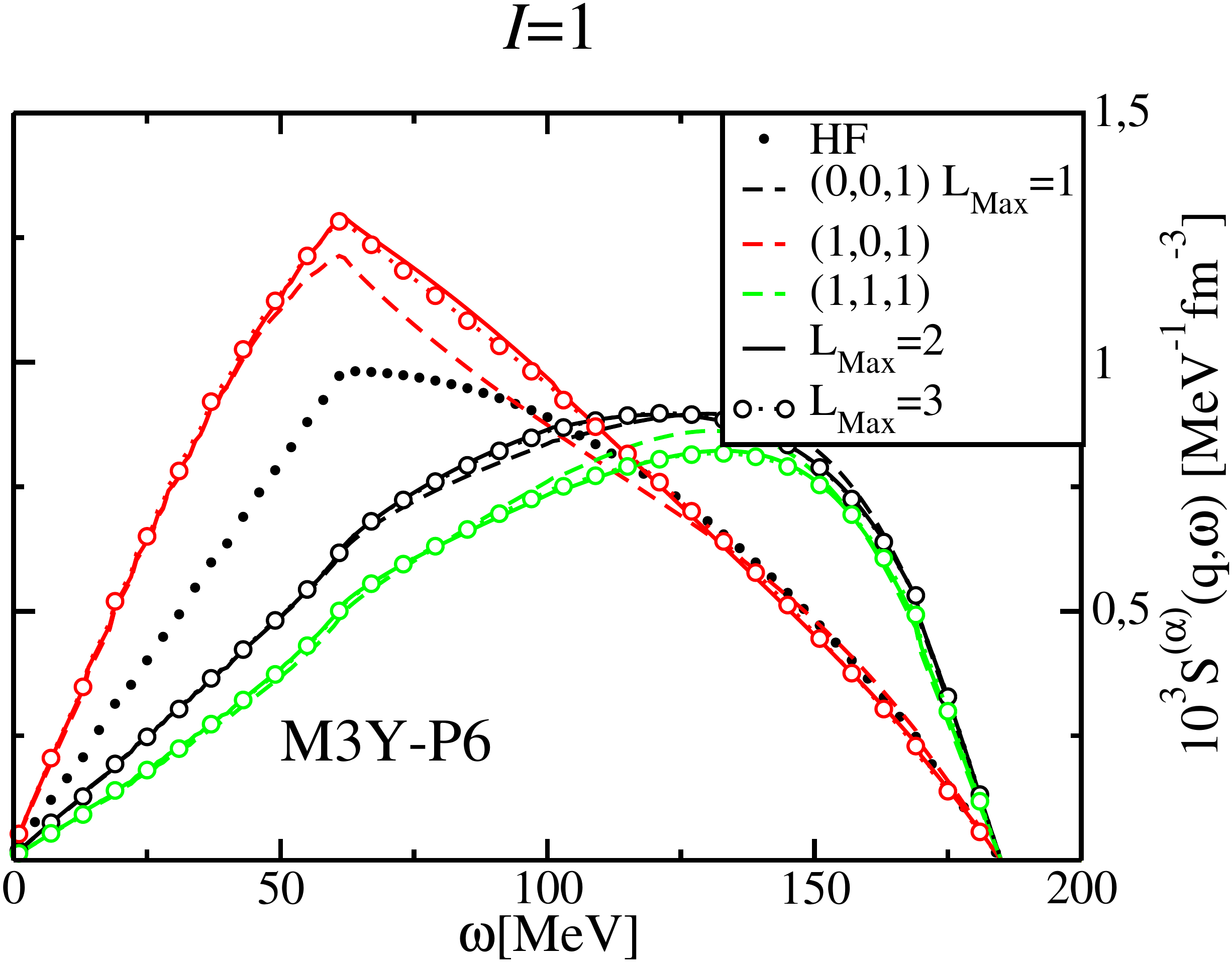}\\
\includegraphics[width=0.3\textwidth,angle=0]{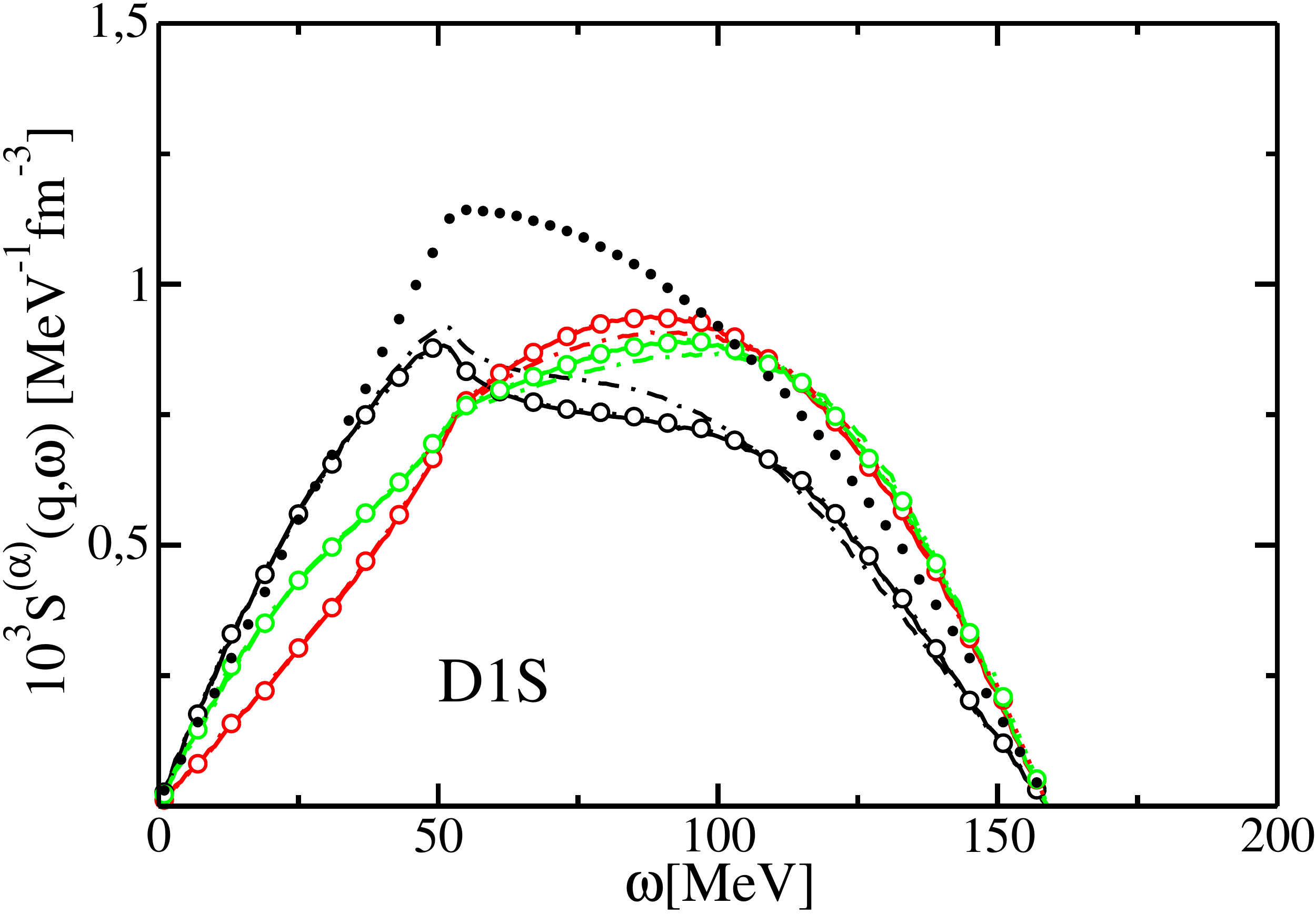}
\includegraphics[width=0.3\textwidth,angle=0]{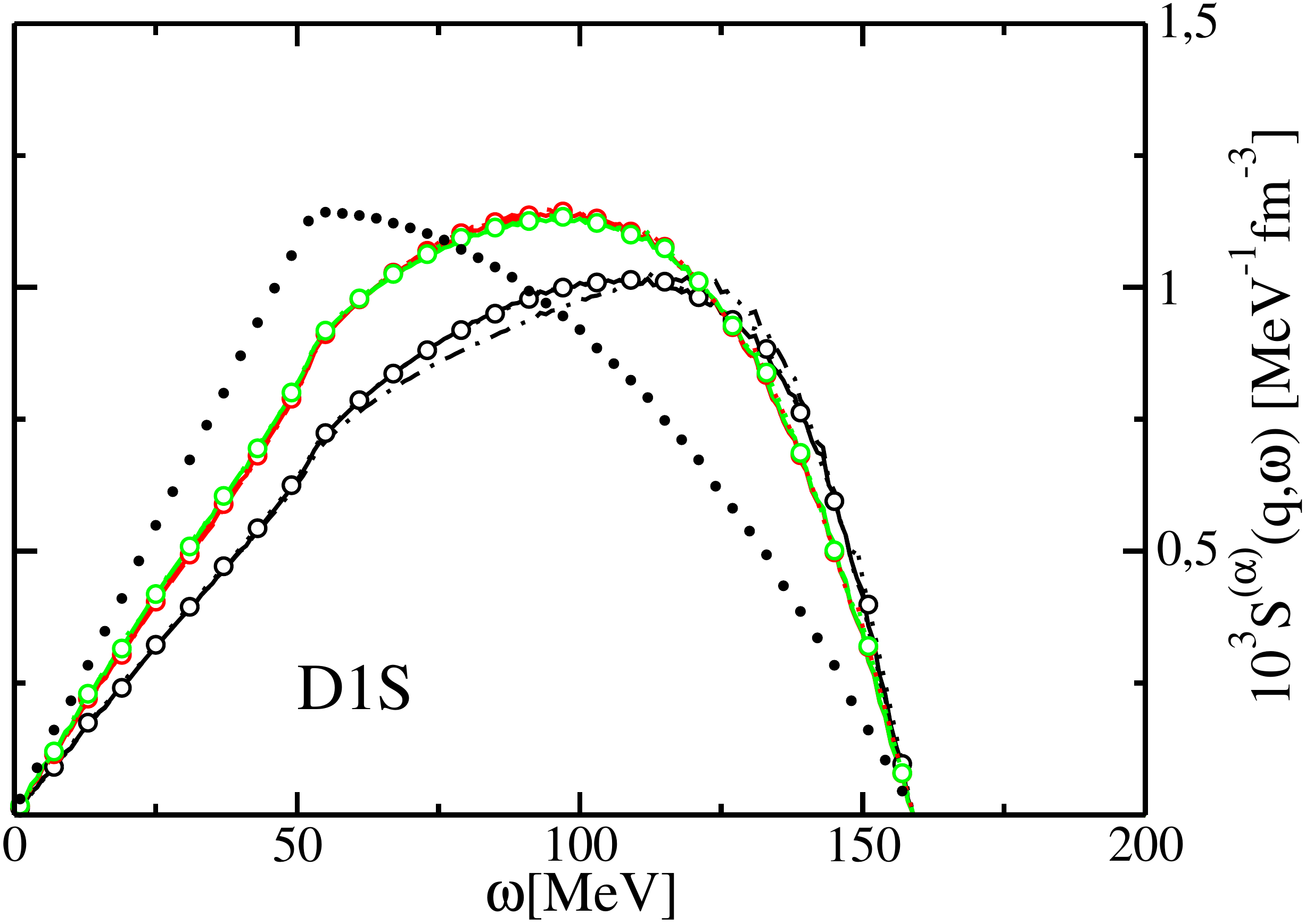}
\end{center}
\caption{Strength functions for the channels $(S,M,I)$ at $\rho=\rho_0$ and  transferred momentum $q/k_F=1$  for  Nakada M3Y-P6 interaction (upper panels)  and Gogny D1S (lower panel) as a function of the cutoff in the multipole expansion. The dots represent the HF response. See text for details.}
\label{Conv:qTOT}
\end{figure}

In Fig.~\ref{Conv:qTOT} are displayed the strength functions calculated for different values of the cutoff parameter $L_{max}$ with interaction D1S, which contains a spin-orbit component, and interaction M3Y-P6, which contains spin-orbit and tensor components. This choice allows one to draw some conclusions independent of the radial form factor and the ranges. The calculations have been done at  saturation density $\rho=\rho_0$ and transferred momentum $q/k_F=1$. 

We can notice that in the spin-isospin channel $(0,0,0)$, both interactions produce quite similar strength functions. In each isospin channels $(1,M,I)$ the $M$-strengths calculated with D1S interaction are very close one to each other ($I=0$) or barely distinguishable ($I=1$). These differences are solely due to the spin-orbit term of the interaction. In contrast, the strength functions in these channels are clearly different in the case of interaction M3Y-P6, due to its tensor component. We clearly see that  for both interactions, $L_{max}=2$ is sufficient to achieve an excellent convergence even in presence of an explicit tensor term. This confirms the findings that we have previously discussed. 

\begin{figure}[!h]
\begin{center}
\includegraphics[width=0.3\textwidth,angle=0]{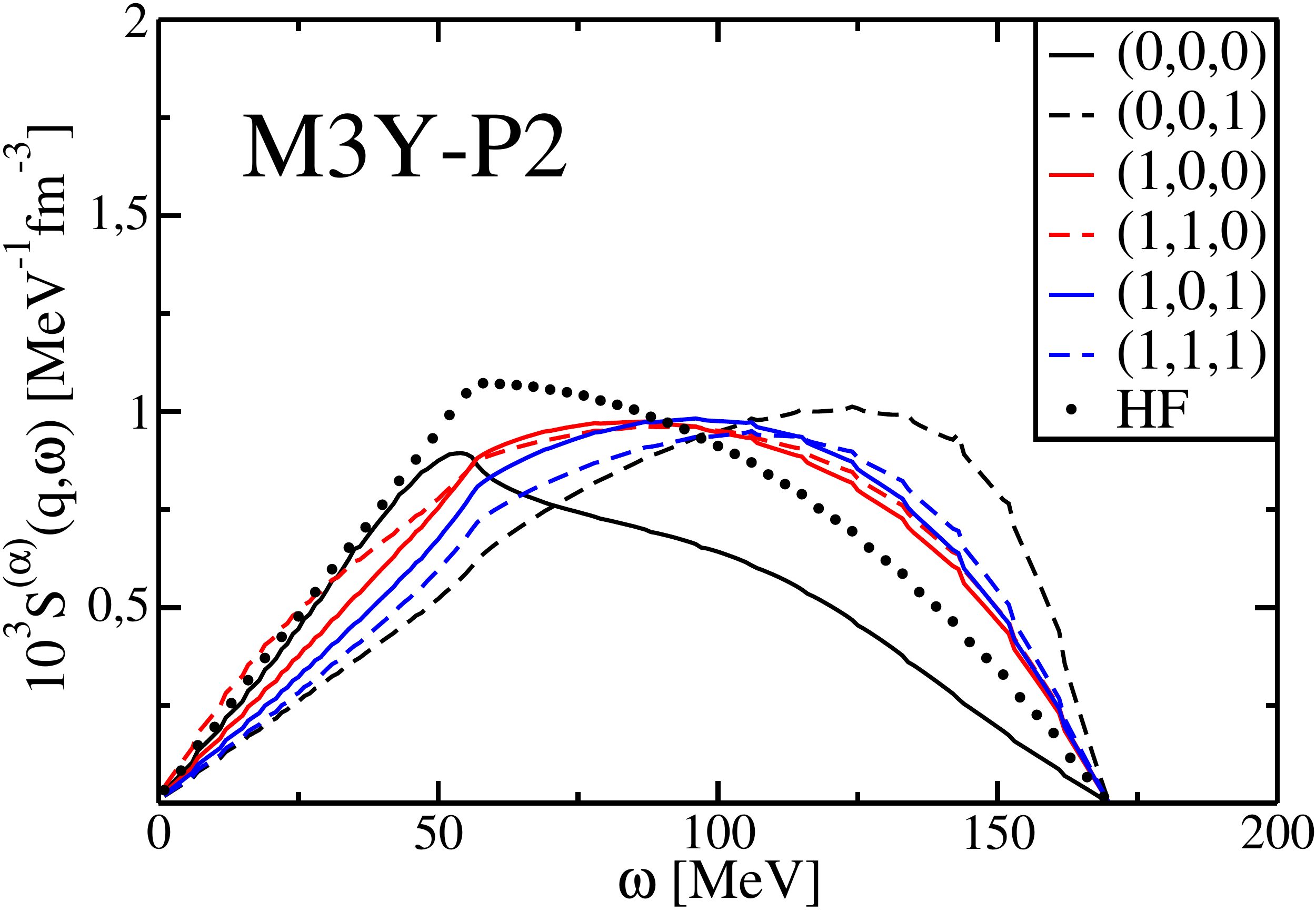}
\includegraphics[width=0.3\textwidth,angle=0]{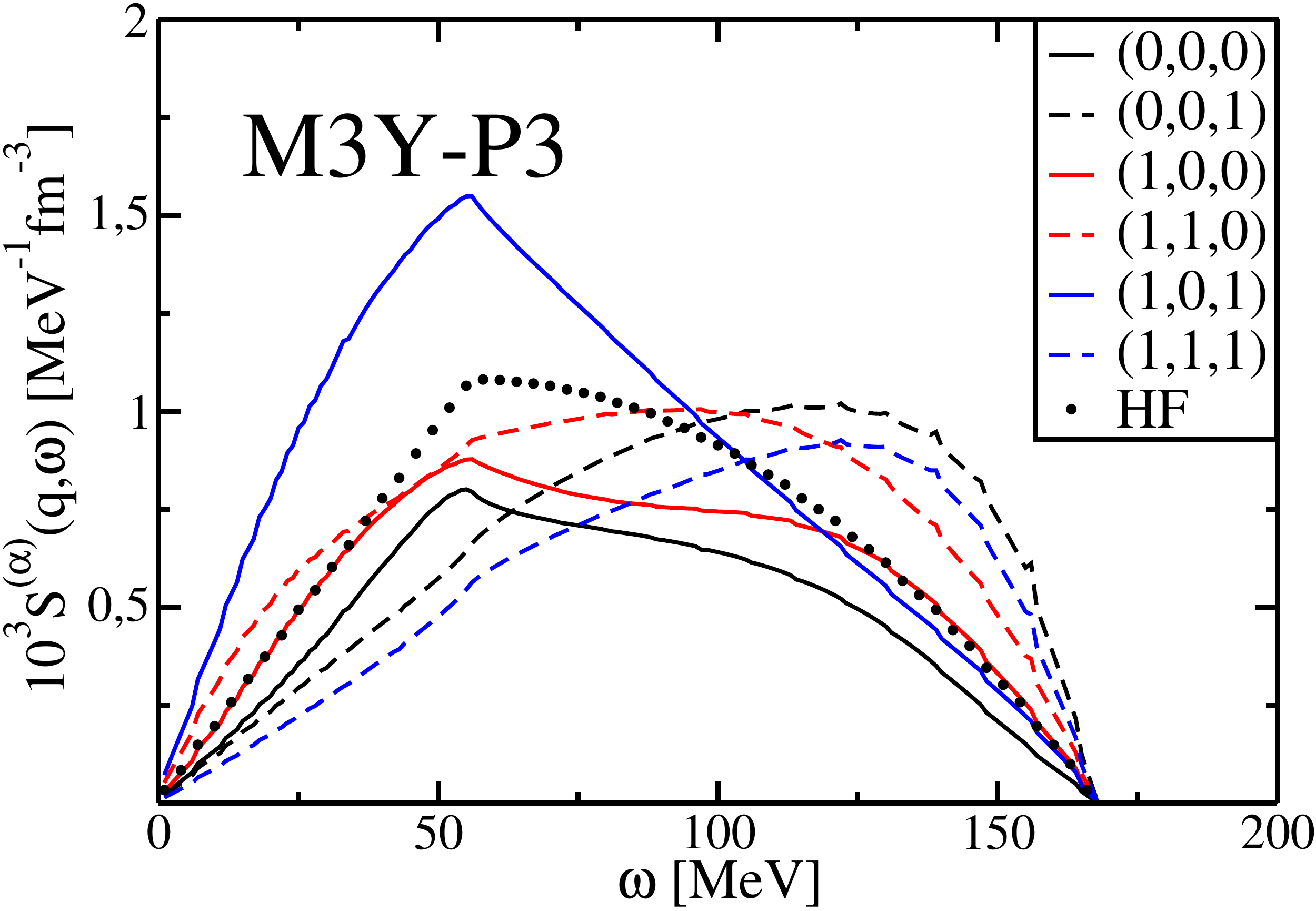}\\
\includegraphics[width=0.3\textwidth,angle=0]{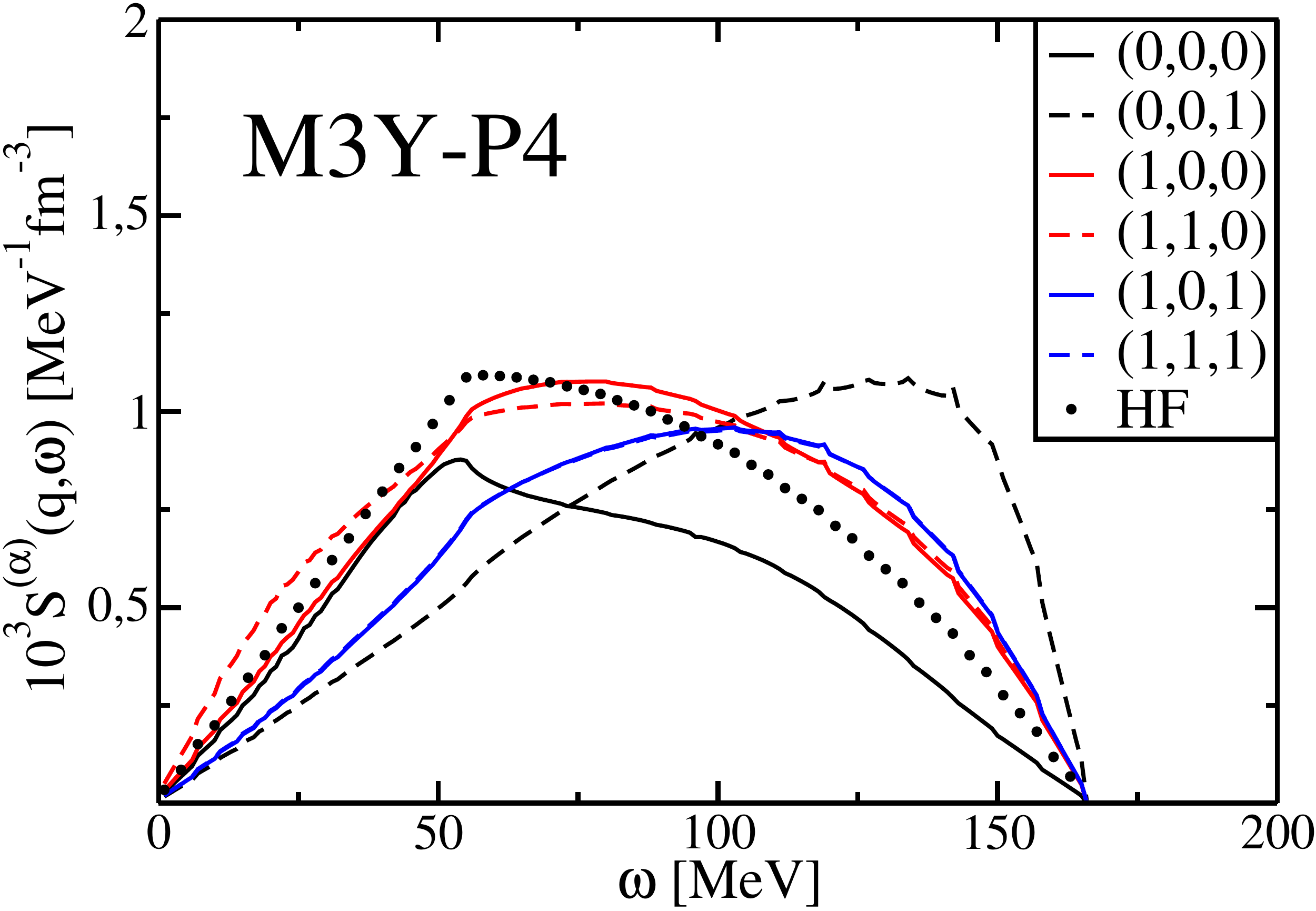}
\includegraphics[width=0.3\textwidth,angle=0]{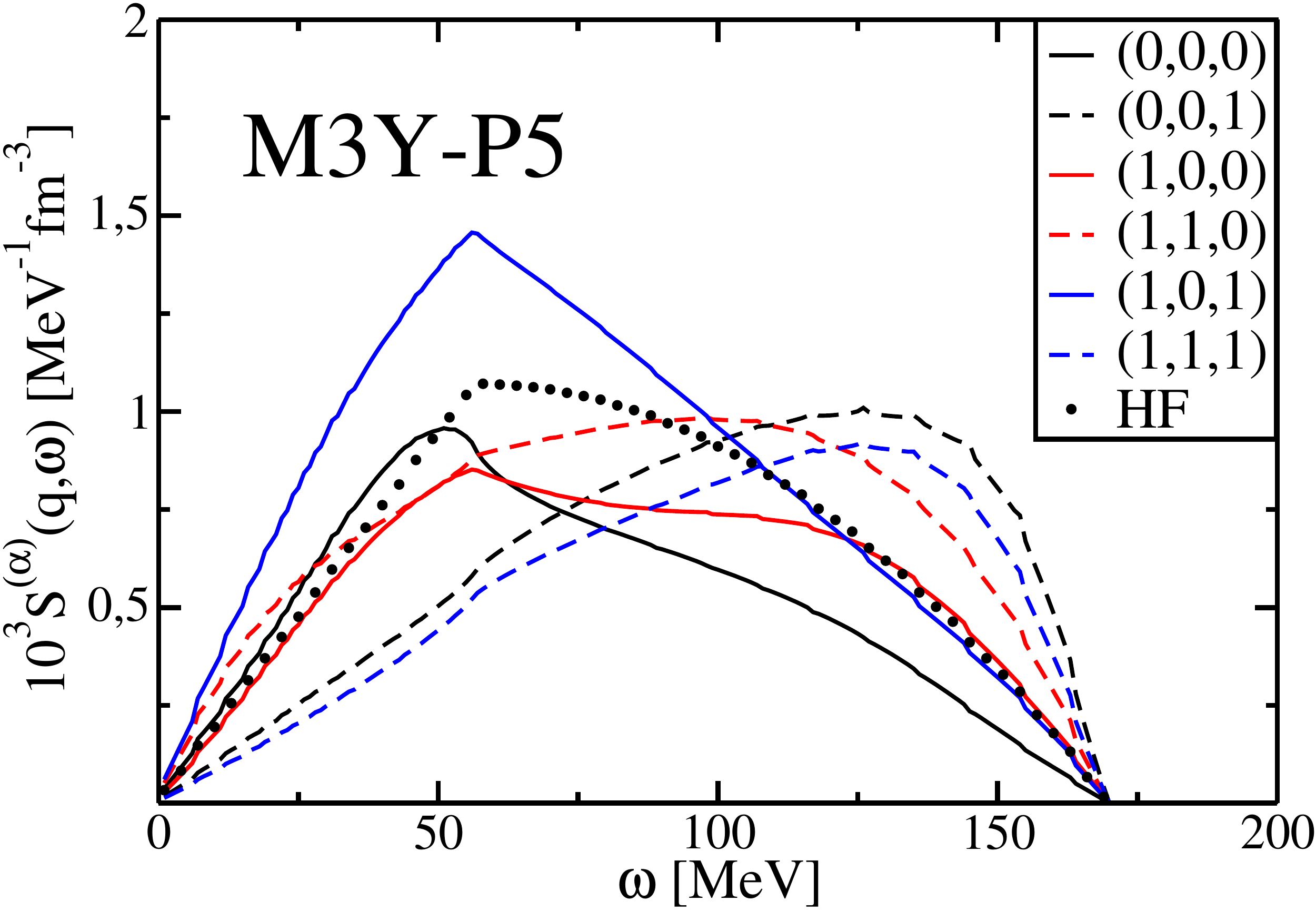}\\
\includegraphics[width=0.3\textwidth,angle=0]{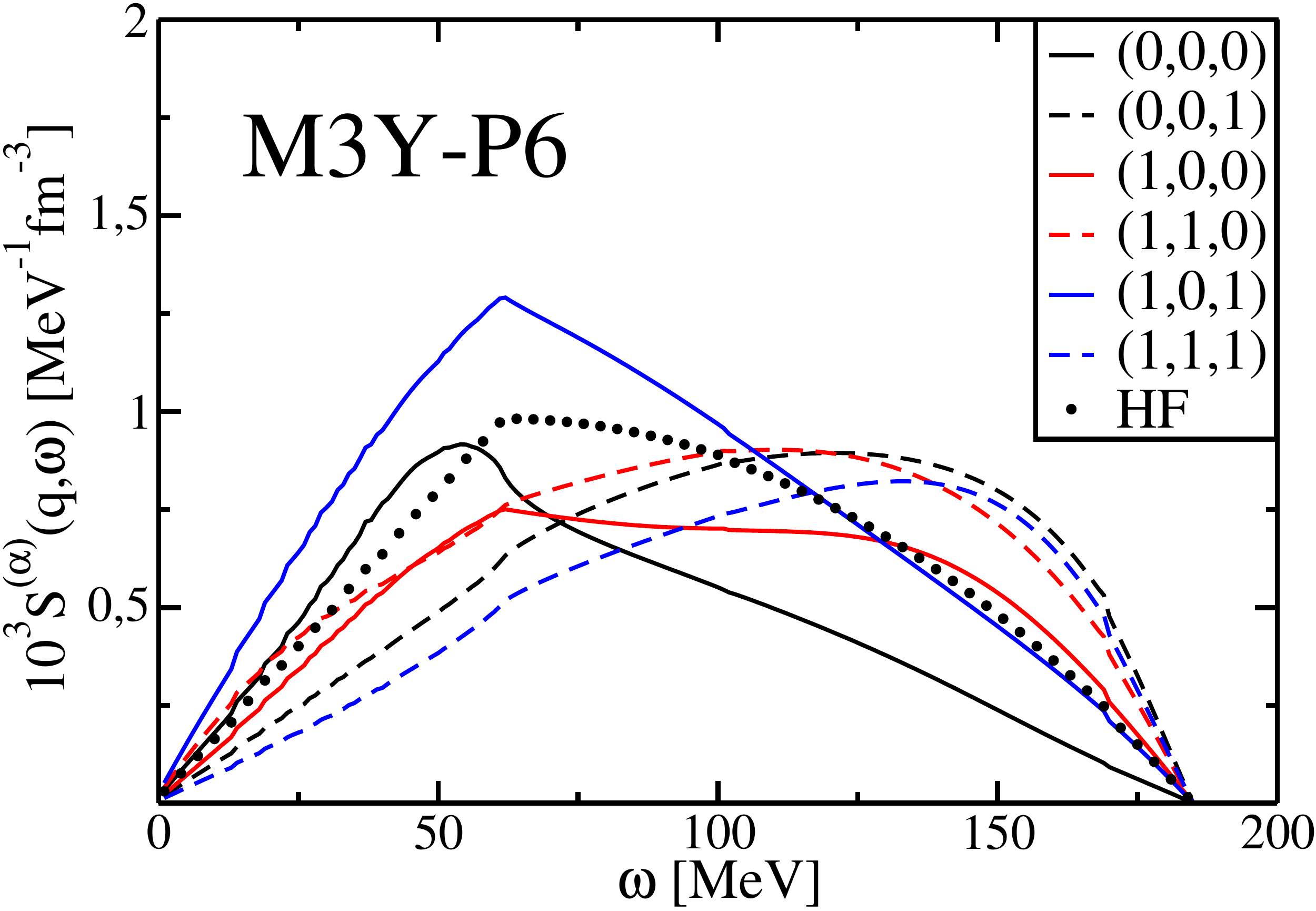}
\includegraphics[width=0.3\textwidth,angle=0]{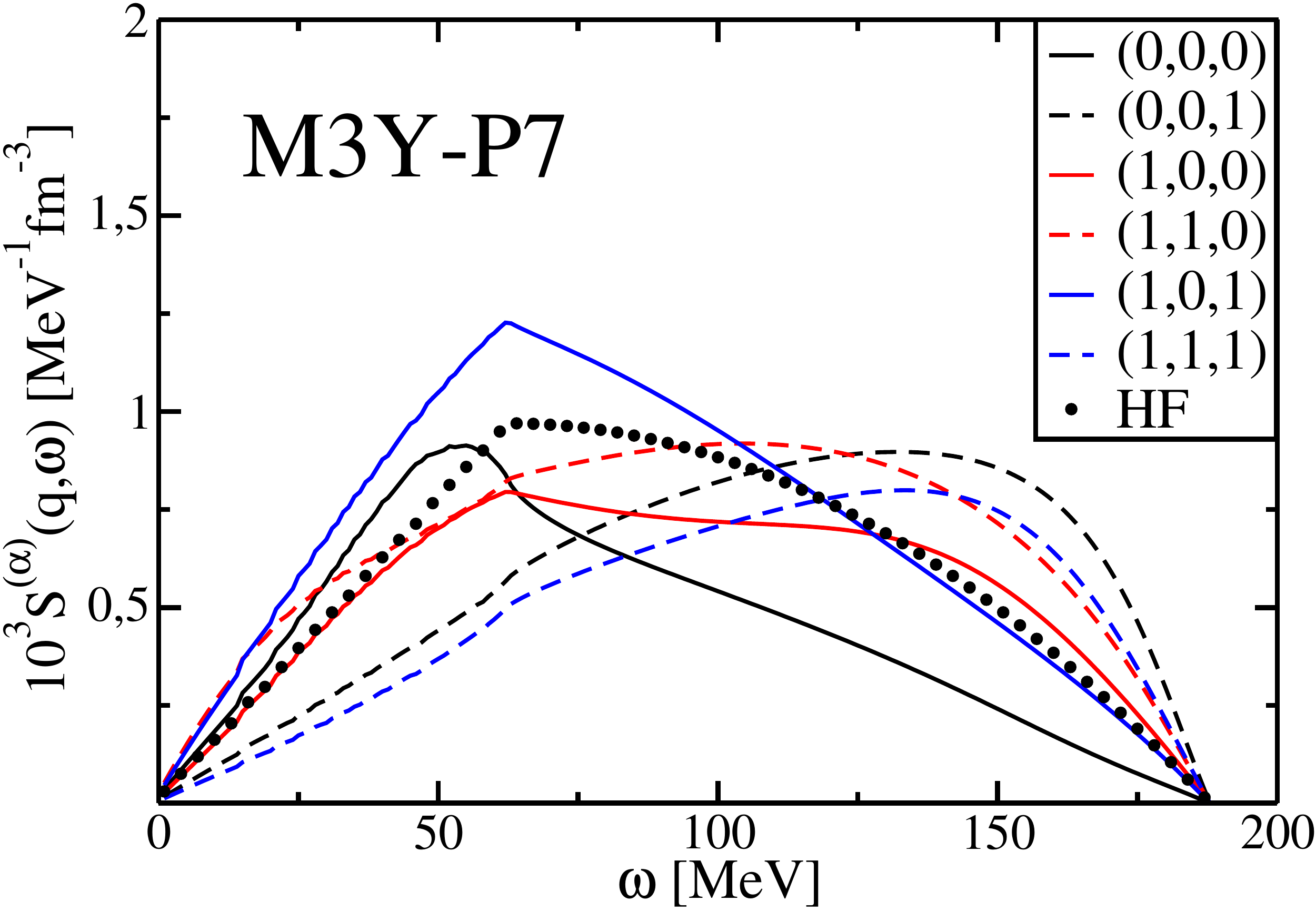}\\
\end{center}
\caption{Strength function for the channels $(S,M,I)$ calculated with the various Nakada interactions at saturation density and $q/k_F=1$.}
\label{M3Y:overview}
\end{figure}

\begin{figure}[!h]
\begin{center}
\includegraphics[width=0.3\textwidth,angle=0]{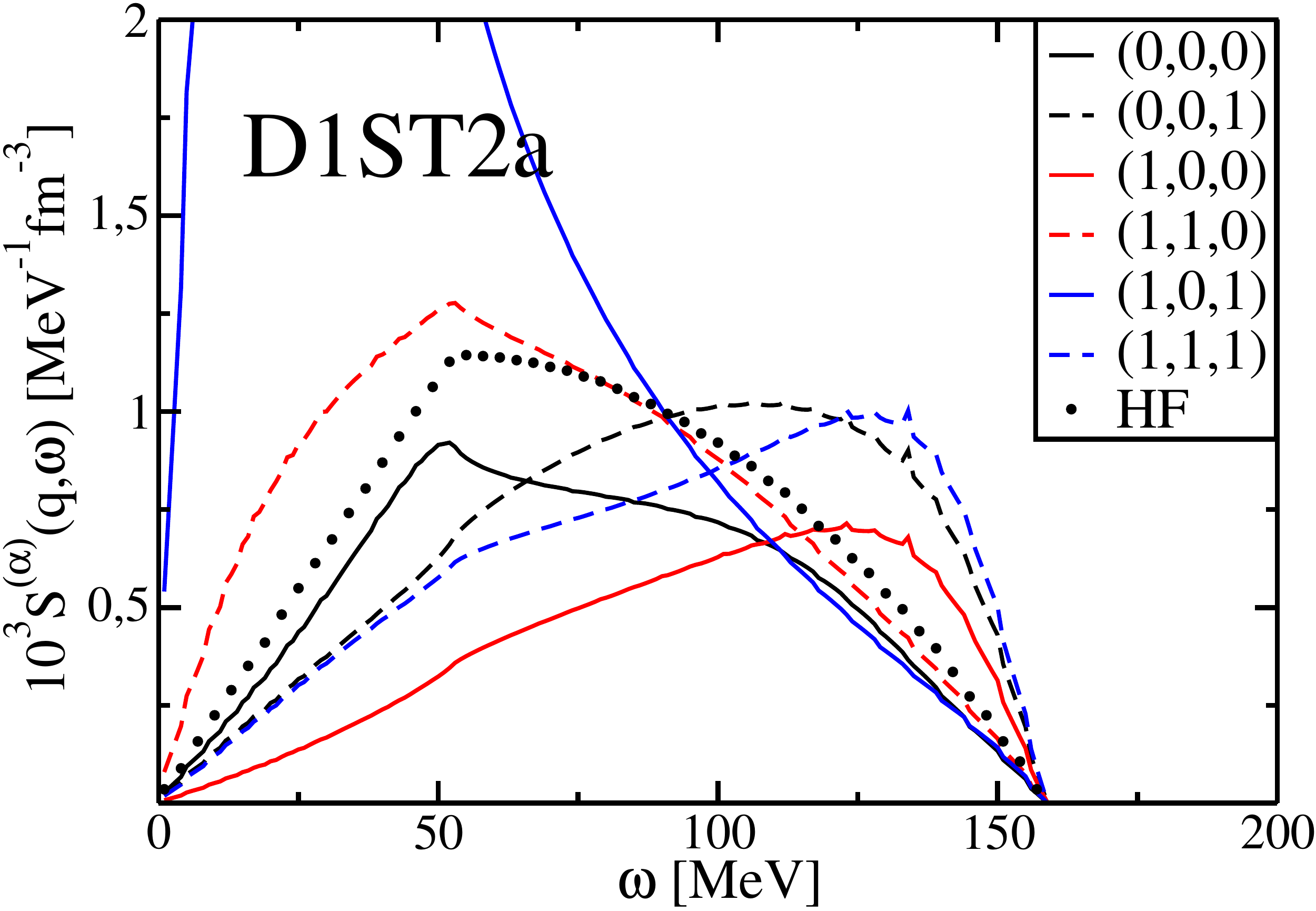}
\includegraphics[width=0.3\textwidth,angle=0]{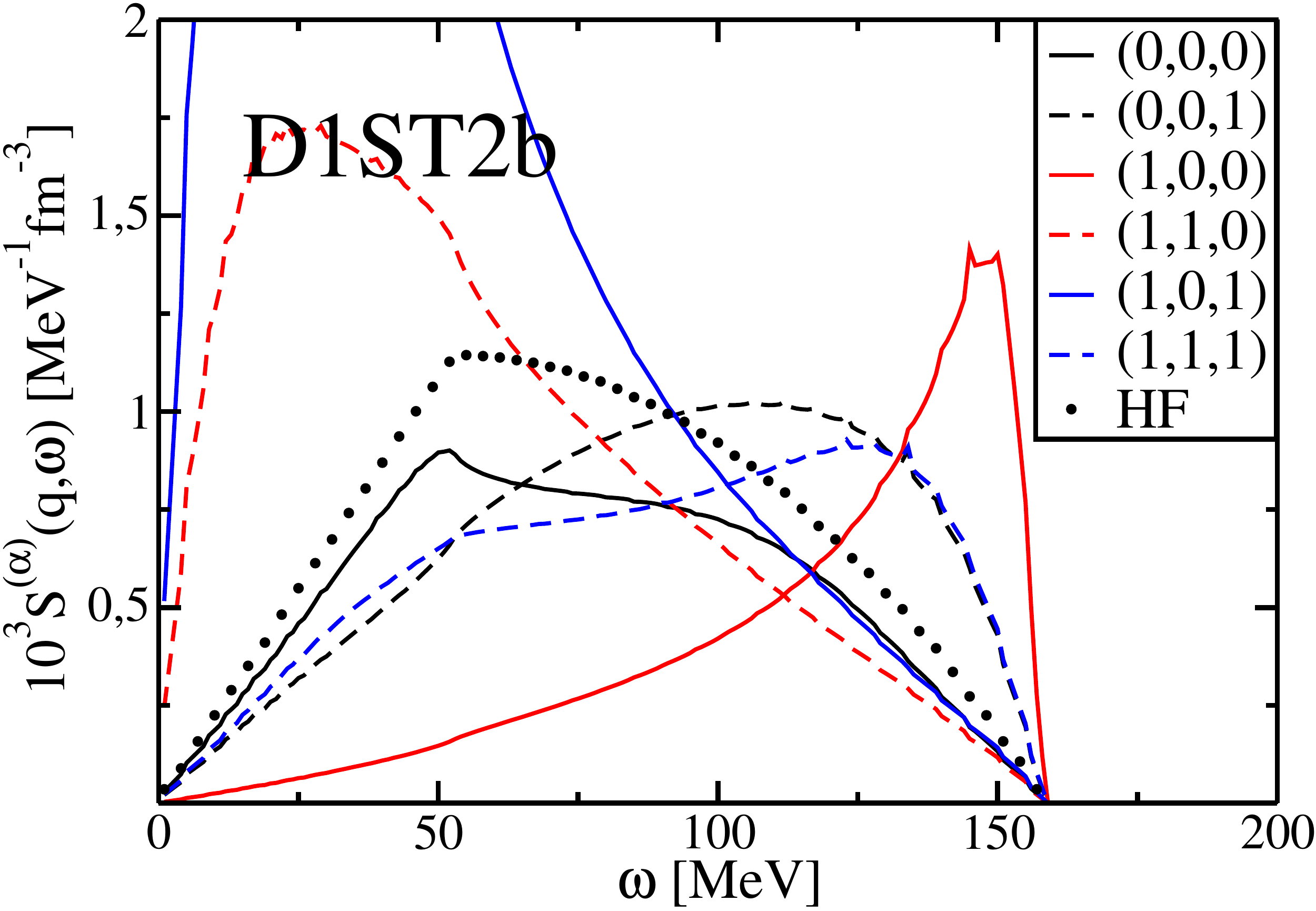}\\
\includegraphics[width=0.3\textwidth,angle=0]{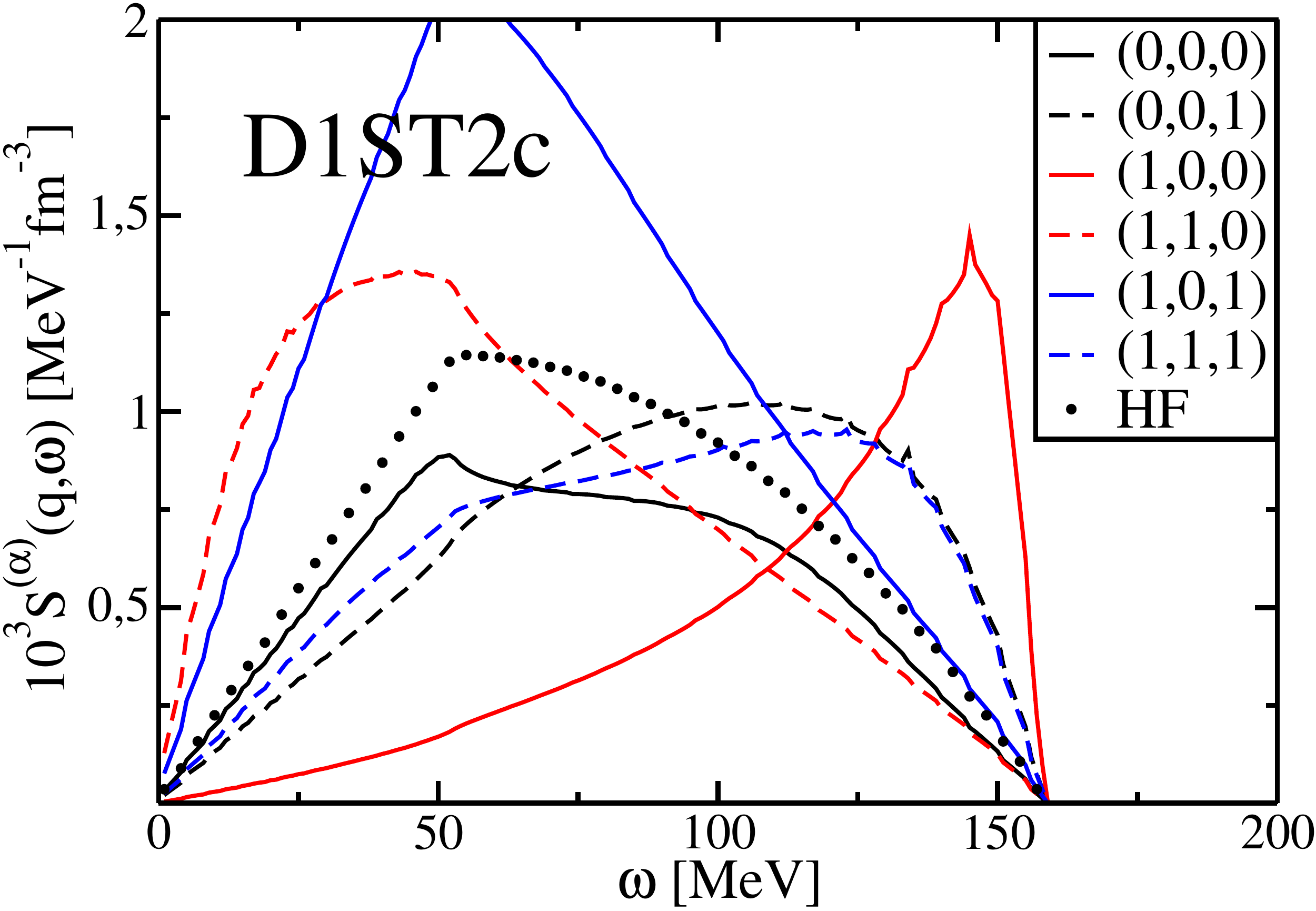}
\includegraphics[width=0.3\textwidth,angle=0]{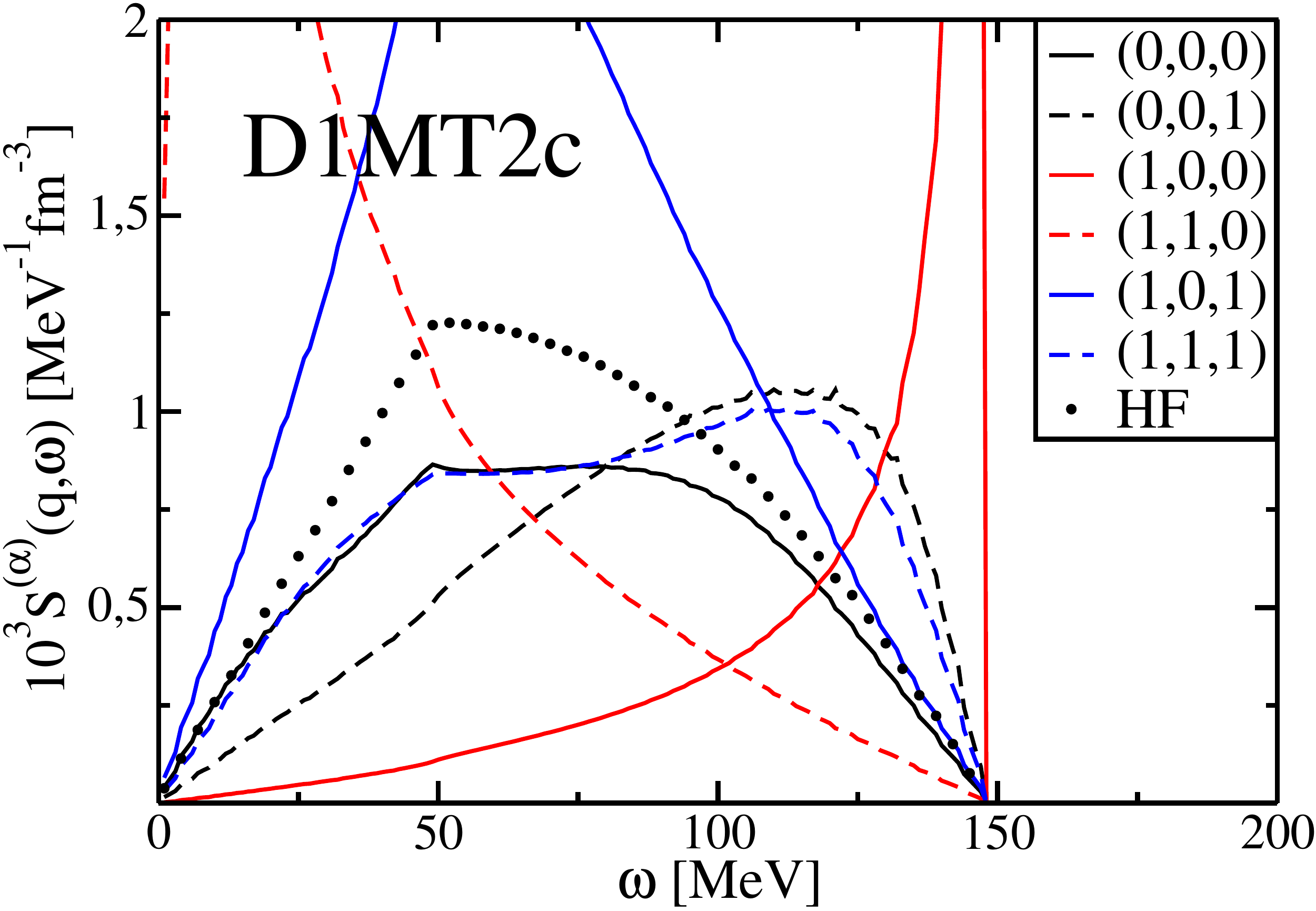}\\
\end{center}
\caption{Strength function for the channels $(S,M,I)$ calculated with the various Gogny interactions plus tensor at saturation density and for $q/k_F=1$.}
\label{gogny:overview}
\end{figure}

To discern the importance of tensor interactions on the strength functions, we compare the results obtained with all the Nakada interactions in Fig.~\ref{M3Y:overview}, and with various Gogny interactions equipped with a tensor 
in Fig.~\ref{gogny:overview}. The strengths are calculated at saturation density and $q/k_F=1$. 

It is interesting to notice that all $S=0$ strength functions are quite similar, while the $S=1$ channel shows more variety, due to the difference in tensor strength adopted in the different interactions. In particular we recognise that the strengths in channels $(1,M,I)$ obtained with interactions M3Y-P2 and M3Y-P4 are nearly the same for $M=0$ and 1, and the same value of $I$. On the contrary, the other interactions produce very strong differences in the $S=1$ channels. This can be seen by the strong separation of different $M$-channels and especially for $I=1$. Of particular interest is to compare the results obtained with the D1ST2a, b and c, since they all have the same central term. The differences observed in the response function arise basically from the tensor term, and less importantly from the modified spin-orbit term. In the isospin $I=1$ channels one can also observe that, as compared with the HF strength, the longitudinal $(M=0)$ and transverse $(M=1)$ strengths are reminiscent of the results of Ref.~\cite{alb82} displayed in Fig.~\ref{alberico-fig3}. 

For all the considered interactions, the multipolar expansion provides converged results with $L_{max}=2$ at saturation density. This small value of $L_{max}$ seems to be  specific to the short-range nuclear interactions. Indeed, the same multipolar expansion method has been employed in Ref.~\cite{sogo2012spontaneous} to calculate the response functions of a two-component dipolar Fermi gas using a (long-range) magnetic dipolar interaction, in the Landau scheme. The convergence required no less than ten multipoles in some channels. 
Keeping in mind that Nakada interactions include a range corresponding to a pion exchange, one thus understand that in Fig.~\ref{Conv:qTOT} $L_{max}=2$ and 3 results are on top of each other in the case of a Gogny interaction, while they are very close but still distinguishable in the case of a Nakada interaction. 

\section{Continued fraction approximation} \label{Sec:CF}

All the previous approximations start from the $G_{HF}$ ph propagator of infinite matter, so that many technical details are not directly transposable to finite nuclei. We discuss now another method to properly deal with the exchange term, whose basic principles can be used in both infinite and finite systems. The idea is to express the RPA response function as a continued fraction (CF), where the full ph interaction appears in the form of multiple momentum averages weighted with $G_{HF}$. Formally, the response function is exact at infinite CF order, but obviously a truncated expansion should be done for practical uses. Calculations in infinite nuclear matter allow one to determine the convergence of the expansion, which can also be assessed by comparing with the results obtained with the previous multipolar expansion method. We expect that the conclusions about the convergence in SNM are also valid for CF calculations in finite nuclei.

Actually, the CF technique was first developed in a different context in Ref.~\cite{len80},  to solve the Lippman Schwinger equation for the transition operator aiming at getting the optical potential. Later on, the quasi-elastic electron scattering on $^{12}$C was analysed in Ref.~\cite{del85} by calculating the Tamm-Dancoff response function at the first CF order, and in Ref.~\cite{bri87} by calculating the RPA response function at the same  order. In both cases, the single-particle wave functions were obtained in a non self-consistent way from a single-particle potential with no density-dependent term. The technique was also employed to calculate the response function of nuclear matter using meson-exchange-type potentials in Refs.~\cite{alb93,bar94,bar96,bar96a}, by truncating the expansion at first order. The calculation was pushed up to second order by De Pace~\cite{pac98}. A slightly different CF expansion was presented in Ref.~\cite{sch89}, using a semiclassical Thomas-Fermi scheme. More recently, another CF expansion was made in Ref.~\cite{mar08}, leading formally to the same results of Ref.~\cite{pac98} up to second order, and differing at next orders.

\subsection{\it The method}

We start with the formalism developed in Refs.~\cite{pac98,pac16}, ignoring for the time being the non-central components of the ph interaction. In that case, its matrix elements are diagonal in the spin-isospin indices and we will omit them for the sake of clarity. Let us write it as 
\be
V_{ph}/n_d = V_D(q) + V_E({\bf k}_{12}) \, ,
\label{Vph-CF}
\ee
where $V_D$ includes the direct and density-dependent parts while $V_E$ is the exchange term. See Eqs.~(\ref{Appb:central}-\ref{Appb:density}) and (\ref{central-Nakada}-\ref{density-Nakada}). 

The RPA response function is written as a CF-like expansion according to the pattern
\be
 \chi_{RPA} = \frac{\chi^{HF}}{1 - A - \cfrac{B}{1-C-\cfrac{D}{1-\dots}}}   \, .
\label{pace-1}
\ee
If the exchange term is dropped out, all quantities $B, C \dots$ are equal to zero while $A= V_D \chi^{HF}$, thus recovering the ring approximation to the response function. The point is to express $A, B, C \dots$ when the exchange term is included, and obtain the successive  orders. To this end, let us go back to the 
Bethe-Salpeter equation~(\ref{bethe-salpeter}), and integrate it with respect to the momentum. One gets
\be
\langle G_{RPA}\rangle = \langle G^{HF}\rangle + \langle G^{HF} V_{ph} G_{RPA} \rangle \,,
\label{iteration}
\ee
where the brackets is a short notation for integrations over chains of momenta as in the examples
\bea
\langle G^{HF} \rangle &=& \int \frac{{\rm d}{\bf k}_1}{(2 \pi)^3} \, G^{HF}({\bf k}_1,{\bf q},\omega)  \nnn
\langle G^{HF} V_{ph} G^{HF} \rangle &=& \int \frac{{\rm d}{\bf k}_1}{(2 \pi)^3} \,
\frac{{\rm d}{\bf k}_2}{(2 \pi)^3} \, G^{HF}({\bf k}_1, {\bf q},\omega) \; V_{ph}({\bf q}, {\bf k}_1, {\bf k}_2) \; G^{HF}({\bf k}_2, {\bf q},\omega) \nn
\eea
Formally, Eq.~(\ref{iteration}) can be solved iteratively thus providing a perturbatively expansion to $\chi_{RPA}$. 
The quantities $A, B, C \dots$ of Eq.~(\ref{pace-1}) are determined so as to  reproduce at $n$th order the perturbative series at the same order. To express the results given in Ref.~\cite{pac98} let us employ a more compact notation, by defining the following quantities related to the exchange ph interaction
\be
E_n = \frac{\langle G^{HF} V_E G^{HF} \dots V_E G^{HF} \rangle}{\langle G^{HF} \rangle^n} \, .
\ee
Then, the first CF order is trivially obtained
\be
A=(V_D+E_1) \chi^{HF} \,.
\label{pace-A}
\ee
The second order was deduced in Ref.~\cite{pac98} by analysing the perturbative series, and is written as 
\be
B= (E_2-E_1^2) \left(\chi^{HF}\right)^2 \,.
\label{pace-B}
\ee
An heuristic derivation for the next order was also given in Ref.~\cite{pac98} as
\be
C= \frac{1}{E_2-E_1^2} \left( E_3-E_1^3-2 E_1 E_2 \right) \chi^{HF} \,.
\label{pace-C}
\ee
Notice that $B$ and $C$, which corresponds to second and third orders respectively, involve only the exchange part of the ph interaction. These expressions were generalized in Ref.~\cite{pac16} to include non-central interactions. 

An alternative CF expansion was given in Ref.~\cite{mar08}. We present it here for a general ph interaction, including non-central terms. The guiding idea is to define an effective interaction $V^{(\alpha)}_{\rm eff}(q, \omega)$ such that the RPA response is written as
\be
\chi_{RPA}^{(\alpha)}(q, \omega) = \frac{\chi^{HF}(q, \omega)}{1- V_{\rm eff}^{(\alpha)}(q, \omega) \chi^{HF}(q, \omega)} \, ,
\label{CF-0}
\ee
that is, with the same form as in the ring approximation given in Eq.~(\ref{ring2}). The effective interaction is 
written as a continued fraction expansion 
\be
V^{(\alpha)}_{\rm eff} = \cfrac{V^{(\alpha)}_1}{ 1 - \cfrac{V^{(\alpha)}_2 \chi^{HF}}{1 - \cfrac{V^{(\alpha)}_3 \chi^{HF}}{1- \dots}}} \, .
\label{CF-1}
\ee
A comparison with the CF expansion (\ref{pace-1}) leads to the identification $A=V_1 \chi^{HF}$, $B=V_1 V_2 \left(\chi^{HF}\right)^2$, and $C=V_3 \left(\chi^{HF}\right)^3$. We will come back to these relations. 
The quantities $V_i^{(\alpha)}$ are obtained by expanding both (\ref{CF-1}) and (\ref{CF-0}) and comparing order by order to the perturbative series for $\chi_{RPA}^{(\alpha)}$ as obtained from the Bethe-Salpeter equation (\ref{bethe-salpeter})
\bea
 \chi^{(\alpha)}_{RPA} &=& \chi^{HF} + n_d \sum_{\alpha'} \langle G^{HF} V^{(\alpha,\alpha')}_{ph} G^{HF} \rangle 
 + n_d \sum_{\alpha',\alpha''} \langle G^{HF} V^{(\alpha,\alpha')}_{ph} G^{HF} V^{(\alpha',\alpha'')}_{ph} G^{HF}\rangle + \dots 
 \label{CF-3}
\eea
The explicit expression for the first three terms are
\bea
V^{(\alpha)}_1 &=& n_d \frac{\sum_{\alpha'} \langle G^{HF} V^{(\alpha,\alpha')}_{ph} G^{HF} \rangle}{(\chi^{HF})^2} 
\label{CF-V1}\;, \\
V^{(\alpha)}_2 &=& n_d \frac{ \sum_{\alpha',\alpha''} \langle G^{HF} V^{(\alpha,\alpha')}_{ph} G^{HF} V^{(\alpha',\alpha'')}_{ph} G^{HF} \rangle}{ V^{(\alpha)}_1 (\chi^{HF})^3} - V^{(\alpha)}_1
\label{CF-V2}\;, \\
V^{(\alpha)}_3 &=& n_d \frac{\sum_{\alpha', \alpha'', \alpha'''} \langle G^{HF} V^{(\alpha,\alpha')}_{ph} G^{HF} V^{(\alpha',\alpha'')}_{ph} G^{HF} V^{(\alpha'',\alpha''')}_{ph} G^{HF} \rangle}{ V^{(\alpha)}_1 V^{(\alpha)}_2 (\chi^{HF})^4} - \frac{(V^{(\alpha)}_1+V^{(\alpha)}_2)^2}{V^{(\alpha)}_2}\;.
\label{CF-V3}
\eea 
The sums in the above equations only concern the non-central parts, because the central and density-dependent parts of $V^{(\alpha,\alpha')}_{ph}$ are diagonal in the indices $\alpha$ and $\alpha'$. The quantities $G^{HF}$ and $\chi^{HF}$ are complex functions of $q$ and $\omega$, and so are the $V^{(\alpha)}_i$ and the effective interaction $V^{(\alpha)}_{\rm eff}$. 

For a ph interaction of the form given in Eq.~(\ref{Vph-CF}) one obtains
\bea
V_1 &=&  V_D + E_1 \, ,
\label{CF-V1-b} \\
V_2 &=& \frac{E_2 - E_1^2}{V_D+E_1} \, ,
\label{CF-V2-b} \\
V_3 &=& \frac{E_3-E_1^3}{E_2-E_1^2} + V_D - 2 E_1 - \frac{E_2 + E_1^2}{V_D+E_1} \, .
\label{CF-V3-b}
\eea
As for the comparison with (\ref{pace-1}) one confirms the previous identifications $A=V_1 \chi^{HF}$ and $B=V_1 V_2 \left(\chi^{HF}\right)^2$, so that the CF first and second orders coincide in both formalisms \cite{pac98} and \cite{mar08}. However, the guess for $C$ only includes the exchange term while $V_3$ depends on both direct and exchange terms.

Concerning the spin-orbit, one can see from Eqs.~(\ref{Appb:so}) and (\ref{so-Nakada}) that a factor $M \left( k_{12}\right)^{1}_{M}$ is present in the multiple momentum integrals entering Eqs.~(\ref{CF-V1}-\ref{CF-V3}). Since $G^{HF}$ does not depend on the azimuthal angle $\varphi$, we conclude that the spin-orbit does not contribute to the first order term $V_1^{(\alpha)}$. This argument does not apply to the tensor terms (see Eqs.~\ref{Appb:ten} and \ref{ten-Nakada}) because it contains instead the factor $\left( k_{12}\right)^{1}_{-M} \left( k_{12}\right)^{1}_{M'}$. The case $M'=M$ leads to a non-vanishing contribution when integrating over $\varphi$, so that the tensor does contribute to $V_1^{(\alpha)}$. At higher orders, there is a similar compensation for the spin-orbit, which thus starts to contribute at second order. In other words, the coupling between $S=0$ and 1 channels induced by the spin-orbit is a second order effect. The tensor term only acts on the $S=1$ channel, but it affects also the $S=0$ channel because of the spin-orbit coupling, and thus the effect of the tensor on the $S=0$ channel is also second order.

\subsection{\it Convergence}
Two tests about the convergence of the CF expansion were analysed in Ref.~\cite{mar08}. The first one is in the particular case of a ph interaction given by Eq.~(\ref{LAN-Vph-ME}), but restricted to the first two central Landau parameters $\ell=0, 1$. The calculation of the functions $V^{(\alpha)}_i$ given in Eqs.~(\ref{CF-V1}-\ref{CF-V3}) is straightforward, and  the effective interaction turns out to be
\bea
V^{(\alpha)}_{\rm eff} &=& f^{(\alpha)}_0 + f^{(\alpha)}_1 \nu^2 \left\{ 1 + \left( -\frac{1}{3} F^{(\alpha)}_1\right) + \left( -\frac{1}{3} F^{(\alpha)}_1\right)^2 + \left( -\frac{1}{3} F^{(\alpha)}_1\right)^3 + \dots \right\} \nnn
&=&f^{(\alpha)}_0 + \frac{f^{(\alpha)}_1 \nu^2}{1+F^{(\alpha)}_1/3}
\eea
We recall that $\nu=\omega m^*/(q k_F)$.  As shown in Sect.~\ref{resp-landau-analytical}, this is precisely the effective interaction deduced from Eq.~(\ref{W-landau}) with Landau parameters $\ell=0, 1$. Typically, the value of these parameters is $|F^{(\alpha)}_1| \le 1$ (see Tab.~\ref{land-param-SNM}), thus guaranteeing the convergence, whose pace will depend on the specific interaction and channel.

\begin{figure}[!h]
\begin{center}
\includegraphics[width=0.5\textwidth,angle=0]{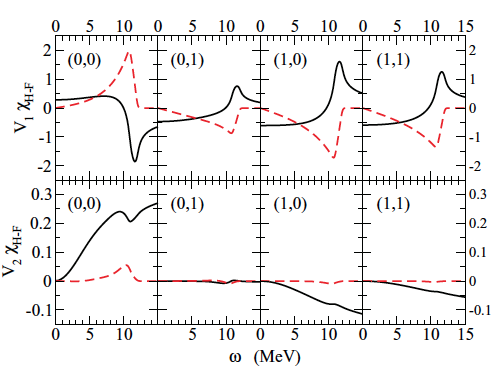}
\end{center}
\caption{Real (solid line) and imaginary (dashed line) part of $V_1 \chi^{HF}$ (top row) and $V_2 \chi^{HF}$ (bottom row) for D1 interaction at saturation density and $q=27$\,MeV. The $(S,T)$ channels are shown in each panel. Taken from Ref.~\cite{mar08}. }
\label{mar08-fig1}
\end{figure}

The second test is numerical. One can have an idea of the convergence rate by comparing the functions $V_1 \chi^{HF}$ and $V_2 \chi^{HF}$, that is, first and second order. This is shown in Fig.~\ref{mar08-fig1} for the Gogny interaction D1 at a momentum transfer $q = 27$\,MeV, roughly $0.1 k_F$. Notice that the scale used to plot $V_2 \chi^{HF}$ is about a factor of ten larger than that of $V_1 \chi^{HF}$. It can be seen that the imaginary parts of $V_2 \chi^{HF}$ are close to zero for the four spin-isospin channels. The real parts are generally small compared to 1, but the situation seems less favourable in the channel $(S, T ) = (0, 0)$. From the behaviour shown in this figure one can expect a rapid convergence of the calculated responses already at the level of $V_2$, although perhaps slower in the case of the $(0, 0)$ channel.

\subsection{\it Results} 
 
\begin{figure}[!h]
\begin{center}
\includegraphics[width=0.4\textwidth,angle=0]{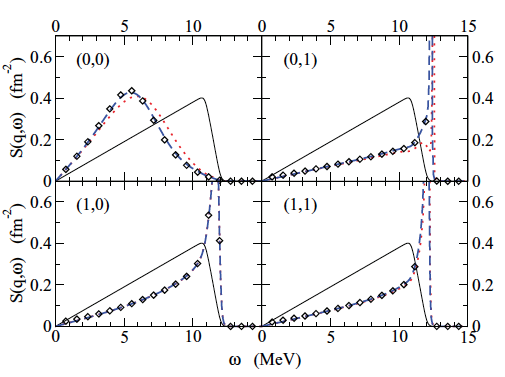}
\end{center}
\caption{Strength functions calculated with D1 interaction at saturation density and momentum transfer $q=27$\,MeV. Thin lines: HF, dotted lines: first order CF, dashed lines: second order CF, open diamonds: converged strength functions from the multipolar expansion. Taken from Ref.~\cite{mar08}.}
\label{mar08-fig2}
\end{figure}

\begin{figure}[!h]
\begin{center}
\includegraphics[width=0.4\textwidth,angle=0]{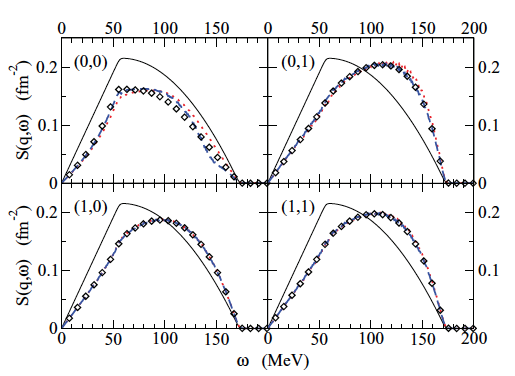}
\end{center}
\caption{Same as Fig.~\ref{mar08-fig2} for $q=270$\,MeV. Taken from Ref.~\cite{mar08}.}
\label{mar08-fig3}
\end{figure}

As an illustration, in Figs.~\ref{mar08-fig2} and \ref{mar08-fig3} are displayed the RPA strength functions  calculated with Gogny D1 interaction at saturation density and for two values of the momentum transfer, at about $k_F/10$ and $k_F$. In the same figures are plotted the converged strengths obtained in a multipolar expansion, which provides the numerically exact results and are taken as benchmark. For comparison, the HF strength is plotted as well. Notice the units: a factor of $\hbar c$ is multiplying Eq.~(\ref{strength2}).

 One can see that for all channels except $(0,0)$, the CF expansion has already converged at first order. For the $(0, 0)$ channel it is necessary to include the second order to get a strength practically superimposed to the benchmark ones. The $(0, 0)$ exception is to be expected from the previous analysis of Fig.~\ref{mar08-fig1}.  Interestingly, the convergence is independent of the value of $q$, as no expansion in powers of $q$ has been done. Indeed, as it can be seen in Eq.~(\ref{CF-1}) the CF convergence of the effective interaction does not rely on $q$ but on the functions $V_i \chi^{HF}$. Instead, as shown in Ref.~\cite{mar08}, the convergence relies on the value of the density, and it deteriorates as the density increases, particularly in the $(0,0)$ channel. 

The convergence in each $(S,T)$ channel depends on the specific interaction employed, as shown in Ref.~\cite{pac16}, where the Gogny D1, D1S, D1N and D1M interactions were considered. These calculations show that in the $(1,1)$ channel, the differences between CF first and second order are clearly evident for interactions D1M and D1N, particularly in the high energy region. We mention here that, at variance with Ref.~\cite{mar08}, the authors of Ref.~\cite{pac16} looked for the presence of collective states in the discrete region near or above the ph continuum. While interactions D1, D1S and D1M predict a collective state isospin channel ($S,1)$ for each value of $S$, the interaction D1N only predicts a collective state in the (1,1) channel. These states are found in both first and second orders. At $q=270$\,MeV these collective states enter into the continuum.

\begin{figure}[!h]
\begin{center}
\includegraphics[width=0.8\textwidth,angle=0]{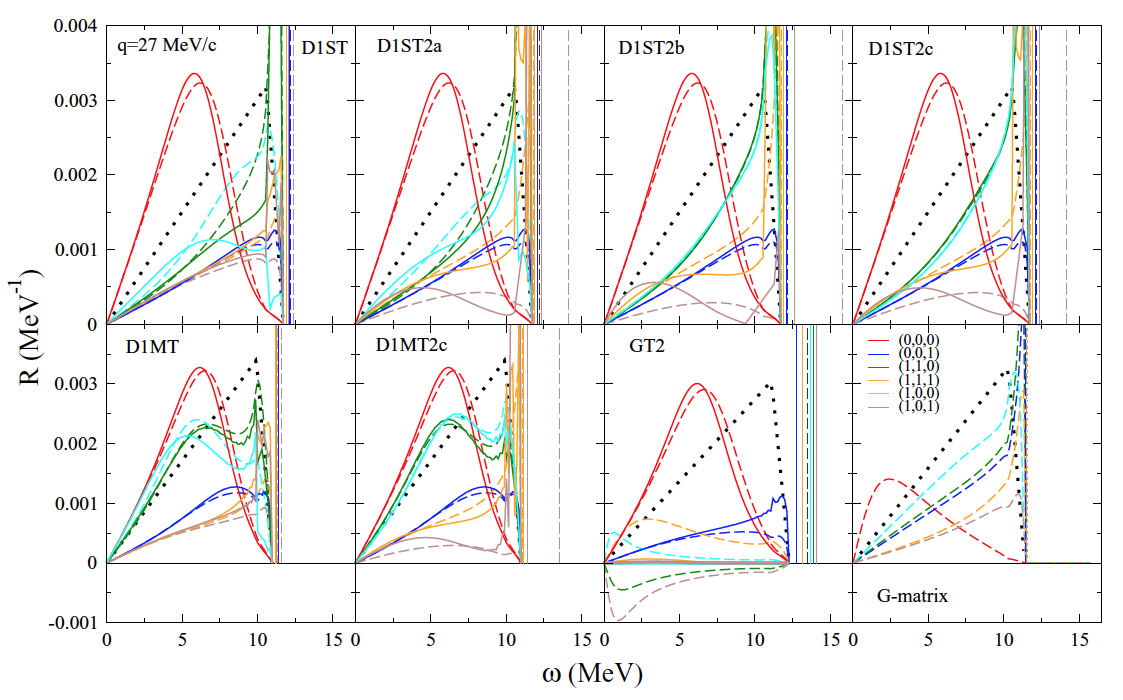}
\end{center}
\caption{Strength functions per nucleon calculated with several Gogny interactions equipped with a tensor component,  at $k_F=270$\,MeV and momentum transfer $q=27$\,MeV. Dashed lines: CF first order, solid lines: CF second order, black dotted: HF strength. For comparison, the strength functions calculated with a $G$-matrix nuclear interaction are also displayed in the last panel. Taken from Ref.~\cite{pac16}.}
\label{marco-fig4}
\end{figure}

\begin{figure}[!h]
\begin{center}
\includegraphics[width=0.8\textwidth,angle=0]{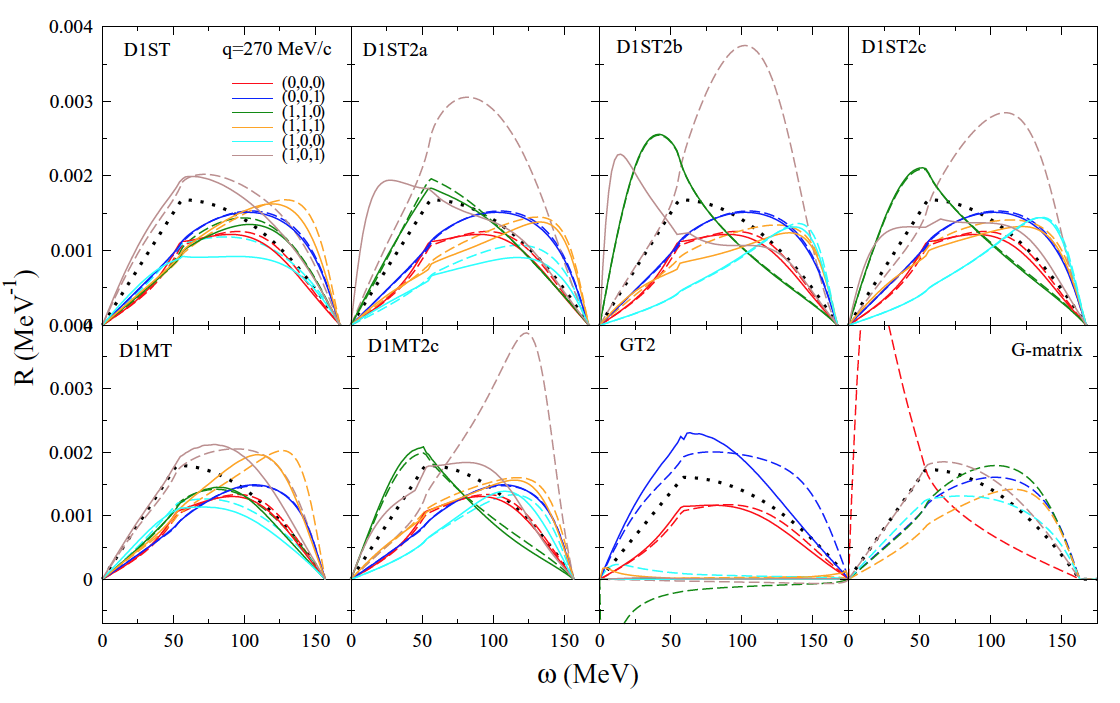}
\end{center}
\caption{Same as Fig.~\ref{marco-fig4} for $q=270$\,MeV. Taken from Ref.~\cite{pac16}.}
\label{marco-fig5}
\end{figure}

We consider now the effect of tensor interaction in the strength functions. In Figs.~\ref{marco-fig4} and \ref{marco-fig5} are displayed the strength functions per particle calculated in Ref.~\cite{pac16} with several Gogny interactions equipped with a tensor component, but ignoring the spin-orbit term. For comparison, the strength functions calculated with an effective interaction~\cite{nak84} based on a $G$-matrix in nuclear matter derived from a one-boson-exchange potential~\cite{hol81}, are also displayed.

We discuss first the results obtained with interactions D1ST, D1STa,b,c, D1MT, and D1MTc. One can see that the $S=0$ strengths are all in qualitative agreement. Results at first and second orders are very close both at $q=27$\,MeV and 270\,MeV. A collective $(0,0,1)$ state appears in the discrete region at $q=27$\,MeV. In both isospin channels, first and second order strengths are very close for all interactions, thus indicating that the CF convergence has been reached. On the contrary, in the $S=1$ channels, the differences between first and second order are clearly visible, very pronounced in some cases, in contrast with the results obtained with central interactions. To this respect, one should keep in mind that these Gogny parametrizations have a stronger tensor interaction as compared to Nakada interactions or even the AV18 interaction (see Fig.~\ref{anguiano-fig3}). Therefore, for these interactions a CF third order calculation is required to get more reliable results. Besides, including spin-orbit interaction is also necessary to analyse the effect on the convergence due to the coupling with the $S=0$ channel. We also notice that a collective state in the discrete region is always present in the longitudinal $(1,0,1)$ channel at $q=27$\,MeV. In the transverse $(1,1,1)$ channel, the collective state can be in the continuum or in the discrete region, depending on the interaction. At $q=270$\,MeV all the strengths are qualitatively very similar. 

Consider now the results obtained with the GT2 interaction. The $S=0$ strengths are qualitatively similar to those obtained with the previous Gogny interaction, although in the $(0,0,1)$ channel, first and second order are much different. The first observation for the $S=1$ strengths is the unphysical results at first order with the GT2 interaction in $(1,1,0)$ and $(1,0,1)$ channels, with negative strengths. At second order, the strengths for channels $(1,0,0)$ and $(1,1,1)$ practically vanishes at both values of $q$, apart from the collective $(1,1,1)$ state at the lower value of $q$. We will see in Sect.~\ref{sec:instabilities} that the unphysical behaviour of interaction GT2 is also reflected in the study of finite-size instabilities. 

Finally, concerning the results obtained with the $G$-matrix, one observes the marked difference in the $(0,0,0)$ channel as compared with those obtained with Gogny interactions. The authors of Ref.~\cite{pac16} stressed that at $q=290, 390$ and 480 MeV, the $K^+$-nucleus quasi-elastic cross section~\cite{pac97}, which is largely dominated by the $(0,0,0)$ channel,  is succesfully reproduced with this  $G$-matrix interaction. Actually, an effective interaction should have the correct OPEP tail to describe inelastic scattering, which is obviously not the case for the Gogny interaction. In that respect, we may hope that the Nakada interactions give better results. 
For the other channels, the results given by the $G$-matrix and the Gogny interactions are in a qualitative agreement.

\section{Phenomenological interactions and finite size instabilities}
\label{sec:instabilities}

We conclude this review by considering an important connection with finite nuclei as is the detection of finite-size instabilities associated with the employed phenomenological interaction. By means of the linear response theory, it is possible to identify the presence of zero-energy modes with infinite strength~\cite{pas12,pas12a}. These modes can be associated with a physical instability as in the case of the spinodal one in the channel (0,0,0). The spinodal instability~\cite{duc07} is associate with the liquid-gas phase transition of the infinite medium and it plays a crucial role in the formation of non-homogeneous phases in the crust of neutron stars~\cite{cha08}.
Other instabilities may also appear, but if not associated with any physical process, we call them spurious. As showed in Refs.~\cite{pas12b,hel13,pas15,les06,fra12,kor14}, these instabilities may manifest in the calculations of properties of atomic nuclei, leading to non-converging results. To avoid the problems of spurious finite-size instabilities, in Ref.\cite{pas13,bec17}, a different optimisation procedure was discussed by making use of explicit results coming from the RPA responses.  

\begin{figure}[!h]
\begin{center}
\includegraphics[width=0.4\textwidth,angle=0]{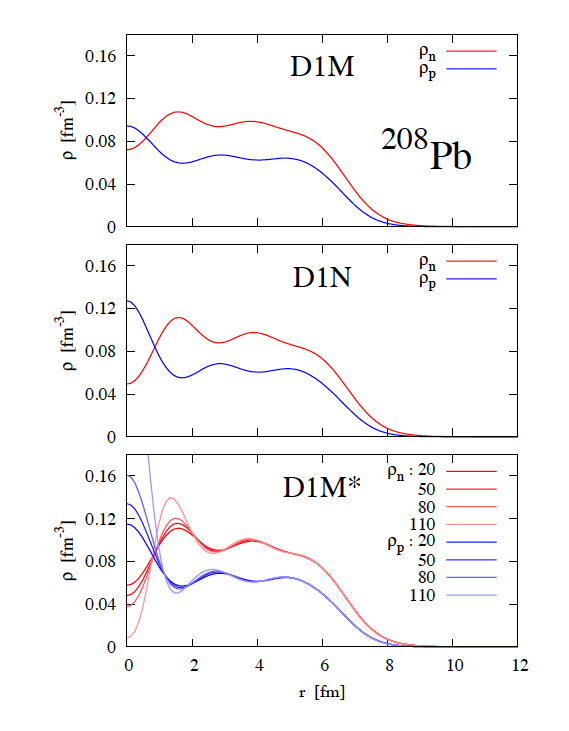}
\end{center}
\caption{ Neutron (in red) and proton (in blue) densities obtained from Hartree-Fock calculations for $^{208}$Pb, with the interactions D1M (top panel), D1N (central panel) and D1M$^*$ (bottom panel). Since the interaction D1M$^*$ does not lead to a self-consistent convergent solution, different levels of red and blue are used to plot the densities after different numbers of iterations as indicated in the figure. Taken from Ref.~\cite{mar19}.}
\label{Fig:instabilities:208pb}
\end{figure}

As an example, we illustrate in Fig.~\ref{Fig:instabilities:208pb}  the proton and neutron densities from an Hartree-Fock calculation with the D1M, D1N and D1M$^*$~\cite{gon18} Gogny interactions for $^{208}$Pb. With the D1M$^*$ interaction, the calculations do not converge and lead to oscillations of the isovector density with very large amplitude, as shown by the behaviour of the densities as the number of HF iterations is increased. For D1M and D1N only the final densities are represented since the calculations are fully converged.

\begin{figure}[!h]
\begin{center}
\includegraphics[width=0.4\textwidth,angle=0]{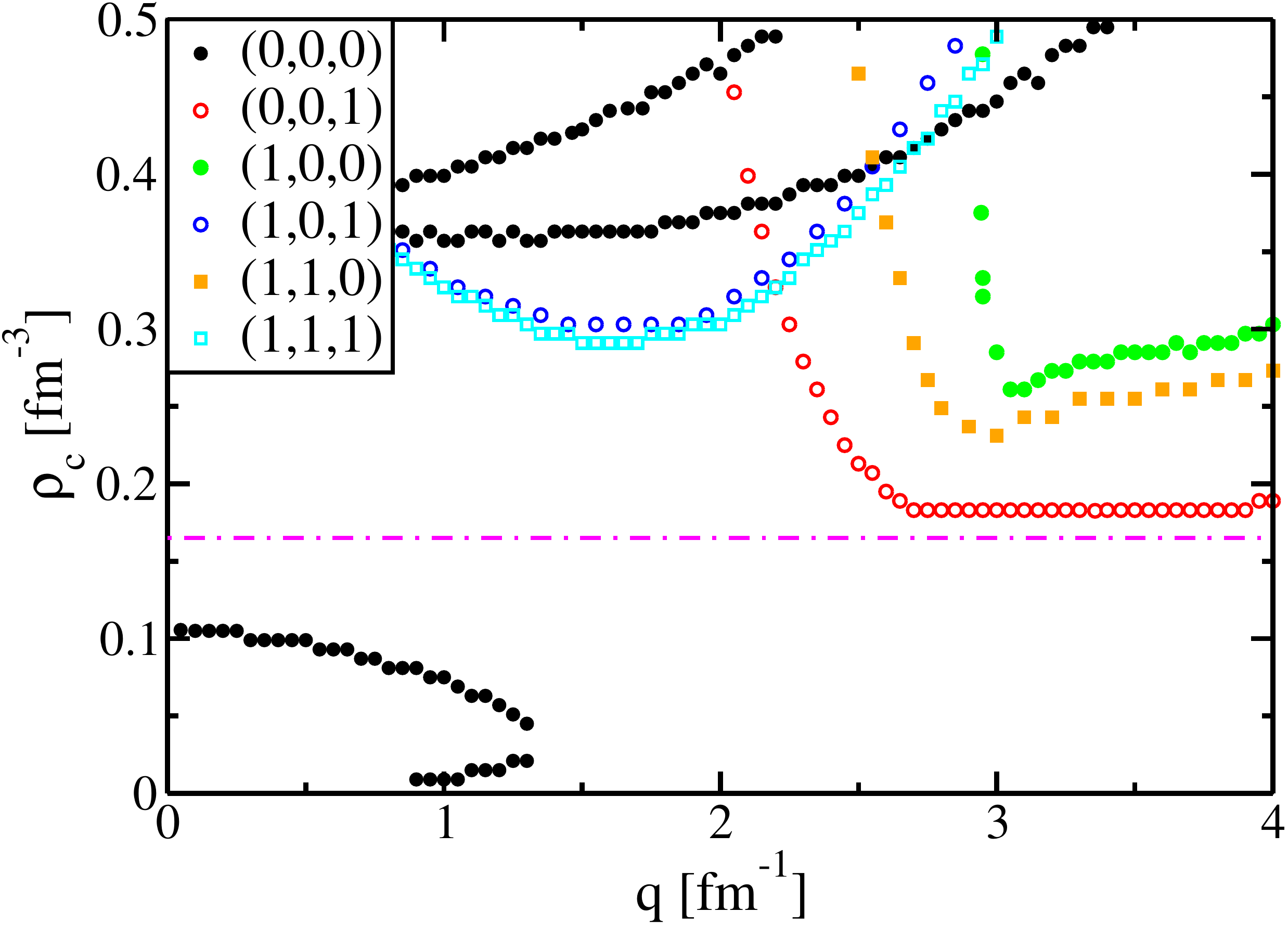}
\end{center}
\caption{Critical densities as a function of the transferred momentum  for the D1M$^*$ interaction for different spin isospin channels (S,M,I). The horizontal dot-dashed line represents the saturation density $\rho_0$. See text for details.}
\label{Fig:instabilities:d1ms}
\end{figure}

Such a behaviour has been also observed in Refs~\cite{les06,hel13} for the case of Skyrme interactions and it has been identified with the presence of spurious zero-energy modes in infinite nuclear matter at densities close to saturation. Using Landau theory of Fermi liquid is it possible to identify all spurious modes associated with zero transferred momentum~\cite{cao10,nav13}, but only thanks to RPA responses it is also possible to identify the instabilities occurring at finite transferred momentum. This happens when
\begin{eqnarray}\label{eq:poles}
1/\chi^{(\alpha)}(\omega=0,q,\rho)=0\;.
\end{eqnarray}
By scanning all values of transferred momentum in a range $q \le 4$\;fm$^{-1}$, we identify the values of the critical densities $\rho_c$ at which Eq.~(\ref{eq:poles}) has a solution. The results are reported in  Fig.~\ref{Fig:instabilities:d1ms}. The same calculation was also performed in Ref.~\cite{mar19}, using CF formalism, but ignoring the spin-orbit ph interaction. Apart from the spinodal instability in the scalar-isoscalar channel, we also observe another instability in the scalar-isovector channel close to the saturation density of the system.
A detailed analysis of finite size instabilities has been performed in Ref.~\cite{hel13} by using Skyrme functionals. The authors have thus identified a simple criterion to establish if the presence of poles may affect or not finite nuclei calculations. In particular, they observed that when an instability appears at densities below 1.2 $\rho_0$ a possible issue in the calculation of finite nuclei may manifest.  This means that at these densities, the HF ground state used for the current calculations is no more the true ground state and the system will undergo a phase-transition. We stress that the findings of Ref.~\cite{hel13} for zero-range interactions are confirmed in Ref.~\cite{mar19} for finite-range interactions.

\begin{figure}[!h]
\begin{center}
\includegraphics[width=0.4\textwidth,angle=0]{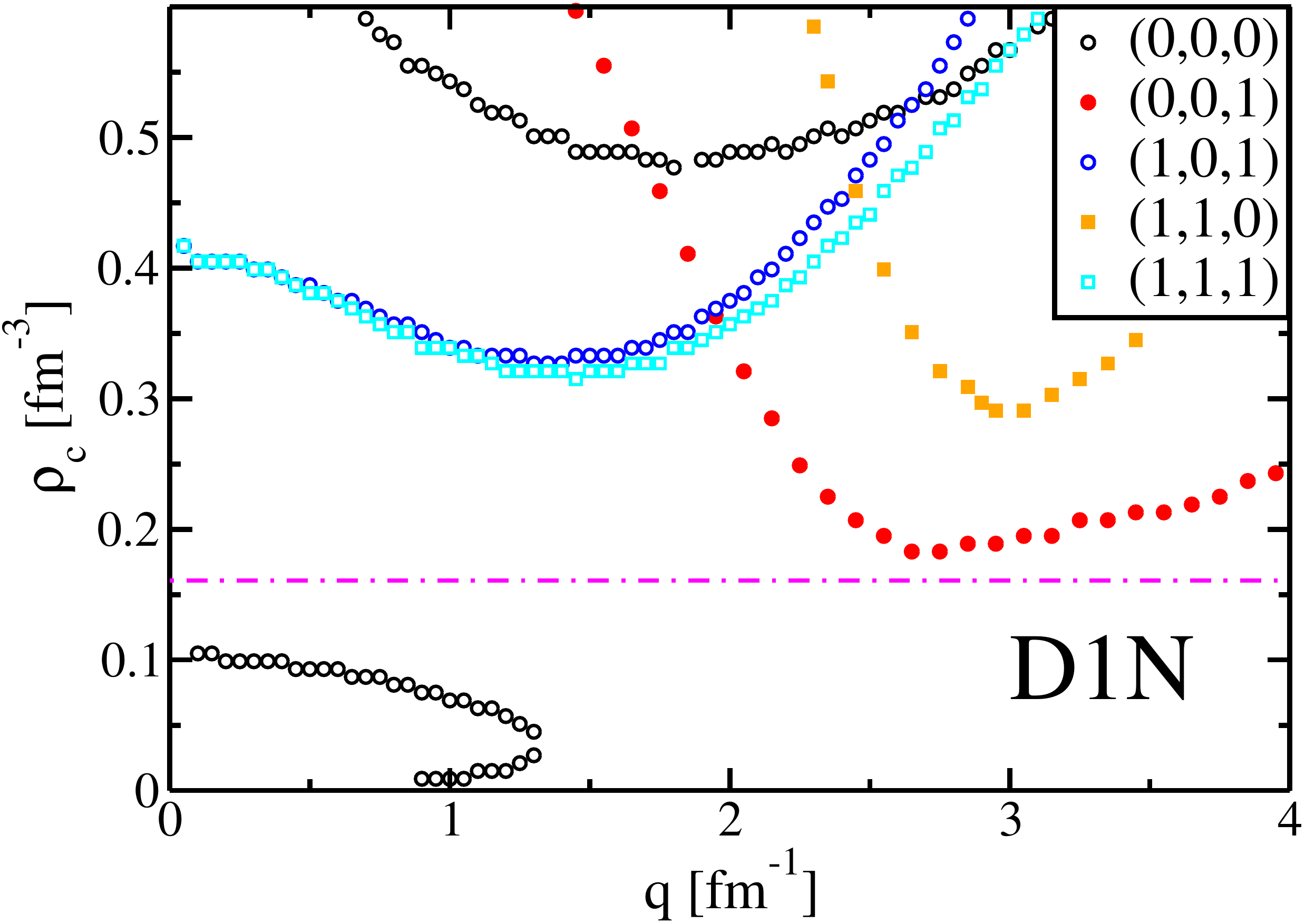}
\includegraphics[width=0.4\textwidth,angle=0]{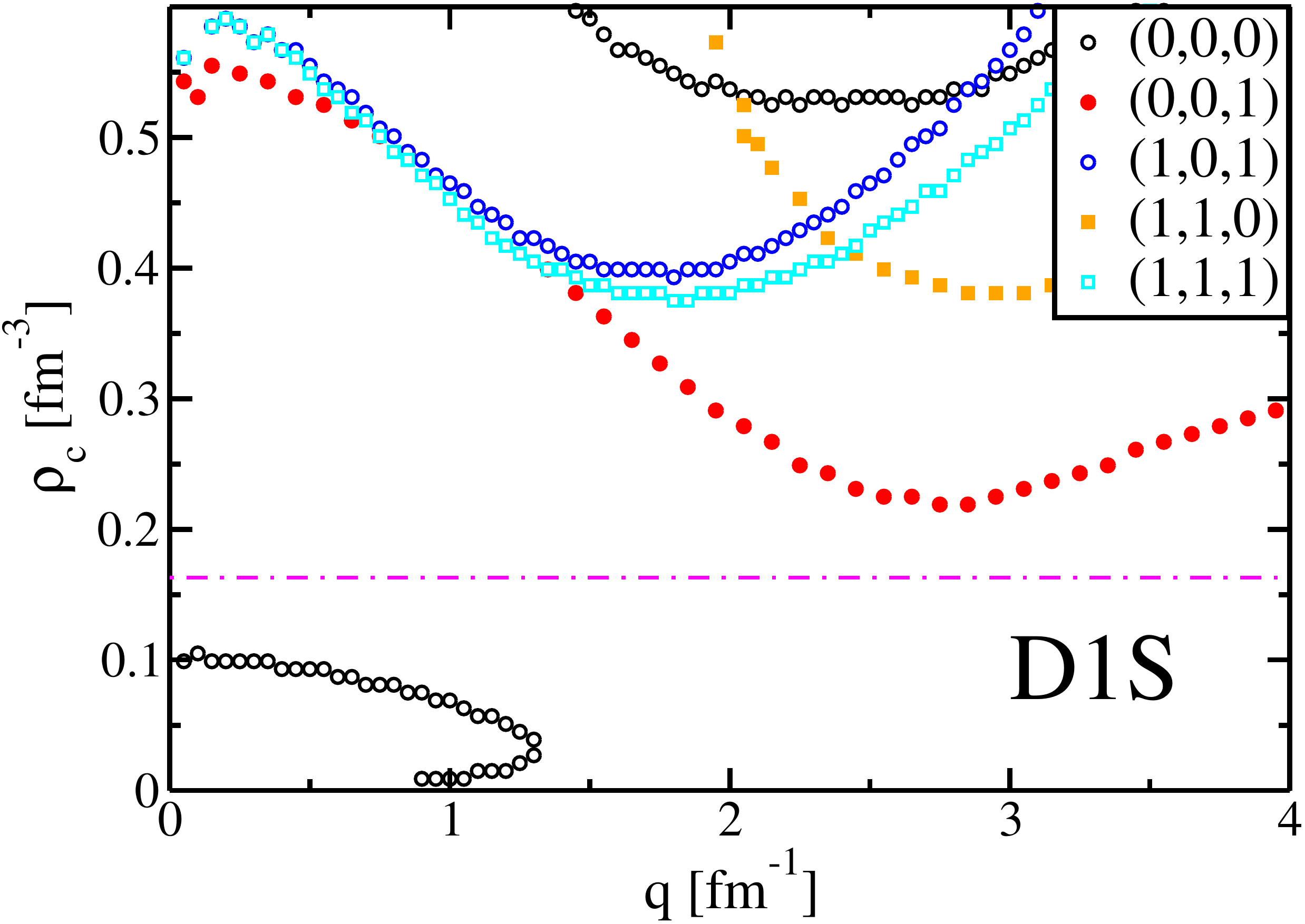}
\end{center}
\caption{Finite-size instabilities for the Gogny D1N (left) and D1S.}
\label{Poles:gogny}
\end{figure}

It is now interesting to explore the position of finite size instabilities for Gogny interactions as done in Ref.~\cite{pac16}. In Fig.~\ref{Poles:gogny}, we illustrate the position of the critical densities for D1S and D1N. 

We observe that the D1S interaction does not present any finite-size instability around saturation density, while D1N presents an instability in the scalar-isovector channel at $\approx1.1\rho_0$. This value falls within the criterion given in Ref.~\cite{hel13} and it means that potentially some finite-nuclei calculations may fail when using this interaction~\cite{mar19}.
The results obtained here using the full multipolar expansion including central and spin-orbit term agrees nicely with the ones given in Ref.\cite{pac16} and based on CF approximation.
It is interesting to note that the D1N fitting protocol includes explicit constraints on Landau parameters~\cite{cha15a}, but these are not enough to prevent the appearance of finite-size instabilities.  As seen from Fig.\ref{Poles:gogny}, the presence of a range is not sufficient to remove such instabilities.

\begin{figure}[!h]
\begin{center}
\includegraphics[width=0.8\textwidth,angle=0]{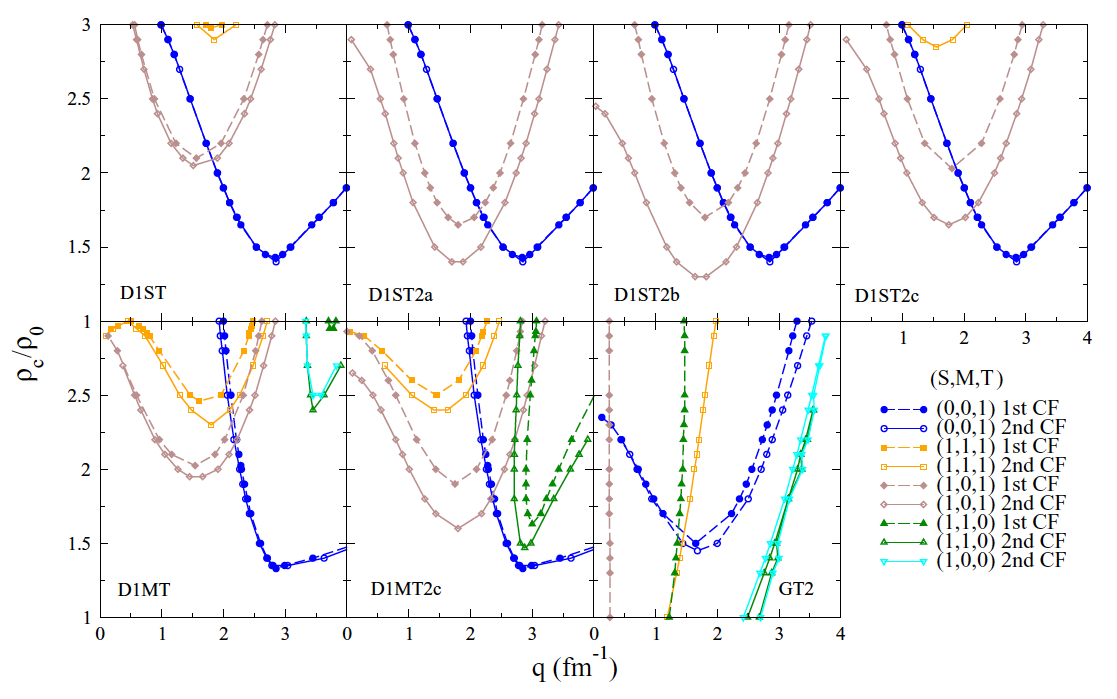}
\end{center}
\caption{Finite-size instabilities for various Gogny interactions including tensor terms. Taken from Ref.\cite{pac16}.}
\label{Poles:gogny:tens}
\end{figure}

In Fig.~\ref{Poles:gogny:tens}, we illustrate the position of the instabilities in the various spin/isospin channels using the CF approximation of Ref.\cite{pac16}. The position of the instabilities have been rescaled for each interaction to the corresponding value of the critical density $\rho_0$ so to make more evident the possible violation of the stability criterion given in Ref.~\cite{hel13}.

Apart from the GT2 interaction that is clearly unstable, as seen in the anomalous behaviour of the response function given in Fig.\ref{marco-fig5}, all the other interactions present instabilities well above saturation density. Since in these calculations there is no spin-orbit term, the instabilities observed in the $S=0$ channel only arise from the central term and thus they are not related to the presence of a tensor term in the ph channel.
Among the various Gogny-tensor interactions, only D1ST2b presents an instability in the (1,0,1) channel that is quite close to the stability limit. In all cases, despite the presence of a strong tensor term  as discussed in Sec.\ref{sec:comp:tens}, we can conclude that these parameterisations are most likely stable.

\begin{figure}[!h]
\begin{center}
\includegraphics[width=0.4\textwidth,angle=0]{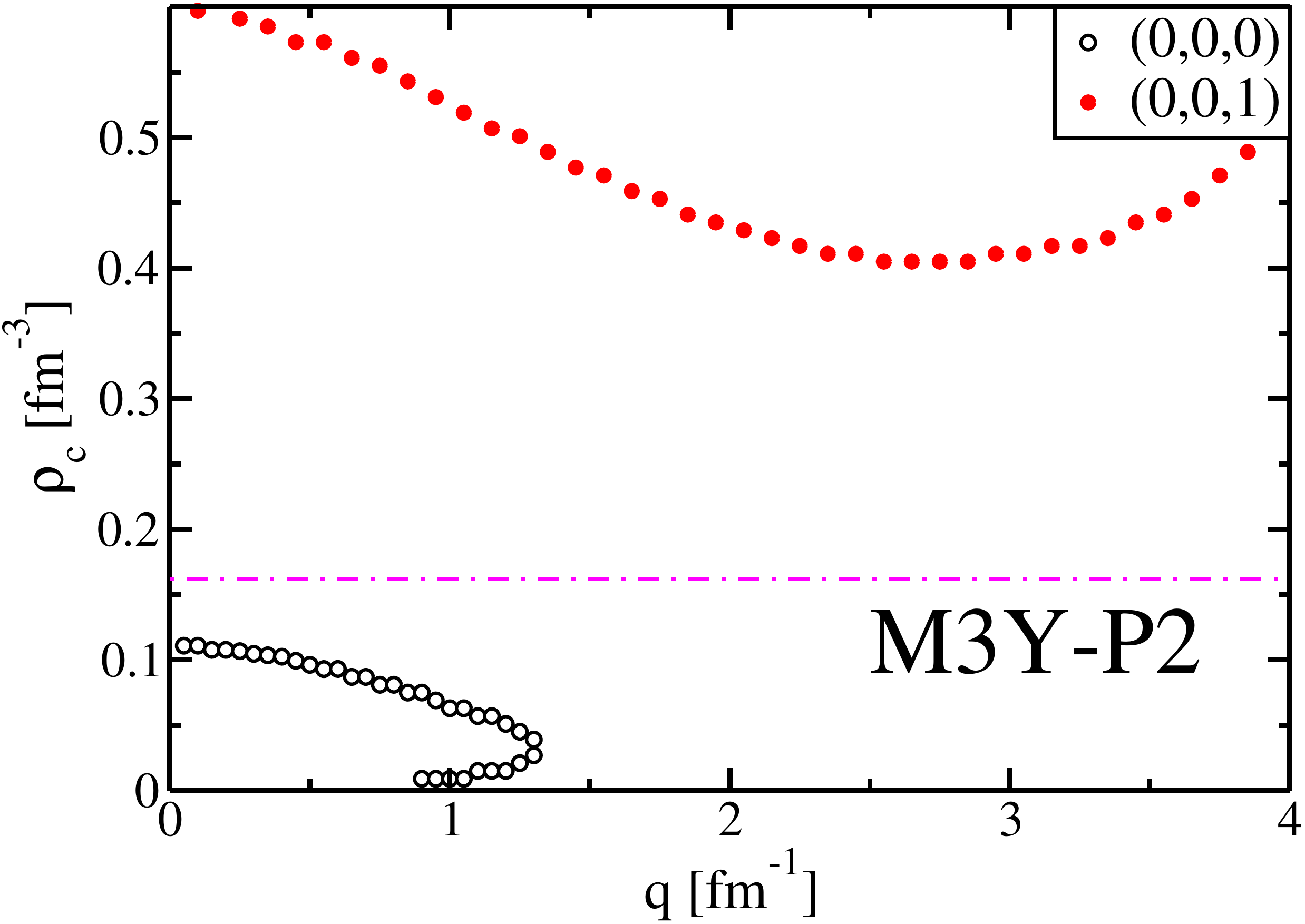}
\includegraphics[width=0.4\textwidth,angle=0]{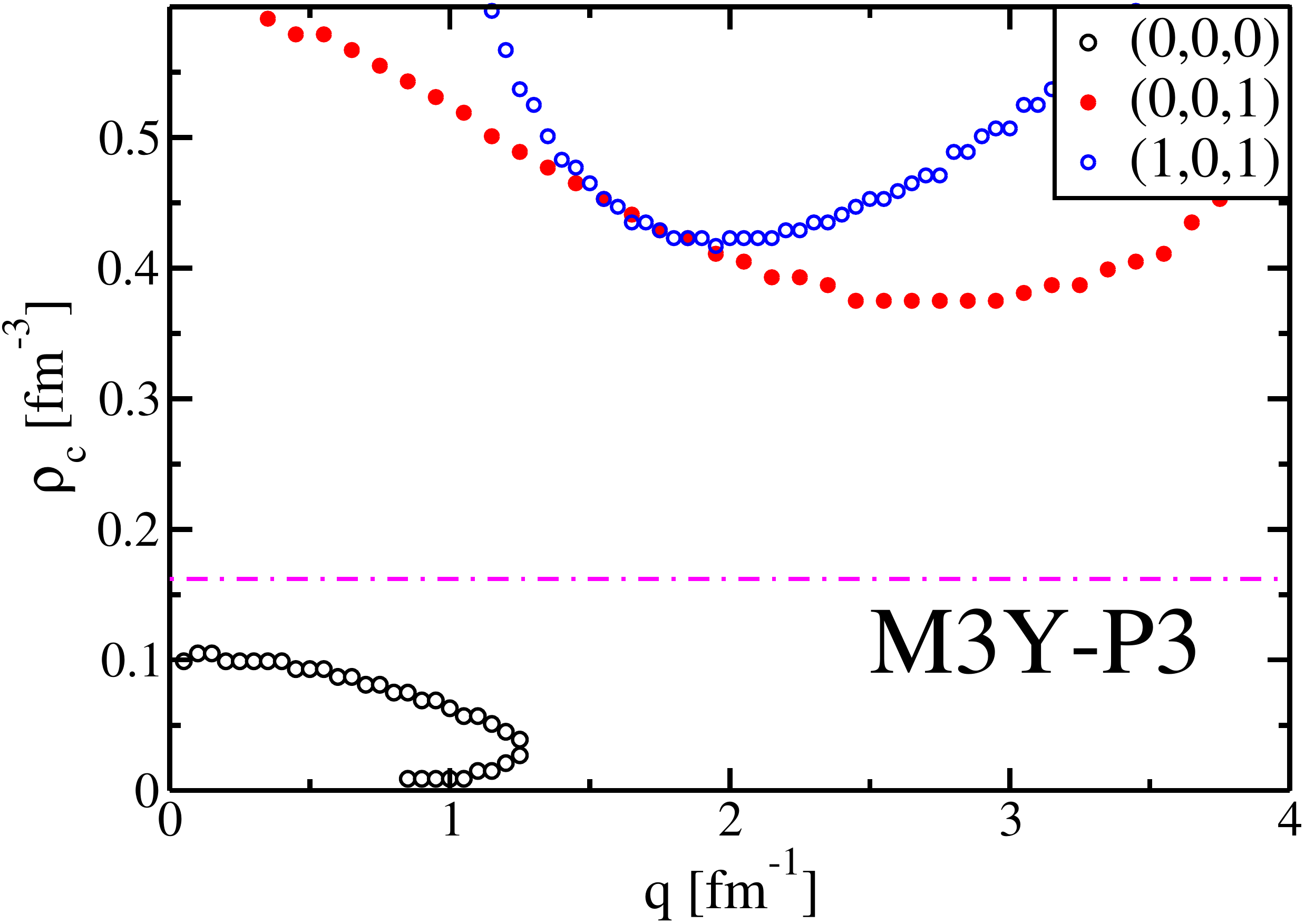}\\
\includegraphics[width=0.4\textwidth,angle=0]{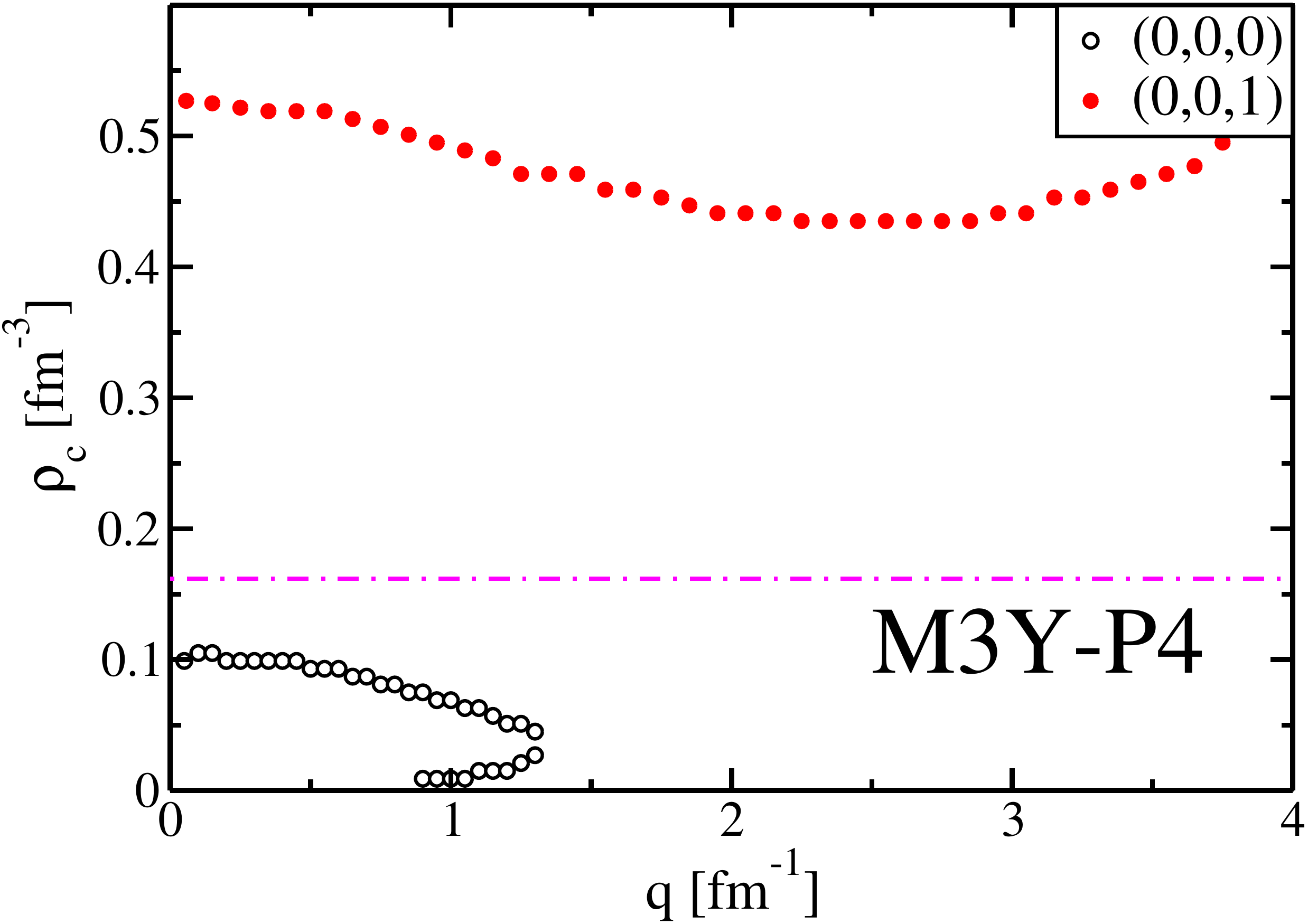}
\includegraphics[width=0.4\textwidth,angle=0]{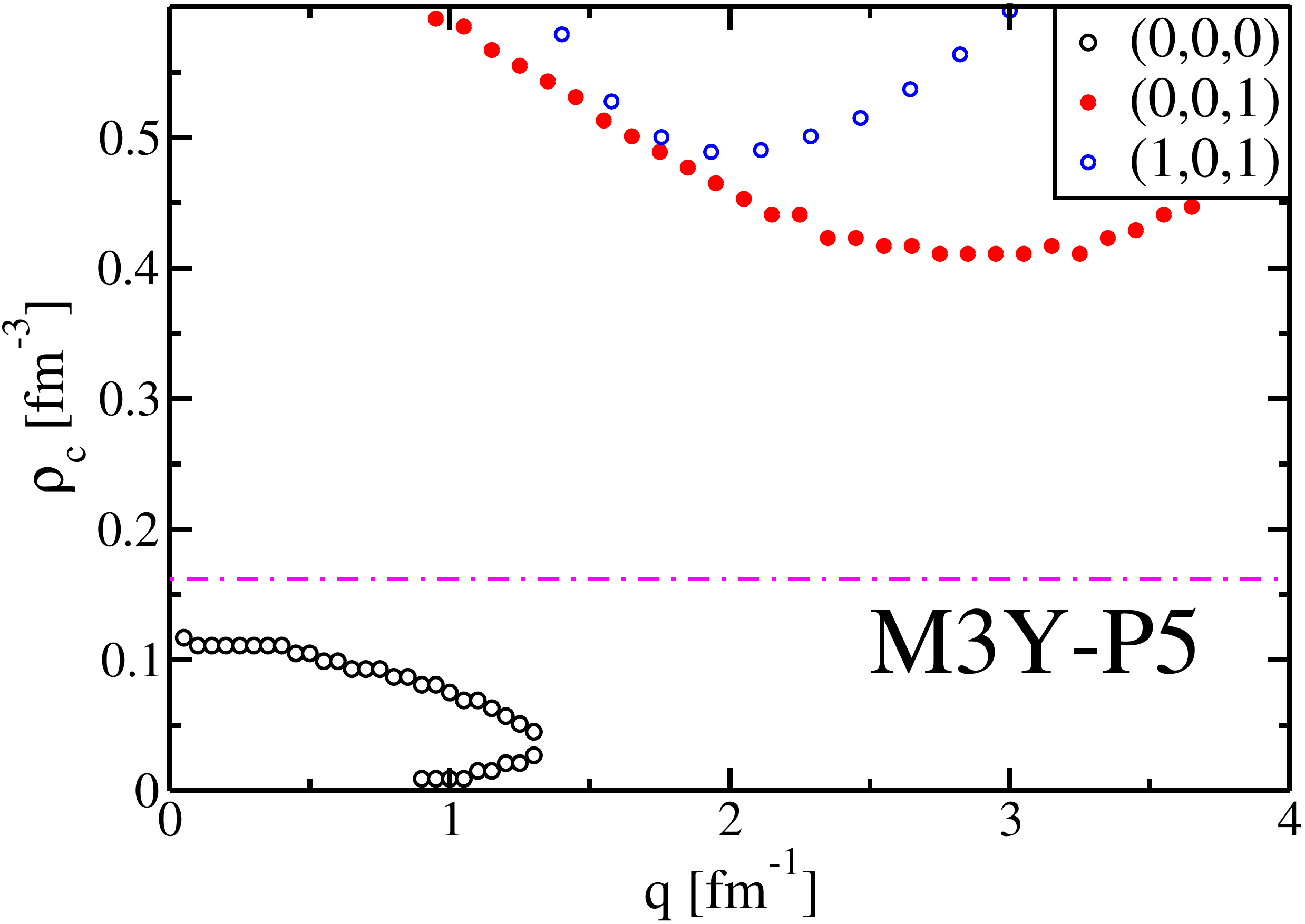}\\
\includegraphics[width=0.4\textwidth,angle=0]{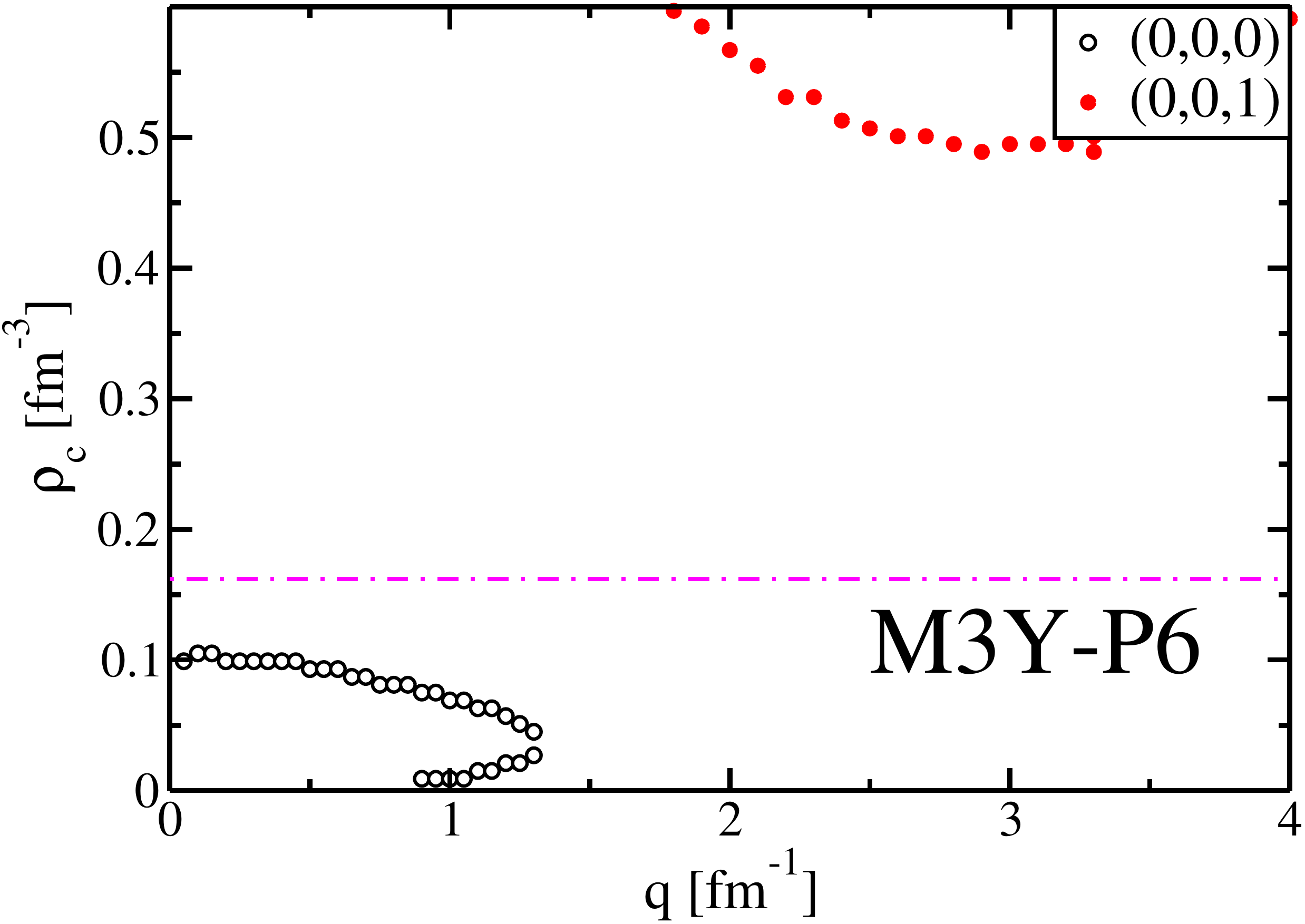}
\includegraphics[width=0.4\textwidth,angle=0]{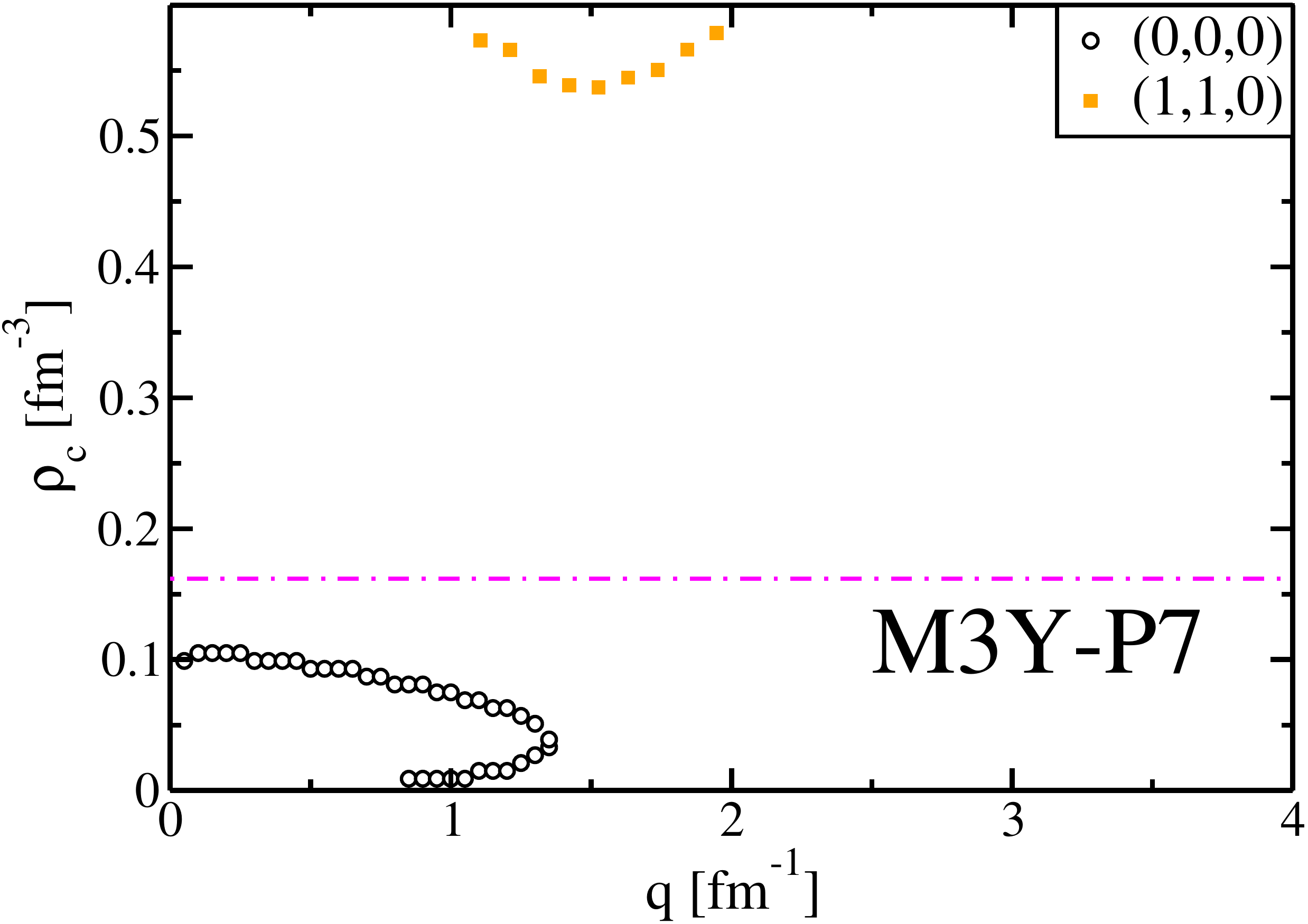}
\end{center}
\caption{Finite-size instabilities for various M3Y interactions}
\label{Poles:m3y}
\end{figure}

In Fig.~\ref{Poles:m3y}, we report the finite-size instabilities of the Nakada interaction. A quite remarkable result is that the Nakada's interactions are essentially free from instabilities up to $\approx2\rho_0$, apart from the physical spinodal instability. This is a very remarkable result since as shown above, it does not apply to Gogny interactions. 
Such a result has been already discussed in Fig.~\ref{fig:Landau:skyrme:instabilities} in the context of the instabilities related to the deformation of the Fermi sphere using Landau parameters. 

These results are remarkably different with the one obtained in the context of the tensor interactions $T_{ij}$~\cite{les07}: in that case all interactions turned out to be unstable. 
The results of Fig.~\ref{Poles:m3y} clearly motivate a much in depth analysis of the properties of the Nakada interaction before making any strong conclusion, but this could be the effect of keeping the long range on the OPEP potential. 

\section{Concluding remarks} \label{Sec:concluding}

This review has focused on the methods for calculating the response function of symmetric nuclear matter based on  the two most common families of phenomenological finite-range interactions, namely Gogny and Nakada. The core of the review has been the discussion of existing approximations and methods to deal with the exchange part of the ph interaction with particular attention to the inclusion of an explicit tensor term.

An analysis of some basic ground states properties of the infinite medium shows that all these interactions  give reasonable results in fair agreement with available {\it ab initio} calculations. This result is of course expected since these properties essentially depend on the central part of the interaction and are directly integrated in the fitting procedure. However, the situation is not so favorable when the tensor terms are included. 
By performing the partial wave decomposition of the energy per particle, it is possible to highlight an inconsistency between the sign of the tensor term added (perturbatively up to now) to the Gogny interaction and the sign of {\it ab initio} one. Moreover, when compared to realistic interactions, the tensor terms in Gogny interactions exhibit a huge discrepancy in magnitude as well. The Nakada interaction, on the contrary, shows good agreement with partial waves determined with {\it ab initio} calculations and the strength, although too weak, is in fair agreement when compared to realistic interactions.

To obtain the response functions of the infinite medium, one has to solve the  Bethe-Salpeter equations. The first ingredient of these equations is the ph propagator. It has been shown explicitly that the usual parabolic approximation on the mean field is reliable up to $q\simeq 2 k_F$ and therefore can be used safely in that range for the results and applications presented in this review. The second important ingredient is the ph interaction, whose exchange part constitutes the most difficult aspect of the calculations. For such a reason, the core of the review has been devoted to the discussion of existing approximations.

The Landau approximation is only reliable for very low-values of transferred momentum, but it gives important informations on the system since the Landau  parameters have a universal character. It is thus important to compare their numerical values calculated from different interactions with the ones produced using various {\it ab-initio} methods. The Landau parameters are linked to macroscopic properties of the infinite medium such as compressibility, effective mass or spin susceptibility. Besides, they must fulfill some general inequalities to guarantee the stability of the system. These inequalities satisfied by Landau parameters must be taken into account at the very first stage of the fitting procedure, in particular for the tensor part, which is known to produce unphysical instabilities at least for zero-range interactions. 

 A possible way to improve the Landau approximation is to keep the explicit momentum dependence on the direct term. In this way the response function calculated at higher momenta is in better agreement with the complete response function. This can be useful for extensive calculations when the exact calculations demand a heavy numerical effort.

A full treatment of the exchange term is necessary to obtain the exact response function. This can achieved using either a multipolar expansion of the ph interaction and the RPA ph propagator or a continued fraction expansion of the RPA response function. In both cases a truncation scheme is required, but the existing calculations show that the convergence is very fast. Regarding the multipolar expansion method, we have provided in the Appendix  detailed expressions of the multipoles of the different terms (central, density-dependent, spin-orbit and tensor) of the finite-range interactions since they can also be used to perform finite-nuclei calculations~\cite{tha11,gra02}. This expansion method is quite general but is time-consuming. Regarding the continued fraction approach, the integrals of the interaction convoluted with the ph propagator are calculated numerically from the start, but the presence of an explicit spin-orbit term makes the calculation quite involved. The approach can be applied to finite nuclei, presumably with a similar convergence rate.

A detailed analysis of nuclear response functions at some characteristic value of the transferred momentum sheds some light on the accuracy of all these approximations, and permits to highlight how the presence of a tensor term can remarkably change the structure of the response functions. These modifications can be connected to the ones observed for the transverse and longitudinal responses in hadronic physics. In particular, we noticed that the strong Gogny tensor leads to a large separation between these two modes, greater that the one observed in hadronic physics. As the Nakada tensor is weaker, it leads to a smaller separation between these two modes and it looks more in qualitative agreement with the results obtained with pionic response function.  

By investigating the poles in the response functions, we have observed that essentially all Gogny interactions with tensor are stable, apart from the GT2 interaction. This is an interesting result since, it was shown in Ref~\cite{report} that all Skyrme interactions having a tensor term are unstable. However, up to now the Gogny tensor has been added perturbatively, {\it i.e.} without refitting all the parameters of the interaction, apart from the GT2 interaction which presents unstabilities in some spin-isospin channels.  All these results point towards the necessity of a more systematic analysis of finite-range tensor based on the complete refit of the parameters in the same spirit of Refs~\cite{les07,hel12}. Considering Nakada interactions, we have shown that all available parametrisation are essentially free from finite-size instabilities up to 2-3 times saturation density. This result makes Nakada interaction a very promising candidate to be used for astrophysical calculations of other important properties~\cite{loa11} involving the nuclear response function. A typical example is the neutrino mean free path (NMFP) in stellar medium~\cite{iwa82,nav99}, which plays a crucial role in the physics of supernovae and neutron stars.  Indeed, as shown in  Refs~\cite{pas12a,pas14}, the presence of a finite-size instability can dramatically change the NMFP by several order of magnitude, thus having a major impact on transport properties of these systems.

\section*{Acknowledgements}

The authors would like to thank Magda Ericson, Alexandros Gezerlis, Marcella Grasso, Marco Martini, Arnau Rios  and Michael Urban for useful comments and discussions. This work has been supported by the FIS2017-84038-C2-1-P and STFC Grants No. ST/M006433/1 and No. ST/P003885/1.  Numerical calculations were  undertaken on the Viking Cluster, which is a high performance computing facility provided by the University of York.


\begin{appendices}


\section{Multipoles of the HF propagator} \label{app:A}

We present here the explicit expressions of the multipoles of the HF propagator $G^{HF}(q,\mathbf{k},\omega)$ (see Eq.~\ref{GHF}). Since this propagator does not depend on the azimuthal angle, its multipoles are simply calculated as
\begin{eqnarray}
G_\ell^{HF}(q,k_1,\omega)=\sqrt{\frac{2\ell+1}{4\pi}}\int d\hat{k}_1 G_{HF}(q,\mathbf{k}_1,\omega)P_\ell (\cos\theta_1).
\label{GLHF}
\end{eqnarray}
For practical reasons, the integral is separated into its real and imaginary parts. 

Consider first the case when the parabolic approximation of the mean field, given in Eq.~(\ref{Uk-parabole}), is used to calculate the difference between single-particle energies entering the denominator of the HF propagator. In that case, the HF propagator has the same form as the free propagator with an effective mass calculated at the Fermi surface. Defining the dimensionless quantities $x=k_1/k_F$, $k=q/(2k_F)$ and $\nu= m^* \omega/(q k_F)$, the imaginary part of (\ref{GLHF}) can be determined analytically with the result
\begin{eqnarray}
\label{ImGHF}
{\rm Im} G^{HF}_\ell(x,k,\nu)= -2\pi^2 \sqrt{\frac{2\ell+1}{4\pi}}\frac{m^*}{2k^2_Fk} \frac{\theta(x-|\nu-k|)}{x}P_\ell\left( \frac{\nu-k}{x}\right)\left[ \theta(1-x)-\theta(1-x^2-4k\nu)\right] \,.
\end{eqnarray}
The real part, on the contrary, has to be determined numerically as
\begin{eqnarray}
\label{ReGHF}
{\rm Re} G^{HF}_\ell(x,k,\nu)= 2\pi  \sqrt{\frac{2\ell+1}{4\pi}}\frac{m^*}{2k^2_Fk}\left\{ \theta(1-x)\int_{-1}^1 du \frac{P_\ell(u)}{\nu-k-xu}-\int_{-1}^1du P_\ell(u) \frac{\theta(1-x^2-4kxu-4k^2)}{\nu-k-xu}\right\}. \nonumber \\
\end{eqnarray}
For a finite-range potential, the effective mass depends explicitly on the momentum. 
For the two cases of interest considered in this review (Gogny and Nakada), one obtains respectively
\begin{eqnarray}
\frac{m}{m^*_G(x)} & = & 1 + \frac{2 m C_E^{(G)}}{\sqrt{\pi} a_G k_F^2 x^3} e^{-a_G^2 (1+x)^2/4} \left[ 2 +x a_G^2 + e^{x a_G^2 }(x a_G^2 -2)\right] \;,\\
\frac{m}{m^*_N(x)} & = & 1 + \frac{m C_E^{(N)}}{\pi a_N k_F^2 x^3}  \left[ -4 x a_N^2 +(1+a_N^2(1+x^2)) \log\left(\frac{1+a_N^2(1+x)^2}{1+a_N^2(1-x)^2}\right)\right]\;,
\end{eqnarray}
where  $a_G = \mu k_F$, $a_N = k_F/\mu$, $C_E^{(G)} = \frac{1}{4}W+\frac{1}{2}B-\frac{1}{2}H-M$ and $C_E^{(N)} = \frac{1}{16}(-3 t^{(SE)}+t^{(SO)}-3t^{(TE)}+9t^{(TO)})$. The effective mass at the Fermi surface is simply obtained from the above equations with $x=1$.

\begin{figure}[!h]
\begin{center}
\includegraphics[width=0.5\textwidth,angle=0]{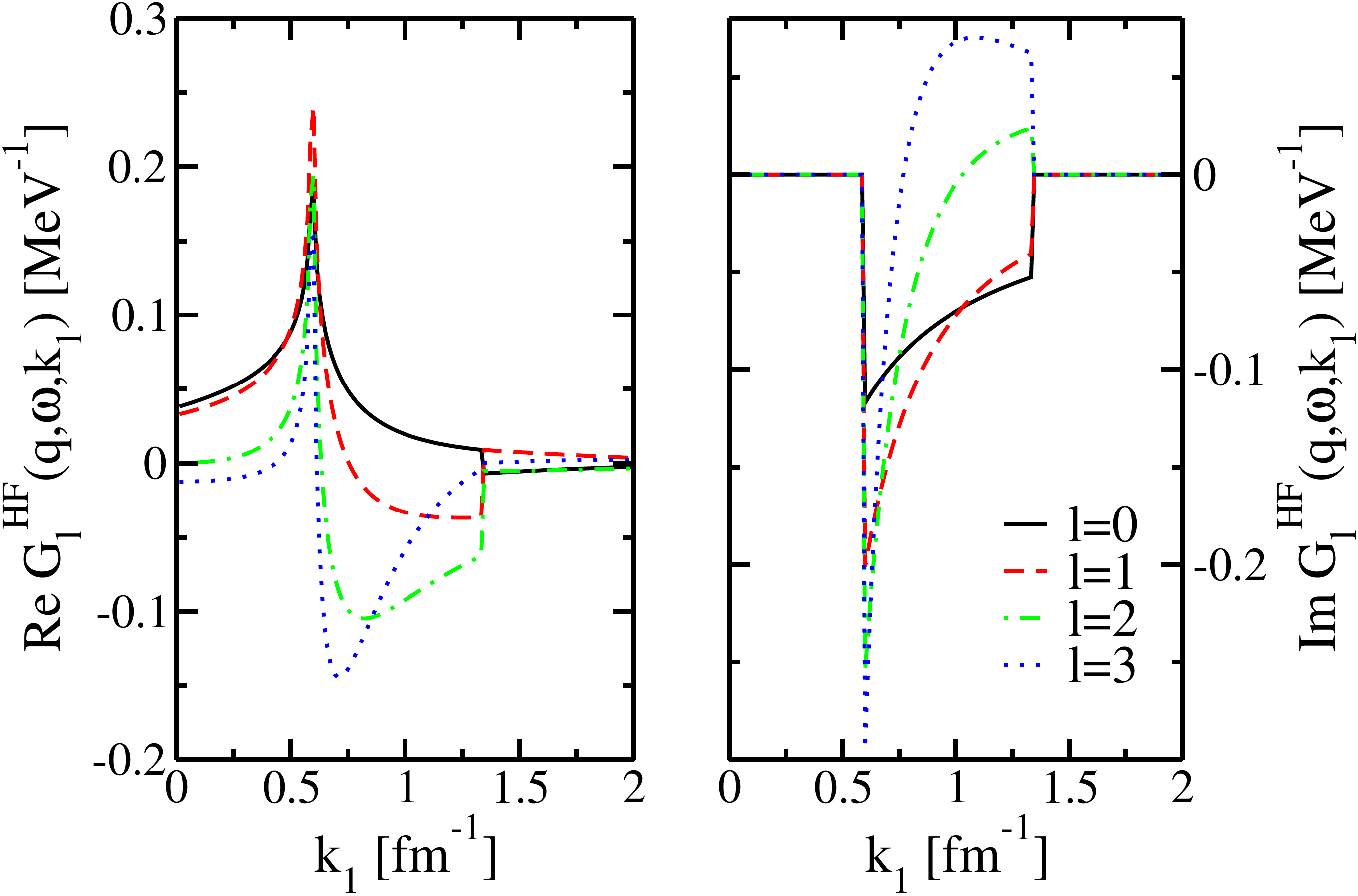}
\end{center}
\caption{Multipole components of the Hartree-Fock propagator for D1S up to $\ell=3$ for $\rho=\rho_0$, $q=k_F$ and $\omega=10$ MeV: real (imaginary) parts are depicted on the left (right) panel.}
\label{Fig:g0}
\end{figure}

One can see that both real and imaginary parts of the multipoles of the HF propagator are piece-wise functions and contain poles. As an example, we show in Fig.~\ref{Fig:g0} the components up to $\ell=3$ of the real and imaginary parts for D1S with $\rho=\rho_0$, $q=k_F$ and $\omega=10$ MeV.

We now turn now to the general case when no approximation on the mean field is made. Instead of the variable $\nu$ we define $\tilde{\nu}=m \omega/(q k_F)$, and obtain from Eq~(\ref{GLHF}) the following expression
\begin{eqnarray}
G_{\ell}^{HF}(q,\omega,k_1) &=& \sqrt{\frac{2\ell +1}{4\pi}} \, \frac{m \pi}{k_F^2 k} \int_{-1}^1 {\rm d}u \, P_\ell(u)
\; \frac{\theta(1-x) - \theta(1-4k^2-x^2-4kxu)}{\tilde{\nu}-k - \tilde U(k,x,u) - x u + i \eta} \;,
\end{eqnarray}
where we have defined $\tilde U(k,x,u) \equiv \frac{m}{2 k_F^2 k}\left[U({\bf k_1}+{\bf q})-U({\bf k_1})\right]$.
The delta function related to the imaginary part implies that the equality $\tilde{\nu}-k - \tilde U(k,x,u) - x u =0$ must be fulfilled. Let $u_0$ be its solution. Then the imaginary part can be written as
\begin{eqnarray}
{\rm Im} \; G_{\ell}^{HF}(q,\omega,k_1) & = & - \sqrt{\frac{2\ell+1}{4\pi}} \; \frac{m\pi^2}{k_F^2 k} \; \frac{P_\ell (u_0)}{| x + (\partial \tilde U/\partial u )_{u_0}|} 
\theta(1-|u_0|) \bigg\{ \theta(1-x) - \theta(1-4k^2-x^2-4kxu_0)\bigg\} \;. \nonumber \\
\end{eqnarray}
For a Gogny interaction, the derivative of $\tilde U$ reads
\begin{eqnarray}
\frac{\partial \tilde U(k,x,u)}{\partial u} & = & \frac{2 m x C_E^{(G)}}{\sqrt{\pi} \mu k_F^3 X^3} e^{-k_F^2\mu^2(1+X)^2/4} \left[  k_F^2 \mu^2 X + 2 + e^{k_F^2 \mu^2 X } \left( k_F^2 \mu^2 X - 2  \right)   \right] \,,
\end{eqnarray}
while for a Nakada interaction, the result is
\begin{eqnarray}
\frac{\partial \tilde U(k,x,u)}{\partial u} & = & -\frac{m x C_E^{(N)}}{\pi \mu k_F^3 X^3}\left[ 4k_F^2X+\left(k_F^2(1+X^2)+\mu^2\right) \log\left[\frac{k_F^2(X-1)^2+\mu^2}{k_F^2(X+1)^2+\mu^2}\right]\right] \,.
\end{eqnarray}
In the above equations, $X$ is a short-hand notation for $X \equiv \sqrt{4 k^2+4k x u + x^2}$.

The real part has to be calculated numerically from the expression
\begin{eqnarray}
{\rm Re} G^{HF}_\ell(x,k,\nu) &=& 2\pi  \sqrt{\frac{2\ell+1}{4\pi}}\frac{m}{2k^2_Fk}\left\{ \theta(1-x)\int_{-1}^1 du \frac{P_\ell(u)}{\tilde{\nu}-k - \tilde U(k,x,u) - x u}  \right. \nonumber \\
&& \left.  \hspace{3cm} -\int_{-1}^1du P_\ell(u) \frac{\theta(1-x^2-4kxu-4k^2)}{\tilde{\nu}-k - \tilde U(k,x,u) - x u}\right\} \,,
\end{eqnarray}
which is similar to Eq.~(\ref{ReGHF}) with the replacements $m^* \to m$ and $\nu \to \tilde{\nu} - \tilde U(k,x,u) $. 

Considering the approximation of the effective mass or the more general case, the HF response function is obtained by an integration of ${\rm Im} \; G_{\ell}^{HF}(q,\omega,k_1)$ over the $k_1$ component. It is thus easy to understand that the presence of poles requires a special treatment. As discussed in Ref.~\cite{mar06}, the simplest solution is to use a dense uniform mesh in k-space. The obvious problem is that a decent accuracy requires a very dense mesh ($\simeq$ 200 points) and consequently the diagonalisation of a large matrix to solve the integro-differential system. 

\section{Multipolar expansion of the ph interactions} \label{app:B}

We present here the explicit expressions of the multipoles of the ph interaction as defined in Eq.~(\ref{eq:dvptV}).
Since a zero-range interaction can be used as a useful test for codes dealing with finite-range interactions, we first show the results for a Skyrme interaction. Then, we give the results for the Gogny and Nakada interactions separately since they differ not only in the radial form factor, but also in the spin-orbit term. The tensor term will be discussed in a specific section.

\subsection{\it Skyrme interaction}\label{sec:skyrme:ph}

The ph matrix elements for Skyrme interaction have already been published in Refs.~\cite{gar92,mar06}. These results have been generalized to extensions of Skyrme interaction as N2LO and N3LO \cite{dav13}. Here we simply provide with the multipoles of the usual Skyrme interaction, which we write as the sum
\be
F^{(\alpha,\alpha')}_{\ell m;\ell' m'} = F^{(\alpha,\alpha')}_{\ell m;\ell' m'; C+DD} + F^{(\alpha,\alpha')}_{\ell m;\ell' m'; SO} \, ,
\ee
where
\bea
F^{(\alpha,\alpha')}_{\ell m;\ell' m'; C+DD} &=& \delta_{\ell \ell'} \delta_{m m'} \delta_{\alpha \alpha'} \left\{ 4\pi\left[ W_1^{(\alpha)}+W_2^{(\alpha)}(k_1^2+k_2^2)\right]\delta_{\ell,0}-\frac{8\pi}{3}W_2^{(\alpha)}k_1k_2\delta_{l,1} \right\}\;, \\
F^{(\alpha,\alpha')}_{\ell m;\ell' m'; SO}  & = & -\frac{4\pi}{\sqrt{3}}qW_0\left[3 \delta_{I,0}+\delta_{I,1}\right]\delta_{I,I'}\delta_{Q,Q'} \left\{ M'\left[k_1\delta_{\ell,1}\delta_{m,M'}\delta_{\ell',0}\delta_{m',0}+k_2\delta_{\ell',1}\delta_{m',-M'}\delta_{\ell,0}\delta_{m,0}\right]\delta_{S',1}\delta_{S,0} \right. \nonumber \\
& & \hskip 4.5 true cm \left. +M\left[k_1\delta_{\ell,1}\delta_{m,-M}\delta_{\ell',0}\delta_{m',0}+k_2\delta_{\ell',1}\delta_{m',M}\delta_{\ell,0}\delta_{m,0}\right] \delta_{S,1}\delta_{S',0}  \right\} \, .\label{AppB:SOSkyrme}
\eea
The coefficients $W^{(\alpha)}_{1,2}$ are combinations of the Skyrme parameters, and are given in Tab.~\ref{skyrme-central}. 

\subsection{\it Gogny interaction}

The multipoles corresponding to the central and density-dependent parts (see Eqs.~(\ref{Appb:central}) and~(\ref{Appb:density})) are written as
\bea
F^{(\alpha,\alpha')}_{\ell m; \ell' m'; C}(k_1,k_2) &=& \delta_{\alpha, \alpha'} \delta_{\ell,\ell'} \delta_{m, m'}
\bigg\{  \delta_{\ell,0} 4 \pi D^{(\alpha)} {\rm e}^{-\frac{1}{4} q^2 \mu^2}  + 2 \pi E^{(\alpha)} {\rm e}^{-\frac{1}{4}(k_1^2+k_2^2) \mu^2} I(\ell) \bigg\} \;,\\
F^{(\alpha,\alpha')}_{\ell m; \ell' m'; DD}(k_1,k_2) &=& \delta_{\alpha, \alpha'} \delta_{\ell,\ell'} \delta_{m, m'} \delta_{\ell,0} 4 \pi R^{(\alpha)} \;,
\eea
while the multipoles of the spin-orbit part coincide with Eq.~(\ref{AppB:SOSkyrme}), since the interactions have the same form. The coefficients $D^{(\alpha)}, E^{(\alpha)}$ have been given in Tab.~\ref{gogny-central}. The functions $I(\ell)$ defined as
\be
I(\ell) = \int_{-1}^1 {\rm d}u P_{\ell}(u) {\rm e}^{a u}\;,
\ee
can be calculated from the recurrence relation
\be
I(\ell+1) = I(\ell-1) - \frac{2\ell+1}{a} I(\ell)\;,
\ee
with 
\bea
I(0) &=& \frac{2}{a} \sinh a\;, \nnn
I(1) &=& \frac{2}{a^2} \left( a \cosh a - \sinh a \right) \;, \nonumber 
\eea
where $a=\frac{1}{2} k_1 k_2 \mu^2$.

\subsection{\it Nakada interaction}

The multipoles corresponding to the central and density-dependent parts (see Eqs.~(\ref{central-Nakada}) and~(\ref{density-Nakada})) are written as
\bea
F^{(\alpha,\alpha')}_{\ell m; \ell' m'; C}(k_1,k_2) &=& \delta_{\alpha, \alpha'} 
\delta_{\ell,\ell'} \delta_{m, m'} \; 
\bigg\{  \delta_{\ell,0}  4 \pi  D^{(\alpha)} \frac{1}{\mu^2 + q^2}  + 2 \pi  E^{(\alpha)} J(\ell)  \bigg\} \;, \\
F^{(\alpha,\alpha')}_{\ell m; \ell' m'; DD}(k_1,k_2) &=& \delta_{\alpha, \alpha'} 
\delta_{\ell,\ell'} \delta_{m, m'}  \delta_{\ell,0}  4 \pi R^{(\alpha)}\;.
\eea
The functions $J(\ell)$ defined as
\be
J(\ell) = \int_{-1}^1  \frac{P_\ell(x)}{b- x} {\rm d}x\;,
\ee
are calculated using the recurrence relation 
\be
J(\ell+1) = \frac{2\ell+1}{\ell+1}  b \, J(\ell) - \frac{\ell}{\ell+1} J(\ell-1) \;,\nn
\ee
where $b=( \mu^2 + k_1^2 + k_2^2) /(2 k_1 k_2) $ and 
\bea
J(0) &=&  \ln \frac{b+1}{b-1} \;,\nnn
J(1) &=&   - 2 + b \ln \frac{b+1}{b-1}. \nonumber
\eea
This kind of integrals will be encountered in the following, but with a different power in the denominator
\be
J_n(\ell) = \int_{-1}^1  \frac{P_\ell(x)}{(b- x)^n} {\rm d}x.
\label{eq:JLnak}
\ee
In that case, the use of the Legendre polynomials standard properties 
\bea
(\ell+1) P_{\ell+1}(x) & = & (2\ell+1) x P_{\ell}(x) - \ell P_{\ell-1}(x)\;, \nnn
P'_{\ell+1}(x) - P'_{\ell-1}(x) & = & (2\ell+1) P_{\ell}(x)\;, \nonumber
\eea
will respectively lead to the following relations
\bea
J_n(\ell+1) & = & -\frac{2\ell+1}{\ell+1} J_{n-1}(\ell) + \frac{2\ell+1}{\ell+1}  b \, J_n(\ell) - \frac{\ell}{\ell+1} J_n(\ell-1)\;, \nnn
J_n(\ell+1) - J_n(\ell-1) & = & -\frac{2\ell+1}{n-1} J_{n-1}(\ell)\;, \nonumber
\eea
which can be combined, when necessary, in 
\bea
\label{AppB:recurrence}
J_n(\ell) & = & \frac{J_n(\ell-1)}{b}  + \left[ 1- \frac{\ell+1}{n-1} \right] \frac{J_{n-1}(\ell)}{b}.
\eea
Therefore, with $J_0(\ell) \equiv J(\ell)$, one can get the other integrals $J_{n>0}(\ell)$.  

The multipoles of the finite-range spin-orbit term given in Eq.~(\ref{so-Nakada}) have a more complex structure and read
\begin{eqnarray*}
F^{(\alpha,\alpha')}_{\ell m; \ell' m'; SO}(k_1,k_2) & = & -\frac{4\pi^2}{\sqrt{3}} q\; \delta_{I,I'}\delta_{Q,Q'}\left[ -\delta_{I,0} \left( t^{(LSE)}+3t^{(LSO)} \right)
 + \delta_{I,1} \left(  t^{(LSE)}- t^{(LSO)} \right)  \right] F_{Y,SO}(q) \nonumber \\
&  & \hskip 1 true cm \left\{ M'\left[k_1\delta_{\ell,1}\delta_{m,M'}\delta_{\ell',0}\delta_{m',0}+k_2\delta_{\ell',1}\delta_{m',-M'}\delta_{\ell,0}\delta_{m,0}\right]\delta_{S',1}\delta_{S,0} \right. \nonumber \\
& & \hskip 1.5 true cm \left. +M\left[k_1\delta_{\ell,1}\delta_{m,-M}\delta_{\ell',0}\delta_{m',0}+k_2\delta_{\ell',1}\delta_{m',M}\delta_{\ell,0}\delta_{m,0}\right] \delta_{S,1}\delta_{S',0}  \right\}  \nonumber \\
&  & - \pi q \; \delta_{I,I'}\delta_{Q,Q'}\left[ \delta_{I,0} \left(  t^{(LSE)}-3t^{(LSO)} \right)
 - \delta_{I,1} \left(  t^{(LSE)}+t^{(LSO)} \right)  \right]  (k_1 F_{SO,\ell'}  - k_2 F_{SO,\ell}) \nonumber \\
&  &  \hskip -1.7 true cm  (-1)^{m} \hat{\ell}\hat{\ell'} \left(\begin{array}{ccc} 
\ell&\ell'&1\\
0&0&0
\end{array} \right)\left[ M' \left(\begin{array}{ccc} 
\ell&\ell'&1\\
-m&m'&M'
\end{array} \right)\delta_{S',1}\delta_{S,0} + M \left(\begin{array}{ccc} 
\ell&\ell'&1\\
-m&m'&-M
\end{array} \right)\delta_{S,1}\delta_{S',0}\right].
\end{eqnarray*}
The functions $F_{SO,\ell}(k_1,k_2)$ are defined by 
\be
\label{AppB:defFSO}
F_{SO,\ell} = \frac{4\pi}{\mu (2k_1k_2)^2} \int_{-1}^1  \frac{P_\ell(x)}{(b- x)^2} {\rm d}x \equiv \frac{4\pi}{\mu (2k_1k_2)^2} J_2(\ell)\;,
\ee
and hence satisfy the following recurrence relation (see Eq.~(\ref{AppB:recurrence}))

\begin{eqnarray*}
F_{SO,\ell} = \frac{F_{SO,\ell-1}}{b} - \frac{4\pi}{\mu (2k_1 k_2)^2} \frac{\ell J(\ell)}{b}\;,
\end{eqnarray*}
with
\begin{eqnarray*}
F_{SO,0}(k_1,k_2) & = & \frac{\pi}{\mu (k_1k_2)^2} \frac{2}{b^2-1}.
\end{eqnarray*}

\subsection{\it Tensor term}\label{app:tens}
In this section, we present the multipoles defined in defined in Eq.~(\ref{eq:dvptV}) for both zero-range and finite-range tensor term.

\subsubsection{\it Zero-range interaction}

For the Skyrme interaction, the tensor term is usually written as
\begin{eqnarray} \label{eq:N3LO:t}
V_{T} &=& \frac{1}{2} t_e T_e({\bf k'},{\bf k})  + \frac{1}{2} t_o T_o({\bf k'},{\bf k})\;,
\end{eqnarray}
where $T_e$ and $T_o$, respectively even and odd under parity transformations, are defined as
\begin{eqnarray}
T_e({\bf k'},{\bf k}) &=& 3 ({\boldsymbol \sigma}_1 \cdot {\bf k'}) ({\boldsymbol \sigma}_2 \cdot {\bf k'}) 
+ 3 ({\boldsymbol \sigma}_1 \cdot {\bf k}) ({\boldsymbol \sigma}_2 \cdot {\bf k}) 
- ({\bf k'}^2 + {\bf k}^2) ({\boldsymbol \sigma}_1 \cdot {\boldsymbol \sigma}_2), \\
T_o({\bf k'},{\bf k}) &=& 3 ({\boldsymbol \sigma}_1 \cdot {\bf k'}) ({\boldsymbol \sigma}_2 \cdot {\bf k}) 
+ 3 ({\boldsymbol \sigma}_1 \cdot {\bf k}) ({\boldsymbol \sigma}_2 \cdot {\bf k'}) 
- 2 ({\bf k'} \cdot {\bf k}) ({\boldsymbol \sigma}_1 \cdot {\boldsymbol \sigma}_2).
\end{eqnarray} 
The ph matrix elements have already been published  in Ref.~\cite{pas12}, and the multipoles read
\bea
F^{(\alpha,\alpha')}_{\ell ' m';\ell m; T}  & = & 2\pi \delta_{S,1}\delta_{S',1}\delta_{I,I'} \delta_{Q,Q'}\left[(t_e+3t_o) \delta_{I,0}+(t_o-t_e) \delta_{I,1}\right]  \nonumber  \\ 
 & &  \left\{k_1k_2\delta_{\ell,1}\delta_{\ell',1} \left[ \frac{2}{3}\delta_{m,m'}\delta_{M,M'} - \delta_{m,M'} \delta_{m',M} - (-1)^{M+M'} \delta_{m,-M} \delta_{m',-M'} \right] \right. \nonumber \\ 
 & &  \left. + \sqrt{6} \left[ (-1)^{m'} k_1^2 \delta_{\ell',0}\delta_{m',0} \delta_{\ell,2} \delta_{m,M'-M} + (-1)^{m} k_2^2 \delta_{\ell,0}\delta_{m,0}  \delta_{\ell',2}\delta_{m',M-M'}\right]
 \left(\begin{array}{@{\hskip0.5mm}c@{\hskip1.3mm}c@{\hskip1.3mm}c@{\hskip0.5mm}} 
1&1&2\\
-M & M' & M-M'
\end{array} \right)    \right\} \nonumber \\
& & \hskip -0.1 true cm + 2 \pi q^2\delta_{S,1}\delta_{S',1}\delta_{I,I'}\delta_{Q,Q'} \left[(t_e-3t_o) \delta_{I,0}-(t_e+t_o) \delta_{I,1}\right]  \left[3\delta_{M,0}\delta_{M',0} -\delta_{M,M'}\right]\delta_{\ell,0}\delta_{\ell',0}\delta_{m,0}\delta_{m',0} 
\eea

\subsubsection{\it Finite-range interaction}

Let us write a general finite-range tensor interaction as
\begin{equation}\label{AppB:T}
V_T = (V_1 + V_2 P_\tau) W_T(r_{12}) S_{T}(\hat{\br}_{12}).
\end{equation}
For the Gogny tensor interaction~(\ref{V-gogny-2}), we made the replacements $V_1 \to V_{T1}$, $V_2 \to V_{T2}$ and $W_T \to e^{-(r/\mu_{\scriptscriptstyle{G}})^2}$, while for the Nakada interaction~(\ref{m3y:interactionT}) we have
\bea
V_{T1} &\to& \frac{1}{2}(t^{(TNE)}+t^{TN0})\;, \\
V_{T2} &\to& \frac{1}{2}(-t^{(TNE)}+t^{TN0})\;, \\
W_T &\to& \frac{e^{-\mu_{\scriptscriptstyle{T}}r}}{\mu_{\scriptscriptstyle{T}}r} r^2 .
\eea
The ph matrix elements of~(\ref{AppB:T}) are written as
\bea
 V_{T\,ph}^{(\alpha,\alpha')} & = & - 8\pi \delta_{S,1}\delta_{S',1}\delta_{I,I'}\delta_{Q,Q'} \left\{  \left[  \delta_{I,0} \left(2V_{1} + V_{2} \right) + \delta_{I,1}  V_{2} \right] F_T (q) q^2 \left[3\delta_{M,0}\delta_{M',0} -\delta_{M,M'}\right] \right. \nonumber \\ 
& & \hskip 2.2 true cm -  \left. \left[ \delta_{I,0} \left(  V_{1} + 2V_{2} \right)  + \delta_{I,1} V_{1}\right] F_{T}(k_{12}) \left[3(-1)^{M}(k_{12})^{1}_{-M}(k_{12})^{1}_{M'} -k_{12}^2\delta_{M,M'}\right] \right\}
\eea
where the function $F_{T}$ is a shorthand notation for $F_{Y,T}, F_{G,T}$ whose expressions are
\bea
F_{Y,T}(k) & = & \frac{8}{\mu_{\scriptscriptstyle{Y}}(k^2+\mu_{\scriptscriptstyle{Y}}^2)^3}\;, \label{eq:FYT} \\
F_{G,T}(k) & = & \frac{1}{4 k^5} \left( 6\pi \hbox{Erf}\left(\frac{k\mu_{\scriptscriptstyle{G}}}{2}\right)-\sqrt{\pi}k\mu_{\scriptscriptstyle{G}} (6+k^2\mu_{\scriptscriptstyle{G}}^2)e^{-k^2\mu_{\scriptscriptstyle{G}}^2/4} \right) \;.  \label{eq:FGT} 
\eea
The final expression for the multipoles reads
\begin{eqnarray*}
F^{(\alpha,\alpha')}_{\ell ' m';\ell m; T} & = & - 8\pi \delta_{S,1}\delta_{S',1}\delta_{Q,Q'} \left\{  \;\; \left[ \delta_{I,0} \left(2V_{1} + V_{2} \right) + \delta_{I,1}  V_{2} \right] 4 \pi F_T (q) q^2 \left[3\delta_{M,0}\delta_{M',0} -\delta_{M,M'}\right]\delta_{\ell,0}\delta_{\ell',0}\delta_{m,0}\delta_{m',0} \right. \\ \nonumber
& & \hskip 2.8 true cm -  \left. \left[ \delta_{I,0} \left(  V_{1} + 2V_{2} \right)  + \delta_{I,1} V_{1}\right] T_{\ell m,\ell' m'} \right\} \;,\\ \nonumber
\end{eqnarray*}
where $T_{\ell m,\ell' m'}$ is defined as
\begin{eqnarray*}
T_{lm,l'm'} &=&3k_1k_2\delta_{M+m,M'+m'} \\ \nonumber
& &  \left\{ \hspace{0.5cm} \sqrt{ \ell' ( \ell'-1)}\delta_{ \ell, \ell'-2}A^T_{ \ell'-1}\left[ (-1)^{m-m'}\left(\begin{array}{@{\hskip0.5mm}c@{\hskip1.3mm}c@{\hskip1.3mm}c@{\hskip0.5mm}} 
1 &  \ell'-1 &  \ell'-2 \\
M' &  m-M' & - m
\end{array} \right)\left(\begin{array}{@{\hskip0.5mm}c@{\hskip1.3mm}c@{\hskip1.3mm}c@{\hskip0.5mm}} 
1 &  \ell'-1 &  \ell' \\
-M & M- m' &  m'
\end{array} \right) \right. \right. \nonumber \\ %
&& \left. \left. \hspace{5cm} +\left(\begin{array}{@{\hskip0.5mm}c@{\hskip1.3mm}c@{\hskip1.3mm}c@{\hskip0.5mm}}  
1 &  \ell'-1 &  \ell'-2 \\
-M & M+ m & - m
\end{array} \right)
\left(\begin{array}{@{\hskip0.5mm}c@{\hskip1.3mm}c@{\hskip1.3mm}c@{\hskip0.5mm}}  
1 &  \ell'-1 &  \ell' \\
M' & -m'-M' & m'
\end{array} \right)\right] \right.\\ \nonumber
& &   \left. + \sqrt{( \ell'+1) ( \ell'+2)}\delta_{ \ell, \ell'+2}A^T_{ \ell'+1}\left[ (-1)^{m-m'}\left(\begin{array}{@{\hskip0.5mm}c@{\hskip1.3mm}c@{\hskip1.3mm}c@{\hskip0.5mm}}  
1 &  \ell'+1 &  \ell' \\
M' & -m'-M' & m'
\end{array} \right)\left(\begin{array}{@{\hskip0.5mm}c@{\hskip1.3mm}c@{\hskip1.3mm}c@{\hskip0.5mm}} 
1 &  \ell'+1 &  \ell'+2 \\
-M & M+ m & - m
\end{array} \right) \right. \right. \nonumber \\ %
&& \left. \left. \hspace{5cm} +\left(\begin{array}{@{\hskip0.5mm}c@{\hskip1.3mm}c@{\hskip1.3mm}c@{\hskip0.5mm}} 
1 &  \ell'+1 &  \ell' \\
-M & M-m' & m'
\end{array} \right)
\left(\begin{array}{@{\hskip0.5mm}c@{\hskip1.3mm}c@{\hskip1.3mm}c@{\hskip0.5mm}}  
1 &  \ell'+1 &  \ell'+2 \\
M' &  m-M' & - m
\end{array} \right)\right] \right. \\ \nonumber 
& &  \left. - \delta_{ \ell, \ell'}  \ell' A^T_{ \ell'-1}\left[ (-1)^{m-m'}\left(\begin{array}{@{\hskip0.5mm}c@{\hskip1.3mm}c@{\hskip1.3mm}c@{\hskip0.5mm}} 
1 &  \ell'-1 &  \ell' \\
-M & M-m' & m'
\end{array} \right)\left(\begin{array}{@{\hskip0.5mm}c@{\hskip1.3mm}c@{\hskip1.3mm}c@{\hskip0.5mm}}  
1 &  \ell'-1 &  \ell' \\
M' &  m-M' & - m
\end{array} \right) \right. \right. \nonumber \\ %
&& \left. \left. \hspace{5cm} +\left(\begin{array}{@{\hskip0.5mm}c@{\hskip1.3mm}c@{\hskip1.3mm}c@{\hskip0.5mm}} 
1 &  \ell'-1 &  \ell' \\
-M & M+ m & - m
\end{array} \right)
\left(\begin{array}{@{\hskip0.5mm}c@{\hskip1.3mm}c@{\hskip1.3mm}c@{\hskip0.5mm}}
1 &  \ell'-1 &  \ell' \\
M' & -m'-M' & m'
\end{array} \right)\right] \right. \\ \nonumber %
& &  \left. - \delta_{ \ell, \ell'} ( \ell'+1) A^T_{ \ell'+1}\left[ (-1)^{m-m'}\left(\begin{array}{@{\hskip0.5mm}c@{\hskip1.3mm}c@{\hskip1.3mm}c@{\hskip0.5mm}} 
1 &  \ell'+1 &  \ell' \\
-M & M-m' & m'
\end{array} \right)\left(\begin{array}{@{\hskip0.5mm}c@{\hskip1.3mm}c@{\hskip1.3mm}c@{\hskip0.5mm}}  
1 &  \ell'+1 &  \ell' \\
M' &  m-M' & - m
\end{array} \right)  \right. \right. \nonumber \\ %
&& \left. \left. \hspace{5cm} +\left(\begin{array}{@{\hskip0.5mm}c@{\hskip1.3mm}c@{\hskip1.3mm}c@{\hskip0.5mm}}  
1 &  \ell'+1 &  \ell' \\
-M & M+ m & - m
\end{array} \right)
\left(\begin{array}{@{\hskip0.5mm}c@{\hskip1.3mm}c@{\hskip1.3mm}c@{\hskip0.5mm}}  
1 &  \ell'+1 &  \ell' \\
M' & -m'-M' & m'
\end{array} \right)\right] \hspace{0.5cm} \right\} \\ \nonumber %
& & + \delta_{M+ m,M'+m'} \sqrt{30}\; \hat{\ell} \hat{\ell'}(-1)^{M'-m} (A^T_{ \ell'}k_1^2+A^T_{ \ell}k_2^2) 
\left(\begin{array}{@{\hskip0.5mm}c@{\hskip1.3mm}c@{\hskip1.3mm}c@{\hskip0.5mm}} 
1 & 1 & 2 \\
-M & M' & M - M'
\end{array} \right)
\left(\begin{array}{@{\hskip0.5mm}c@{\hskip1.3mm}c@{\hskip1.3mm}c@{\hskip0.5mm}}  
\ell' &  \ell & 2 \\
0 & 0 & 0
\end{array} \right)
\left(\begin{array}{@{\hskip0.5mm}c@{\hskip1.3mm}c@{\hskip1.3mm}c@{\hskip0.5mm}}  
\ell' &  \ell & 2 \\
m' & - m &  m-m'
\end{array} \right) 
\nonumber \\ 
& &+  \delta_{\ell,\ell'} \delta_{m,m'}  \delta_{M,M'} 2k_1 k_2
\left( \frac{\ell}{2\ell+1}A^T_{ \ell-1}+\frac{\ell+1}{2\ell+1}A^T_{ \ell+1} \right)
.\nonumber 
 \end{eqnarray*}
The coefficients $A^T_{ \ell}$ are defined as
\be
A^T_{ \ell} = 2 \pi \int_{-1}^1 {\rm d}x  F_T(\sqrt{k_1^2+k_2^2-2 k_1 k_2 x}) P_\ell(x) \,,
\ee
where $F_T$ is given in Eqs.~(\ref{eq:FYT}) and (\ref{eq:FGT}). 

For Nakada interaction, they can be written as
\begin{eqnarray*}
A^{Y,T}_{\ell} & = & 
\frac{16\pi}{\mu (2k_1k_2)^3} J_3(\ell) \;,\nonumber
\end{eqnarray*}
where the function $J_3$ is defined in Eq.~(\ref{eq:JLnak}). From Eq.~(\ref{AppB:recurrence}) a useful recursion relation can be obtained as
\begin{eqnarray*}
A^{Y,T}_{\ell} & = & \frac{A^{Y,T}_{ \ell-1}}{b} - \frac{16\pi}{\mu (2k_1 k_2)^3} \frac{(\ell-1) J_2(\ell)}{2b}\;, \nonumber
\end{eqnarray*}
with
\begin{eqnarray*}
A^{Y,T}_{0} =  \frac{\pi}{\mu(k_1k_2)^3} \frac{4b}{(b^2-1)^2}\;, \nonumber
\end{eqnarray*}
where we used the abridged notation $b=\frac{\mu^2+k_1^2+k_2^2}{2k_1k_2}$. 

For the Gogny interaction, we have not been able to obtain a  closed form as the one obtained for Nakada. We give the explicit expression for the first coefficients
\begin{eqnarray*}
\pi^{-3/2} \mu_{\scriptscriptstyle{G}}^{-5} A^{G,T}_{0}  & = &  
\frac{e^{-a} [a \sinh (b)+b \cosh (b)]}{4 b \left(b^2-a^2\right)}+\frac{\sqrt{\pi}  
    \text{Erf}\left(\sqrt{a-b}\right)}{16 b (a-b)^{3/2}}-\frac{\sqrt{\pi}  \text{Erf}\left(\sqrt{a+b}\right)}{16
   b (a+b)^{3/2}} \\ \nonumber
\pi^{-3/2} \mu_{\scriptscriptstyle{G}}^{-5} A^{G,T}_{1}  & = &    
   \frac{ e^{-a}   [a \cosh (b)+b \sinh (b)]}{4 b \left(b^2-a^2\right)}-\frac{\sqrt{\pi} 
    (2 a-3 b)  \text{Erf}\left(\sqrt{a-b}\right)}{16 b^2 (a-b)^{3/2}}+\frac{\sqrt{\pi}   (2 a+3 b)
    \text{Erf}\left(\sqrt{a+b}\right)}{16 b^2 (a+b)^{3/2}} \\ \nonumber
\pi^{-3/2}    \mu_{\scriptscriptstyle{G}}^{-5} A^{G,T}_{2}  & = & 
    -\frac{ e^{-a}   [ \left(6 a^2+(a-6)
  b^2\right) \sinh (b)+b^3 \cosh (b) ]}{4 b^3 (a-b) (a+b)} \\ \nonumber
& &   -\frac{\sqrt{\pi}   \left(12 a^2-18 a b+5 b^2\right)
    \text{Erf}\left(\sqrt{a-b}\right)}{16 b^3 (a-b)^{3/2}}+\frac{\sqrt{\pi}   \left(12 a^2+18 a b+5 b^2\right)
    \text{Erf}\left(\sqrt{a+b}\right)}{16 b^3 (a+b)^{3/2}}\\ \nonumber
\pi^{-3/2}  \mu_{\scriptscriptstyle{G}}^{-5} A^{G,T}_{3}  & = & 
    -\frac{ e^{-a}   [ b \left(10
   a^2+(a-10) b^2\right) \cosh (b)+\left(10 a^2 (2 a-1)+10 (1-2 a) b^2+b^4\right) \sinh (b) ]}{4 b^4 (a-b)
   (a+b)} \\ \nonumber
   & &   -\frac{\sqrt{\pi}   \left(40 a^3-60 a^2 b+12 a b^2+7 b^3\right)  \text{Erf}\left(\sqrt{a-b}\right)}{16
   b^4 (a-b)^{3/2}} \\ \nonumber
   & &   +\frac{\sqrt{\pi}   \left(40 a^3+60 a^2 b+12 a b^2-7 b^3\right)
    \text{Erf}\left(\sqrt{a+b}\right)}{16 b^4 (a+b)^{3/2}} \\ \nonumber
\pi^{-3/2}    \mu_{\scriptscriptstyle{G}}^{-5} A^{G,T}_{4}  & = & 
   -\frac{ e^{-a}   [ 14 a^2
   \left(4 a^2-2 a+3\right)+(a-6) b^4-2 (a (25 a-14)+21) b^2  \sinh (b) ]}{4 b^5 (a-b) (a+b)}  \\ \nonumber 
   & &   -\frac{ e^{-a} \, b \left[ 14 a^2 (2 a-3)+14 (3-2 a)
   b^2+b^4\right] \cosh (b) }{4 b^5 (a-b) (a+b)}  \\ \nonumber 
   & &   -\frac{\sqrt{\pi}   \left(112 a^4-168 a^3 b+12 a^2 b^2+52
   a b^3-9 b^4\right)  \text{Erf}\left(\sqrt{a-b}\right)}{16 b^5 (a-b)^{3/2}} \\ \nonumber
   & &   +\frac{\sqrt{\pi}   \left(112
   a^4+168 a^3 b+12 a^2 b^2-52 a b^3-9 b^4\right)  \text{Erf}\left(\sqrt{a+b}\right)}{16 b^5 (a+b)^{3/2}}\\ \nonumber 
\pi^{-3/2}   \mu_{\scriptscriptstyle{G}}^{-5} A^{G,T}_{5}  & = & 
     -\frac{ e^{-a} \!  \left[18 a^2 (6 a - 4 a^2 + 8 a^3\!-\!15 ) \!-\! 
 2 (  2 a (27 + a (7 + 40 a))\!-\!135) b^2\! + 4 (25 + 4 a) b^4 + b^6\right] \sinh (b)}{4 b^6 (a-b) (a+b)}  \\ \nonumber 
 & &    -\frac{ e^{-a}  \, b \left[18 a^2 (15 - 6 a + 4 a^2) - 
   2 (135 + a (-54 + 31 a)) b^2 + (-10 + a) b^4 \right] \cosh (b)}{4 b^6 (a-b) (a+b)}  \\ \nonumber 
   & &  +\frac{\sqrt{\pi}   \left(-288 a^5 + 432 a^4 b + 32 a^3 b^2 - 228 a^2 b^3 + 42 a b^4 + 11 b^5\right)  \text{Erf}\left(\sqrt{a-b}\right)}{16 b^6 (a-b)^{3/2}}  \\ \nonumber 
   & &   +\frac{\sqrt{\pi}   \left(288 a^5 + 432 a^4 b - 32 a^3 b^2 - 228 a^2 b^3 - 42 a b^4 + 11 b^5\right)  \text{Erf}\left(\sqrt{a+b}\right)}{16 b^6 (a+b)^{3/2}} \;. 
\end{eqnarray*}
where we have defined $a=\frac{(k_1^2+k_2^2)\mu_{\scriptscriptstyle{G}}^2}{4}$ and $b=\frac{k_1k_2\mu_{\scriptscriptstyle{G}}^2}{2}$. 

\end{appendices} 

\bibliographystyle{elsarticle-num}
\bibliography{biblio-PPNP}

\begin{thebibliography}{100}
\expandafter\ifx\csname url\endcsname\relax
  \def\url#1{\texttt{#1}}\fi
\expandafter\ifx\csname urlprefix\endcsname\relax\def\urlprefix{URL }\fi
\expandafter\ifx\csname href\endcsname\relax
  \def\href#1#2{#2} \def\path#1{#1}\fi

\bibitem{fet71}
A.~L. Fetter, J.~D. Walecka, Quantum Theory of Many-Particle Systems,
  McGraw-Hill, New York, 1971.

\bibitem{rin80}
P.~Ring, P.~Schuck, The Nuclear Many-Body Problem, Springer-Verlag Berlin
  Heidelberg, 1980.

\bibitem{pin66}
D.~Pines, P.~Nozi\`eres, The Theory of Quantum Liquids, Benjamin, New York,
  1966.

\bibitem{lip08}
E.~Lipparini, Modern Many-Particle Physics, World Scientific, Singapore, 2008.

\bibitem{dic08}
W.~Dickhoff, D.~V. Neck, Many-Body Theory Exposed, World Scientific, Singapore,
  2008.

\bibitem{din18}
P.~M. Dinh, J.~Navarro, E.~Suraud, An Introduction to Quantum Fluids, CRC
  Press, Boca Raton, 2018.

\bibitem{cen90}
M.~Centelles, M.~Pi, X.~Vi\~nas, F.~Garcias, M.~Barranco, Nucl. Phys. A 510
  (1990) 397.

\bibitem{bal99}
M.~Baldo, Nuclear methods and the nuclear equation of state, World Scientific,
  Singapore, 1999.

\bibitem{cen07}
M.~Centelles, P.~Schuck, X.~Vi\~nas, Ann. Phys. (NY) 322 (2007) 363.

\bibitem{cha08}
N.~Chamel, P.~Haensel, Living Reviews in Relativity 11 (2008) 10.

\bibitem{wes75}
G.~B. West, Phys. Rep. 18 (1975) 263.

\bibitem{bur00}
A.~Burrows, T.~Young, Phys. Rep. 333 (2000) 63.

\bibitem{vol00}
C.~Volpe, N.~Auerbach, G.~Colo, T.~Suzuki, N.~Van~Giai, Phys. Rev. C 62~(1)
  (2000) 015501.

\bibitem{bortignon2019giant}
P.~F. Bortignon, A.~Bracco, R.~A. Broglia, Giant Resonances, Routledge, 2019.

\bibitem{suh07}
J.~Suhonen, From Nucleons to Nucleus, Springer-Verlag Berlin Heidelberg, 2007.

\bibitem{har01}
M.~N. Harakeh, A.~Woude, Vol.~24, Oxford University Press on Demand, 2001.

\bibitem{kha05}
E.~Khan, N.~Sandulescu, N.~Van~Giai, Phys. Rev. C 71~(4) (2005) 042801.

\bibitem{bar10}
S.~Baroni, A.~Pastore, F.~Raimondi, F.~Barranco, R.~A. Broglia, E.~Vigezzi,
  Phys. Rev. C 82 (2010) 015807.

\bibitem{cha13}
N.~Chamel, D.~Page, S.~Reddy, Phys. Rev. C 87 (2013) 035803.

\bibitem{car15}
J.~Carlson, S.~Gandolfi, F.~Pederiva, S.~Pieper, R.~Schiavila, K.~Schmidt,
  R.~Wiringa, Rev. Mod. Phys. 87 (2015) 1067.

\bibitem{hag14}
G.~Hagen, T.~Papenbrock, A.~Ekström, K.~Wendt, G.~Baardsen, S.~Gandolfi,
  M.~Hjorth-Jensen, C.~Horowitz, Phys. Rev. C 89 (2014) 014319.

\bibitem{bog10}
S.~Bogner, R.~Furnstahl, A.~Schwenk, Prog. Part. Nucl. Phys. 65 (2010) 94.

\bibitem{dic04}
W.~Dickhoff, C.~Barbieri, Prog. Part. Nucl. Phys. 52 (2004) 377.

\bibitem{zuo02}
W.~Zuo, A.~Lejeune, U.~Lombardo, J.~Mathiot, Nucl. Phys. A706 (2002) 418.

\bibitem{akm98}
A.~Akmal, V.~Pandharipande, D.~Ravenhall, Phys. Rev. C 58 (1998) 1804.

\bibitem{ter87}
B.~ter Haar, R.~Malfliet, Phys. Rep. 149 (1987) 207.

\bibitem{ben03}
M.~Bender, P.-H. Heenen, P.-G. Reinhard, Rev. Mod. Phys. 75 (2003) 121.

\bibitem{gra19}
M.~Grasso, Prog. Part. Nucl. Phys. 106 (2019) 256.

\bibitem{duc07}
C.~Ducoin, P.~Chomaz, F.~Gulminelli, Nucl. Phys. A 781 (2007) 407.

\bibitem{bar73}
S.~Barshay, G.~Brown, Phys. Lett. B 47 (1973) 107.

\bibitem{alb80}
W.~Alberico, M.~Ericson, A.~Molinari, Phys. Lett. B 92 (1980) 153.

\bibitem{vid84}
A.~Vidaurre, J.~Navarro, J.~Bernabeu, Astron. Astroph. 135 (1984) 361.

\bibitem{pas12b}
A.~Pastore, K.~Bennaceur, D.~Davesne, J.~Meyer, Int. J. Mod. Phys. E 21~(05)
  (2012) 1250040.

\bibitem{hel13}
V.~Hellemans, A.~Pastore, T.~Duguet, K.~Bennaceur, D.~Davesne, J.~Meyer,
  M.~Bender, P.-H. Heenen, Phys. Rev. C 88 (2013) 064323.

\bibitem{pas15}
A.~Pastore, D.~Tarpanov, D.~Davesne, J.~Navarro, Phys. Rev. C 92 (2015) 024305.

\bibitem{mar19}
{Martini, M.}, {De Pace, A.}, {Bennaceur, K.}, Eur. Phys. J. A 55 (2019) 150.

\bibitem{les06}
T.~Lesinski, K.~Bennaceur, T.~Duguet, J.~Meyer, Phys. Rev. C 74 (2006) 044315.

\bibitem{fra12}
S.~Fracasso, E.~B. Suckling, P.~D. Stevenson, Phys. Rev. C 86 (2012) 044303.

\bibitem{pas13}
A.~Pastore, D.~Davesne, K.~Bennaceur, J.~Meyer, V.~Hellemans, Phys. Scripta
  2013~(T154) (2013) 014014.

\bibitem{sad12}
J.~Sadoudi, M.~Bender, K.~Bennaceur, D.~Davesne, R.~Jodon, et~al., Phys.
  Scripta T154 (2013) 014013.

\bibitem{bec17}
P.~Becker, D.~Davesne, J.~Meyer, J.~Navarro, A.~Pastore, Phys. Rev. C 96 (2017)
  044330.

\bibitem{bro06}
B.~Brown, T.~Duguet, T.~Otsuka, D.~Abe, T.~Suzuki, Phys. Rev. C 74~(6) (2006)
  061303.

\bibitem{les07}
T.~Lesinski, M.~Bender, K.~Bennaceur, T.~Duguet, J.~Meyer, Phys. Rev. C 76
  (2007) 014312.

\bibitem{ben09B}
M.~Bender, K.~Bennaceur, T.~Duguet, P.~H. Heenen, T.~Lesinski, J.~Meyer, Phys.
  Rev. C 80 (2009) 064302.

\bibitem{sag14}
H.~Sagawa, G.~Col{\`o}, Progr. Part. Nucl. Phys. 76 (2014) 76.

\bibitem{sky59}
T.~H.~R. Skyrme, Nucl. Phys. 9 (1959) 615.

\bibitem{report}
A.~Pastore, D.~Davesne, J.~Navarro, Phys. Rep. 563 (2015) 1.

\bibitem{gog75}
D.~Gogny, Proc. Int. Conf. Nuclear Self-consistent Fields, Edited by G. Ripka
  and M. Porneuf, Noth-Holland, Amsterdam, 1975.

\bibitem{dec80}
J.~Decharg\'e, D.~Gogny, Phys. Rev. C 21 (1980) 1568--1593.

\bibitem{rob19}
L.~Robledo, T.~Rodr\'{\i}guez, R.~Rodr\'{\i}guez-Guzm\'an, J. Phys. G 46 (2019)
  1.

\bibitem{nak03}
H.~Nakada, Phys. Rev. C 68 (2003) 014316.

\bibitem{nak20}
H.~Nakada, Int. J. Mod. Phys. E 29 (2020) 1930008.

\bibitem{rai14}
F.~Raimondi, K.~Bennaceur, J.~Dobaczweski, J. Phys. G 41 (2014) 055112.

\bibitem{ben17}
K.~Bennaceur, A.~Idini, J.~Dobaczewski, P.~Dobaczewski, M.~Kortelainen,
  F.~Raimondi, J. Phys. G 44 (2017) 045106.

\bibitem{vau72}
D.~Vautherin, D.~Brink, Phys. Rev. C 5 (1972) 626.

\bibitem{gar92}
C.~Garc\'{\i}a-Recio, J.~Navarro, V.~G. Nguyen, L.~Salcedo, Ann. Phys. (NY) 214
  (1992) 293.

\bibitem{bra94}
F.~Braghin, D.~Vautherin, Phys. Lett. B 333 (1994) 289.

\bibitem{bra95}
F.~L. Braghin, D.~Vautherin, A.~Abada, Phys. Rev. C 52 (1995) 2504.

\bibitem{her96}
E.~S. Hern\'andez, J.~Navarro, A.~Polls, J.~Ventura, Nucl. Phys. A 597 (1996)
  1.

\bibitem{her97}
E.~S. Hern\'andez, J.~Navarro, A.~Polls, Nucl. Phys. A 627 (1997) 460.

\bibitem{her99}
E.~Hern{\'a}ndez, J.~Navarro, A.~Polls, Nucl. Phys. A 658 (1999) 327.

\bibitem{nav99}
J.~Navarro, E.~S. Hern\'andez, D.~Vautherin, Phys. Rev. C 60 (1999) 045801.

\bibitem{mar03}
J.~Margueron, J.~Navarro, N.~Van~Giai, Nucl. Phys. A 719 (2003) 169c.

\bibitem{sta77}
F.~Stancu, D.~Brink, H.~Flocard, Phys. Lett. B 68 (1977) 108.

\bibitem{bri07}
D.~M. Brink, F.~Stancu, Phys. Rev. C 75 (2007) 064311.

\bibitem{col07}
G.~Col\`o, H.~Sagawa, S.~Fracasso, P.~Bortignon, Phys. Lett. B 646 (2007) 227.

\bibitem{bai09}
C.~Bai, H.~Sagawa, H.~Zhang, X.~Zhang, G.~Col\`o, F.~Xu, Phys. Lett. B 675
  (2009) 28.

\bibitem{bai09a}
C.~L. Bai, H.~Q. Zhang, X.~Z. Zhang, F.~R. Xu, H.~Sagawa, G.~Col\`o, Phys. Rev.
  C 79 (2009) 041301.

\bibitem{ton83}
F.~Tondeur, Phys. Lett. B 123 (1983) 139.

\bibitem{liu91}
K.-F. Liu, H.~Luo, Z.~Ma, Q.~Shen, S.~Moszkowski, Nucl. Phys. A 534 (1991) 1.

\bibitem{she19}
S.~Shen, G.~Col{\`o}, X.~Roca-Maza, et~al., Phys. Rev. C 99~(3) (2019) 034322.

\bibitem{sam04}
M.~Samyn, S.~Goriely, M.~Bender, J.~M. Pearson, Phys. Rev. C 70 (2004) 044309.

\bibitem{cen09}
M.~Centelles, X.~Roca-Maza, X.~Vi\~nas, M.~Warda, Phys. Rev. Lett. 102 (2009)
  122502.

\bibitem{dan09}
J.~Danielewicz, J.~Lee, Int. J. Mod. Phys. E 18 (2009) 892.

\bibitem{roc11}
X.~Roca-Maza, M.~Centelles, X.~Vi\~nas, M.~Warda, Phys. Rev. Lett. 106 (2011)
  252501.

\bibitem{rep13}
A.~Repko, P.-G. Reinhard, V.~O. Nesterenko, J.~Kvasil, Phys. Rev. C 87 (2013)
  024305.

\bibitem{blo13}
J.~P. Blocki, A.~G. Magner, P.~Ring, A.~A. Vlasenko, Phys. Rev. C 87 (2013)
  044304.

\bibitem{kut89}
M.~Kutschera, W.~W\'ojcik, Phys. Lett. B 223 (1989) 11.

\bibitem{rio05}
A.~Rios, A.~Polls, I.~Vida\~na, Phys. Rev. C 71 (2005) 055802.

\bibitem{mar02}
J.~Margueron, J.~Navarro, V.~G. Nguyen, Phys. Rev. C 66 (2002) 014303.

\bibitem{bul02}
A.~Bulgac, Y.~Yu, Phys. Rev. Lett. 88 (2002) 042504.

\bibitem{gra03}
M.~Grasso, M.~Urban, Phys. Rev. A 68 (2003) 033610.

\bibitem{bor06}
P.~Borycki, J.~Dobaczewski, W.~Nazarewicz, M.~Stoitsov, Phys. Rev. C 73 (2006)
  044319.

\bibitem{ber07}
G.~Bertsch, D.~Dean, W.~Nazarewicz, SciDAC Rev 6 (2007) 42.

\bibitem{kor10}
M.~Kortelainen, T.~Lesinski, J.~Mor\'e, W.~Nazariewicz, J.~Sarich, N.~Schunck,
  M.~Stoitsov, S.~Wild, Phys. Rev. C 82 (2010) 024313.

\bibitem{kor14}
M.~Kortelainen, J.~McDonnell, W.~Nazariewicz, E.~Olsen, P.~Reinhard, J.~Sarich,
  N.~Schunck, M.~Stoitsov, S.~Wild, D.~Davesne, J.~Erler, A.~Pastore, Phys.
  Rev. C 89 (2014) 054314.

\bibitem{car10}
B.~Carlsson, J.~Dobaczewski, Phys. Rev. Lett. 105 (2010) 122501.

\bibitem{car08}
B.~G. Carlsson, J.~Dobaczewski, M.~Kortelainen, Phys. Rev. C 78 (2008) 044326.

\bibitem{rai11}
F.~Raimondi, B.~G. Carlsson, J.~Dobaczewski, Phys. Rev. C 83 (2011) 054311.

\bibitem{dav13}
D.~Davesne, A.~Pastore, J.~Navarro, J. Phys. G 40 (2013) 095104.

\bibitem{bec15}
P.~Becker, D.~Davesne, J.~Meyer, A.~Pastore, J.~Navarro, J. Phys. G 42 (2015)
  034001.

\bibitem{pas13b}
A.~Pastore, D.~Davesne, J.~Navarro, J. Phys. G 41 (2014) 055103.

\bibitem{mar05}
J.~Margueron, N.~Van~Giai, J.~Navarro, Phys. Rev. C 72 (2005) 034311.

\bibitem{ose82}
E.~Oset, H.~Toki, W.~Weise, Phys. Rep. 83 (1982) 281.

\bibitem{del85}
A.~Dellafiore, F.~Lenz, F.~Brieva, Phys. Rev. C 31 (1985) 1088.

\bibitem{bri87}
F.~Brieva, A.~Dellafiore, Phys. Rev. C 36 (1987) 899.

\bibitem{sch89}
P.~Schuck, R.~Hasse, J.~Jaenicke, C.~Gr\'egoire, B.~R\'emaud, F.~S\'ebille,
  E.~Suraud, Prog. Part. Nucl. Phys. 22 (1989) 181.

\bibitem{pac98}
A.~De~Pace, Nucl. Phys. A 635 (1998) 163.

\bibitem{mar08}
J.~Margueron, J.~Navarro, N.~Van~Giai, P.~Schuck, Phys. Rev. C 77 (2008)
  064306.

\bibitem{pac16}
A.~De~Pace, M.~Martini, Phys. Rev. C 94 (2016) 024342.

\bibitem{mig67}
A.~B. Migdal, The Theory of Finite Fermi Systems, Wiley, New York, 1967.

\bibitem{vau96}
D.~Vautherin, Adv. Nucl. Phys. 22 (1996) 123.

\bibitem{lin54}
J.~Lindhard, Kgl. Danske Videnskab. Selkab, Mat-Fys. Medd. 28 (1954).

\bibitem{alb82b}
W.~Alberico, A.~Molinari, R.~Cenni, M.~B. Johnson, Ann. Phys. (NY) 138 (1982)
  178.

\bibitem{mar06}
J.~Margueron, J.~Navarro, N.~Van~Giai, Phys. Rev. C 74 (2006) 015805.

\bibitem{dav09}
D.~Davesne, M.~Martini, K.~Bennaceur, J.~Meyer, Phys. Rev. C 80 (2009) 024314.

\bibitem{pas12}
A.~Pastore, D.~Davesne, Y.~Lallouet, M.~Martini, K.~Bennaceur, J.~Meyer, Phys.
  Rev. C 85 (2012) 054317.

\bibitem{rod07}
T.~R. Rodr\'{\i}guez, J.~L. Egido, Phys. Rev. Lett. 99 (2007) 062501.

\bibitem{per14}
S.~P\'eru, M.~Martini, Eur. Phys. J. A 50 (2014) 88.

\bibitem{egi16}
L.~Egido, Phys. Scripta 91 (2016) 073003.

\bibitem{ber77}
G.~Bertsch, J.~Borysowicz, H.~McManus, W.~Love, Nucl. Phys. A 284 (1977) 399.

\bibitem{ana83}
N.~Anantaraman, H.~Toki, G.~Bertsch, Nucl. Phys. A 398 (1983) 269.

\bibitem{cha08b}
F.~Chappert, M.~Girod, S.~Hilaire, Phys. Lett. B 668 (2008) 420.

\bibitem{gor09a}
S.~Goriely, S.~Hilaire, M.~Girod, S.~P{\'e}ru, Phys. Rev. Lett. 102 (2009)
  242501.

\bibitem{ots06}
T.~Otsuka, T.~Matsuo, D.~Abe, Phys. Rev. Lett. 97 (2006) 162501.

\bibitem{co11}
G.~Co', V.~De~Donno, M.~Anguiano, A.~Lallena, J. Phys. Conf. Series 267 (2011)
  012022.

\bibitem{ang11}
M.~Anguiano, G.~Co', V.~De~Donno, A.~Lallena, Phys. Rev. C 83 (2011) 064306.

\bibitem{wir95}
R.~Wiringa, V.~Stoks, R.~Schiavila, Phys. Rev. C 51 (1995) 38.

\bibitem{ang12}
M.~Anguiano, M.~Grasso, G.~Co', V.~De~Donno, A.~Lallena, Phys. Rev. C 86 (2012)
  054302.

\bibitem{ang16}
M.~Anguiano, A.~Lallena, G.~Co', V.~De~Donno, M.~Grasso, R.~Bernard, Eur. Phys.
  J. A 52 (2016) 183.

\bibitem{oni78}
N.~Onishi, J.~Negele, Nucl. Phys. A301 (1978) 336.

\bibitem{cha15a}
F.~Chappert, N.~Pillet, M.~Girod, J.-F. Berger, Phys. Rev. C 91 (2015) 034312.

\bibitem{gon18}
C.~Gonzalez-Boquera, M.~Centelles, X.~Vi{\~n}as, L.~M. Robledo, Phys. Lett. B
  779 (2018) 195.

\bibitem{bal97}
M.~Baldo, G.~Burgio, I.~Bombaci, Astron. Astrophys. 328 (1997) 274.

\bibitem{dav16b}
D.~Davesne, P.~Becker, A.~Pastore, J.~Navarro, Ann. Phys. (NY) 375 (2016) 288.

\bibitem{sel14}
R.~Sellahewa, A.~Rios, Phys. Rev. C 90 (2014) 054327.

\bibitem{dav17b}
D.~Davesne, P.~Becker, A.~Pastore, J.~Navarro, Acta Phys. Pol. B 48 (2017) 256.

\bibitem{vid11}
I.~Vida\~na, A.~Polls, C.~Provid{\^e}ncia, Phys. Rev. C 84 (2011) 062801.

\bibitem{dav15a}
D.~Davesne, J.~Navarro, P.~Becker, R.~Jodon, J.~Meyer, A.~Pastore, Phys. Rev. C
  91 (2015) 064303.

\bibitem{dav16c}
D.~Davesne, P.~Becker, A.~Pastore, J.~Navarro, Phys. Rev. C 93~(6) (2016)
  064001.

\bibitem{nak10}
H.~Nakada, Phys. Rev. C 81 (2010) 027301.

\bibitem{nak08}
H.~Nakada, Phys. Rev. C 78 (2008) 054301.

\bibitem{dav18}
D.~Davesne, J.~Navarro, J.~Meyer, K.~Bennaceur, A.~Pastore, Phys. Rev. C 97
  (2018) 044304.

\bibitem{dut12}
M.~Dutra, O.~Louren{\c{c}}o, J.~S. Martins, A.~Delfino, J.~R. Stone,
  P.~Stevenson, Phys. Rev. C 85 (2012) 035201.

\bibitem{nak15}
H.~Nakada, T.~Inakura, Phys. Rev. C 91 (2015) 021302.

\bibitem{nak13}
H.~Nakada, Phys. Rev. C 87 (2013) 014336.

\bibitem{gra13}
M.~Grasso, M.~Anguiano, Phys. Rev. C 88 (2013) 054328.

\bibitem{bai10}
C.~Bai, H.~Zhang, H.~Sagawa, X.~Zhang, G.~Col{\`o}, F.~Xu, Phys. Rev. Lett.
  105~(7) (2010) 072501.

\bibitem{cao10}
L.-G. Cao, G.~Col{\`o}, H.~Sagawa, et~al., Phys. Rev. C 81 (2010) 044302.

\bibitem{min13}
F.~Minato, C.~Bai, Phys. Rev. Lett. 110 (2013) 122501.

\bibitem{per04}
E.~Perli\ifmmode~\acute{n}\else \'{n}\fi{}ska, S.~G.
  Rohozi\ifmmode~\acute{n}\else \'{n}\fi{}ski, J.~Dobaczewski, W.~Nazarewicz,
  Phys. Rev. C 69 (2004) 014316.

\bibitem{dav14}
D.~Davesne, A.~Pastore, J.~Navarro, J. Phys. G 41 (2014) 065104.

\bibitem{dav16}
D.~Davesne, A.~Pastore, J.~Navarro, Astron. Astrophys. 585 (2016) A83.

\bibitem{kor13}
M.~Kortelainen, J.~McDonnell, W.~Nazarewicz, E.~Olsen, P.-G. Reinhard,
  J.~Sarich, N.~Schunck, S.~M. Wild, D.~Davesne, J.~Erler, A.~Pastore, Phys.
  Rev. C 89 (2014) 054314.

\bibitem{bla95}
J.~Blaizot, J.~Berger, J.~Decharg\'e, M.~Girod, Nucl. Phys. A 591 (1995) 435.

\bibitem{jerome}
J.~Margueron, Effet du milieu sur la propagation de neutrinos dans la mati\`ere
  nucl\'eaire, Ph.D thesis, University of Paris XI, 2001.

\bibitem{lan57a}
L.~Landau, Sov. Phys. JETP 3 (1957) 920.

\bibitem{lan57b}
L.~Landau, Sov. Phys. JETP 5 (1957) 101.

\bibitem{lan59}
L.~Landau, Sov. Phys. JETP 8 (1959) 70.

\bibitem{abr59}
A.~Abrikosov, I.~Khalatnikov, Rep. Prog. Phys. 22 (1959) 329.

\bibitem{bay91}
G.~Baym, C.~Pethick, Landau Fermi-Liquid Theory, Wiley, New York, 1991.

\bibitem{kha02}
E.~Khan, N.~Sandulescu, M.~Grasso, N.~Van~Giai, Phys. Rev. C 66 (2002) 024309.

\bibitem{gog77}
D.~Gogny, R.~Padjen, Nucl. Phys. A293 (1977) 365.

\bibitem{bac79}
S.~Backman, O.~Sjoberg, A.~Jackson, Nucl. Phys. A 321 (1979) 10.

\bibitem{bac80}
S.-O. B{\"a}ckman, G.~Brown, V.~Klemt, J.~Speth, Nucl. Phys. A 345 (1980) 202.

\bibitem{fri81}
B.~Friman, P.~Haensel, Phys. Lett. B 98 (1981) 323.

\bibitem{hae82}
P.~Haensel, A.~Jerzak, Phys. Lett. B 112 (1982) 285.

\bibitem{fuj87}
T.~Fujita, K.~F. Quader, Phys. Rev. B 36 (1987) 5152.

\bibitem{dab76}
J.~Dabrowski, P.~Haensel, Ann. Phys. (NY) 97 (1976) 452.

\bibitem{ols04}
E.~Olsson, P.~Haensel, C.~Pethick, Phys. Rev. C 70 (2004) 025804.

\bibitem{sch04}
A.~Schwenk, B.~Friman, Phys. Rev. Lett. 92 (2004) 082501.

\bibitem{ben13}
O.~Benhar, A.~Cipollone, A.~Loreti, Phys. Rev. C 87 (2013) 014601.

\bibitem{hol13}
J.~W. Holt, N.~Kaiser, W.~Weise, Phys. Rev. C 87 (2013) 014338.

\bibitem{dav15}
D.~Davesne, J.~Holt, A.~Pastore, J.~Navarro, Phys. Rev. C 91 (2015) 014323.

\bibitem{hol11}
J.~W. Holt, N.~Kaiser, W.~Weise, Nucl. Phys. A870-871 (2011) 1.

\bibitem{hol12}
J.~W. Holt, N.~Kaiser, W.~Weise, Nucl. Phys. A876 (2012) 61.

\bibitem{pas14a}
A.~Pastore, D.~Davesne, J.~Navarro, J. Phys. G 41 (2014) 055103.

\bibitem{nav13}
J.~Navarro, A.~Polls, Phys. Rev. C 87 (2013) 044329.

\bibitem{gia81}
N.~V. Giai, H.~Sagawa, Phys. Lett. B 106 (1981) 379.

\bibitem{ben02}
M.~Bender, J.~Dobaczewski, J.~Engel, W.~Nazarewicz, Phys. Rev. C 65~(5) (2002)
  054322.

\bibitem{roc12}
X.~Roca-Maza, G.~Col{\`o}, H.~Sagawa, Phys. Rev. C 86~(3) (2012) 031306.

\bibitem{mey81}
J.~Meyer-ter Vehn, Physics Reports 74 (1981) 323--378.

\bibitem{ericson-weise}
T.~Ericson, W.~Weise, Pions and Nuclei, Clarendon Press, Oxford, 1988.

\bibitem{mig78}
A.~Migdal, Rev. Mod. Phys. 50 (1978) 107.

\bibitem{gyu77}
M.~Gyulassi, W.~Greiner, Ann. Phys. (NY) 109 (1977) 485.

\bibitem{eri78}
M.~Ericson, J.~Delorme, Phys. Lett. B 76 (1978) 182.

\bibitem{tok79}
H.~Toki, W.~Weise, Phys. Rev. Lett. 42 (1979) 1034.

\bibitem{alb82}
W.~Alberico, M.~Ericson, A.~Molinari, Nucl. Phys. A379 (1982) 429.

\bibitem{spe80}
J.~Speth, V.~Klemt, J.~Wambach, G.~Brown, Nucl. Phys. A 343 (1980) 382.

\bibitem{dic83}
W.~Dickhoff, Nucl. Phys. A359 (1983) 287.

\bibitem{ich06}
M.~Ichimura, H.~Sakai, T.~Wasaka, Prog. Part. Nucl. Phys. 56 (2006) 446.

\bibitem{pin87}
D.~Pines, Can. J. Phys. 65 (1987) 1357.

\bibitem{wei91}
S.~Weisgerber, P.~Reinhard, Phys. Lett. A 158 (1991) 407.

\bibitem{bar93}
M.~Barranco, D.~Jezek, E.~Hern\'andez, J.~Navarro, L.~Serra, Z. Phys. D 28
  (1993) 257.

\bibitem{barr96}
M.~Barranco, E.~Hern\'andez, J.~Navarro, Phys. Rev. B 54 (1996) 7394.

\bibitem{sogo2012spontaneous}
T.~Sogo, M.~Urban, P.~Schuck, T.~Miyakawa, Phys. Rev. A 85 (2012) 031601.

\bibitem{len80}
F.~Lenz, E.~Moniz, K.~Yakazi, Ann. Phys. (NY) 129 (1980) 84.

\bibitem{alb93}
W.~Alberico, M.~Barbaro, A.~De~Pace, T.~Donnelly, A.~Molinari, Nucl. Phys. A563
  (1993) 605.

\bibitem{bar94}
M.~Barbaro, A.~De~Pace, T.~Donnelly, A.~Molinari, Nucl. Phys. A569 (1994) 701.

\bibitem{bar96}
M.~Barbaro, A.~De~Pace, T.~Donnelly, A.~Molinari, Nucl. Phys. A598 (1996) 503.

\bibitem{bar96a}
M.~Barbaro, A.~De~Pace, T.~Donnelly, A.~Molinari, Nucl. Phys. A596 (1996) 553.

\bibitem{nak84}
K.~Nakayama, S.~Krewald, J.~Speth, G.~Love, W, Nucl. Phys. A431 (1984) 419.

\bibitem{hol81}
K.~Holinde, Phys. Rep. 68 (1981) 121.

\bibitem{pac97}
A.~De~Pace, C.~Garc\'{\i}a-Recio, E.~Oset, Phys. Rev. C 55 (1997) 1394.

\bibitem{pas12a}
A.~Pastore, M.~Martini, V.~Buridon, D.~Davesne, K.~Bennaceur, J.~Meyer, Phys.
  Rev. C 86 (2012) 044308.

\bibitem{tha11}
H.~S. Than, E.~Khan, N.~Van~Giai, J. Phys. G 38~(2) (2011) 025201.

\bibitem{gra02}
M.~Grasso, N.~Van~Giai, N.~Sandulescu, Phys. Lett. B 535~(1-4) (2002) 103.

\bibitem{hel12}
V.~Hellemans, P.-H. Heenen, M.~Bender, Phys. Rev. C 85 (2012) 014326.

\bibitem{loa11}
D.~T. Loan, N.~H. Tan, D.~T. Khoa, J.~Margueron, Phys. Rev. C 83~(6) (2011)
  065809.

\bibitem{iwa82}
N.~Iwamoto, C.~J. Pethick, Phys. Rev. D 25 (1982) 313.

\bibitem{pas14}
A.~Pastore, M.~Martini, D.~Davesne, J.~Navarro, S.~Goriely, N.~Chamel, Phys.
  Rev. C 90 (2014) 025804.

\end{thebibliography}

\end{document}